\documentclass[11pt]{report}

\usepackage{amstext,chapterbib}
\usepackage{amssymb}

\newcommand{\R}{\mathbb R} 
\newcommand{\C}{\mathbb C}
\newcommand{\A}{\mathbb A}
\newcommand{\Pbb}{\mathbb P}
\newcommand{\D}{\mathbb D}
\newcommand{\I}{\mathbb I}
\newcommand{\G}{\mathbb G}
\newcommand{\gbb}{\mathbf{g}}
\newcommand{\E}{\mathbb E}
\newcommand{\F}{\mathbb F}
\newcommand{\Z}{\mathbb Z}
\newcommand{\Poincare}{Poincar\'{e}}

\begin{document}

\begin{center}

\vskip 50mm

\LARGE{M.Sc. thesis}

\vskip 10mm

\Large{\textbf{Symmetries, higher order symmetries and supersymmetries}}

\vskip 10mm


Andrea Ferrantelli

\vskip 2mm
{\it Dipartimento di Fisica Teorica, Universit\'a di Torino,\\
via P. Giuria 1, I-10125 Torino, Italy}


\vskip 20mm

Torino, July 11$^{\rm{th}}$ 2002

\vskip 10mm


\end{center}



\date{July, 11th 2002}


%

\newpage

\tableofcontents

\newpage

\section*{Abstract}

This thesis deals with an alternative (geometrical)
formulation for the study of symmetries and supersymmetries. It is called \textsl{
Gauge-Natural}, and it has been developed mainly by the Mathematical Physics groups in Turin (Italy) and in Brno (Czech Republic).

	Gauge-Natural field theories are a generalization of the so-called \textsl{
natural theories}, i.e. field theories where all space-time diffeomorphisms
are symmetries. In natural theories, the Lagrangian of the system is
required to be covariant with respect to all space-time diffeomorphisms
which act on the fields.
However, physicists realized very soon that natural field theories
were not enough to describe physical phenomenology. If a gauge symmetry is
involved, a more general framework is needed; its mathematical bases have to
deal with the notions of principal and associated bundles. Gauge-Natural
field theories regard interactions between natural and gauge fields.

	In this framework, the group of automorphisms of some suitable principal
bundle $\mathcal{P}$, the \textsl{structure bundle}, acts on fields by means
of gauge transformations. At this point, one requires such a group to contain the symmetries of the
theory. Since the fields do not carry any representation of space-time
diffeomorphisms, every consideration about symmetries is moved from the
space-time manifold to this bundle.

	This is one of the most innovative features of the Gauge-Natural formalism,
since actually the concept of symmetry in physics is mainly related to that
of manifold. The model will be exposed in details in Chapter \ref{chapt:gn}.

	Chapter \ref{chapt:structure} deals with the mathematical basics of the model, i.e. we define principal,
associated and jet bundles.
	In Chapter \ref{chapt:variational} these are used to define the Lagrangian formulation of
field theories (by introducing the \Poincare-Cartan form).
	Chapter \ref{chapt:spin} introduces spin structures on the Gauge-Natural bundles thus described. Spin structures overcome the problems encountered when defining
spinors in a curved space.
	Chapter \ref{chapt:susy} deals with the Wess-Zumino model. Here Supersymmetry (SUSY) is global, since the SUSY transformations are point-independent. The generator of supersymmetries is indeed covariantly conserved.
	When this no longer holds, the corresponding theory is Supergravity (SUGRA), namely local SUSY. As an example, we consider the
Rarita-Schwinger model in Chapter \ref{chapt:rs}. We calculate the on-shell covariance of the Lagrangian and the on-shell closure of
the SUSY algebra. We show that this is problematic for the formalism which is exposed in this thesis, and introduce a particular
model for Supergravity, to which the Gauge-Natural framework might be applied.

	Finally, in the Appendix we give some background on supergroups,
supermanifolds and other mathematical tools which are widely used in theories of Supersymmetry and of Supergravity.

\addcontentsline{toc}{section}{Abstract}


\chapter{Structure group bundles}\label{chapt:structure}

\section{Introduction}

The concept of \textsl{field} as a physical entity endowed with energy,
impulse and angular momentum follows from Faraday's and Maxwell's
works about electromagnetism.

	A field, by definition, provides each space-time point with several
quantities. If we denote with $F$ the set of values that can be taken, the 
\textsl{configurations} in an open subset $U$ of space-time are locally
described by the functions $\varphi :U\longrightarrow F.$ After assigning a
topology and a differential structure to $F$, one can require $\varphi $
to be continue and differentiable.

	The above object can be used to introduce a geometrical point of view of field theories. The function 
$\varphi $ can be equivalently defined by means of its graph $\left\{ \left( x,\varphi \left(
x\right) \right) \in U\times F\right\} $. One therefore defines the
function $\widehat{\varphi }:U\longrightarrow U\times F$ that is given by $\widehat{%
\varphi }:x\longmapsto \left( x,\varphi \left( x\right) \right) .$ Hence $U\times F$ can be regarded as the local model of a bundle with fiber $F$.
As we shall see, in this formalism $\widehat{\varphi }$ is the local expression of a section. It follows immediately that the field
configurations are the bundle sections, and that such a bundle can be called \textsl{configuration bundle}.

	At a first sight, this way of introducing bundles in field theory may seem
forced and not useful. One may think that only trivial bundles (i.e. Cartesian products like $M\times F$) are essential to physics.
	However, the formulation of field theories on fiber bundles is not only motivated by some principle of generalization. On the contrary, it
is an \textsl{empirical} consequence of physical situations that we can find in
nature.

	For example, if we want to study the motion of a particle on a
sphere, we must give the position $x \in S^{2}$, a tangent
velocity $\upsilon$ in $T_{x}S^{2}$ and finally a point $\left( x,\upsilon
\right) \in TS^{2}$ of the bundle $TS^{2}$ that is tangent to the sphere $S^{2}$. It can be
easily proven that $TS^{2}$, as a bundle on $S^{2}$, is non trivial.
When looking for the solutions of the equations of motion (e.o.m.) on the
sphere, one has first to choose a local frame on it. On this open subset the
bundle $TS^{2}$ can be trivialized, and therefore the e.o.m. are locally
written on $\R^{2}\times \R^{2}$. Together with these equations, some
conditions which guarantee the regularity of their solutions at every point of $S^2$ must be also given. This is non trivial e.g. at the points
which are excluded from the open subset previously chosen, like the north pole.

	Moreover, General Relativity can provide us with a possibly crucial argument
in favor of a geometric framework for physics. This theory assumes that space-time
is a generic differentiable manifold, not flat
Minkowski. Then, to choose a trivial bundle $M\times F$
as the configuration bundle of a field theory is often both mathematically
incorrect and physically wrong, since this arbitrary choice regards just one among many possibilities. In this chapter, we will introduce (in a basic and not complete manner) some concepts of the theory of fiber bundles. This will be useful for the discussion of the Gauge-Natural formalism in Chapter 3.

\section{Structure group bundles}

A fiber bundle on a manifold $M$, intuitively speaking, is a space whose
local topology is that of a Cartesian product $U_{\alpha }\times F$, where $%
\left\{ U_{\alpha }\right\} $ form a covering of $M$. These local models can
be \textsl{glued} together in a non trivial way, and consequently the
bundle, as a whole, may not be a Cartesian product.

	The gluing procedure is done by choosing suitable automorphisms on the
standard fiber $F$. In general, the group $\hbox{Diff}(F)$ we have chosen is
infinite dimensional.

	It happens however for the automorphisms to be chosen only in a finite subgroup $G\subset \hbox{Diff}(M)$. In this case we say
that $G$ is the \textsl{structure} \textsl{group} of the bundle.
Hereafter we will be interested only in bundles admitting a structure group,
thence we give the following definition.\medskip \newline 
\textbf{Definition (1.2.1.1):}\thinspace a bundle $\mathcal{B}=\left(
B,M,\pi ,F,\lambda ,G\right) $ with structure group $G$ is an object such that\newline
\textbf{(a) }$B,M,F$ are differentiable manifolds (paracompact) called respectively \textsl{total} \textsl{space}, \textsl{base} and \textsl{standard fiber}. The map $\pi :B\longrightarrow M$ surjective and of maximum rank is called \textsl{projection}. The \textsl{structure group} $G$ is a Lie group and $\lambda $ is a left action of $G$ on the standard fiber $F$.\newline
\textbf{(b)} there exists an open covering $\left\{ U_{\alpha }\right\}
_{\alpha \in I}$ of the base $M$ such that for each $\alpha \in I$ there exists a diffeomorphism
\begin{equation}
t_{\left( \alpha \right) }:\pi ^{-1}\left( U_{\alpha }\right)
\longrightarrow U_{\alpha }\times F
\end{equation}
The pair $\left( U_{\alpha },t_{\left( \alpha \right) }\right) $ (but often only $t_{\left( \alpha \right) }$) is called \textsl{local trivialization} of $B$.
The set of all local trivializations $\left\{ \left( U_{\alpha },t_{\left(
\alpha \right) }\right) \right\} _{\alpha \in I}$ is simply labeled as the \textsl{trivialization} of $B$.\newline
\textbf{(c)} if $U_{\alpha \beta }=U_{\alpha }\cap U_{\beta }$ and $e_{G}$ is the group identity, there exists a collection of maps $g_{\left( \alpha \beta \right) }:U_{\beta \alpha}\longrightarrow G$ satisfying the relations:
\begin{equation}
\left\{
\begin{tabular}{l}
$g_{\left( \alpha \alpha \right) }\left( x\right) =e_{G}$ \\ 
$g_{\left( \alpha \beta \right) }\left( x\right) =\left[ g_{\left( \beta
\alpha \right) }\left( x\right) \right] ^{-1}$ \\ 
$g_{\left( \alpha \beta \right) }\left( x\right) \cdot g_{\left( \beta
\gamma \right) }\left( x\right) \cdot g_{\left( \gamma \alpha \right)
}\left( x\right) =e_{G}$%
\end{tabular}
\right.
\end{equation}
such that: 
\begin{equation}
\left\{
\begin{tabular}{l}
\vspace{0.1cm}$id_{U_{\alpha \beta }}\times \widehat{g}_{\left( \alpha \beta \right)
}:U_{\beta \alpha }\times F\longrightarrow U_{\alpha \beta }\times F$ \\ 
\vspace{0.1cm}$t_{\left( \beta \right) }:\pi ^{-1}\left( U_{\beta \alpha }\right)
\longrightarrow U_{\alpha \beta }\times F$ \\ 
$t_{\left( \alpha \right) }:\pi ^{-1}\left( U_{\alpha \beta }\right)
\longrightarrow U_{\beta \alpha }\times F$%
\end{tabular}
\right.
\end{equation}
where $\widehat{g}_{\left( \alpha \beta \right) }:U_{\beta \alpha }\times
F\longrightarrow F$ is defined by: 
\begin{equation}
\widehat{g}_{\left( \alpha \beta \right) }:\left( x,\varphi \right)
\longmapsto \lambda \left( g_{\left( \alpha \beta \right) }\left( x\right)
,\varphi \right)
\end{equation}
The functions $g_{\left( \alpha \beta \right) }$ are called \textsl{transition functions} and depend on the trivialization.
The bundles $\left( M\times F,M,p_{1,}F,\lambda ,\left\{ e\right\} \right) $%
, where $p_{1}:M\times F\longrightarrow M$ is the projection on the first factor, are called \textsl{trivial} \textsl{bundles}.\medskip\newline
\textbf{Definition (1.2.1.2):}\thinspace a \textsl{morphism} between the bundles $
\mathcal{B}=(B,M,\pi ,F)$ and $\mathcal{B}^{\prime }=(B^{\prime },M^{\prime },\pi ^{\prime },F^{\prime})$ is the pair of maps $\Phi =(\phi ,f)$ with $\phi :B\longrightarrow $ $B^{\prime }$ and $f:M\longrightarrow M^{\prime }$ making the following diagram be commutative:
\begin{equation}
\begin{tabular}{lcl}
$\;\;B$ & $\stackrel{\phi}{\longrightarrow}$ & $B^{\prime }$ \\ 
$\pi \downarrow $ & \multicolumn{1}{l}{} & $\downarrow \pi ^{\prime }$ \\ 
$\;\;M$ & $\stackrel{f}{\longrightarrow}$ & $M^{\prime }$
\end{tabular}
\end{equation}
A morphism $\Phi =(\phi ,f)$ is called \textsl{strong} if $f:M\longrightarrow M^{\prime }$ is a diffeomorphism.
To the whole of structure bundles one can apply the usual terminology of
morphisms: the notions of bundles epimorphisms, isomorphisms, endomorphisms,
automorphisms are well defined.

	We remark that the only difference between the bundles analyzed above and
those more commonly mentioned in literature (which do not necessarily have a
structure group), is that axiom \textbf{(c)} in definition (1.2.1.1) needs
not to be satisfied in the latter class of bundles, which we will call 
\textsl{geometrical} \textsl{bundles}. In fact, no additional structure has
been added to them.

	In the following we will deal only with structure bundles.\medskip\newline
\textbf{Proposition (1.2.1.1):}\thinspace let $M$ be a manifold, $\left\{U_{\alpha }\right\} _{\alpha \in I}$ one of its open coverings and $g_{\left( \alpha \beta \right) }:U_{\alpha\beta }\longrightarrow G$ satisfying the conditions
\begin{equation}
\left\{
\begin{tabular}{l}
$g_{\left( \alpha \alpha \right) }\left( x\right) =e_{G}$ \\ 
$g_{\left( \alpha \beta \right) }\left( x\right) \cdot g_{\left( \beta
\gamma \right) }\left( x\right) \cdot g_{\left( \gamma \alpha \right)
}\left( x\right) =e_{G}$%
\end{tabular}
\right.
\end{equation}
Moreover an action $\lambda :G\times F$\textbf{\ }$\longrightarrow F$ of $G$ on a manifold $F$ is given: so there exists a bundle $\mathcal{B}=(B,M,\pi ,F,\lambda,G)$ unique up to isomorphisms which admits $g_{\left( \alpha \beta \right)}$ as transition functions.

\subsection{Sections of bundles}

\textbf{Definition (1.2.2.1): }given a bundle $\mathcal{B}=(B,M,\pi ,F)$, a
map $\rho :U\longrightarrow \pi ^{-1}(U)$ such that $\pi \circ \rho =id_{U}$
is called \textsl{local} \textsl{section}. If $U=M$, $\rho $ is a \textsl{%
global} \textsl{section}.\medskip\newline
The existence of local sections is guaranteed by local trivializations;
indeed if a local trivialization is $t_{\left( \alpha \right) }:\pi
^{-1}(U_{\alpha })\longrightarrow U_{\alpha }\times F$, fixing a map $%
\varphi :U_{\alpha }\longrightarrow F$ one can define the local section $%
\rho _{\varphi }:x\longmapsto t_{\left( \alpha \right) }^{-1}\left(
x,\varphi \left( x\right) \right) $.
On the contrary, global sections of a bundle may not exist in general.

\subsection{Fibered coordinates and local expressions}

Let us consider a point $p\in B$ and fix a local trivialization giving $%
t_{\left( \alpha \right) }(p)=(x,\varphi )$; let also be $p=\left[ x,\varphi
\right] _{\alpha }$. There can be chosen a chart $\left\{ x^{\mu }\right\} $
in $U_{\alpha }$, neighborhood of $x$, and another chart $\left\{ \varphi
_{a}\right\} $ in F, neighborhood of $\varphi $.

	This way a system of coordinates $\left\{ x^{\mu },\varphi _{a}\right\} $
has been defined over $B$; such coordinate systems are named\textsl{\
fibered coordinate systems}.
Consequently, a morphism between two bundles $\mathcal{B}$ and $\mathcal{B}%
^{\prime }$ in fibered coordinates has the following local expression over
the trivialization domain $U_{\alpha }$: 
\begin{equation}
\left\{ 
\begin{tabular}{l}
\vspace{0.1cm}$x^{\prime \mu }=f_{\left( \alpha \right) }^{\mu }\left( x\right) $ \\ 
$\varphi ^{\prime a}=\Phi _{\left( \alpha \right) }^{a}\left( x,\varphi
\right) $%
\end{tabular}
\right.
\end{equation}

\subsection{Particular classes of bundles}

There are several subclasses within the class of fibered bundles. We discuss them briefly here.
A bundle is called \textbf{vector bundle} if it has a vector space $V$ as
standard fiber and $GL\left( n,\R\right) $ ($n=\dim (V)$) represented
on $V$ through the standard representation (therefore the transition
functions are linear). A morphism between vector bundles is \textsl{linear}
if it acts linearly on the fibers.

	An \textbf{affine bundle} has an affine space $A$ as standard fiber and the
affine group as structure group, represented on $A$ with the standard
representation. A morphism is \textsl{affine} if, restricted on each fiber,
it is an affine map.

	A \textbf{principal bundle} has a Lie group $G$ as fiber and as structure
group; $G$ is represented over itself with the left translation $%
L_{g}:G\longrightarrow G:h\longmapsto g\cdot h$. Consequently, we will
indicate a principal bundle with $\mathcal{P=}\left( P,M,\pi ,G\right) $.\medskip\newline
On every principal bundle $\mathcal{P=}\left( P,M,\pi ,G\right) $, a \textsl{right action} of the structure group $G$ is locally defined by 
\begin{equation}
\widetilde{R}_{g}:P\longrightarrow P:\left[ x,h\right] _{\alpha }\longmapsto
\left[ x,h\cdot g\right] _{\alpha }
\end{equation}
which is independent of the used trivialization (see section 1.4.2).
If now $\mathcal{P=}\left( P,M,\pi ,G\right) $ and $\mathcal{P}^{\prime }%
\mathcal{=}\left( P^{\prime },M^{\prime },\pi ^{\prime },G^{\prime }\right) $
are two principal bundles and $\theta :G\longrightarrow G^{\prime }$ a Lie
groups homomorphism, a fiber bundle morphism defined by $\Phi =\left( \phi ,f\right) :%
\mathcal{P}\longrightarrow \mathcal{P}^{\prime }$ is a \textsl{principal
morphism with respect to }$\theta $ if:
\begin{equation}
\begin{tabular}{lll}
$\;\hspace{0.4cm}P$ & $\stackrel{\phi}{\longrightarrow}$ & $P^{\prime }$ \\ 
$\widetilde{R}_{g}\downarrow $ &  & $\downarrow \widetilde{R}_{\theta \left(
g\right) }$ \\ 
$\hspace{0.4cm}\;P$ & $\stackrel{\phi}{\longrightarrow}$ & $P^{\prime }$%
\end{tabular}
\end{equation}
In the case where $G=G^{\prime }$ and $\theta =id_{G}$, $\Phi $ is simply
called \textsl{principal morphism}.

\subsection{The Lie derivative}

Let $\mathcal{B}=(B,M,\pi ,F)$ be a bundle. We can consider the space $TB$
tangent to the total space $B$.\medskip\newline
\textbf{Definition (1.2.5.1): }a vector $v\in TB$ is called \textsl{vertical} if $T\pi \left( v\right) =0$.
The set of all vertical vectors is a subbundle of $TB$ which we will denote by $V\left( \pi \right) \longrightarrow B$.
\begin{equation}
\begin{tabular}{lll}
$\hspace{-0.1cm}V\left( \pi \right) $ & $\longrightarrow $ & $\hspace{-0.1cm}TB$ \\ 
$\downarrow $ &  & $\downarrow \tau _{B}$ \\ 
$\hspace{-0.05cm}B$ &  & $\hspace{-0.05cm}B$ \\ 
&  & $\downarrow \pi $ \\ 
&  & $\hspace{-0.1cm}M$
\end{tabular}
\end{equation}
By composing the projections $V\left( \pi \right) \longrightarrow
B\longrightarrow M$, we obtain a bundle over $M$; the notion of \textsl{%
vertical vector fields on a submanifold }$U\subset M$ \textsl{of the base
manifold }$M$ is thence well defined. Analogously, for a section $\sigma
:M\longrightarrow B$ there exist \textsl{vertical fields over the section }$%
\sigma $; they are defined only over $\sigma \left( M\right) \subset B$.

	If a vector field $\Xi $ over $B$ is considered, it can happen that there
can exist over $M$ a field $\xi $ such that: 
\begin{equation}
T\pi \left( \Xi \left( b\right) \right) =\xi \left( x\right) ,\;\hspace{3cm}%
\forall b\in \pi ^{-1}\left( x\right) \hspace{2cm}\!
\end{equation}
This vector field is called \textsl{projectable} and $\xi $ is its \textsl{%
projection}. For instance, a vertical vector field is projectable and its
projection is the null vector field over $M$.

	If $\Xi $ is a projectable field over $B$ vanishing in $\pi ^{-1}\left(
U\right) $, with abuse of notation we say that $\Xi $ vanishes over $%
U\subset M$.\medskip\newline
\textbf{Definition (1.2.5.2):}\thinspace let $\rho :M\longrightarrow B$ be a section
of $\mathcal{B}$, $\Xi $ a projectable vector field over $\mathcal{B}$ and $\xi $ its projection.

	We define the \textsl{Lie derivative of the section }$\rho $ \textsl{along
the field }$\Xi $ with the following expression: 
\begin{equation}
\pounds _{\Xi }\rho =T\rho \left( \xi \right) -\Xi \circ \rho
\end{equation}
It is easy to prove that $\pounds _{\Xi }\rho $ is a vertical field over $%
\rho .$

\section{Jet bundles}

\subsection{The prolongation of order $k$ of a bundle}

Let $\mathcal{B}=(B,M,\pi ,F)$ be a bundle over $M$ and $\Gamma _{x}\left( 
\mathcal{B}\right) $ the set of all sections of $\mathcal{B}$ locally
defined in a neighborhood of $x\in M$; consider the equivalence relation $%
\sim _{x}^{k}$ defined on $\Gamma _{x}\left( \mathcal{B}\right) $ by 
\begin{equation}
\rho\sim _{x}^{k}\sigma \Leftrightarrow \forall f:B\longrightarrow \R \:\text{and}\: \forall \gamma
:\R\longrightarrow M \:\text{such that}\: \gamma
\left( 0\right) =x\label{getti}
\end{equation}
Defining now the functions 
\begin{equation}
\left\{ 
\begin{tabular}{l}
$\left( f\circ \rho \circ \gamma \right) :\R\longrightarrow \R$ \\ 
$\left( f\circ \sigma \circ \gamma \right) :\R\longrightarrow \R
\hspace{0.1cm}$
\end{tabular}
\right.
\end{equation}
the equivalence relation (\ref{getti}) is equivalent to: 
\begin{equation}
t_{0}^{k}\left( f\circ \rho \circ \gamma \right) =t_{0}^{k}\left( f\circ
\sigma \circ \gamma \right)
\end{equation}
i.e., they have the same Taylor expansion up to the order $k$ in $t=0$.

	If we choose a local trivialization in $\mathcal{B}$ and a fibered
coordinate system $\left( x^{\mu },\varphi ^{a}\right) $, a local section $%
\rho :U\longrightarrow \pi ^{-1}\left( U\right) $ is given in coordinates by 
$\rho :x\longmapsto \left( x,\rho _{\left( \alpha \right) }\left( x\right)
\right) $ (with $\rho _{\left( \alpha \right) }:U\longrightarrow F$). The
equivalence relation (\ref{getti}) reduces to require the local expressions
of $\rho _{\left( \alpha \right) }$ and $\sigma _{\left( \alpha \right) }$
to have the same Taylor expansion up to the order $k$ around the point $x\in
M$. This equivalence relation is obviously independent of the trivialization
and the fibered coordinates.

Here and hereafter we denote by $j_{x}^{k}\rho $ the equivalence class (with
respect to $\sim _{x}^{k}$) pointed out by the representative $\rho \in
\Gamma _{x}\left( \mathcal{B}\right) $. Moreover, the space formed by these
equivalence classes is called $J_{x}^{k}B$ and $J^{k}B=\amalg _{x\in
M}\left( J_{x}^{k}B\right) $ the disjoint union of all the spaces $%
J_{x}^{k}B $.

We have thence the local fibered coordinate system: 
\begin{equation}
\left( x^{\mu },\varphi ^{a},\varphi _{\mu _{1}}^{a},\varphi _{\mu
_{1}\ldots\mu _{k}}^{a}\right)
\end{equation}
named \textsl{natural coordinates over }${J}^{k}{B}$. Please
note that the $\varphi _{\mu _{1}\ldots\mu _{k}}^{a}$ are symmetric in the
indices $\left( \mu _{1},\ldots,\mu _{k}\right) $, representing the values of
the partial derivations of $\varphi ^{a}$ with respect to $x^{\mu }$.

\subsection{The transition functions of $J^{k}B$}

In the case $k=1$ we can easily compute the transition functions by changing
the trivialization. In the new coordinates, the local expression for the
same representative $\rho $ is given by $\rho _{\left( \alpha \right) }$
such that: 
\begin{equation}
\rho _{\left( \alpha \right) }^{a}\left( x\right) =g_{\left( \alpha \beta
\right) }^{a}\left( x,\rho _{\left( \beta \right) }\left( x\right) \right)
\end{equation}
where $g_{\left( \alpha \beta \right) }$ are the transition functions over $%
\mathcal{B}$. By derivative with respect to $x\in M$, one obtains the
transition functions 
\begin{equation}
\left\{ 
\begin{tabular}{l}
\vspace{0.1cm}$\varphi ^{\prime a}=g_{\left( \alpha \beta \right) }^{a}\left( x,\varphi
\right) $ \\ 
$\varphi _{\mu }^{\prime a}=\partial _{\mu }g_{\left( \alpha \beta \right)
}^{a}\left( x,\varphi \right) +\partial _{b}g_{\left( \alpha \beta \right)
}^{a}\left( x,\varphi \right) \varphi _{\mu }^{b}$%
\end{tabular}
\right.
\end{equation}
which are in the form 
\begin{equation}
\varphi _{\mu }^{\prime a}=Y_{\mu }^{a}\left( x,\varphi \right)
+Y_{b}^{a}\left( x,\varphi \right) \varphi _{\mu }^{b}
\end{equation}
In the case $k=2$, one can add to the previous one 
\begin{eqnarray}
\varphi _{\mu \nu }^{\prime a} &=&\partial _{\mu \nu }g_{\left( \alpha \beta
\right) }^{a}\left( x,\varphi \right) +\partial _{\mu b}g_{\left( \alpha
\beta \right) }^{a}\left( x,\varphi \right) \varphi _{\nu }^{b}+\partial
_{\nu b}g_{\left( \alpha \beta \right) }^{a}\left( x,\varphi \right) \varphi
_{\mu }^{b}+  \nonumber \\
&&+\partial _{bc}g_{\left( \alpha \beta \right) }^{a}\left( x,\varphi
\right) \varphi _{\mu }^{b}\varphi _{\nu }^{c}+\partial _{b}g_{\left( \alpha
\beta \right) }^{a}\left( x,\varphi \right) \varphi _{\mu \nu }^{b}
\end{eqnarray}
The structure is given again by 
\begin{equation}
\varphi _{\mu \nu }^{\prime a}=Y_{\mu \nu }^{a}\left( x^{\lambda },\varphi
^{c},\varphi _{\lambda }^{c}\right) +Y_{b}^{a}\left( x,\varphi \right)
\varphi _{\mu \nu }^{b}
\end{equation}
The coefficients $Y_{\mu \nu }^{a}$ are polynomial functions of degree $2$
in $\varphi _{\lambda }^{c}$.

In general, to the $k$-order, this structure is preserved being 
\begin{equation}
\varphi _{\mu _{1}\ldots\mu _{k}}^{\prime a}=Y_{\mu _{1}\ldots\mu _{k}}^{a}\left(
x^{\lambda },\varphi ^{c},\varphi _{\lambda }^{c},\varphi _{\lambda
_{1}\ldots\lambda _{k-1}}^{c}\right) +Y_{b}^{a}\left( x,\varphi \right) \varphi
_{\mu _{1}\ldots\mu _{k}}^{b}
\end{equation}
and the $Y_{\mu _{1}\ldots\mu _{k}}^{a}$ are polynomial functions of degree $2$
in $\varphi _{\lambda _{1}\ldots\lambda _{k-1}}^{c}$.

	The natural coordinates define therefore an atlas of class $\mathcal{C}%
^{\infty }$ providing a differentiable structure to $J^{k}\mathcal{B}$.
Moreover, one can define the projections 
\begin{equation}
\pi _{k-1}^{k}:J^{k}B\longrightarrow J^{k-1}B
\end{equation}
with 
\begin{equation}
\pi _{k-1}^{k}:\left( x^{\mu },\varphi ^{a},\varphi _{\mu
_{1}}^{a},\ldots,\varphi _{\mu _{1}\ldots\mu _{k}}^{a}\right) \longmapsto \left(
x^{\mu },\varphi ^{a},\varphi _{\mu _{1}}^{a},\ldots,\varphi _{\mu _{1}\ldots\mu
_{k-1}}^{a}\right)
\end{equation}
Now, the transition functions are affine maps in the variables on the fiber $%
\varphi _{\mu _{1}\ldots\mu _{k}}^{a}$; consequently, we can build a family of
affine bundles defined by the projections $\pi _{k-1}^{k}$. A sequence of
maps is obtained: 
\begin{equation}
M\stackrel{\pi}{\longleftarrow}B=J^{0}B\stackrel{\pi_{0}^1}{\longleftarrow}J^{1}B\stackrel{\pi_{1}^2}{\longleftarrow}J^{2}B\longleftarrow \ldots
\end{equation}
each map defines an affine bundle $J^{k+1}B$ over $J^{k}B$. In general we
define for composition the projections 
\begin{equation}
\pi ^{k+h}:J^{k+h}B\longrightarrow J^{k}B
\end{equation}
and we denote simply by $\pi ^{k}:J^{k}B\longrightarrow M$ the projections
on the base $M$.
In the following the bundle $\left( J^{k}B,M,\pi ^{k},J_{0}^{k}\left( \R
^{m}\times F\right) \right) $ will be defined only by $J^{k}\mathcal{B}$.

\subsection{The prolongation of a fibered morphism}

Let now $\mathcal{B}$ and $\mathcal{B}^{\prime }$ be two bundles and $\Phi
=\left( \phi ,f\right) :\mathcal{B}\longrightarrow \mathcal{B}^{\prime }$ a
strong bundle morphism; one can define the \textsl{prolongation of }$k$%
\textsl{-order of the (strong) morphism }$\Phi $ in the following way: 
\begin{equation}
J^{k}\Phi :J^{k}\mathcal{B}\longrightarrow J^{k}\mathcal{B}^{\prime
}:j_{x}^{k}\rho \longmapsto j_{f\left( x\right) }^{k}\left( \phi \circ \rho
\circ f^{-1}\right)
\end{equation}
which turns out to be a (strong) bundle morphism.

	Consequently, we have the\medskip\newline
\textbf{Proposition (1.3.3.1): }given two morphisms\textbf{\ }$\Phi :\mathcal{B}\longrightarrow \mathcal{C}$ and $\Psi :\mathcal{C}\longrightarrow \mathcal{D}$ we have: 
\begin{equation}
\left\{ 
\begin{tabular}{l}
$J^{k}\left( \Psi \circ \Phi \right) =J^{k}\left( \Psi \right) \circ
J^{k}\left( \Phi \right) $ \\ 
$J^{k}\left( id_{\mathcal{B}}\right) =id_{J^{k}\mathcal{B}}$%
\end{tabular}
\right.
\end{equation}

\subsection{The prolongation of sections}

If $\rho $ is a section of $\mathcal{B}$ we can define a bundle morphism, to
which $J^{k}$ can be applied, giving:
\begin{equation}
\begin{tabular}{lll}
$\hspace{-0.05cm}M$ & $\stackrel{\rho}{\longrightarrow} $ & $\hspace{-0.05cm}B$ \\ 
$\Vert $ &  & $\downarrow \pi $ \\ 
$\hspace{-0.05cm}M$ &  & $\hspace{-0.05cm}M$
\end{tabular}
\hspace{1cm}  \stackrel{J^{k}}{\longrightarrow}  \hspace{1cm} 
\begin{tabular}{lll}
$\hspace{-0.07cm}M$ & $\stackrel{J^{k}\rho}{\longrightarrow} $ & $\hspace{-0.1cm}J^{k}B$ \\ 
$\Vert $ &  & $\downarrow \pi $ \\ 
$\hspace{-0.07cm}M$ &  & $\hspace{-0.07cm}M$%
\end{tabular}
\end{equation}
The map $J^{k}\rho $ can be reinterpreted as a section of $J^{k}\mathcal{B}$
called the \textsl{prolongation to the }$k$-\textsl{order of the section }$%
\rho $; by convention, such a section is denoted $j^{k}\rho $ instead of $%
J^{k}\rho $. Among the sections of $J^{k}\mathcal{B}$, those which can be
obtained as prolongations of sections of $\mathcal{B}$ are regarded as 
\textsl{holonomic sections.}

	Given a local section of $\mathcal{B}$ represented by $\rho _{\left( \alpha
\right) }:U\longrightarrow F:x^{\mu }\longmapsto \rho _{\left( \alpha
\right) }\left( x\right) $, the prolongation assumes the form: 
\begin{equation}
j^{k}\rho :x^{\mu }\longmapsto \left( x^{\mu },\rho _{\left( \alpha \right)
}\left( x\right) ,\partial _{\mu _{1}}\rho _{\left( \alpha \right)
}^{a}\left( x\right) ,\ldots,\partial _{\mu _{1}\ldots\mu _{k}}\rho _{\left(
\alpha \right) }^{a}\left( x\right) \right)
\end{equation}

\subsection{Contact forms}

\textbf{Definition (1.3.5.1): }a form $\omega $ over $J^{k}\mathcal{B}$ is a
c\textsl{ontact form} if it vanishes on all the holonomic sections, i.e. 
\begin{equation}
\left( j^{k}\rho \right) ^{*}\omega =0
\end{equation}
The contact forms of $J^{k}\mathcal{B}$ create an ideal of the external
algebra over $J^{k}\mathcal{B}$. A set of generators is formed by
\begin{equation}
\left\{ 
\begin{tabular}{l}
\vspace{0.1cm}$\omega ^{a}=d\varphi ^{a}-\varphi _{\sigma }^{a}dx^{\sigma }$ \\ 
\vspace{0.1cm}$\omega _{\mu }^{a}=d\varphi _{\mu }^{a}-\varphi _{\mu \sigma
}^{a}dx^{\sigma }$ \\ 
\vspace{0.1cm}$\ldots$ \\ 
\vspace{0.1cm}$\omega _{\mu _{1}\ldots\mu _{k-1}}^{a}=d\varphi _{\mu _{1}\ldots\mu
_{k-1}}^{a}-\varphi _{\mu _{1}\ldots\mu _{k-1}\sigma }^{a}dx^{\sigma }$ \\ 
$d\omega _{\mu _{1}\ldots\mu _{k-1}}^{a}=-\varphi _{\mu _{1}\ldots\mu _{k-1}\sigma
}^{a}\wedge dx^{\sigma }$%
\end{tabular}
\right.
\end{equation}

\subsection{Total derivatives}

We introduce a family of operators $d_{\mu }$, called \textsl{total
derivatives}, acting on the functions of $J^{k}\mathcal{B}$ to give
functions over $J^{k+1}\mathcal{B}$ such that: 
\begin{equation}
\forall \rho :M\longrightarrow B,\left( d_{\mu }F\right) \circ j^{k+1}\rho
\left( x\right) =\partial _{\mu }\left( F\circ j^{k}\rho \left( x\right)
\right)
\end{equation}
It is easy to show that the operators $d_{\mu }$ act as follows: 
\[
d_{\mu }F\left( x^{\mu },\varphi ^{a},\varphi _{\mu _{1}}^{a},\ldots,\varphi
_{\mu _{1}\ldots\mu _{k}}^{a}\right) =\partial _{\mu }F+\partial _{a}F\cdot
\varphi _{\mu }^{a}+ 
\]
\begin{equation}
+\partial _{a}^{\mu _{1}}F\cdot \varphi _{\mu \mu _{1}}^{a}+\ldots+\partial
_{a}^{\mu _{1}\ldots\mu _{k}}F\cdot \varphi _{\mu \mu _{1}\ldots\mu _{k}}^{a}
\end{equation}
where $\partial _{a}$ denotes the derivative with respect to $\varphi ^{a}$, 
$\partial _{a}^{\mu _{1}}$ those w.r. to $\varphi _{\mu _{1}}^{a}$ and so on.

	If a change of fibered coordinates (or an automorphism) over $\mathcal{B}$
is considered, i.e. 
\begin{equation}
\left\{ 
\begin{tabular}{l}
$x^{\prime \mu }=f^{\mu }\left( x\right) $ \\ 
$\varphi ^{\prime a}=\phi ^{a}\left( x,\varphi \right) $%
\end{tabular}
\right.
\end{equation}
the prolongation takes the form:
\begin{equation}
\left\{ 
\begin{tabular}{l}
\vspace{0.1cm}$x^{\prime \mu }=f^{\mu }\left( x\right) \hspace{1cm}\overline{J}_{\mu
}^{\nu }=\partial _{\mu }\left( f^{-1}\right) ^{\nu }\left( x\right) $ \\ 
\vspace{0.1cm}$\varphi ^{\prime a}=\phi ^{a}\left( x,\varphi \right) $ \\ 
\vspace{0.1cm}$\varphi _{\mu _{1}}^{\prime a}=\overline{J}_{\mu _{1}}^{\nu _{1}}d_{\nu
_{1}}\phi ^{a}\left( x,\varphi \right) $ \\ 
\vspace{0.1cm}$\varphi _{\mu _{1}\mu _{2}}^{\prime a}=\overline{J}_{\mu _{2}}^{\nu
_{2}}d_{\nu _{2}}\left( \overline{J}_{\mu _{1}}^{\nu _{1}}d_{\nu _{1}}\phi
^{a}\left( x,\varphi \right) \right) $ \\ 
\vspace{0.1cm}$\ldots$%
\end{tabular}
\right.
\end{equation}
which represents also the transition functions of $J^{k}\mathcal{B}$.

\section{Principal fiber bundles}

\subsection{The right action}

	Let $\mathcal{P}=\left( P,M,\pi ,G\right) $ be a principal bundle and fix a
local trivialization on the open subset $U_{\alpha }$. There are locally
four maps:
\begin{eqnarray}
\widetilde{R}_{g}^{\left( \alpha \right) } &:&\pi ^{-1}\left( U_{\alpha
}\right) \longrightarrow \pi ^{-1}\left( U_{\alpha }\right) :\left[
x,h\right] _{\alpha }\longmapsto \left[ x,h\cdot g\right] _{\alpha } 
\nonumber \\
\widetilde{L}_{g}^{\left( \alpha \right) } &:&\pi ^{-1}\left( U_{\alpha
}\right) \longrightarrow \pi ^{-1}\left( U_{\alpha }\right) :\left[
x,h\right] _{\alpha }\longmapsto \left[ x,h\cdot g\right] _{\alpha } 
\nonumber \\
\widetilde{L}_{p}^{\left( \alpha \right) } &:&G\longrightarrow \pi
^{-1}\left( U_{\alpha }\right) :g\longmapsto \widetilde{R}_{g}^{\left(
\alpha \right) }\left( p\right) \\
\widetilde{R}_{p}^{\left( \alpha \right) } &:&G\longrightarrow \pi
^{-1}\left( U_{\alpha }\right) :g\longmapsto \widetilde{L}_{g}^{\left(
\alpha \right) }\left( p\right)  \nonumber
\end{eqnarray}
	An important theorem states that there exists a global action $\widetilde{R}%
_{g} $ of $G$ over $P$ which is vertical, free and fiber-transitive, whose
local expressions are given by $\widetilde{R}_{g}^{\left( \alpha \right) }$.
As a corollary, we say that \textbf{it is possible to
associate to each local section} $\sigma^{\left( \alpha \right) }:U_{\alpha }\longrightarrow P$ \textbf{a local trivialization} $%
t_{\left( \alpha \right) }:\pi ^{-1}\left( U_{\alpha }\right)
\longrightarrow U_{\alpha }\times G$, \textbf{canonically and in one-to-one correspondence}.

	For this reason, on principal bundles, with an abuse of notation, a
trivialization is often assigned by defining a set of local sections $\left\{
\sigma ^{\left( \alpha \right) }\right\} $ whose domains form a covering of $%
M.$
This corollary proves (but the proof will be omitted here) that principal
bundles are completely identified by their right action. In fact the
following is derived:\medskip\newline
\textbf{Property (1.4.1.1):} given a manifold $P$ and a free right action
of $G$ so that the quotient space is a manifold, $P$ is the total space of a
principal bundle.

\subsection{The bundle of $s$-frames}

As an example of principal bundle let us consider 
\begin{equation}
L^{s}\left( M\right) =\left\{ j_{0}^{s}\epsilon \mid \epsilon :\R
^{m}\longrightarrow M\right\}
\end{equation}
with $\epsilon $ locally invertible around the origin. The projection $\pi
:L^{s}\left( M\right) \longrightarrow M:j_{0}^{s}\epsilon \longmapsto
\epsilon \left( 0\right) $ takes values in all the codomain.

	As standard fiber let us define: 
\begin{equation}
GL^{s}\left( M\right) =\left\{ j_{0}^{s}\alpha \mid \alpha :\R
^{m}\longrightarrow \R^{m}\right\}
\end{equation}
where $\alpha $ is locally invertible around the origin and $\alpha \left(
0\right) =0$.
The product is: 
\begin{equation}
j_{0}^{s}\alpha \cdot j_{0}^{s}\beta :=j_{0}^{s}\left( \alpha \circ \beta
\right)
\end{equation}
Therefore we have that $\left( L^{s}\left( M\right) ,M,\pi ,GL^{s}\left(
M\right) \right) $ is a principal bundle; in fact one can define the right
action 
\begin{equation}
j_{0}^{s}\epsilon \cdot j_{0}^{s}\alpha :=j_{0}^{s}\left( \epsilon \circ
\alpha \right)
\end{equation}
which turns out to be free.

	Given a morphism between manifolds $f:M\longrightarrow M^{\prime },$ 
\begin{equation}
L^{s}\left( f\right) :L^{s}\left( M\right) \longrightarrow L^{s}\left(
M^{\prime }\right) :j_{0}^{s}\epsilon \longmapsto j_{0}^{s}\left( f\circ
\epsilon \right)
\end{equation}
can be defined.

	With these hypotheses $\left( L^{s}\left( f\right) ,f\right) $ is a
principal morphism. The bundle $L^{s}\left( M\right) $ is called \textsl{%
bundle of the }$s$\textsl{-frames} while $L^{s}\left( f\right) $ is regarded
as the \textsl{natural lift to the }$s$\textsl{-frames.}

\section{Canonical constructions of fiber bundles}

\subsection{Associated bundles}

Suppose to have a principal bundle $\mathcal{P}=\left( P,M,\pi ,G\right) $
and $\lambda :G\times F\longrightarrow F$ a left action of the group $G$
over a manifold $F$. The associated bundle $\mathcal{P}\times _{\lambda }F$
is the bundle having as total space the quotient of $P\times F$ with respect
to the equivalence relation 
\begin{equation}
\left( p,\varphi \right) \sim \left( p^{\prime },\varphi ^{\prime }\right)
\Longleftrightarrow \exists g\in G\mid \widetilde{R}_{g}p=p^{\prime } \text{and} \hspace{0.1cm}\varphi =\lambda \left( g,\varphi ^{\prime }\right)  \label{equivalenza}
\end{equation}
The equivalence classes will be denoted with $\left[ p,\varphi \right]
_{\lambda }$. Indeed, given a trivialization $\sigma ^{\left( \alpha \right) }$ of $%
\mathcal{P}$ we have the trivialization: 
\begin{equation}
t_{\lambda }^{\left( \alpha \right) }:\pi _{\lambda }^{-1}\left( U_{\alpha
}\right) \longrightarrow U_{\alpha }\times F:\left[ p,\varphi \right]
_{\lambda }\longmapsto \left( x,\lambda \left( g,\varphi \right) \right) 
\hspace{1cm}p=\sigma ^{\left( \alpha \right) }\left( x\right) \cdot g
\end{equation}
which is one-to-one because of (\ref{equivalenza}).

\subsection{Structure bundles}

Let $\mathcal{B}=\left( B,M,\pi ,F,\lambda ,G\right) $ be a bundle with
structure group; let us fix a trivialization $t_{\left( \alpha \right) }$
and let $g_{\left( \alpha \beta \right) }$ be the corresponding transition
functions. By applying Proposition (1.2.1.1), we can define a bundle having $%
G$ as standard fiber with the use of $g_{\left( \alpha \beta \right) }$ and
of the left translation $L_{g}:G\longrightarrow G$. The bundle $\mathcal{P}%
=\left( P,M,\pi ,G\right) $ we obtain is a principal bundle called \textsl{%
structure bundle of }$\mathcal{B}$.

	Now, if we start from $\mathcal{P}$, choose $F$ as standard fiber and build
the associated bundle with the action $\lambda :G\times F\longrightarrow F$,
we obtain $\mathcal{B}$ again. Thus every bundle with structure group can be
viewed as associated to some structure bundle. Though in physics associated
bundles have, as we shall see, the precise meaning of configuration bundles,
while structure bundles do not have a direct physical interpretation, it is
actually of use building up the configuration bundles by starting from their
structure bundles. Taking into account this equivalence between structure
and associated bundles, we will adopt systematically this habit.

\subsubsection{An example}

We will show that the tangent bundle $TM$ is associated to the frame bundle $%
L\left( M\right) $.
The frame bundle is a principal bundle with group $GL\left( m,\R\right) 
$ and we can choose the natural representation: 
\begin{equation}
\lambda :GL\left( m,\R\right) \times \R^{m}\longrightarrow \R%
^{m}:\left( J_{\nu }^{\mu },\upsilon ^{\nu }\right) \longmapsto J_{\nu
}^{\mu }\upsilon ^{\nu }
\end{equation}
Therefore we build the associated bundle $L\left( M\right) \times _{\lambda
}^{m}\R$. We shall show that this bundle is isomorphic to $TM$.

	We begin by fixing a chart $\varphi _{\left( \alpha \right) }:U_{\alpha
}\longrightarrow \R^{m}$ of the base $M$ and by denoting with $%
\overline{\varphi }_{\left( \alpha \right) }:\varphi _{\left( \alpha \right)
}\left( U_{\alpha }\right) \longrightarrow U_{\alpha }$ its inverse map;
hence a local section of $L\left( M\right) $ is 
\begin{equation}
e^{\left( \alpha \right) }:x\longmapsto j_{\overline{x}}^{s}\varphi _{\left(
\alpha \right) }\hspace{1cm}\overline{x}=\varphi _{\left( \alpha \right)
}\left( x\right)
\end{equation}
and consequently a trivialization of $L\left( M\right) $.

	Moreover, the choice of $\varphi _{\left( \alpha \right) }$ induces locally
the natural base $\partial _{\mu }^{\left( \alpha \right) }$ of the vectors
tangent to $M$. A point of $L\left( M\right) \times _{\lambda }\R^{m}$
is thus in the form $\left[ e^{\left( \alpha \right) }\cdot \Vert a_{\nu
}^{\mu }\Vert ,\upsilon ^{\nu }\right] _{\lambda }=\left[ e^{\left( \alpha
\right) },a_{\nu }^{\mu }\cdot \upsilon ^{\nu }\right] _{\lambda }$ and we
can locally define the isomorphisms: 
\begin{equation}
\psi _{\left( \alpha \right) }:L\left( M\right) \times _{\lambda }\R
^{m}\longrightarrow TM:\left[ e^{\left( \alpha \right) },\upsilon _{\left(
\alpha \right) }^{\mu }\right] _{\lambda }\longmapsto \widetilde{\upsilon }%
_{\left( \alpha \right) }^{\mu }\partial _{\mu }^{\left( \alpha \right) }
\end{equation}
where $\widetilde{\upsilon }_{\left( \alpha \right) }^{\mu }=a_{\nu }^{\mu
}\cdot \upsilon _{\left( \alpha \right) }^{\nu }$.

	The above local morphisms point out a single global bundle isomorphism $\psi :L\left(
M\right) \times _{\lambda }\R^{m}\longrightarrow TM$ because the
morphisms $\psi _{\left( \alpha \right) }$ satisfy to the required
compatibility conditions . In fact, if we choose another chart $\varphi
_{\left( \beta \right) }$ and define over $\varphi _{\left( \alpha \right)
}\left( U_{\alpha \beta }\right) \subset \R^{m}$, where both the charts
are defined, the map $f=( \varphi _{\left( \beta \right) }\circ \varphi
_{\left( \alpha \right) }^{-1}) :\varphi _{\left( \alpha \right)
}\left( U_{\alpha \beta }\right) \longrightarrow \varphi _{\left( \beta
\right) }\left( U_{\alpha \beta }\right) $, we get: 
\begin{equation}
\left\{ 
\begin{tabular}{l}
$e^{\left( \alpha \right) }=e^{\left( \beta \right) }\cdot J$ \\ 
$\partial _{\mu }^{\left( \alpha \right) }=J_{\mu }^{\nu }\partial _{\nu
}^{\left( \alpha \right) }$%
\end{tabular}
\right.
\end{equation}
the compatibility condition is therefore:
\begin{equation}
\left\{ 
\begin{tabular}{l}
\vspace{0.1cm}$\psi _{\left( \alpha \right) }:\left[ e^{\left( \alpha \right) },\upsilon
_{\left( \alpha \right) }^{\mu }\right] \longrightarrow \widetilde{\upsilon }%
_{\left( \alpha \right) }^{\mu }\partial _{\mu }^{\left( \alpha \right) }$
\\ 
\vspace{0.1cm}$\psi _{\left( \beta \right) }:\left[ e^{\left( \beta \right) },\upsilon
_{\left( \beta \right) }^{\mu }\right] \longrightarrow \widetilde{\upsilon }%
_{\left( \beta \right) }^{\mu }\partial _{\mu }^{\left( \beta \right) }$ \\ 
\vspace{0.1cm}$\left[ e^{\left( \alpha \right) },\upsilon _{\left( \alpha \right) }^{\mu
}\right] =\left[ e^{\left( \beta \right) },\upsilon _{\left( \beta \right)
}^{\mu }\right] $ \\ 
\vspace{0.1cm}$\widetilde{\upsilon }_{\left( \alpha \right) }^{\mu }\partial _{\mu
}^{\left( \alpha \right) }=\widetilde{\upsilon }_{\left( \beta \right)
}^{\mu }\partial _{\mu }^{\left( \beta \right) }$ \\ 
\vspace{0.1cm}$\partial _{\mu }^{\left( \alpha \right) }=J_{\mu }^{\nu }\partial _{\nu
}^{\left( \beta \right) }$ \\ 
$\widetilde{\upsilon }_{\left( \beta \right) }^{\mu }=J_{\nu }^{\mu }%
\widetilde{\upsilon }_{\left( \alpha \right) }^{\nu }$%
\end{tabular}
\right.
\end{equation}
Note also that the isomorphism $TM\simeq L\left( M\right) \times
_{\lambda }\R^{m}$ created in this way does not depend on any
additional structure and it is by consequence \textsl{canonical}.

\subsection{Natural bundles}

\textbf{Definition (1.5.4.1): }a fiber bundle is \textsl{natural} if it is
canonically isomorphic to a bundle associated to $L^{s}\left( M\right) $,
for some $s\geq 1$.\medskip\newline
Hence we have just shown that $TM$ is a natural bundle. Over each natural
bundle $L^{s}\left( M\right) \times _{\lambda }F$ one can define, for any
manifold diffeomorphism $f:M\longrightarrow M$, the bundle morphism: 
\begin{equation}
f_{\lambda }:L^{s}\left( M\right) \times _{\lambda }F\longrightarrow
L^{s}\left( M\right) \times _{\lambda }F:\left[ j_{0}^{s}\epsilon ,\varphi
\right] _{\lambda }\longrightarrow \left[ L^{s}\left( f\right) \left(
j_{0}^{s}\epsilon \right) ,\varphi \right] _{\lambda }
\end{equation}
called \textsl{natural lift of }$f$.

\section{Infinitesimal generators of principal automorphisms}


Let $\mathcal{P}=\left( P,M,\pi ,G\right) $ be a principal bundle. In the
following the group $Aut\left( \mathcal{P}\right) 
$ of the principal automorphisms of $\mathcal{P}$ will be particularly relevant. In analogy with generic transformation
groups, each one-parameter subgroup is the flow of a vector field on $P$
called the \textsl{infinitesimal generator }of the subgroup. Throughout this
section we fix a trivialization $\sigma ^{\left( \alpha \right) }$ of $%
\mathcal{P}$ and a set of right invariant vector fields $\rho _{A}.\medskip $\newline
\textbf{Proposition (1.6.1.1): }an infinitesimal generator of automorphisms
of $\mathcal{P}$ has locally the following form:
\begin{equation}
\Xi =\xi ^{\mu }\left( x\right) \partial _{\mu }+\xi ^{A}\left( x\right)
\rho _{A}
\end{equation}
\textbf{Proof:} a one-parameter family of automorphisms of $\mathcal{P}$, $%
\left( \Phi _{t},f_{t}\right) \in Aut\left( \mathcal{P}\right) $; in a trivialization we have:
\begin{equation}
\left\{ 
\begin{tabular}{l}
\vspace{0.1cm}$\Phi _{t}:P\longrightarrow P$ \\ 
\vspace{0.1cm}$f_{t}:M\longrightarrow M$ \\ 
$\Phi _{t}\left[ x,h\right] _{\alpha }=\left[ f_{t}\left( x\right) ,\Phi
_{t}\left( x\right) \cdot h\right] _{\alpha }$%
\end{tabular}
\right.
\end{equation}
The generator of this automorphism is therefore 
\begin{eqnarray*}
\Xi =\xi ^{\mu }\left( x\right) \partial _{\mu }+\xi ^{A}\left( x\right)
\rho _{A} 
\end{eqnarray*}
with
\begin{equation}
\left\{ 
\begin{tabular}{l}
\vspace{0.05cm}$\xi ^{\mu }\left( x\right) =\dot{f}_{0}\left( x\right) $ \\ 
$\xi ^{A}\left( x\right) =\left( T^{-1}\right) _{a}^{A}\dot{\Phi}%
_{0}^{a}\left( x\right) $%
\end{tabular}
\right.
\end{equation}
\textbf{Theorem (1.6.1.1): }if every infinitesimal isomorphism is in the form 
\begin{equation}
\Xi =\xi ^{\mu }\left( x\right) \partial _{\mu }+\xi ^{A}\left( x\right) 
\widehat{\rho }_{A}
\end{equation}
$\widehat{\rho }_{A}$ are right invariant fields.\medskip\newline
\textbf{Proof:} the infinitesimal generators of automorphisms are right invariant: 
\begin{eqnarray}
T_{p}\widetilde{R}_{g}\Xi _{p} &=&\Xi _{p\cdot g}\Longrightarrow \xi
^{A}\left( x\right) T_{p}\widetilde{R}_{g}\widehat{\rho }_{A}\left( p\right)
=\xi ^{A}\left( x\right) \widehat{\rho }_{A}\left( p\cdot g\right)
\Longrightarrow  \nonumber \\
&\Longrightarrow &T_{p}\widetilde{R}_{g}\widehat{\rho }_{A}\left( p\right) =%
\widehat{\rho }_{A}\left( p\cdot g\right)
\end{eqnarray}
The group $Aut\left( P\right) $ has a subgroup $Aut_{\left( V\right) }\left(
P\right) $ consisting of vertical automorphisms, that is automorphisms
projecting on the identity in $M$. The infinitesimal generators of vertical
automorphisms are: 
\begin{equation}
\Xi _{p}=\xi ^{A}\left( x\right) \rho _{A}\left( p\right) \hspace{1cm}\pi
\left( p\right) =x
\end{equation}

\subsection{The bundle of vertical infinitesimal automorphisms}

Let us call $\gbb$ the Lie algebra of the group $G$ and indicate with $%
Ad:G\times \gbb\longrightarrow \gbb$ the adjoint representation of
the group over the algebra. We can build the associated bundle $P\times _{Ad}%
\gbb$, called the \textsl{bundle of vertical infinitesimal automorphisms} because
the following holds:\medskip\newline
\textbf{Proposition (1.6.2.1): }there is a one-to-one correspondence between infinitesimal generators of vertical automorphisms and sections of $P\times _{Ad}
\gbb.$

\section{Principal connections}

\subsection{Definition of a principal connection}

Let $\mathcal{P}=\left( P,M,\pi ,G\right) $ be a principal bundle, $%
\sigma ^{\left( \alpha \right) }$ be a trivialization and $\rho _{A}$ a base
of right invariant vertical vectors.

	Note that the vertical vectors of $\mathcal{P}$ have an intrinsic
meaning, independent of any additional structure. Over each bundle the
vectors in $\ker \left( T\pi \right) $ form indeed a subbundle $V\left( \pi
\right) $ of the tangent bundle $TP$ to the total space $P$. On the
contrary, there is no canonical notion of horizontal vector; if in a local
trivialization we consider the vectors $\partial _{\mu }^{\left( \alpha
\right) }$ these, by changing local trivialization, acquire also a vertical
component. In other terms, such a concept of horizontal vectors should
depend on the trivialization, and therefore would be non canonical.

	The notion of connection allows to preserve the globality, taking into
account the impossibility of a canonical choice.\medskip\newline
\textbf{Definition (1.7.1.1): }a \textsl{connection }is a family $%
H_{p}\subset T_{p}P$ such that:\newline
\textbf{(a)} the family is \textsl{smooth} with respect to the point $p$: $H=\left\{ \upsilon \in TP\mid p=\tau _{P}\left( \upsilon \right) ,\upsilon\in H_{p}\right\} $ must be a subbundle of $TP$\newline
\textbf{(b) }$\forall p\in P,T_{p}P=H_{p}\oplus V_{p}\left(\pi \right)$ where $V_{p}\left(\pi \right) $ is the set of the vertical vectors in $p\in P.$\newline
The connection is \textsl{principal} if the family $H_{p}$ is invariant
under the right action of $G$ on $P$, i.e.:\newline
\textbf{(c) }$T\widetilde{R}_{g}H_{p}=H_{p\cdot g}$\newline
A vector $\upsilon \in T_{p}P$ is \textsl{horizontal} if $\upsilon \in H_{p}$.

	Notice that \textbf{(c)} determines $H_{p\cdot g}$ once given $H_{p}$;
hence, with the transitivity of the right action over the fibers, condition 
\textbf{(c)} states that for giving a principal connection it is enough to
assign properly a subspace $H_{p}$ for each fiber.

\subsection{Equivalent definitions of connection}

Connections are sometimes introduced in the literature in a way equivalent
to Definition (1.7.1.1): a connection induces a lift $\omega
:TM\longrightarrow TP$ which associates to each vector $\upsilon \in TM$ the
only vector in $H_{p}$ projecting, through $T_{\pi }$, over $\upsilon$. This vector may be locally written in the form 
\begin{equation}
\omega \left( \xi ^{\mu }\partial _{\mu }\right) =\xi ^{\mu }\left( \partial
_{\mu }+\omega _{\mu }^{A}\left( p\right) \rho _{A}\right)
\end{equation}
\textbf{Proposition (1.7.2.1): }a connection\textbf{\ }$\omega $ over\textbf{%
\ }$\mathcal{P}$ is principal if and only if it is in the form 
\begin{equation}
\omega =dx^{\mu }\otimes \left( \partial _{\mu }+\omega _{\mu }^{A}\left(
x\right) \rho _{A}\right)
\end{equation}
\textbf{Proof: }if $\omega $ is in the required form, $\omega \left( \xi
\right) $ is right invariant and thence the connection is principal. Conversely, if the connection $\omega =dx^{\mu }\otimes \left( \partial _{\mu }+\omega _{\mu }^{A}\left(
x,g\right) \rho _{A}\right) $ is principal, $\omega _{\mu }^{A}\left(
x,g\right) $ is constant on the fibers and as such depends only on $x$.\medskip\newline 
If we choose a trivialization $\sigma ^{\left( \alpha \right) }$ in $%
\mathcal{P}$, a trivialization $\partial _{\mu }^{\left( \alpha \right) }$
in $L\left( M\right) $ and we denote by $dx_{\left( \alpha \right) }^{\mu }$
the dual base of the 1-forms of $M$, the local expression of the principal
connection is thence: 
\begin{equation}
\omega =dx^{\mu }\otimes \left( \partial _{\mu }^{\left( \alpha \right)
}+\omega _{\mu }^{\left( \alpha \right) A}\rho _{A}^{\left( \alpha \right)
}\right)
\end{equation}
This is well defined also under a change of trivializations.

Another representation of principal connections, used mainly in
Mathematical physics, is that using the differential forms on $P$ valued in
the Lie algebra $\gbb$ of the group $G$. This representation considers
the projection of a vector $\Xi \in TP$ over the vertical part $\Xi _{\left(
v\right) }\in V\left( \pi \right) $, depending on the connection $\omega $
even if the set of vertical vectors is independent of $\omega $. At this
point, if one uses the isomorphism $\sim $ between the algebra $\gbb$
and the fiber of $V_{p}\left( \pi \right) $, there can be defined a $\gbb
$-valued form $\omega $ over $P$. In this way, $\omega \left( \Xi \right) $
is the element of $\gbb$ which, through the isomorphism, corresponds to $%
\Xi _{\left( v\right) }$.

	The form $\omega $ thus introduced has the following properties:
\begin{equation}
\left\{
\begin{tabular}{l}
\vspace{0.1cm}$\left( \mathbf{a}\right) \hspace{1cm}\omega \left( H_{p}\right) =0$ \\ 
\vspace{0.1cm}$\left( \mathbf{b}\right) \hspace{1cm}\omega \left( \Xi _{\left( v\right)
}\right) \sim \Xi _{\left( v\right) }$ \\ 
$\left( \mathbf{c}\right) \hspace{1cm}\overline{R}_{g}^{*}\omega =Ad\left(
g^{-1}\right) \omega $
\end{tabular}
\right.
\end{equation}
Conversely, if a $\gbb$-valued form $\omega $ over $P$ satisfies the
conditions \textbf{(b)} and \textbf{(c)}, then \textbf{(a)} defines a family
of subspaces $H_{p}\subset TP$ pointing out a principal connection. Locally,
if $\theta ^{A}$ is the base of the right invariant 1-forms dual to the base
of $\rho _{A}$, the form representing the principal connection $\omega $ is: 
\begin{equation}
\omega =\left( \theta ^{A}+A_{\mu }^{A}\left( x\right) dx^{\mu }\right) T_{A}%
\hspace{1cm}A_{\mu }^{A}\left( x\right) =-\omega _{\mu }^{A}\left( x\right)
\end{equation}
The components $A_{\mu }^{A}\left( x\right) $ are called \textsl{vector
potentials }of the connection.

\subsection{Induced connections on associated bundles}

If we choose a principal connection $\omega =dx^{\mu }\otimes \left(
\partial _{\mu }+\omega _{\mu }^{A}\left( x\right) \rho _{A}\right) $ over a
principal bundle $\mathcal{P}=\left( P,M,\pi ,G\right) ,$ we can build in a
canonical way a connection on every bundle $\mathcal{P}\times _{\lambda }F$
associated to $\mathcal{P}$ named the \textsl{induced connection}.

	In fact, chosen $\varphi \in F$, one can define the map 
\begin{equation}
\Phi _{\varphi }:P\longrightarrow P\times _{\lambda }F:p\longmapsto \left[
p,\varphi \right] _{\lambda }
\end{equation}
Let now $\left( p,\varphi \right) $ be a representative of the point $\left[
p,\varphi \right] _{\lambda }\in P\times _{\lambda }F$; the tangent map $%
T\Phi _{\varphi }:T_{p}P\longrightarrow T_{\left[ p,\varphi \right]
_{\lambda }}\left( P\times _{\lambda }F\right) $ defines the horizontal
subspaces in the associated bundle: 
\begin{equation}
\widehat{H}_{\left[ p,\varphi \right] _{\lambda }}:=T_{p}\Phi _{\varphi
}\left( H_{p}\right)
\end{equation}
One can check that $\widehat{H}_{\left[ p,\varphi \right] _{\lambda }}$ does
not depend on any chosen representative for the point $\left[ p,\varphi
\right] _{\lambda }\in P\times _{\lambda }F.$

	It can be also proven that $\widehat{H}_{\left[ p,\varphi \right] _{\lambda
}}$ is a connection over $P\times _{\lambda }F$: for instance, the
connection induced on $\mathcal{P\times }_{Ad \gbb}$ by a principal
connection $\omega $ on $\mathcal{P}$ is in the form: 
\begin{equation}
\widehat{\omega }_{\left( x^{\mu },\upsilon ^{a}\right) }=dx^{\mu }\otimes
\left( \partial _{\mu }+c_{\cdot BC}^{A}\omega _{\mu }^{B}\upsilon
^{C}\partial _{A}\right)
\end{equation}
where $c_{\cdot BC}^{A}$ are the structure constants of the group $G$ with
respect to the generators $T_{A}$ chosen in its Lie algebra $\gbb$.

\subsection{The covariant derivative of a section}

A bundle $\mathcal{B}=\left( B,M,\pi ,F\right) $ (associated to $\mathcal{P}$%
) and a connection $\widehat{\omega }$ in $\mathcal{B}$ (induced by a
principal connection $\omega $ in $\mathcal{P}$) are given.\medskip\newline
\textbf{Definition (1.7.4.1): }let $\xi $ be a vector field on the base $M$
and $\rho :M\longrightarrow B$ a section. The \textsl{covariant derivative of }$%
\rho $ along $\xi $ is the vector 
\begin{equation}
\nabla _{\xi }\rho \left( x\right) =T_{x}\rho \left( \xi \left( x\right)
\right) -\widehat{\omega }_{\rho \left( x\right) }\left( \xi \left( x\right)
\right)
\end{equation}
If we choose a chart on $M$ and the natural base of tangent vectors $%
\partial _{\mu }^{\left( \alpha \right) }$, the covariant derivative of $%
\rho $ along $\partial _{\mu }^{\left( \alpha \right) }$ is $\nabla _{\mu
}^{\left( \alpha \right) }\rho \left( x\right) $.\medskip\newline
\textbf{Proposition (1.7.4.1): }the vector $\nabla _{\xi }\rho \left(
x\right) $ is a vertical vector in\textbf{\ }$\rho \left( x\right) $.\textbf{%
\medskip }\newline
\textbf{Proof: }$T_{x}\rho \left( \xi \left( x\right) \right) $ and $%
\widehat{\omega }_{\rho \left( x\right) }\left( \xi \left( x\right) \right) $
are tangent vectors of $B$ in the point $\rho \left( x\right) $ and they project both on $\xi \left( x\right)$.\medskip\newline
If $\left( x^{\mu },\varphi ^{i}\right) $ are fibered coordinates on $B$, $%
\widehat{\omega }=dx^{\mu }\otimes \left( \partial _{\mu }+\omega _{\mu
}^{i}\left( x,\varphi \right) \partial _{i}\right) $ is the connection, $\xi
=\xi ^{\mu }\partial _{\mu }$ is the vector field on the base and $\rho
:x\longmapsto \left( x,\rho ^{i}\left( x\right) \right) $ is the section,
the covariant derivative is then given by 
\begin{equation}
\nabla _{\xi }\rho \left( x\right) =\xi ^{\mu }\left( \partial _{\mu }\rho
^{i}-\omega _{\mu }^{i}\left( x,\rho \right) \right) \partial _{i}
\end{equation}

\subsection{The stress tensor of a principal connection}

If $\xi =\xi ^{A}\rho _{A}$ is an infinitesimal generator of vertical
automorphisms, it corresponds to a section in $\mathcal{P\times }_{Ad}\gbb$ 
of which the covariant
derivatives can be defined: \medskip\newline
\textbf{Proposition (1.7.5.1): }called $F_{\mu \nu }^{A}:=d_{\nu }\omega
_{\mu }^{A}-d_{\mu }\omega _{\nu }^{A}+c_{BC}^{A}\omega _{\mu }^{B}\omega
_{\nu }^{C}$ the stress tensor of the principal connection $\omega $, we get 
\begin{equation}
\left[ \nabla _{\mu },\nabla _{\nu }\right] \xi ^{A}=c_{BC}^{A}F_{\mu \nu
}^{B}\xi ^{C}
\end{equation}
\textbf{Proof: }the local expression of the section representing $\xi $ is 
\begin{equation}
x\longmapsto \left( x,\xi ^{A}\right)
\end{equation}
the covariant derivatives of this section are thence: 
\begin{equation}
\nabla _{\mu }\xi ^{A}=\partial _{\mu }\xi ^{A}-c_{BC}^{A}\omega _{\mu
}^{B}\xi ^{C}
\end{equation}
the iteration of indices gives the thesis.

\section{Gauge-Natural prolongations of principal bundles}


Let $\mathcal{P}=\left( P,M,\pi ,G\right) $ be a principal bundle. The
standard fiber of the prolongation $J^{r}P$ is the group $J^{r}G=\left\{
j_{0}^{r}a\mid a:\R^{m}\longrightarrow G\right\} $. Let us denote by $%
W^{\left( s,r\right) }G=GL^{s}\left( m\right) \times J^{r}G$ the semidirect
product defined by the product law 
\begin{equation}
\left[ j_{0}^{s}\alpha ,j_{0}^{r}a\right] \odot \left[ j_{0}^{s}\beta
,j_{0}^{r}b\right] =\left[ j_{0}^{s}\left( \alpha \circ \beta \right)
,j_{0}^{r}\left( \left( a\circ \beta \right) \cdot b\right) \right]
\end{equation}
where $\odot $ denotes the product in $W^{\left( s,r\right) }G$. Such a
product is well defined provided $s\geq r$ (otherwise, $j_{0}^{r}\left(
\left( a\circ \beta \right) \cdot b\right) $ would depend on the
representative of $j_{0}^{s}\beta $ chosen). The semidirect product $%
W^{\left( s,r\right) }G$ will be also denoted by $GL^{s}\left( m\right)
\odot J^{r}G$ and it will be called \textsl{Gauge-Natural prolongation of
order }$\left( s,r\right) $\textsl{\ of }$G$.\medskip\newline
\textbf{Theorem (1.8.1.1): }$W^{\left( s,r\right) }\mathcal{P}=L^{s}\left(
M\right) \times _{M}J^{r}\mathcal{P}$ is a principal bundle called\textbf{\ }\textsl{Gauge-Natural prolongation of order }$\left(s,r\right) $\textsl{\ of the principal bundle } $\mathcal{P}$.\medskip\newline
\textbf{Proof: }$W^{\left( s,r\right) }\mathcal{P}$ is a bundle with fiber $%
W^{\left( s,r\right) }G$. A point in $W^{\left( s,r\right) }\mathcal{P}$ has the form $\left[ j_{0}^{s}\epsilon ,j_{x}^{r}\sigma \right] $ with $\epsilon
:\R^{m}\longrightarrow M$, locally invertible in a neighborhood of the origin and such that $\alpha
\left( 0\right) =x,$ while $\sigma:M\longrightarrow P$ is a local section defined around $x\in M$.
To show that it is principal, we define the right action: 
\begin{equation}
\left[ j_{0}^{s}\epsilon ,j_{x}^{r}\sigma \right] \odot \left[
j_{0}^{s}\alpha ,j_{0}^{r}a\right] =\left[ j_{0}^{s}\left( \epsilon \circ
\alpha \right) ,j_{x}^{r}\left( \sigma \left( a\circ \alpha ^{-1}\circ
\epsilon ^{-1}\right) \right) \right]
\end{equation}
where $\odot $ denotes now the right action of $W^{\left( s,r\right) }G$ on $%
W^{\left( s,r\right) }\mathcal{P}$.

\subsection{Gauge-Natural bundles}


We define here the Gauge-Natural bundles of finite order associated to a
principal bundle $\mathcal{P}$. It can be shown that these objects are
actually characterized by the action of $Aut\left( \mathcal{P}\right) ,$
whose expression will be given immediately below. Furthermore Gauge-Natural
bundles of infinite order do not exist. For these results we refer to the
literature (see Ref.[2]), since the according proofs need an advanced and
more abstract formalism which is beyond the scope of this work.\medskip\newline
\textbf{Definition (1.9.1.1): }let $\mathcal{P}=\left( P,M,\pi ,G\right) $ a
principal bundle. A \textsl{Gauge-Natural bundle} \textsl{of order }$\left( s,r\right) $\textsl{\
associated to} $\mathcal{P}$ is a bundle $\mathcal{C}$ isomorphic to an associated bundle $W^{\left( r,s\right) }\mathcal{P}\times _{\lambda}F $ through an action $\lambda $ of $W^{\left( r,s\right) }G$ on the manifold $F$.\medskip\newline
If now $\left( \Phi ,f\right) \in Aut\left( \mathcal{P}\right) $ is an
automorphism of $\mathcal{P}$, it induces an automorphism of $\mathcal{C}$
in the following way: 
\begin{eqnarray}
W_{\lambda }\Phi &:&W^{\left( r,s\right) }\mathcal{P}\times _{\lambda
}F\longrightarrow W^{\left( r,s\right) }\mathcal{P}\times _{\lambda }F 
\nonumber \\
&:&\left[ \left( j_{0}^{s}\epsilon ,j_{x}^{r}\sigma \right) ,\widehat{%
\varphi }\right] _{\lambda }\longmapsto \left[ W^{\left( r,s\right) }\Phi
\left( j_{0}^{s}\epsilon ,j_{x}^{r}\sigma \right) ,\widehat{\varphi }\right]
_{\lambda }  \label{automorphism}
\end{eqnarray}

\subsection{Local expressions}

If $\mathcal{C}$ is a Gauge-Natural bundle, it admits local coordinates $%
\left( x^{\mu },\varphi ^{a}\right) $. These are chosen to parametrize the point $%
\left[ \left( j_{0}^{s}\epsilon ^{\left( \alpha \right) },j_{x}^{r}\sigma
^{\left( \alpha \right) }\right) ,\varphi \right] _{\lambda }\in \mathcal{C}$
according to the conventions fixed in 1.6. The local expression of an
automorphism of $\mathcal{P}$ represented over $\mathcal{C}$ is: 
\begin{equation}
\left\{ 
\begin{tabular}{l}
\vspace{0.1cm}$x^{\prime \mu }=f_{t}^{\mu }\left( x\right) $ \\ 
$g^{\prime }=\Phi _{t}\left( x\right) \cdot g$%
\end{tabular}
\right.
\end{equation}
its infinitesimal generator is in the form: 
\[
\Xi =\xi ^{\mu }\left( x\right) \partial _{\mu }+\xi ^{A}\left( x\right)
\rho _{A} 
\]
where 
\begin{equation}
\left\{ 
\begin{tabular}{l}
\vspace{0.1cm}$\xi ^{\mu }\left( x\right) =\dot{f}_{0}^{\mu }\left( x\right) $ \\ 
$\xi ^{A}\left( x\right) =\left( T^{-1}\right) _{a}^{A}\dot{\phi}%
_{0}^{a}\left( x\right) $%
\end{tabular}
\right.
\end{equation}
and the components of the infinitesimal generator of the automorphism $%
\left( \ref{automorphism}\right) $ induced on $\mathcal{C}$ (and therefore
the Lie derivative $\pounds _{\Xi }\varphi =T\varphi \left( \xi \right) -\Xi
_{\lambda }\circ \varphi $) depend linearly on $\xi ^{\mu }\left( x\right)
,\partial _{\sigma }\xi ^{\mu }\left( x\right) ,\ldots,\partial _{\sigma
_{1}\ldots\sigma _{s}}\xi ^{\mu }\left( x\right) $ and $\xi ^{A}\left( x\right)
,\partial _{\sigma }\xi ^{A}\left( x\right) ,\ldots,\partial _{\sigma
_{1}\ldots\sigma _{r}}\xi ^{A}\left( x\right) $.

\subsubsection{An example: the principal connections bundle}

Let us define the Gauge-Natural bundle whose sections are in one-to-one
correspondence to the principal connections of $\mathcal{P}=\left( P,M,\pi
,G\right) .$

	We choose the vector space $\left( \left( \R^{m}\right) ^{*}\otimes 
\gbb\right) $and we fix a basis $\left( \partial ^{\mu }\otimes
T_{A}\right) $. Let us define the following representation: 
\begin{eqnarray}
\lambda &:&\left( GL\left( m\right) \odot J^{1}G\right) \times \left( \left( 
\R^{m}\right) ^{*}\otimes \gbb\right) \longrightarrow \left( \left( 
\R^{m}\right) ^{*}\otimes \gbb\right) :  \nonumber \\
&:&\left( \left( a_{\mu }^{\nu },g^{a},g_{\mu }^{a}\right) ,\omega _{\mu
}^{A}\right) \longmapsto \overline{a}_{\mu }^{\nu }\left( Ad_{B}^{A}\left(
g\right) \omega _{\nu }^{B}+\overline{R}_{a}^{A}\left( g\right) g_{\nu
}^{a}\right)
\end{eqnarray}
where the bar denotes, as usual, the inverse matrix. The Gauge-Natural
bundle $W^{\left( 1,1\right) }\mathcal{P}\times _{\lambda }\left( \left( 
\R^{m}\right) ^{*}\otimes \gbb\right) $ will be indicated with $%
\mathcal{C}_{\mathcal{P}}$.\medskip\newline
\textbf{Proposition(1.9.3.1): }there is a one-to-one relation between
sections of $\mathcal{C}_{\mathcal{P}}$\ and principal connections of $\mathcal{P}$
.\medskip\newline
\textbf{Proof: }chosen a local trivialization, let 
\begin{equation}
\omega =dx_{\left( \alpha \right) }^{\mu }\otimes \left( \partial _{\mu
}^{\left( \alpha \right) }+\omega _{\mu }^{A\left( \alpha \right) }\left(
x\right) \rho _{A}^{\left( \alpha \right) }\right)
\end{equation}
be the principal connection on $\mathcal{P}$. It induces a section of $%
\mathcal{C}_{\mathcal{P}}$ (which we denote again by $\omega $): 
\begin{equation}
\omega :M\longrightarrow \mathcal{C}_{\mathcal{P}}:x\longmapsto \left[
\partial _{\mu }^{\left( \alpha \right) },j^{1}\sigma ^{\left( \alpha
\right) },\omega _{\mu }^{A\left( \alpha \right) }\left( x\right) \partial
^{\mu \left( \alpha \right) }\otimes T_{A}\right] _{\lambda }
\end{equation}
it is well defined because of the form of the representation $\lambda$ and
the transformation rules of principal connections.

\subsection{The Lie derivative of connection and curvature}

Now that we are able to regard the principal connections with the sections
of a bundle, we can evaluate the Lie derivatives with respect to a
projectable field $\left( \Xi ,\xi \right) $ of $\mathcal{P}$. It can be
easily proven that: 
\begin{equation}
\pounds _{\Xi }\omega _{\mu }^{A}=F_{\mu \nu }^{A}\xi ^{\nu }-\nabla _{\mu
}\xi _{\left( \upsilon \right) }^{A}
\end{equation}
being $\xi _{\left( \upsilon \right) }^{A}=\xi ^{A}-\omega _{\mu }^{A}\xi
^{\mu }$ the vertical part of the generator $\Xi $.

	For the curvature, we get: 
\begin{equation}
\pounds _{\Xi }F_{\mu \nu }^{A}=\nabla _{\sigma }F_{\mu \nu }^{A}\xi
^{\sigma }+F_{\mu \sigma }^{A}\nabla _{\nu }\xi ^{\sigma }+F_{\sigma \nu
}^{A}\nabla _{\mu }\xi ^{\sigma }+c_{\cdot BC}^{A}F_{\mu \nu }^{B}\xi
_{\left( \upsilon \right) }^{C}
\end{equation}
where the covariant derivative of the field strength (curvature) $F_{\mu \nu
}^{A}$ is 
\begin{equation}
\nabla _{\sigma }F_{\mu \nu }^{A}=d_{\sigma }F_{\mu \nu }^{A}-F_{\rho \nu
}^{A}\Gamma _{\mu \sigma }^{\rho }-F_{\mu \rho }^{A}\Gamma _{\nu \sigma
}^{\rho }+c_{\cdot BC}^{A}F_{\mu \nu }^{B}\omega _{\sigma }^{C}
\label{curvatura}
\end{equation}


Now, by using the form (\ref{curvatura}) of the covariant derivative of
the curvature, it is easy to verify the following:\medskip\newline
\textbf{Proposition: }for the stress tensor the \textsl{Bianchi identities}
hold: 
\begin{equation}
\nabla _{\sigma }F_{\mu \nu }^{A}+\nabla _{\mu }F_{\nu \sigma }^{A}+\nabla
_{\nu }F_{\sigma \mu }^{A}=0 \label{bianchi}
\end{equation}
(notice that the connection $\Gamma _{\mu \sigma }^{\rho }$ used is torsion
free).


\chapter{Geometrical formalism for variational calculus}\label{chapt:variational}

\section{Introduction}

	This chapter regards the study of variational calculus from a
geometrical point of view, that has been developed in order to extend the
results found in $\R^{n}$ to general manifolds. However, it assumes a
quite strong notion of regularity (i.e. we shall use $\mathcal{C}^{\infty }$
objects). We are not usually interested in searching for solutions: this is
a \textsl{local} problem, and as such it can be treated by the powerful
functional analysis.

	Despite at a first sight the regularity hypotheses may seem very restrictive, these are typically satisfied in the applications in 
fundamental physics. Moreover, the geometrical framework is very effective
in searching for the fundamental structures of field theories. Indeed, it
often simplifies the work made in a local framework: the need to introduce
an enough general structure, which has to be well defined from the global
point of view, limits the possible choices in a local framework.

	As an example, we will analyze in details symmetries and conserved
quantities. It is clear from physics that they play a fundamental role in
model building. Yet, from a mathematical point of view,
there exist many ways to implement this concept. The characteristic property
of symmetries is to preserve the space of solutions of the field equations,
but depending on the case they can be more or less generalized. In this
chapter we choose a characterization of the concept of symmetry which preserves the Lagrangian structure of the field equations (i.e.,
it leaves the Poincar\'{e}-Cartan form invariant). Though this is not the most general characterization, it
will be enough for our purposes.

	Within the geometrical framework, then, some
key-concepts such as the variation of the action
functional can be obtained. The definition we give for this object is deeply geometric. Though coinciding substantially with the usual definition of variation used in the local paradigm, this is conceptually much simpler. More precisely, the
functional derivative is normally defined in functional analysis as
a directional derivative of the action functional. namely, of a function on
the infinite dimensional space of sections defined over $D$, with boundary
conditions fixed on the boundary $\partial D$. Consequently, when dealing with variation one has to specify first of all this functional space, and then define derivatives, theorems of derivation and so on. After this, one can proceed with the
variation. The result is the same when using the geometrical formalism,
where function spaces or other infinite dimensional spaces are absent. The functional variation we will introduce simply coincides
with the total derivative of a real function of a real variable.

	Besides, there are phenomena, such as monopoles and instantons, whose
properties are hidden in the boundary
conditions, thus remain quite obscure from a local perspective.
Since the chosen function space is infinite dimensional, we
usually do not know almost anything of it. This occurs if one uses the local point of view. Using the geometrical formalism,
on the contrary, instantons and monopoles can be described in terms of
sections of non trivial bundles. This fact is a further evidence that field
theories should be formulated on bundles instead of Cartesian
products (i.e. the local models), simply because the world does work this way.

	However, although we prefer a geometrical perspective, we do not think local
results are of minor interest: on the contrary, they are often the
foundation of the geometrical viewpoint. Variational calculus is, among the
branches of Mathematics, maybe the discipline where the interaction between
the analytic and the geometric formalisms have been developed more
extensively.

In this chapter we provide a rapid summary of the geometrical formulation
of field theories, embedding the structures introduced in the previous chapter
in a physical environment. The resulting Lagrangian formalism is
provided with techniques of variational calculus over bundles. The topic of conserved quantities is also briefly addressed.

\section{Lagrangian formalism}

As anticipated in the introduction, the configurations of a field theory
are, by definition, the smooth sections of the configuration bundle $%
\mathcal{C}=\left( C,M,\pi ,F\right) $ over the space-time $M$. A section of 
$\mathcal{C}$ describes the values of the fields in $F$ at each point of the
space-time $M$, and it represents indeed the entire evolution of the system.
In other words, the Euler-Lagrange equations will have to single out
particular sections of $\mathcal{C}$.

	A Lagrangian of order $k$ is a bundle morphism $L:J^{k}\mathcal{C}%
\longrightarrow \Lambda _{m}^{0}\left( M\right) $:

\begin{equation}
\begin{tabular}{lll}
$\;\;\,J^{k}\mathcal{C}$ & $\longrightarrow $ & $\hspace{-0.1cm}\Lambda _{m}^{0}\left(
M\right) $ \\ 
$\pi ^{k}\downarrow $ &  & $\downarrow \tau _{m}^{0}$ \\ 
\ \hspace{0.28cm}$M$ &  & $\hspace{-0.1cm}M$%
\end{tabular}
\end{equation}
where $\Lambda _{m}^{0}\left( M\right) $ is the $m$-forms bundle over $M$ $%
\left( m=\dim \left( M\right) \right) .$ The bundle $J^{k}\mathcal{C}$ is
called \textsl{(lagrangian) phases bundle}.

	Such a definition could seem odd, but it leads in a direct manner to the
formulation of \textsl{Hamilton's principle of stationary action}. In
fact, if $D\subset M$ is a region of $M$ (that is, a compact submanifold of
dimension $m$ with a boundary $\partial D$ which is a compact submanifold of
dimension $m-1$) and $\rho :M\longrightarrow \mathcal{C}$ is a
configuration, we can evaluate the Lagrangian $L$ on the $k$-order
prolongation of the section. One then obtains an $m$-form $L\circ
j^{k}\rho $ over $M$; this is the correct object to be integrated over $D$. We
define the action in $D$ of the section $\rho $ as follows: 
\begin{equation}
A_{D}\left( \rho \right) =\int_{D}L\circ j^{k}\rho
\end{equation}
In the case of Mechanics $J^{1}\left( \R\times Q\right) \simeq \R%
\times TQ$ and a first order Lagrangian is locally described by 
\begin{equation}
L:\R\times TQ\longrightarrow \Lambda _{1}^{0}\left( \R\right)
:\left( t,q^{\alpha },u^{\alpha }\right) \longmapsto \mathcal{L}\left(
t,q^{\alpha },u^{\alpha }\right) dt
\end{equation}
which coincides with the notion of Lagrangian given in (time-dependent)
Mechanics.

Let us now choose a vertical field $X$ with compact support $D$ such that $%
j^{k-1}X$ vanishes over $\partial D$; this field is called \textsl{%
deformation} \textsl{(with fixed values at the boundary) over }$D$. Now we
can \textsl{drag} a section $\rho :M\longrightarrow \mathcal{C}$ in order to
define a one-parameter family of sections: 
\begin{equation}
\rho _{s}=\phi _{s}\circ \rho
\end{equation}
and calculate the action for each element of the family: 
\begin{equation}
A_{D}^{s}\left( \rho \right) =\int_{D}L\circ j^{k}\rho _{s}
\end{equation}
Being $\rho $ and $X$ fixed, $A_{D}^{s}\left( \rho \right) $ is a function
associating to $s\in \R$ the value of the action in the region $D$ of
the section $\rho $ dragged along the field $X.$ It is thence a function $%
A_{D}^{s}\left( \rho \right) :\R\longrightarrow \R$ and one can
consider the derivative at $s=0$. Therefore we define: 
\begin{equation}
\delta _{X}A_{D}^{s}\left( \rho \right) =\left[ \frac{d}{ds}A_{D}^{s}\left(
\rho \right) \right] _{s=0}
\end{equation}
and enunciate \textsl{Hamilton's principle of stationary action}:\medskip\newline
\textbf{Definition (2.2.1): }a configuration $\rho :M\longrightarrow 
\mathcal{C}$ is a \textsl{section of motion } (or a \textsl{critical section}\ or a\textbf{\ }\textsl{shell}) if for every
region $D$ and for every deformation $X$ over $D$ the following holds: 
\begin{equation}
\delta _{X}A_{D}^{s}\left( \rho \right) =0
\end{equation}
Since there is not any \textsl{a priori} physical reason to accept
this rule, it is very important to explain its meaning.
	From a mathematical point of view, it is relatively simple to show that this
axiom implies that the section $\rho $ is a solution of the Euler-Lagrange
equations (see the next section). It remains anyway obscure the \textsl{%
physical} reason why some phenomena (and among them, all the fundamental
physics) have to obey this prescription. A partial clarification comes by observing that the Lagrangian is not an
observable, and that it is not determined by the system.

	The only reason to be of the Lagrangian is indeed to determine canonically (through
variational calculus and Euler-Lagrange equations) and universally (at
least for a huge class of phenomena) the correct equations which select the
observed evolution.

	Accordingly, the principle of stationary action may be viewed as an implicit definition of
the Lagrangian: it is the object which gives the correct field equations through the Euler-Lagrange equations.

\section{Euler-Lagrange equations}

The variation of the action introduced in the previous paragraph can be
expanded as follows: 
\begin{eqnarray}
\delta _{X}A_{D}^{s}\left( \rho \right) &=&\left[ \frac{d}{ds}\int_{D}L\circ
j^{k}\phi _{s}\circ j^{k}\rho \right] _{s=0}=  \nonumber \\
&=&\int_{D}\left[ \frac{d}{ds}\left( L\circ j^{k}\phi _{s}\circ j^{k}\rho
\right) \right] _{s=0}=  \nonumber \\
&:=&\int_{D}\left\langle \delta L\circ j^{k}\rho \mid
j^{k}X\right\rangle
\end{eqnarray}
where $\delta L:J^{k}\mathcal{C}\longrightarrow V^{*}\left( J^{k}\mathcal{C}%
\right) \otimes \Lambda _{m}^{0}\left( M\right) $ is a global bundle
morphism (see Ref.[2]).

	For each Lagrangian of order $k$ over $\mathcal{C}$ there exists a unique
Euler-Lagrange morphism 
\begin{equation}
\E\left( L\right) :J^{2k}\mathcal{C}\longrightarrow V^{*}\left( 
\mathcal{C}\right) \otimes \Lambda _{m}^{0}\left( M\right)
\end{equation}
and a family of Poincar\'{e}-Cartan morphisms parametrized by a linear
fibered connection $\gamma $ 
\begin{equation}
\F\left( L,\gamma \right) :J^{2k-1}\mathcal{C}\longrightarrow
V^{*}\left( J^{k-1}\mathcal{C}\right) \otimes \Lambda _{m-1}^{0}\left(
M\right)
\end{equation}
such that, for each section $\rho :M\longrightarrow \mathcal{C}$ and for
each vertical vector field $X$ over $\mathcal{C}$, the \textsl{\ first
variation formula }holds: 
\begin{equation}
\left\langle \delta L\circ j^{k}\rho \mid j^{k}X\right\rangle =\left\langle 
\E\left( L\right) \circ j^{2k}\rho \mid X\right\rangle +d\left[
\left\langle \F\left( L,\gamma \right) \circ j^{2k-1}\rho \mid
j^{k-1}X\right\rangle \right]
\label{first variation}
\end{equation}
where $\left\langle \cdot \mid \cdot \right\rangle $ denotes the duality
between $V\left( \mathcal{C}\right) $ and $V^{*}\left( \mathcal{C}\right) $
and even its prolongations, for instance between $V\left( J^{k-1}\mathcal{C}%
\right) $ and $V^{*}\left( J^{k-1}\mathcal{C}\right) $. With this convention 
$\F\left( L,\gamma \right) $ $\circ j^{2k-1}\rho $ is an element of $%
V^{*}\left( J^{k-1}\mathcal{C}\right) \otimes \Lambda _{m-1}^{0}\left(
M\right) $ and therefore the object $\left\langle \F\left( L,\gamma \right) \circ
j^{2k-1}\rho \mid j^{k-1}X\right\rangle $ is an $\left( m-1\right) $ form
over $M.$

	Although the fibered connection $\gamma $ is useful to prove the
globality of these morphisms, it has been shown that the Poincar\'{e}-Cartan
morphism depends only on a connection over the base. However, in the case we
are going to study $\left( k=1\right) $ this dependence does not exist, and
both the morphisms are uniquely defined.

	The first variation formula contains all the information of variational
calculus. Indeed, for deducing the Euler-Lagrange equations from the
principle of stationary action, it is enough to operate in the following
way:
\begin{eqnarray}
&&\delta _{X}A_{D}\left( \rho \right) =\int_{D}\left\langle \delta L\circ
j^{k}\rho \mid j^{k}X\right\rangle = \nonumber\\
&&=\int_{D}\left\langle \E\left( L\right) \circ j^{2k}\rho \mid
X\right\rangle +\int_{D}d\left[ \left\langle \F\left( L,\gamma \right)
\circ j^{2k-1}\rho \mid j^{k-1}X\right\rangle \right] = \nonumber\\
&&=\int_{D}\left\langle \E\left( L\right) \circ j^{2k}\rho \mid
X\right\rangle +\int_{\partial D}\left[ \left\langle \F\left( L,\gamma
\right) \circ j^{2k-1}\rho \mid j^{k-1}X\right\rangle \right]
\end{eqnarray}
Now it is easy to note that the second integral gives no contribution
because the integrand is calculated over $\partial D$ where $j^{k-1}X=0.$

So we obtain the following,
\begin{equation}
\int_{D}\left\langle \E\left( L\right) \circ j^{2k}\rho \mid
X\right\rangle =0
\end{equation}
in every region $D$ and for each deformation $X$. Since $D$ is arbitrary,
the integrand $\left\langle \E\left( L\right) \circ j^{2k}\rho \mid
X\right\rangle $ has to vanish for any $X$, hence: 
\begin{equation}
\E\left( L\right) \circ j^{2k}\rho =0  
\label{euler}
\end{equation}
which are the Euler-Lagrange equations. These are partial differential
equations of order $2k$ in the section $\rho $.\medskip\newline
\textbf{Definition (2.3.1): }a \textsl{formal divergence }is the operator
associating uniquely a morphism $Div\theta :J^{h+1}\mathcal{C}\longrightarrow \Lambda
\left( M\right) $ to a morphism $\theta :J^{h}\mathcal{C}\longrightarrow $ $\Lambda \left( M\right) $ such that 
\begin{equation}
\forall \rho :M\longrightarrow \mathcal{C},\hspace{1cm}\left( Div\theta
\right) \circ j^{h+1}\rho =d\left( \theta \circ j^{h}\rho \right)
\end{equation}\medskip 
It is also easy to prove:\medskip\newline
\textbf{Theorem (2.3.1):} if the Lagrangian is a formal $m$-divergence (i.e., if there exists a morphism $\theta :J^{k-1}\mathcal{C}\longrightarrow \Lambda
_{m-1}^{0}\left( M\right) $ such that $L=Div\theta $), then $\forall \rho :M\longrightarrow \mathcal{C},$ $\E\left( L\right)\circ j^{2k}\rho =0.$\medskip\newline
\textbf{Proof: }notice that the variation of the action is identically zero: 
\begin{eqnarray}
\delta _{X}A_{D}\left( \rho \right) &=&\delta _{X}\int_{D}L\circ j^{k}\rho =
\nonumber \\
&=&\delta _{X}\int_{D}\left( Div\theta \right) \circ j^{k}\rho =\delta
_{X}\int_{\partial D}\theta \circ j^{k-1}\rho \equiv 0
\end{eqnarray}
being $j^{k-1}X\mid _{\partial D}=0$.

	The above property justifies the previous claim that the Lagrangian of a system
is not uniquely determined. Let $L$ be a Lagrangian of order $k$, $\forall
\theta :J^{h-1}\mathcal{C}\longrightarrow \Lambda _{m-1}^{0}\left( M\right) $.
The new Lagrangian $L+Div\theta $ is (by abuse of notation) another
Lagrangian (of order less than $\max \left( k,h\right) $) which gives the
same Euler-Lagrange equations. It is therefore equivalent to $L$.

\section{Poincar\'{e}-Cartan form and symmetries}

To each Lagrangian $L$ of order $k$, once we choose a linear connection $%
\gamma $, we can associate an $m$-form $\Theta \left( L,\gamma \right) $
over $J^{2k-1}\mathcal{C}$ called \textsl{Poincar\'{e}-Cartan form}, whose
local expression is 
\begin{equation}
\Theta \left( L,\gamma \right) =\mathcal{L}\mathbf{ds}+\left[ \widehat{\mathbf{f}}%
_{\alpha }^{\mu _{1}}\omega ^{\alpha }+\widehat{\mathbf{f}}_{\alpha }^{\mu
_{1}\mu _{2}}\omega _{\mu _{2}}^{\alpha }+\ldots+\widehat{\mathbf{f}}_{\alpha
}^{\mu _{1}\mu _{2}\ldots\mu _{k}}\omega _{\mu _{2}\ldots\mu _{k}}^{\alpha
}\right] \wedge \mathbf{ds}_{\mu _{1}}  \label{poincare}
\end{equation}
where 
\begin{equation}
\left\{ 
\begin{tabular}{l}
\vspace{0.1cm}$\widehat{\mathbf{f}}_{\alpha }^{\mu _{1}}:=\F^{\prime }\left( L,\gamma
\right) _{\alpha }^{\mu _{1}}$ \\ 
\vspace{0.1cm}$\widehat{\mathbf{f}}_{\alpha }^{\mu _{1}\mu _{2}}:=\F^{\prime }\left(
L,\gamma \right) _{\alpha }^{\mu _{1}\mu _{2}}$ \\ 
$\widehat{\mathbf{f}}_{\alpha }^{\mu _{1}\mu _{2}\ldots\mu _{k}}:=\F%
^{\prime }\left( L,\gamma \right) _{\alpha }^{\mu _{1}\mu _{2}\ldots\mu _{k}}$%
\end{tabular}
\right.
\end{equation}
with 
\begin{equation}
\left\{ 
\begin{tabular}{l}
\vspace{0.1cm}$\F^{\prime }\left( L,\gamma \right) _{\alpha }^{\mu _{1}}:=\hat{p}%
_{\alpha }^{\mu _{1}}-\nabla _{\mu _{2}}\hat{p}_{\alpha }^{\mu _{1}\mu
_{2}}+\ldots+\left( -1\right) ^{k-1}\nabla _{\mu _{2}\ldots\mu _{k}}\hat{p}%
_{\alpha }^{\mu _{1}\mu _{2}\ldots\mu _{k}}$ \\ 
$\F^{\prime }\left( L,\gamma \right) _{\alpha }^{\mu _{1}\mu _{2}\ldots\mu
_{k}}:=\hat{p}_{\alpha }^{\mu _{1}\mu _{2}\ldots\mu _{k}}$%
\end{tabular}
\right.
\end{equation}
and the $\hat{p}_{\alpha }$, called \textsl{covariant momenta }of the
Lagrangian $L$, are defined by the following identity: 
\begin{equation}
\begin{tabular}{l}
$p_{\alpha }X^{\alpha }+p_{\alpha }^{\mu _{1}}X_{\mu _{1}}^{\alpha
}+\ldots+p_{\alpha }^{\mu _{1}\mu _{2}\ldots\mu _{k}}X_{\mu _{1}\mu _{2}\ldots\mu
_{k}}^{\alpha }=$ \\ 
$=\hat{p}_{\alpha }\hat{X}^{\alpha }+\hat{p}_{\alpha }^{\mu _{1}}\hat{X}%
_{\mu _{1}}^{\alpha }+\ldots+\hat{p}_{\alpha }^{\mu _{1}\mu _{2}\ldots\mu _{k}}%
\hat{X}_{\mu _{1}\mu _{2}\ldots\mu _{k}}^{\alpha }$%
\end{tabular}
\end{equation}
with the vertical field $X=X^{\alpha }\partial _{\alpha }.$\newline
In Eq.(\ref{poincare}), $\omega ^{\alpha },\omega _{\mu _{2}}^{\alpha
},\ldots,\omega _{\mu _{2}\ldots\mu _{k}}^{\alpha }$ are the contact forms defined
on $J^{k+1}\mathcal{C}$. If $k=0,1,2,$ $\Theta \left( L,\gamma \right) $
does not depend on the connection.\newline
Being $\Theta \left( L,\gamma \right) -\mathcal{L}\mathbf{ds}$ a contact form, the
action functional can be rewritten as: 
\begin{equation}
A_{D}\left( \rho \right) =\int_{D}\left( j^{2k-1}\rho \right) ^{*}\Theta
\left( L,\gamma \right)
\end{equation}
The Euler-Lagrange equations can be thence formulated in terms of the
Poincar\'{e}-Cartan form with the variation 
\begin{eqnarray*}
&&\delta _{X}A_{D}\left( \rho \right) =\left[ \frac{d}{ds}\int_{D}\left(
j^{2k-1}\rho \right) ^{*}\left( j^{2k-1}\phi _{s}\right) ^{*}\Theta \left(
L,\gamma \right) \right] _{s=0} =\\
&&=\int_{D}\left( j^{2k-1}\rho \right) ^{*}\left( i_{j^{2k-1}X}\circ d+d\circ
i_{j^{2k-1}X}\right) \Theta \left( L,\gamma \right) =\\
&&=\int_{D}\left( j^{2k-1}\rho \right) ^{*}i_{j^{2k-1}X}\circ d\Theta \left(
L,\gamma \right) +\int_{\partial D}\left( j^{2k-1}\rho \right)
^{*}i_{j^{2k-1}X}\Theta \left( L,\gamma \right)
\end{eqnarray*}
The last term gives no contribution because it depends linearly on $j^{k-1}X$
and is evaluated on $\partial D$ (where $j^{k-1}X=0$); $D$ and $X$ are
arbitrary, therefore: 
\begin{equation}
\E\left( L\right) \circ j^{2k}\rho =\left( j^{2k-1}\rho \right)
^{*}i_{j^{2k-1}X}\circ d\Theta \left( L,\gamma \right) =0
\end{equation}
are equivalent to the field equations.\medskip\newline
\textbf{Definition (2.4.1): }a \textsl{(Lagrangian) symmetry }for the
Lagrangian $L$ is an automorphism of the configuration bundle:
\begin{equation}
\begin{tabular}{lll}
$\hspace{0.3cm}\mathcal{C}$ & $\stackrel{\Phi }{\longrightarrow}$ & $\mathcal{C}$ \\ 
$\pi \downarrow $ &  & $\downarrow \pi $ \\ 
$\hspace{0.25cm}M$ & $\stackrel{f}{\longrightarrow}$ & $\hspace{-0.07cm}M$%
\end{tabular}
\end{equation}
such that $\left( j^{2k-1}\Phi \right) ^{*}\Theta \left( L,\gamma \right)
=\Theta \left( L,\gamma \right) $, i.e. such that it leaves the Poincar\'{e}-Cartan form invariant.\medskip\newline
\textbf{Theorem (2.4.1): }if\textbf{\ }$\left( \Phi ,f\right) $\textbf{\ }is
a symmetry for\textbf{\ }$L,$ it sends solutions into solutions.\medskip\newline
\textbf{Proof:} let $\rho :M\longrightarrow \mathcal{C}$ be a solution of
the Euler-Lagrange equations, i.e. such that $j^{k-1}X=0$ on $\partial D$, we have: 
\begin{equation}
\left( j^{2k-1}\rho \right) ^{*}i_{j^{2k-1}X}d\Theta \left( L,\gamma \right)
=0
\end{equation}
We can use $\left( \Phi ,f\right) $ to define a new section $\rho ^{\prime
}=\Phi \circ \rho \circ f^{-1}$; then $\rho ^{\prime }$ is itself a solution: in fact, 
\begin{eqnarray}
&&\left( j^{2k-1}\rho ^{\prime }\right) ^{*}i_{j^{2k-1}X}d\Theta \left(
L,\gamma \right) = \nonumber\\
&&=\left( f^{-1}\right) ^{*}\left( j^{2k-1}\rho \right) ^{*}\left(
j^{2k-1}\Phi \right) ^{*}i_{j^{2k-1}X}d\Theta \left( L,\gamma \right) =\nonumber\\ 
&&=\left( f^{-1}\right) ^{*}\left( j^{2k-1}\rho \right) ^{*}i_{j^{2k-1}\left(
\Phi _{*}X\right) }\left( j^{2k-1}\Phi \right) ^{*}d\Theta \left( L,\gamma
\right) = \nonumber\\
&&=\left( f^{-1}\right) ^{*}\left( j^{2k-1}\rho \right) ^{*}i_{j^{2k-1}\left(
\Phi _{*}X\right) }d\Theta \left( L,\gamma \right)
\end{eqnarray}
from which it follows that: 
\begin{eqnarray}
&&\left( j^{2k-1}\rho ^{\prime }\right) ^{*}i_{j^{2k-1}X}d\Theta \left(
L,\gamma \right) = \nonumber\\ 
&&=\left( f^{-1}\right) ^{*}\left( j^{2k-1}\rho \right) ^{*}i_{j^{2k-1}\left(
\Phi _{*}X\right) }d\Theta \left( L,\gamma \right) =0\Longleftrightarrow \nonumber\\
&&\Longleftrightarrow \left( j^{2k-1}\rho \right) ^{*}i_{j^{2k-1}X}d\Theta
\left( L,\gamma \right) =0
\end{eqnarray}
This is not the most general definition of symmetry, but it is enough to develop a theory of conserved quantities that covers most of cases of physical interest.\medskip\newline
\textbf{Theorem (2.4.2): }$\left( \Phi ,f\right) $ is a symmetry for $L$ if
and only if\textbf{\ }$\mathcal{J\cdot L}\circ j^{k}\Phi =$ $\mathcal{L}$, where\textbf{\ }$\mathcal{J}=\det \left( J\right) ,$ $L=$ $%
\mathcal{L}\mathbf{ds}$ (so that $\mathcal{L}$ will be called \textsl{Lagrangian } \textsl{density})
and
$J$\ is the Jacobian of the diffeomorphism\textbf{\ }$f:M\longrightarrow M$.

\section{Covariant Lagrangians and the N\"{o}ther theorem}

Let $G\subset Aut\left( \mathcal{C}\right) $ be a subgroup of automorphisms
of the configuration bundle.\medskip\newline
\textbf{Definition (2.5.1): }a Lagrangian of order $k$ is $G$\textsl{%
-covariant }if every automorphism $\left( \phi ,f\right) \in G$ is a symmetry.\medskip\newline
Let $H\subset G$ be a 1-parameter subgroup $\left( \phi _{s},f_{s}\right) $
of symmetries: 
\begin{equation}
\varphi ^{\prime a}=\phi _{s}^{a}\left( x,\varphi \right) ,\hspace{1cm}s\in 
\R
\end{equation}
\begin{equation}
x^{\prime \mu }=f_{s}^{\mu }\left( x\right)
\end{equation}
and let us denote with $\Xi =\xi ^{\mu }\partial _{\mu }+\xi ^{a}\partial
_{a}$ its infinitesimal generator. The following holds:\medskip\newline
\textbf{Theorem (2.5.2): }if the Lagrangian is $G$-covariant, 
\begin{equation}
\left\langle \delta L\mid j^{k}\pounds _{\Xi }\varphi \right\rangle
=Div\left( i_{\xi }L\right)  \label{identità fondamentale}
\end{equation}
which is called \textsl{fundamental identity}.\medskip\newline
\textbf{Proof: }$\left( \phi _{s},f_{s}\right) $ are symmetries, therefore it must be $\mathcal{L}=\mathcal{J\cdot L}\circ j^{k}\Phi _{s}$ which implies the infinitesimal condition:
\begin{equation}
0=\left( \partial _{\mu }\xi ^{\mu }\right) \mathcal{L}+\xi ^{\mu }\left(
\partial _{\mu }\mathcal{L}\right) +p_{a}\xi ^{a}+p_{a}^{\mu }\xi _{\mu
}^{a}+\ldots+p_{a}^{\mu _{1}\ldots\mu _{k}}\xi _{\mu _{1\ldots\mu _{k}}}^{a}
\end{equation}
that can be recast as
\begin{equation}
d_{\mu }\left( \mathcal{L}\xi ^{\mu }\right) =p_{a}\left( \pounds _{\Xi
}\varphi ^{a}\right) +p_{a}^{\mu }\left( \pounds _{\Xi }\varphi _{\mu
}^{a}\right) +\ldots+p_{a}^{\mu _{1}\ldots\mu _{k}}\left( \pounds _{\Xi }\varphi
_{\mu _{1}\ldots\mu _{k}}^{a}\right)  \label{identità in coordinate}
\end{equation}
This is the local expression of (\ref{identità fondamentale}) in natural
fibered coordinates.

	Note that the intrinsic expression (\ref{identità fondamentale}) is
much more useful than the one in coordinates (\ref{identità in coordinate}).
It can happen for instance that a Lagrangian (supposed here of the first
order, for simplicity) depends on the fields' derivatives
just through some combination: 
\begin{equation}
L=\mathcal{L}\left( x^{\mu },\varphi ^{a},R^{A}\left( \varphi ^{b},\varphi
_{\nu }^{b}\right) \right) \mathbf{ds}
\end{equation}
In this case, the fundamental identity can be written as 
\begin{equation}
d_{\mu }\left( \mathcal{L}\xi ^{\mu }\right) =\left( \partial _{a}\mathcal{L}%
\right) \left( \pounds _{\Xi }\varphi ^{a}\right) +\left( \partial _{A}%
\mathcal{L}\right) \left( \pounds _{\Xi }R^{A}\right)  \label{okoo}
\end{equation}
where $\pounds _{\Xi }R^{A}$ is expressed in terms of $\pounds _{\Xi
}\varphi ^{a}$ and $\pounds _{\Xi }\varphi _{\mu }^{a}.$ When the function $R^{A}$ is complicated, it is simpler to use Eq.(\ref{okoo}) than the version
in local natural coordinates Eq.(\ref{identità in coordinate}).

	Using now the fundamental identity and the first variation formula, we can
obtain the following property: 
\begin{equation}
Div\left( i_{\xi }L\right) =\left\langle \E\left( L\right) \mid \pounds
_{\Xi }\varphi \right\rangle +Div\left\langle \F\left( L,\gamma \right)
\mid j^{k-1}\pounds _{\Xi }\varphi \right\rangle
\end{equation}
which can be easily recast in the form: 
\begin{equation}
Div\mathcal{E}\left( L,\Xi \right) =\mathcal{W}\left( L,\Xi \right)
\label{divergenza}
\end{equation}
with 
\begin{equation}
\left\{ 
\begin{tabular}{l}
\vspace{0.1cm}$\mathcal{E}\left( L,\Xi \right) =\left\langle \F\left( L,\gamma \right)
\mid j^{k-1}\pounds _{\Xi }\varphi \right\rangle -i_{\xi }L  \label{n1}$ \\ 
$\mathcal{W}\left( L,\Xi \right) =-\left\langle \E\left( L\right) \mid
\pounds _{\Xi }\varphi \right\rangle$
\end{tabular}
\right.
\end{equation}
	Equations (\ref{divergenza}) and (\ref{n1}) are one of the forms
of the \textsl{N\"{o}ther theorem}, which gives a prescription to
construct the $\left( m-1\right) $-form $\mathcal{E}\left( L,\Xi \right) $
on $j^{2k-1}\mathcal{C}$ starting from the infinitesimal generator of
generalized symmetries $\left( \Xi ,\xi \right) $ for the Lagrangian $L$\footnote{see Section 3.6 for the definition of generalized symmetries.}.
Such $\left(m-1\right) $-form associates, by pull-back, an $\left( m-1\right) $-form
of the base $M$ to each section $\rho :M\longrightarrow \mathcal{C}$,
\begin{equation}
\mathcal{E}\left( L,\Xi ,\rho \right) =\left( j^{2k-1}\rho \right) ^{*}%
\mathcal{E}\left( L,\Xi \right)
\end{equation}
which is closed if $\rho $ is a solution, because in that case 
\begin{equation}
\mathcal{W}\left( L,\Xi ,\rho \right) =\left( j^{2k}\rho \right) ^{*}%
\mathcal{W}\left( L,\Xi \right) =0
\end{equation}
\textbf{Definition (2.5.2): }the form $\mathcal{E}\left( L,\Xi \right) $ is
called \textsl{conserved current (on-shell)}, while $\mathcal{W}\left( L,\Xi \right) $ is the \textsl{
work form}.\medskip\newline
\textbf{Definition (2.5.3): }let us consider the conserved current $\mathcal{E}\left( L,\Xi \right) $; we
define the corresponding \textsl{conserved quantity}: 
\begin{equation}
Q_{D}\left( L,\Xi ,\rho \right) =\int_{D}\mathcal{E}\left( L,\Xi ,\rho
\right)
\end{equation}
If the configuration bundle is a natural bundle, each diffeomorphism of the
base $M$ can be lifted to an automorphism of the configuration bundle. Thus we can (improperly) identify the group $\hbox{Diff}\left( M\right) $ with a
subgroup of $Aut\left( \mathcal{C}\right) .$ The Lagrangians $\hbox{Diff}\left(
M\right) $-covariant are called \textsl{natural Lagrangians}. A theory
described by a natural configuration bundle and by a natural Lagrangian is
called \textsl{natural theory}.

\subsection{The superpotential}

As we have seen above, to evaluate the conserved quantities one has to
integrate the closed form $\mathcal{E}\left( L,\Xi ,\rho \right) $ in a
region $D$. It is therefore interesting to know if $\mathcal{E}$ is exact, in view of a
possible use of the Stokes theorem.\medskip\newline
\textbf{Definition (2.6.1): }if the conserved current $\mathcal{E}\left(L,\Xi \right) $ can be written in the form 
\begin{equation}
\mathcal{E}\left( L,\Xi \right) =\widetilde{\mathcal{E}}\left( L,\Xi \right)
+Div\left( \mathcal{U}\left( L,\Xi \right) \right)
\end{equation}
where $\widetilde{\mathcal{E}}\left( L,\Xi ,\rho \right) :=\left(
j^{2k-1}\rho \right) ^{*}\widetilde{\mathcal{E}}\left( L,\Xi \right) =0$ for
every $\rho :M\longrightarrow \mathcal{C}$ which is a solution, the $\left( m-1\right) $-form $\widetilde{\mathcal{E}}\left(
L,\Xi \right) $ is the \textsl{reduced current,} while the $\left( m-2\right) $-form $\mathcal{U}\left( L,\Xi \right) $ is the \textsl{
superpotential}. If the reduced current and the superpotential exist for each 1-parameter subgroup of symmetries, we
say that \textsl{the theory admits a superpotential}.\medskip\newline
\textbf{Proposition (2.6.1):} if a theory admits a superpotential, the
conserved quantities are rewritten as follows: 
\begin{equation}
Q_{D}\left( \Xi ,\rho \right) =\int_{D}\mathcal{E}\left( L,\Xi ,\rho \right)
=\int_{D}d\mathcal{U}\left( L,\Xi ,\rho \right) =\int_{\partial D}\mathcal{U}%
\left( L,\Xi ,\rho \right)
\end{equation}
	The conserved current $\mathcal{E}\left( L,\Xi \right) $ is, in general,
closed only \textsl{on-shell} (namely, along the solutions). We say
that it is subject to a \textsl{weak conservation law.}
However, the existence of the superpotential allows to define a quantity, 
$\mathcal{E}\left( L,\Xi \right) -\widetilde{\mathcal{E}}\left( L,\Xi
\right) $, which is conserved also \textsl{off-shell} (i.e., even for sections which are not a solution). In this case, we say that it is
subject to a \textsl{strong conservation law.}

	In view of the applications to quantum field theories, it may happen that
the strong conserved currents play a different role with respect to the
weakly conserved currents. Indeed, in a \textsl{path-integrals formulation}
of quantum theories one integrates not only along the classical solutions,
but over \textsl{all} the configurations. In this context we can reasonably
expect quantum effects depending strongly on the specific Hamiltonian, that is chosen in the
class of the classically equivalent Hamiltonians. The strong conservation
law can be a way to select the \textsl{true} quantum Hamiltonian, or at least
to select a class of Hamiltonians which are physically reasonable. This could be the topic of a future
work.

	More generally, it has been proven that all natural theories admit superpotentials. An
analogous proof has been given for gauge theories coupled with bosonic and
fermionic matter. In the next chapter, we will define Gauge-Natural theories, which encompass
the natural theories on one side and the gauge theories on the other.

\nopagebreak


\chapter{Gauge-Natural formalism}\label{chapt:gn}

\section{Introduction}

Contemporary physics has provided with some insights about the world that
will be hardly disproved in the near future. First of all, most of what we know to be fundamental in nature can be
stated in terms of the structure groups of the theories. Since there exists
a \textsl{duality} between symmetries and conserved quantities given by
N\"{o}ther's theorem, we may also claim that conserved quantities are a way
for analyzing the fundamental structure of physical theories.

	Secondly, it is clear that some conserved quantities (energy, momentum and
angular momentum) should have a meaning in any theory, while others are
conserved only within certain theories and not in others. In natural
theories this is easily implemented, because there are symmetries which are
the lift of space-time diffeomorphisms. These symmetries, consistently with
General Relativity, must exist in every theory.

	Moreover, each theory can
have further own symmetries, and therefore more conserved quantities. In
this framework, a canonical notion of energy, momentum and angular momentum
can be given. This is connected to the symmetry under space-time
diffeomorphisms. Within the natural theories, the lift of diffeomorphisms allows to identify a class of \textsl{horizontal}
symmetries to which these conserved quantities are associated. This is
possible because the fields are natural objects.

	We also know that, in general, physical fields are not natural objects. In other words, there is not a canonical way to associate a transformation of the
configuration bundle to a diffeomorphism of the space-time. In such theories,
a notion of horizontal symmetry is absent. If we want to define
energy, momentum and angular momentum, we must require supplementary
structures. These can replace the natural lift and
define at least the concept of 1-parameter flow of horizontal symmetries,
that are needed in N\"{o}ther's theorem.

	We believe that these ingredients are a motivation for Gauge-Natural
theories as they will be presented in this chapter. Moreover, it has been
shown that all the theories in use in particle physics and in General
Relativity can be obtained with this formalism as well. This is encouraging for
our future investigations.

\section{Gauge-Natural field theories}

\textbf{Definition (3.2.1): }a \textsl{Gauge-Natural field theory} consists of:\newline
\textbf{(a)} a \textsl{structure bundle} $\mathcal{P}$ which is a principal
bundle with fiber $G$;\newline
\textbf{(b) }a \textsl{configuration bundle }$\mathcal{C}$ which is a
Gauge-Natural bundle of order $\left( s,r\right) $ associated to the structure bundle;\newline
\textbf{(c) }a \textsl{Lagrangian} $L$ of order $k$ over $\mathcal{C}$ which
is $Aut\left( \mathcal{P}\right) $-covariant, where $Aut\left( \mathcal{P}\right) $ acts on $\mathcal{C}$ with the canonical action defined in Eq.(\ref{automorphism});\newline
\textbf{(d) }two morphisms $\omega :J^{k}\mathcal{C}\longrightarrow \mathcal{%
C}_{\mathcal{P}}$ and $\Gamma :J^{k}\mathcal{C}\longrightarrow \mathcal{C}%
_{L\left( M\right) }$ associating to each configuration $\rho :M\longrightarrow C$ a principal connection $\omega
\circ j^{k}\rho $ of the structure bundle $\mathcal{P}$ and a principal connection $\Gamma \circ
j^{k}\rho $ of the frame bundle $L\left( M\right)$.

	The structure bundle encodes the structure of the symmetries of the theory.
The configuration bundle is supposed to be a Gauge-Natural bundle associated
to the structure bundle, in order to have the canonical action (\ref
{automorphism}) of $Aut\left( \mathcal{P}\right) $ over $\mathcal{C}$ which
partly replaces the lift of the natural theories. The morphism $\omega $
defines on the structure bundle a principal connection, called \textsl{
dynamical connection}, to which connections on the configuration bundle are
associated. This means that a flow of \textsl{horizontal} symmetries can be now defined.

	In this chapter we will define the conserved currents
and the superpotentials for a generic Gauge-Natural theory. Using the morphisms $\omega :J^{k}\mathcal{C}\longrightarrow \mathcal{C}_{%
\mathcal{P}}$ and $\Gamma :J^{k}\mathcal{C}\longrightarrow \mathcal{C}%
_{L\left( M\right) }$, one can define the symmetrized covariant
derivatives: 
\begin{equation}
\left\{ 
\begin{tabular}{ll}
$\nabla _{\sigma }\xi ^{\mu }=d_{\sigma }\xi ^{\mu }+\Gamma _{\rho \sigma
}^{\mu }\xi ^{\rho };$ & $\nabla _{\sigma }\xi ^{A}=d_{\sigma }\xi
^{A}-\omega _{B\rho }^{A}\xi ^{B}$ \\ 
$\nabla _{\sigma _{1}\sigma _{2}}\xi ^{\mu }=\nabla _{(\sigma _{1}}\nabla
_{\sigma _{2})}\xi ^{\mu };$ & $\nabla _{\sigma _{1}\sigma _{2}}\xi
^{A}=\nabla _{(\sigma _{1}}\nabla _{\sigma _{2})}\xi ^{A}$ \\ 
$\ldots$ & $\ldots$
\end{tabular}
\right.
\end{equation}
The conserved currents can be expanded on the basis of these covariant
derivatives: 
\begin{eqnarray}
\mathcal{E}\left( L,\Xi \right) &=&\left\langle \F\left( L,\gamma
\right) \mid j^{k-1}\pounds _{\Xi }\varphi \right\rangle -i_{\xi }\circ L= 
\nonumber \\
&=&(T_{\mu }^{\lambda }\xi ^{\mu }+T_{\mu }^{\lambda \sigma }\nabla _{\sigma
}\xi ^{\mu }+\ldots+T_{\mu }^{\lambda \sigma _{1}\ldots\sigma _{s+k-1}}\nabla
_{\sigma _{1}\ldots\sigma _{s+k-1}}\xi ^{\mu }+ \\
&&+T_{A}^{\lambda }\xi ^{A}+T_{A}^{\lambda \sigma }\nabla _{\sigma }\xi
^{A}+\ldots+T_{A}^{\lambda \sigma _{1}\ldots\sigma _{r+k-1}}\nabla _{\sigma
_{1}\ldots\sigma _{r+k-1}}\xi ^{A})\mathbf{ds}_{\lambda }  \nonumber
\end{eqnarray}
Accordingly the work form is: 
\begin{eqnarray}
\mathcal{W}\left( L,\Xi \right) &=&-\left\langle \E\left( L\right) \mid
\pounds _{\Xi }\varphi \right\rangle =  \nonumber \\
&=&(W_{\mu }\xi ^{\mu }+W_{\mu }^{\sigma }\nabla _{\sigma }\xi ^{\mu
}+\ldots+W_{\mu }^{\sigma _{1}\ldots\sigma _{s}}\nabla _{\sigma _{1}\ldots\sigma
_{s}}\xi ^{\mu }+ \\
&&+W_{A}\xi ^{A}+W_{A}^{\sigma }\nabla _{\sigma }\xi ^{A}+\ldots+W_{A}^{\sigma
_{1}\ldots\sigma _{r}}\nabla _{\sigma _{1}\ldots\sigma _{r}}\xi ^{A})ds  \nonumber
\end{eqnarray}
The quantities $T_{\mu }^{\lambda },T_{\mu }^{\lambda \sigma },\ldots,T_{\mu
}^{\lambda \sigma _{1}\ldots\sigma _{s+k-1}}$ and $T_{A}^{\lambda
},T_{A}^{\lambda \sigma },\ldots,T_{A}^{\lambda \sigma _{1}\ldots\sigma _{r+k-1}}$
are tensor densities symmetric in the upper indices (except $\lambda $) and
are called \textsl{canonical tensors}, while the quantities $W_{\mu },W_{\mu
}^{\sigma },\ldots,W_{\mu }^{\sigma _{1}\ldots\sigma _{s}}$ and $%
W_{A},W_{A}^{\sigma },\ldots,W_{A}^{\sigma _{1}\ldots\sigma _{r}}$ are symmetric
in the upper indices and they are called the \textsl{stress tensors}.

	The numbers $\left( s,r\right) $ are the \textsl{geometrical orders }of the
theory, while the numbers $\left( \alpha ,\beta \right) =\left( s+k-1,r+k-1\right) $ are
the \textsl{effective orders}. Moreover, $n=\max \left( \alpha ,\beta
\right) $ is the \textsl{total order}.

\section{The generalized Bianchi identities}

We can consider a Gauge-Natural field theory where the conserved current and
the work form assume the following expression: 
\begin{eqnarray}
\mathcal{E}\left( L,\Xi \right) &=&(T_{\mu }^{\lambda }\xi ^{\mu }+T_{\mu
}^{\lambda \sigma }\nabla _{\sigma }\xi ^{\mu }+\ldots+T_{\mu }^{\lambda \sigma
_{1}\ldots\sigma _{s+k-1}}\nabla _{\sigma _{1}\ldots\sigma _{s+k-1}}\xi ^{\mu }+ \\
&&+T_{A}^{\lambda }\xi ^{A}+T_{A}^{\lambda \sigma }\nabla _{\sigma }\xi
^{A}+\ldots+T_{A}^{\lambda \sigma _{1}\ldots\sigma _{r+k-1}}\nabla _{\sigma
_{1}\ldots\sigma _{r+k-1}}\xi ^{A})ds_{\lambda}  \nonumber
\end{eqnarray}
\begin{eqnarray}
\mathcal{W}\left( L,\Xi \right) &=&(W_{\mu }\xi ^{\mu }+W_{\mu }^{\sigma
}\nabla _{\sigma }\xi ^{\mu }+\ldots+W_{\mu }^{\sigma _{1}\ldots\sigma _{s}}\nabla
_{\sigma _{1}\ldots\sigma _{s}}\xi ^{\mu }+  \label{forma lavoro} \\
&&+W_{A}\xi ^{A}+W_{A}^{\sigma }\nabla _{\sigma }\xi ^{A}+\ldots+W_{A}^{\sigma
_{1}\ldots\sigma _{r}}\nabla _{\sigma _{1}\ldots\sigma _{r}}\xi ^{A})ds  \nonumber
\end{eqnarray}
with some of these vector densities possibly vanishing. Integrating by parts (\ref{forma lavoro}) and recalling the symmetry one
obtains: 
\begin{equation}
\mathcal{W}=\left( B_{\rho }\xi ^{\rho }+B_{A}\xi ^{A}\right) ds+Div\left( 
\widetilde{\mathcal{E}}\left( L,\Xi \right) \right) =Div\left( \mathcal{E}%
\left( L,\Xi \right) \right)  \label{lavoro}
\end{equation}
where: 
\begin{equation}
\left\{ 
\begin{tabular}{l}
\vspace{0.1cm}$B_{\rho }=W_{\rho }-\nabla _{\sigma _{1}}W_{\rho }^{\sigma _{1}}+\ldots+\left(
-1\right) ^{\alpha }\nabla _{\sigma _{1}\ldots\sigma _{\alpha }}W_{\rho
}^{\sigma _{1}\ldots\sigma _{\alpha }}$ \\ 
$B_{A}=W_{A}-\nabla _{\sigma _{1}}W_{A}^{\sigma _{1}}+\ldots+\left( -1\right)
^{\alpha }\nabla _{\sigma _{1}\ldots\sigma _{\alpha }}W_{A}^{\sigma
_{1}\ldots\sigma _{\alpha }}$%
\end{tabular}
\right.  \label{a}
\end{equation}
and the reduced current is
\begin{eqnarray*}
\widetilde{\mathcal{E}}\left( L,\Xi \right) &=&(\widetilde{T}_{\mu
}^{\lambda }\xi ^{\mu }+\widetilde{T}_{\mu }^{\lambda \sigma }\nabla
_{\sigma }\xi ^{\mu }+\ldots+\widetilde{T}_{\mu }^{\lambda \sigma _{1}\ldots\sigma
_{s+k-1}}\nabla _{\sigma _{1}\ldots\sigma _{s+k-1}}\xi ^{\mu }+ \\
&&+\widetilde{T}_{A}^{\lambda }\xi ^{A}+\widetilde{T}_{A}^{\lambda \sigma
}\nabla _{\sigma }\xi ^{A}+\ldots+\widetilde{T}_{A}^{\lambda \sigma
_{1}\ldots\sigma _{r+k-1}}\nabla _{\sigma _{1}\ldots\sigma _{r+k-1}}\xi
^{A})ds_{\lambda }
\end{eqnarray*}
with
\begin{equation}
\left\{ 
\begin{tabular}{l}
\vspace{0.1cm}$\widetilde{T}_{\rho }^{\sigma }=W_{\rho }^{\sigma }-\nabla _{\sigma
_{2}}W_{\rho }^{\sigma \sigma _{2}}+\ldots+\left( -1\right) ^{\alpha -1}\nabla
_{\sigma _{2}\ldots\sigma _{\alpha }}W_{\rho }^{\sigma \sigma _{2}\ldots\sigma
_{\alpha }}$ \\ 
\vspace{0.1cm}$\widetilde{T}_{\rho }^{\sigma \sigma _{2}}=W_{\rho }^{\sigma \sigma
_{2}}-\nabla _{\sigma _{3}}W_{\rho }^{\sigma \sigma _{2}\sigma
_{3}}+\ldots+\left( -1\right) ^{\alpha -2}\nabla _{\sigma _{3}\ldots\sigma
_{\alpha }}W_{\rho }^{\sigma \sigma _{2}\sigma _{3}\ldots\sigma _{\alpha }}$ \\ 
\vspace{0.1cm}$\ldots$ \\ 
$\widetilde{T}_{\rho }^{\sigma \sigma _{2}\ldots\sigma _{\alpha }}=W_{\rho
}^{\sigma \sigma _{2}\ldots\sigma _{\alpha }}$%
\end{tabular}
\right.  \label{b}
\end{equation}
\begin{equation}
\left\{ 
\begin{tabular}{l}
\vspace{0.1cm}$\widetilde{T}_{A}^{\sigma }=W_{A}^{\sigma }-\nabla _{\sigma
_{2}}W_{A}^{\sigma \sigma _{2}}+\ldots+\left( -1\right) ^{\beta -1}\nabla
_{\sigma _{2}\ldots\sigma _{\beta }}W_{A}^{\sigma \sigma _{2}\ldots\sigma _{\beta
}}$ \\ 
\vspace{0.1cm}$\widetilde{T}_{A}^{\sigma \sigma _{2}}=W_{A}^{\sigma \sigma _{2}}-\nabla
_{\sigma _{3}}W_{A}^{\sigma \sigma _{2}\sigma _{3}}+\ldots+\left( -1\right)
^{\beta -2}\nabla _{\sigma _{3}\ldots\sigma _{\beta }}W_{A}^{\sigma \sigma
_{2}\sigma _{3}\ldots\sigma _{\beta }}$ \\ 
\vspace{0.1cm}$\ldots$ \\ 
$\widetilde{T}_{A}^{\sigma \sigma _{2}\ldots\sigma _{\beta }}=W_{A}^{\sigma
\sigma _{2}\ldots\sigma _{\beta }}$%
\end{tabular}
\right.  \label{c}
\end{equation}
Notice that the coefficients of $\widetilde{\mathcal{E}}\left( L,\Xi \right) 
$ are totally symmetric in the upper indices.\medskip\newline
\textbf{Proposition (3.3.1): }with the introduced notations, $B_{\rho }=0$
and $B_{A}= 0.$\medskip\newline
\textbf{Proof: }let us integrate Eq.(\ref{lavoro}) over a region $\Omega
\subset J^{2k}\mathcal{C}$ and for a field $\Xi$ with compact support contained in $\Omega $; recalling (\ref{divergenza}) we obtain: 
\begin{equation}
\int_{\Omega }\left( B_{\rho }\xi ^{\rho }+B_{A}\xi ^{A}\right)
ds+\int_{\partial \Omega }\widetilde{\mathcal{E}}\left( L,\Xi \right)
=\int_{\partial \Omega }\mathcal{E}\left( L,\Xi \right) =0
\end{equation}
which, being $\Omega $ and $\Xi $ arbitrary, implies $B_{\rho }=0$ and $%
B_{A}=0;$ these are called \textsl{generalized Bianchi identities.}

\section{Existence of the superpotentials}

The following results are very important to compute the conserved currents
within a Gauge-Natural theory of any effective order $\left( \alpha ,\beta
\right) $. We refer to the literature for some of the proofs.
First of all, we present a theorem which computes explicitly the
superpotentials for a Gauge-Natural theory of effective order $\left(
2,2\right)$. This will then be extended to the general case.

	It is always possible to write the conserved current in
the form 
\begin{equation}
\mathcal{E}\left( L,\Xi \right) =\widehat{\mathcal{E}}\left( L,\Xi \right)
+Div\left( \mathcal{U}\left( L,\Xi \right) \right)  \label{superpotenziale}
\end{equation}
with the coefficients of $\widehat{\mathcal{E}}\left( L,\Xi \right) $
completely symmetric in the upper indices. Before proving the above, let us study the unicity:\medskip\newline
\textbf{Lemma (3.4.1): }there exists a unique form $\widehat{\mathcal{E}}%
\left( L,\Xi \right) $ with symmetric coefficients, called reduced current, satisfying Eq.(\ref{superpotenziale}).%
\textbf{\medskip}\newline
\textbf{Corollary (3.4.1):} if it exists, the reduced current $\widehat{\mathcal{E}}\left( L,\Xi \right) $ coincides with $\widetilde{\mathcal{E}}\left( L,\Xi \right) $, i.e. it is given by (\ref{a}),\ (\ref{b}), (\ref{c}).\medskip\newline
\textbf{Proof:} by differentiating both sides of (\ref{superpotenziale}) we obtain, for the generalized Bianchi identities: 
\begin{equation}
Div\left( \widehat{\mathcal{E}}\left( L,\Xi \right) \right) =\mathcal{W}%
\left( L,\Xi \right) =Div\left( \widetilde{\mathcal{E}}\left( L,\Xi \right)
\right)
\end{equation}
and from the theorem above it follows that $\widehat{\mathcal{E}}\left(
L,\Xi \right) =\widetilde{\mathcal{E}}\left( L,\Xi \right)$ .\medskip\newline
\textbf{Lemma (3.4.2): }these identities hold: 
\begin{eqnarray}
T_{\alpha }^{\lambda \rho \sigma }\nabla _{\sigma \rho }\xi ^{\alpha }
&=&\nabla _{\sigma }\left[ \frac{4}{3}T_{\alpha }^{\left[ \lambda \sigma
\right] \rho }\nabla _{\rho }\xi ^{\alpha }\right] -\frac{4}{3}\nabla
_{\sigma }T_{\alpha }^{\left[ \lambda \sigma \right] \rho }\nabla _{\rho
}\xi ^{\alpha }+  \nonumber \\
&&+T_{\alpha }^{\left( \lambda \rho \sigma \right) }\nabla _{\sigma \rho
}\xi ^{\alpha }+\frac{1}{3}T_{\beta }^{\left[ \sigma \rho \right] \lambda
}R_{\alpha \sigma \rho }^{\beta }\xi ^{\alpha }
\end{eqnarray}
\begin{eqnarray}
T_{A}^{\lambda \rho \sigma }\nabla _{\sigma \rho }\xi ^{A} &=&\nabla
_{\sigma }\left[ \frac{4}{3}T_{A}^{\left[ \lambda \sigma \right] \rho
}\nabla _{\rho }\xi ^{A}\right] -\frac{4}{3}\nabla _{\sigma }T_{A}^{\left[
\lambda \sigma \right] \rho }\nabla _{\rho }\xi ^{A}+  \nonumber \\
&&+T_{A}^{\left( \lambda \rho \sigma \right) }\nabla _{\sigma \rho }\xi ^{A}+%
\frac{1}{3}T_{B}^{\left[ \sigma \rho \right] \lambda }c_{CA}^{B}F_{\sigma
\rho }^{C}\xi ^{A}
\end{eqnarray}
\textbf{Lemma (3.4.3):} if the theory is of effective order $\left(
1,1\right) $, namely $T_{\alpha }^{\lambda \rho \sigma }=0$ and $T_{A}^{\lambda \rho \sigma }=0$,
\begin{equation}
\begin{tabular}{l}
$\mathcal{E}\left( L,\Xi \right) =\widetilde{\mathcal{E}}\left( L,\Xi
\right) +Div\left( \mathcal{U}\left( L,\Xi \right) \right) $ \\ 
\\ 
$\mathcal{U}\left( L,\Xi \right) =\frac{1}{2}\left( T_{A}^{\left[ \lambda
\sigma \right] }\xi ^{A}+T_{\alpha }^{\left[ \lambda \sigma \right] }\xi
^{\alpha }\right) $%
\end{tabular}
\end{equation}
\textbf{Proof:} integrating by parts we obtain: 
\begin{eqnarray*}
\mathcal{E}\left( L,\Xi \right) &=&\left[ T_{\mu }^{\lambda }\xi ^{\mu
}+T_{\mu }^{\lambda \sigma }\nabla _{\sigma }\xi ^{\mu }+T_{A}^{\lambda }\xi
^{A}+T_{A}^{\lambda \sigma }\nabla _{\sigma }\xi ^{A}\right] ds_{\lambda }=
\\
&=&[\left( T_{\mu }^{\lambda }-\nabla _{\sigma }T_{\mu }^{\left[ \lambda
\sigma \right] }\right) \xi ^{\mu }+\left( T_{A}^{\lambda }-\nabla _{\sigma
}T_{A}^{\left[ \lambda \sigma \right] }\right) \xi ^{A}ds_{\lambda }+ \\
&&+T_{\mu }^{\left( \lambda \sigma \right) }\nabla _{\sigma }\xi ^{\mu
}+T_{A}^{\left( \lambda \sigma \right) }\nabla _{\sigma }\xi ^{A}+\nabla
_{\sigma }\left( T_{\mu }^{\left[ \lambda \sigma \right] }\xi ^{\mu
}+T_{A}^{\left[ \lambda \sigma \right] }\xi ^{A}\right) ]ds_{\lambda }
\end{eqnarray*}
By the condition of weak conservation for the current:
\begin{equation}
\left\{ 
\begin{tabular}{ll}
\vspace{0.1cm}$\nabla _{\lambda }T_{\alpha }^{\lambda }+\frac{1}{2}T_{\beta }^{\lambda
\sigma }R_{\alpha \lambda \sigma }^{\beta }=W_{\alpha }$ & $\nabla _{\lambda
}T_{A}^{\lambda }+\frac{1}{2}T_{B}^{\lambda \sigma }c_{CA}^{B}F_{\lambda
\sigma }^{C}=W_{A}$ \\ 
\vspace{0.1cm}$T_{\alpha }^{\sigma }+\nabla _{\lambda }T_{\alpha }^{\lambda \sigma
}=W_{\alpha }^{\sigma }$ & $T_{A}^{\sigma }+\nabla _{\lambda }T_{A}^{\lambda
\sigma }=W_{A}^{\sigma }$ \\ 
\vspace{0.1cm}$T_{\alpha }^{\left( \lambda \sigma \right) }=W_{\alpha }^{\lambda \sigma
}\equiv 0$ & $T_{A}^{\left( \lambda \sigma \right) }=W_{A}^{\lambda \sigma
}\equiv 0$%
\end{tabular}
\right.
\end{equation}
we can easily obtain: 
\begin{eqnarray}
\mathcal{E}\left( L,\Xi \right) &=&[W_{\mu }^{\lambda }\xi ^{\mu }+W_{\mu
}^{\lambda \sigma }\nabla _{\sigma }\xi ^{\mu }+W_{A}^{\lambda }\xi
^{A}+W_{A}^{\lambda \sigma }\nabla _{\sigma }\xi ^{A}+  \nonumber \\
&&+\nabla _{\sigma }\left( T_{\mu }^{\left[ \lambda \sigma \right] }\xi
^{\mu }+T_{A}^{\left[ \lambda \sigma \right] }\xi ^{A}\right)]ds_{\lambda }
\end{eqnarray}
We can finally enunciate the following theorem:\medskip\newline
\textbf{Theorem (3.4.1): }every Gauge-Natural theory of effective order (2,2) admits a superpotential: 
\begin{eqnarray}
\mathcal{E}\left( L,\Xi \right) &=&\widetilde{\mathcal{E}}\left( L,\Xi
\right) +Div\left( \mathcal{U}\left( L,\Xi \right) \right)  \nonumber \\
\mathcal{U}\left( L,\Xi \right) &=&\frac{1}{2}[\left( T_{A}^{\left[ \lambda
\sigma \right] }-\frac{2}{3}\nabla _{\rho }T_{A}^{\left[ \lambda \sigma
\right] \rho }\right) \xi ^{A}+\frac{4}{3}T_{A}^{\left[ \lambda \sigma
\right] \rho }\nabla _{\rho }\xi ^{A}+ \\
&&+\left( T_{\mu }^{\left[ \lambda \sigma \right] }-\frac{2}{3}\nabla _{\rho
}T_{\mu }^{\left[ \lambda \sigma \right] \rho }\right) \xi ^{\mu }+\frac{4}{3%
}T_{\mu }^{\left[ \lambda \sigma \right] \rho }\nabla _{\rho }\xi ^{\mu
}]ds_{\lambda \sigma } \nonumber
\end{eqnarray}

\section{Existence of the superpotentials: the general case}

\textbf{Theorem (3.5.1):} every Gauge-Natural theory of effective order ($%
\alpha ,\beta $) admits a superpotential.\medskip\newline
The proof is given by induction up to the effective order $\left( \alpha
,\beta \right).$

	In conclusion, we can write each current in the form: 
\begin{equation}
\mathcal{E}\left( L,\Xi \right) =\widetilde{\mathcal{E}}\left( L,\Xi \right)
+Div\left( \mathcal{U}\left( L,\Xi \right) \right)
\end{equation}
with the coefficients of $\widetilde{\mathcal{E}}\left( L,\Xi \right) $
completely symmetric and vanishing along the solutions of the field
equations.

	We stress that currents and superpotentials have a physical meaning, because
they give the conserved quantities when integrated. The
physically relevant objects are therefore the values of the integrals which depend
only on the cohomology class, not on the chosen representative. In
simpler terms, the currents are defined \textsl{up to a divergence} while
the superpotentials are defined \textsl{up to a form with vanishing
divergence.}

\section{Generalized symmetries}

In many applications, symmetries of a more general nature than the standard
Lagrangian symmetries are often used. One can consider transformations which preserve the Lagrangian not \textsl{exactly}, but modulo contact forms
and exact differentials. Both contact forms and exact forms are, in fact,
irrelevant to the N\"{o}ther theorem.

	One can generally consider a transformation on the infinite jet
prolongation of the configuration bundle. The infinitesimal generators of
such transformations are the \textsl{generalized vector fields}.

\begin{equation}
\begin{tabular}{lll}
$\left( \pi _{0}^{k}\right) ^{*}\left( TC\right) $ & $\stackrel{\Phi
^{*}}{\longrightarrow} $ & $\hspace{-0.05cm}TC$ \\ 
$\pi ^{*}\downarrow $ &  & $\downarrow \tau _{C}$ \\ 
$\hspace{0.2cm}J^{k}C$ & $\stackrel{\pi _{0}^{k}}{\longrightarrow} $ & $C$ \\ 
$\pi ^{k}\downarrow $ &  & $\downarrow \pi $ \\ 
$\hspace{0.4cm}M$ &  & $\hspace{-0.05cm}M$
\end{tabular}
\end{equation}
The bundle $\left( \pi _{0}^{k}\right) ^{*}\left( TC\right) $ is the
pull-back of the bundle $TC$ along the map $\pi
_{0}^{k}:J^{k}C\longrightarrow C$ and $\Phi ^{*}$is uniquely defined; a
point in $\left( \pi _{0}^{k}\right) ^{*}\left( TC\right) $ is a pair $%
\left( p,\upsilon \right) $ where $p\in J^{k}\mathcal{C}$, $\upsilon \in TC$
and $\tau _{C}\left( \upsilon \right) =\pi _{0}^{k}\left( p\right) $, i.e. $%
\upsilon \in T_{\pi _{0}^{k}\left( p\right) }C.$ Let us consider a section $%
\Xi $ of the bundle $\left( \pi _{0}^{k}\right) ^{*}\left( TC\right) $ $%
\longrightarrow J^{k}C$; its local expression is 
\begin{equation}
\Xi :J^{k}C\longrightarrow \left( \pi _{0}^{k}\right) ^{*}\left( TC\right)
:p\longmapsto \left( p,\xi ^{\mu }\left( p\right) \partial _{\mu }+\xi
^{i}\left( p\right) \partial _{i}\right)
\end{equation}
The section $\Xi $ is called \textsl{generalized vector field }and by an
abuse of notation is denoted simply by 
\begin{equation}
\Xi =\xi ^{\mu }\left( x^{\mu },y^{i},y_{\mu }^{i},\ldots,y_{\mu _{1}\ldots\mu
_{k}}^{i}\right) \partial _{\mu }+\xi ^{i}\left( x^{\mu },y^{i},y_{\mu
}^{i},\ldots,y_{\mu _{1}\ldots\mu _{k}}^{i}\right) \partial _{i}
\end{equation}
Notice that a generalized vector field $\Xi $ is not a vector field on $C$
(unless $k=0$) since its components depend on the derivatives of fields.
Furthermore, it can be shown that no jet prolongation $j^{r}\Xi$ of a
generalized vector field $\Xi$ is an ordinary vector field on any $J^{s}%
\mathcal{C}$. So, even if one can drag sections along $\Xi$, this
object does not define a transformation on any finite jet bundle.
Generalized vector fields can in fact be regarded as infinitesimal
generators of transformations on $J^{\infty }C$, as shown by Saunders. We
do not discuss this further, and we remand to the literature. Here we are only interested in defining the generalized
symmetries:\medskip\newline
\textbf{Definition (3.6.1): }a \textsl{generalized symmetry} of a Lagrangian 
$L$ is a generalized vector field $\Xi $ over $C$ such that the Poincar\'{e}-Cartan
form of $L$ satisfies the following 
\begin{equation}
\pounds _{j^{1}\Xi }\Theta _{L}=\omega +d\alpha  \label{gen sym}
\end{equation}
where $\omega $ is any contact form and $\alpha $ is any $\left( m-1\right) $%
-form (both possibly depending on the derivatives of fields).

	Since the sum of horizontal and contact forms is direct, condition (\ref
{gen sym}) is equivalent to the following: 
\begin{equation}
hor\left( \pounds _{j^{1}\Xi }\Theta _{L}\right) =hor \:d\alpha
\end{equation}
which, since the Lie derivative of a contact form is again a contact form,
is equivalent to 
\begin{equation}
hor\left( \pounds _{j^{1}\Xi }\left( hor \Theta _{L}\right) \right)
=hor\left( \pounds _{j^{1}\Xi }\left( \mathcal{L}ds\right) \right) =hor \:
d\alpha
\end{equation}
By expanding this last equation in local fibered coordinates $\left( x^{\mu
},y^{i},y_{\mu }^{i}\right) $ over $J^{1}C$, for the jet prolongation of a
generalized vector field $j^{1}\Xi =\xi ^{\mu }\partial _{\mu }+\xi
^{i}\partial _{i}+\xi _{\mu }^{i}\partial _{i}^{\mu }$ projecting on $\xi
=\xi ^{\mu }\partial _{\mu }$, one obtains 
\begin{equation}
d_{\mu }\left( \xi \mathcal{L}\right) -p_{i}\left( \xi ^{\mu }y_{\mu
}^{i}-\xi ^{i}\right) -p_{i}^{\mu }\left( \xi ^{\nu }y_{\mu \nu }^{i}-\xi
_{\mu }^{i}\right) =d_{\mu }\alpha ^{\mu }  \label{sym}
\end{equation}
One can also regard this condition as global, expressed directly in terms of
the Lagrangian (i.e. not involving the Poincar\'{e}-Cartan form); in the
notation used in the previous sections, Eq.(\ref{sym}) can be intrinsically
expressed as 
\begin{equation}
\pounds _{\xi }L-<\delta L|\pounds _{\Xi }\sigma >=Div\left( \alpha \right)
\end{equation}
where $\sigma :M\longrightarrow C$ is a section of $C.$

	The conclusion is that a generalized symmetry leaves the Lagrangian
invariant up to a pure divergence plus terms depending on the first
variation $\delta L.$ Now the \textsl{generalized N\"{o}ther theorem} follows: \medskip\newline
\textbf{Theorem (3.6.1):} for any generalized symmetry generator (of order $s$) $\Xi$ the following holds true: 
\begin{equation}
\pounds _{j^{1}\Xi }\Theta _{L}=\omega +d\alpha
\end{equation}
Expanding the Lie derivatives and collecting terms this can be recast as 
\begin{equation}
d\left( i_{j^{1}\Xi }\Theta _{L}-\alpha \right) =-i_{j^{1}\Xi }d\Theta
_{L}+\omega  \label{form}
\end{equation}
If we define now the quantities 
\begin{equation}
\left\{ 
\begin{tabular}{l}
\vspace{0.1cm}$\mathcal{E}=\left( j^{2k-1}\sigma \right) ^{*}\left( i_{j^{1}\Xi }\Theta
_{L}-\alpha \right)$\\
$\mathcal{W}=-\left( j^{2k}\sigma \right) ^{*}i_{j^{1}\Xi }d\Theta _{L}$
\end{tabular}
\right.
\end{equation}
the identity (\ref{form}), pulled back on $M$ along a section $\sigma $, can
be expressed as 
\begin{equation}
d\mathcal{E}=\mathcal{W}
\end{equation}
Therefore, whenever $\sigma $ is a solution of field equations then $%
\mathcal{W}=0$ and thence the N\"{o}ther current $\mathcal{E}$ is conserved: 
$d\mathcal{E}=0.$


\chapter{Spin structures}\label{chapt:spin}

\section{Introduction}

It is evident from the observed phenomenology that there exists a set of
elementary particles, called \textsl{spinors }or \textsl{fermions}, which
are not described by \textsl{natural} objects. They show some peculiar
transformation characteristics with respect to changes of frame. In Minkowski
space, for instance, if the frame is rotated by an angle $\theta =2\pi $,
the spinor changes sign, remaining invariant only for rotations of $%
\theta =4\pi $. Despite these odd transformation properties of spinors, one can anyway define a lift of the 1-parameter group of Minkowski \textsl{isometries} (i.e. of the Poincar\'{e} group).

	The problem arises when describing spinors over a curved
space, namely if one wants to describe the interaction between the
gravitational field and the spinors. In this case, the analogous of the
infinitesimal flows of the Poincar\'{e} transformations, that is the Killing
vectors, may not exist depending on the metric structure of the space-time.
Moreover, a natural way to define the lift of
an arbitrary space-time transformation, does not seem to exist.

	Similarly, in Minkowski space-time we cannot
define a lift for a transformation that is not an isometry. Global
topological problems can also make it impossible to solve the Dirac equations. An ambiguity in the sign arises from the particular behavior of
spinors under transformations of the Poincar\'{e} group. These problems are
faced by substituting the Lorentz group with its universal covering, that is
the group $Spin\left( 1,m-1\right) $. This operation has hence no effect on
infinitesimal transformations.

	The results here reported are valid for $Spin\left( \eta \right)$, where $\eta =\left(r,s\right)$ is a signature fixed from the beginning, and for its connected component to the identity, $Spin_{0}\left( \eta \right)$. For
this reason and for simplicity, we indicate the component
connected to the identity only when it is important.

	All the above fundamental problems are circumvented by introducing the concept of \textsl{%
spin structure}. It is a principal bundle morphism $\Lambda :\Sigma
\longrightarrow SO\left( M,g\right) $, where $\Sigma $ is a principal bundle
with structure group $Spin\left( \eta \right) $ called \textsl{spin bundle}. $SO\left( M,g\right) $ is the orthonormal frame bundle of the
pseudo-Riemannian manifold $\left( M,g\right), $ with signature $\eta =\left(
r,s\right)$. Spin structures define the spin fields, which are indeed the
sections of a Gauge-Natural vector bundle $E_{\sigma }=W\left( 1,0\right)
=\Sigma \times _{\sigma }V$ associated to the spin bundle. More precisely,
the spin structure is necessary to overcome the problems in defining the
global Dirac equations in curved spaces, possibly endowed with a non
trivial topology.

	The standard theory of spin structures, as it will be exposed in section
(4.2), has unfortunately some disadvantages. There exists in fact a clear
link between metric and spin structure: one cannot define the latter without
a background metric fixed \textsl{a priori}. When a metric on the manifold
is fixed, the spin bundle $\Sigma $ may not exist at all. It may be uniquely
defined or there can be many different nonequivalent spin bundles.

	The problem is that one cannot give a physical interpretation of a non-dynamical
background metric, since the only metric \textsl{physically} defined over
the space-time is related to the gravitational field and it must be
dynamical. This is true for the influence of spin fields on gravity, a fact
that is expected for physical reasons. We can then formulate the
following principle:\medskip\newline
\textbf{Axiom 0:} a physically meaningful theory of spin structures must not
have non-dynamical \textsl{background} fields.\medskip\newline
In other terms, we should create a theory in the category of manifolds (and
not in that of pseudo-Riemannian manifolds). Moreover, if the metric \textsl{%
g} is dynamical, the spin structure itself must be dynamical, since the standard construction of the spin structure
depends on the metric. Being
every dynamical field a section of some bundle, we state the following
principle:\medskip\newline
\textbf{Axiom 1: }there must exist a bundle whose sections are in one-to-one
correspondence to the spin structures.\medskip\newline
The second problem arises from the fact that the relation between metrics
and spin structures is \textsl{one-to-many}. This means that if the metric
varies, such a deformation does not induce canonically a
corresponding variation of the spin structure.

	A solution can be to choose as fundamental field variables the spin
structures (modifying their definition, to make them formally independent of
the metric background). The new spin structures (which will be called 
\textsl{spin frames}) determine uniquely a metric, called \textsl{%
associated metric}. Hence a deformation of the spin structures (which is
canonically defined by axiom 1) induces a variation in the metric. In this
context, spin frames are the true variables which describe the gravitational
field, while metric and geometry of space-time are derived structures.

	The theory defined this way is Gauge-Natural: a canonical treatment of the conserved quantities is now possible. Besides, this is
easily generalized to fermionic matter in interaction with a dynamical
gravitational field. In other terms, we find a Lagrangian theory which
describes the interaction between gravitational field and fermionic matter. This allows not only to determine the way a gravitational field
influences fermions, but also to evaluate the gravitational field generated
by fermions themselves.

\section{Spin structures}

We shall assume the reader to be familiar with the Clifford algebra
framework. We refer to the references and to the literature for details and
notation.

	Let $\left( M,g\right) $ be a pseudo-Riemannian manifold of signature $\eta
=\left( r,s\right)$ .\medskip\newline
\textbf{Definition (4.2.1): }a \textsl{spin structure on }$\left( M,g\right) 
$ is a pair $\left( \Sigma ,\Lambda \right) $ such that
\textbf{(a) }$\Sigma $ is a principal bundle with structure group $%
Spin\left( \eta \right) $ called \textsl{spin bundle}.\newline
\textbf{(b) }Let us denote by $SO\left( M,g\right) $ the orthonormal frame
bundle of $\left( M,g\right) $; we define with $\Lambda :\Sigma \longrightarrow SO\left( M,g\right) $ a principal
bundle morphism with respect to the Lie groups morphism given by $\ell :Spin\left( \eta \right)
\longrightarrow SO\left( \eta \right) $, namely
\begin{equation}
\begin{tabular}{lll}
$\hspace{-0.05cm}\Sigma $ & $\stackrel{\Lambda}{\longrightarrow} $ & $\hspace{-0.1cm}SO\left(
M,g\right)
$
\\ 
$\downarrow $ &  & $\downarrow $ \\ 
$\hspace{-0.1cm}M$ &  & $\hspace{-0.1cm}M$
\end{tabular}
\end{equation}
\textbf{Definition (4.2.2): }let us hereafter denote by $g_{\left(
\alpha \beta \right) }$ the transition functions of $SO\left( M,g\right) $;\textbf{\ }the functions $G_{\left( \alpha
\beta \right) }:U_{\alpha \beta }\longrightarrow Spin\left( \eta \right) $
are a \textsl{lift }of $g_{\left( \alpha \beta \right) }:$ $U_{\alpha \beta }\longrightarrow
SO\left( M,g\right) $ if $\ell \circ G_{\left( \alpha \beta \right)
}=g_{\left( \alpha \beta \right) }.$ They are a \textsl{cocycle }if $\forall x\in U_{\alpha \beta }$ the following holds: 
\begin{equation}
\left\{
\begin{tabular}{l}
$G_{\left( \alpha \alpha \right) }=\I$ \\ 
$G_{\left( \alpha \beta \right) }\circ G_{\left( \beta \gamma \right) }\circ
G_{\left( \gamma \alpha \right) }=\I$%
\end{tabular}
\right.
\end{equation}
\textbf{Proposition (4.2.1): }if $\left\{ G_{\left( \alpha \beta \right)
}\right\} $ are a cocycle and a lift of $\left\{ g_{\left( \alpha \beta
\right) }\right\} $, then $\left( M,g\right) $ admits a spin structure.\medskip\newline
A theorem (the proof, which is not shown here, can be found in Ref.[3])
claims that a lift of the $\left\{ g_{\left( \alpha \beta \right) }\right\}$ always exists as a consequence of the properties of the covering map $\ell$.

	Let us finally state an important result by Geroch [4]:\medskip\newline
\textbf{Theorem (4.2.1): }in 4 dimensions and signature $\left( 1,3\right) $
a non-compact pseudo-Riemannian manifold $\left( M,g\right) $ admits spin structures if
and only if it admits an orthonormal global frame.\medskip\newline
This result is important because its hypotheses are satisfied by all \textsl{
physically} admissible space-times. The fact that $\left( M,g\right)$ is not compact implies the possibility of a well-defined Cauchy problem for the Einstein
equations. Moreover, problems with causality (closed
time-like or light-like geodesics) can be avoided. It can indeed be proven that in compact space-times closed causal curves always exist.

\section{Spin frames}

\subsection{The structure bundle}

Fix a manifold $M$, oriented which admits pseudo-Riemannian metrics of a
fixed signature $\eta =\left( r,s\right)$.

	Let $\Sigma $ be a principal bundle with group $Spin\left( \eta \right) $%
:\medskip\newline
\textbf{Definition (4.3.1.1): }a \textsl{spin
frame} on$\Sigma $ is a morphism $\Lambda :\Sigma \longrightarrow L\left(
M\right) $, where $L\left( M\right) $ denotes the frame bundle of $M$ and such that
\begin{equation}
\begin{tabular}{lll}
$\hspace{-0.025cm}\Sigma $ & $\stackrel{\Lambda }{\longrightarrow} $ &
$\hspace{-0.1cm}L\left(M\right)
$
\\ 
$\downarrow $ &  & $\downarrow $ \\ 
$\hspace{-0.05cm}M$ &  & $\hspace{-0.05cm}M$%
\end{tabular}
\hspace{0.5cm} 
\begin{tabular}{lll}
$\hspace{0.55cm}\Sigma $ & $\stackrel{\Lambda }{\longrightarrow} $ & $\hspace{-0.1cm}L\left(
M\right) $ \\ 
$R_{\mathcal{S}}\downarrow $ &  & $\downarrow R_{\hat{\ell}\left( \mathcal{S}%
\right) }$ \\ 
\hspace{0.5cm}\ $\Sigma $ & $\stackrel{\Lambda }{\longrightarrow}$ & $\hspace{-0.1cm}L\left(
M\right) $%
\end{tabular}
\end{equation}
\begin{equation}
\left\{
\begin{tabular}{l}
\vspace{0.1cm}$\hat{\ell} = i\circ \ell :Spin\left( \eta \right) \longrightarrow GL\left(
n\right) $\\
$i:SO\left( \eta \right) \hookrightarrow GL\left( n\right)$
\end{tabular}
\right.
\end{equation}
The analogy with the spin structures defined in the previous section is
clear. The main differences are that a background metric is not needed and
that the bundle $\Sigma $ is defined once for all. In this context $\Sigma $
will be called \textsl{structure bundle}.

	The codomain of spin frames is the entire frame bundle, instead of the bundle
of orthonormal frames of the background metric. This happens first of all
for avoiding any reference to such a metric. Secondly, if we deform
a spin frame we modify its image within $L\left( M\right) .$ The image of $%
\Lambda $ is a principal subbundle of $L\left( M\right) $ with group $%
SO\left( \eta \right) ,$ and therefore it singles out uniquely an \textsl{%
associated metric} $g\left( \Lambda \right) $ such that 
\begin{equation}
Im\left( \Lambda \right) =SO\left( M,g\left( \Lambda \right) \right)
\end{equation}
Depending on $\Sigma$, there may (or may not) exist spin frames on $\Sigma$. If
there are no spin frames, a meaningful application to field theories is anyway
impossible. Therefore we require the following:\medskip\newline
\textbf{Axiom 2:} the structure bundle $\Sigma$ must be chosen such that
there exists at least one spin frame.\medskip\newline
If the (oriented) manifold $M$ admits spin structures and a metric $g$ of
signature $\eta $, there always exists a structure bundle $\Sigma $
verifying axiom 2. It is easy to prove, indeed, that if $i_{g}:SO\left(
M,g\right) \hookrightarrow L\left( M\right) $ is the canonical immersion in
the frame bundle and $\left( \Sigma ,\Lambda \right) $ is one of the spin
structures which can be considered over $\left( M,g\right) $, $i_{g}\circ
\Lambda :\Sigma \longrightarrow L\left( M\right) $ is a spin frame over $%
\Sigma $; the spin bundle defined in this way can be taken as structure
bundle because it satisfies axiom 2.

	If more than one structure bundle $\Sigma $ is available on $M$, let us fix
a representative $\Sigma $ and consider every spin frame on $\Sigma $. To
each such spin frame we can associate the induced metric, and obtain a whole
class of metrics on $M$. Depending on the topology of $M$, it may happen that
all metrics on $M$ can be obtained by spin frames over $\Sigma $ (these
are called $\Sigma $-admissible metrics). But it can also happen that the
classes of $\Sigma $-admissible metrics are just a subset of the set of all
metrics over $M$. In this case, the set of metrics is disconnected into
classes of $\Sigma $-admissible metrics for $\Sigma $ varying over all
possible structure bundles. In any case, $\Sigma $ can be fixed once for all. By the variation of the spin frame one can get all the $\Sigma $%
-admissible metrics on $M$. The only difference between the metric
and the spinor theories is that the latter allows to describe the
fermionic matter canonically. As long as the gravitational field is considered in the vacuum (or with bosonic matter), the two theories are equivalent.

We remark that in two important cases all metrics are $\Sigma $%
-admissible:\medskip\newline
\textbf{Proposition (4.3.1.1): }in a strictly Riemannian signature, if $M$
is parallelizable all metrics are $\Sigma $-admissible.\medskip\newline
\textbf{Proof:} the fact that $M$ is parallelizable implies the existence of
a global frame, which induces, by the Gram-Schmidt procedure, a global orthonormal frame with respect to any metric $g$.

	The orthonormal frame bundle $SO\left( M,g\right) \simeq SO\left( M,g\right) $ is trivial. If we choose $%
\Sigma $ to be the trivial bundle, every metric is $\Sigma $-admissible.\medskip\newline
\textbf{Proposition (4.3.1.2): }in 4 dimensions and signature $\left(
1,3\right)$, for each non-compact space-time admitting spin structures all the metrics are $\Sigma $-admissible (by Geroch theorem).

\subsection{Infinitesimal generators of automorphisms of $\Sigma $}

We can represent the group $Spin\left( \eta \right) $ as a subgroup of $%
GL\left( k,\mathbb K\right) $ defined by 
\begin{equation}
S\cdot \gamma _{a}\cdot S^{-1}=\gamma _{b}a_{a}^{b}\hspace{1cm}a_{a}^{b}\in
SO\left( \eta \right)
\end{equation}
where $\left\{ \gamma _{a}\right\} $ is a set of Dirac matrices. This
provides the way to give an expression for the infinitesimal generators of
automorphisms of $\Sigma$.

	Let $\Phi _{t}=\left( \phi _{t},f_{t}\right) $ be a 1-parameter subgroup of
automorphisms of $\Sigma $ given locally by $\phi _{t}:\left[ x,S\right]
_{\alpha }\longmapsto \left[ f\left( x\right) ,\varphi \left( x\right)
S\right] _{\alpha }$.

	The infinitesimal generator has the form 
\begin{equation}
\Xi =\xi ^{\mu }\partial _{\mu }+\dot{\varphi}_{0\beta }^{\alpha }S_{\gamma
}^{\beta }\frac{\partial }{\partial S_{\gamma }^{\alpha }}=\xi ^{\mu
}\partial _{\mu }+\xi ^{ab}\sigma _{ab}
\end{equation}
where we set
\begin{equation}
\left\{ 
\begin{tabular}{l}
\vspace{0.1cm}$\xi ^{\mu }=\dot{f}_{0}^{\mu }\left( x\right) $ \\ 
\vspace{0.1cm}$\xi ^{ab}=\partial _{\alpha }^{\beta }\ell _{c}^{[a}\left( \I\right)
\eta ^{b]c}\dot{\varphi}_{0\beta }^{\alpha }$ \\ 
$\sigma _{ab}=\frac{1}{8}\left( \left[ \gamma _{a},\gamma _{b}\right]
S\right) _{\beta }^{\alpha }\frac{\partial }{\partial S_{\beta }^{\alpha }}$%
\end{tabular}
\right.
\end{equation}
For theorem (1.6.1.1) the fields $\sigma _{ab}$ are right-invariant and they
form a local basis of right-invariant vertical fields of $\Sigma$.

\section{The spin frames bundle}

\subsection{Definition of the spin frames bundle}

For what has been said in the previous paragraphs, axiom 0 is always
verified, axiom 2 is true for all the manifolds $M$ admitting spin
structures.

	It remains to prove that axiom 1 always holds. We begin by defining the
action of the group over the manifold $GL\left( m\right) $: 
\begin{eqnarray}
\lambda &:&Spin\left( \eta \right) \times GL\left( m\right) \times GL\left(
m\right) \longrightarrow GL\left( m\right)  \nonumber \\
&:&\left( S,J,e\right) \longmapsto J\cdot e\cdot \ell \left( S^{-1}\right)
\label{azione}
\end{eqnarray}
and we give the following\medskip\newline
\textbf{Definition (4.4.1.1): }we choose as \textsl{spin frames bundle }the
bundle associated to $W^{\left( 1,0\right) }\Sigma =L\left( M\right) \times \Sigma $
through the representation $\lambda $ defined by (\ref{azione}), namely 
\begin{equation}
\Sigma _{\lambda }=W^{\left( 1,0\right) }\Sigma \times _{\lambda }GL\left(
n\right)
\end{equation}
The bundle $\Sigma _{\lambda }$ is by construction a Gauge-Natural bundle of
order $\left( 1,0\right) $

associated to $\Sigma$ .\medskip\newline
\textbf{Proposition (4.4.1.1): }there is a one-to-one correspondence between
sections of $\Sigma _{\lambda }$ and spin frames on $\Sigma $.\medskip\newline
\textbf{Proof: }$s^{\left( \alpha \right) }$ is a trivialization of $\Sigma $
and $\partial ^{\left( \alpha \right) }$ a trivialization of $L\left(
M\right) $.

	A section of $\Sigma _{\lambda }$ is locally given by 
\begin{equation}
\rho :x\longmapsto \left[ s^{\left( \alpha \right) },\partial ^{\left(
\alpha \right) },e_{i}^{\left( \alpha \right) \mu }\right] _{\lambda }
\end{equation}
with the compatibility condition 
\begin{equation}
e^{\left( \alpha \right) }=g_{\left( \alpha \beta \right) }\cdot e^{\left(
\beta \right) }\cdot \ell \left( G_{\left( \beta \alpha \right) }\right)
\label{comp}
\end{equation}
where $g_{\left( \alpha \beta \right) }$ are the transition functions of $%
L\left( M\right) $ and $G_{\left( \beta \alpha \right) }$ are the transition functions of $\Sigma$.

	We can associate to the section $\rho $ the spin frame locally defined by 
\begin{equation}
\Lambda _{\rho }:s^{\left( \alpha \right) }\cdot S\longmapsto \partial
^{\left( \alpha \right) }\cdot e^{\left( \alpha \right) }\cdot \ell \left(
S\right)
\end{equation}
For the compatibility condition (\ref{comp}), $\Lambda _{\rho }$ is a global
spin frame.

	The application associating $\rho $ to $\Lambda _{\rho }$ is one-to-one
because we can define its inverse which takes a spin frame $\Lambda :\Lambda _{\rho }:s^{\left(
\alpha \right) }\cdot S\longmapsto \partial ^{\left( \alpha \right) }\cdot
u^{\left( \alpha \right) }$ and associates it to the section 
\begin{equation}
\rho _{\lambda }:x\longmapsto \left[ s^{\left( \alpha \right) },\partial
^{\left( \alpha \right) },u^{\left( \alpha \right) }\cdot \ell \left(
S^{-1}\right) \right] _{\lambda }
\end{equation}
which is global, being $\Lambda $ global.\medskip\newline
The existence of the metric associated to a spin frame can be stated now by
saying that there exists a bundle epimorphism
\begin{equation}
\begin{tabular}{lll}
$\hspace{-0.07cm}\Sigma _{\lambda }$ & $\longrightarrow $ & $\hspace{-0.1cm}Met\left( M,\eta \right) $ \\ 
$\downarrow $ &  & $\downarrow $ \\ 
$\hspace{-0.07cm}M$ &  & $\hspace{-0.05cm}M$%
\end{tabular}
\end{equation}
where $Met\left( M,\eta \right) $ denotes the bundle of metrics with
signature $\eta $. This epimorphism is canonical, not depending on the
choice of any background. Moreover the bundle $\Sigma _{\lambda }$ is not a
bundle of groups, because the transition functions do not fix the identity.
Besides, axiom 2 ensures that $\Sigma _{\lambda }$ admits at least one
global section.

	Admitting each point in $\Sigma _{\lambda }$ a representative in the form $%
\left[ s^{\left( \alpha \right) },\partial ^{\left( \alpha \right)
},e_{i}^{\left( \alpha \right) \mu }\right] _{\lambda }$, $\left( x^{\mu
},e_{i}^{\mu }\right) $ are coordinates on $\Sigma _{\lambda }.$ The
associated metric is in the form 
\begin{equation}
g_{\mu \nu }=\bar{e}_{\mu }^{a}\eta _{ab}\bar{e}_{\nu }^{b}
\end{equation}
where $\bar{e}_{\mu }^{a}$ is the inverse matrix of $e_{a}^{\mu }$.

In the following we will analyze in details analogies and differences
between spin frames and \textsl{vielbein}. In analogy with the usual
convention about \textsl{vielbein,} we use the matrix $\eta _{ab}$ to lower
the Latin indices, its inverse $\eta ^{ab}$ to raise them and the induced
metric $g_{\mu \nu }$ and its inverse $g^{\mu \nu }$ to lower and raise the
Greek indices. For this reason we can omit everywhere the bar for the
inverse, because the position of indices identifies unambiguously the matrix. For instance, $e_{a\mu }$ can be obtained by lowering an
index in $e_{a}^{\mu }$ or in $\bar{e}_{\mu }^{a}$; in both cases the result
is the same.

\subsection{The spin connection over $\Sigma $}

Let $\Lambda :\Sigma \longrightarrow L\left( M\right) $ be a spin frame over 
$\Sigma $ and $g$ is its associated metric. Let us choose natural
coordinates $\left( x^{\mu },u_{a}^{\mu }\right) $ on $L\left( M\right) .$

	The metric $g$ defines a Levi-Civita connection 
\begin{equation}
\Gamma =dx^{\mu }\otimes \left( \partial _{\mu }-\Gamma _{\sigma \mu
}^{\lambda }\rho _{\lambda }^{\sigma }\right)
\end{equation}
where $\rho _{\lambda }^{\sigma }=u_{a}^{\sigma }\frac{\partial }{\partial
u_{a}^{\lambda }}$ are right-invariant fields over $L\left( M\right) $ and $%
\Gamma _{\sigma \mu }^{\lambda }$ are the Christoffel symbols of the metric $%
g.$

	The spin frame $\Lambda $ allows to define, by pull-back, a connection over $%
\Sigma $, called \textsl{spin connection}, given by 
\begin{equation}
\omega =dx^{\mu }\otimes \left( \partial _{\mu }-\Gamma _{\mu }^{ab}\sigma
_{ab}\right) \hspace{1cm}\Gamma _{\mu }^{ab}=e_{\lambda }^{a}\left( \Gamma
_{\sigma \mu }^{\lambda }e^{\sigma b}+d_{\mu }e^{\lambda b}\right)
\label{spin connection}
\end{equation}
This implies the following\medskip\newline
\textbf{Theorem (4.4.2.1):} there exist two morphisms
\begin{equation}
\left\{
\begin{tabular}{l}
$\Gamma :J^{1}\Sigma _{\lambda }\longrightarrow \mathcal{C}_{L\left(
M\right) }$ \\ 
$\omega :J^{1}\Sigma _{\lambda }\longrightarrow \mathcal{C}_{\Sigma }$%
\end{tabular}
\right.
\end{equation}
which allow to construct principal connections on $L\left( M\right) $\ and $%
\Sigma $\ starting from dynamical fields (and their first derivatives).\medskip\newline
As a consequence of the definition of the spin connection $\omega $, the
following holds:\medskip\newline
\textbf{Theorem (4.4.2.2): }$\nabla _{\mu }e_{a}^{\nu }=d_{\mu }e_{a}^{\nu
}+\Gamma _{\sigma \mu }^{\nu }e_{a}^{\sigma }-\Gamma _{a\mu }^{b\cdot
}e_{b}^{\nu }=0$\medskip \newline
Indeed, for the definition of the spin connection (\ref{spin connection}), we recover 
\begin{equation}
\nabla _{\mu }e_{a}^{\nu }=e_{c}^{\nu }e_{\sigma }^{c}\left( d_{\mu
}e_{a}^{\sigma }+\Gamma _{\rho \mu }^{\sigma }e_{a}^{\rho }\right) -\Gamma
_{a\mu }^{c\cdot }e_{c}^{\nu }=0
\end{equation}

\subsection{Vielbein}

\textbf{Definition (4.4.3.1): }if $M$ is a manifold, a \textsl{system of
(local) vielbein} is a family of local sections $e^{\left( \alpha \right)
}:U_{\alpha }\longrightarrow L\left( M\right) $ where $\left\{ U_{\alpha
}\right\} $ is an open covering of $M$, such that the transition functions,
defined by $e^{\left( \beta \right) }=e^{\left( \alpha \right) }\cdot
a_{\left( \alpha \beta \right) }$, are orthogonal group-valued, namely $%
a_{\left( \alpha \beta \right) }:U_{\alpha \beta }\longrightarrow SO\left(
\eta \right)$ .\medskip\newline
\textbf{Proposition (4.4.3.1):} on every (oriented) manifold $M$\
admitting a metric $g$\ of signature $\eta $ there exists a system of local vielbein.\medskip\newline
\textbf{Proof: }consider a family of local sections $\partial ^{\left(\alpha \right) }:U_{\alpha }\longrightarrow SO\left( M,g\right) $ associated to a trivialization of the bundle of the oriented orthonormal frame.
Using the canonical immersion $i_{g}:SO\left( M,g\right) \longrightarrow L\left( M\right) $ we obtain a set
of local sections of $L\left( M\right) $ which we indicate again with $
\partial ^{\left( \alpha \right) }:U_{\alpha }\longrightarrow $
$L\left( M\right) .$ The transition functions of the trivialization on $%
L\left( M\right) $ associated to this set are orthogonal group-valued, and therefore the $\partial
^{\left( \alpha \right) }$ are vielbein.\medskip\newline
\textbf{Proposition (4.4.3.2): }the choice of a spin frame induces (in a non
canonical way) a system of local vielbein.\medskip\newline
\textbf{Proof: }fix a spin frame $\Lambda :\Sigma \longrightarrow L\left(
M\right) $. If we define a trivialization $\sigma ^{\left( \alpha \right) }$ of $\Sigma $ it induces a system of
vielbein $e^{\left( \alpha \right) }=\Lambda \left( \sigma ^{\left( \alpha
\right) }\right).$

	If we change trivialization $\hat{s}^{\left( \alpha \right) }=s^{\left(
\alpha \right) }\cdot S^{\left( \alpha \right) }$ of $\Sigma $, with
transition functions $S^{\left( \alpha \right) }:U_{\left( \alpha \right) }\longrightarrow
Spin\left( \eta \right) $, we obtain another vielbein 
\begin{equation}
\hat{e}^{\left( \alpha \right) }=e^{\left( \alpha \right) }\cdot \ell \left(
S^{\left( \alpha \right) }\right)
\end{equation}
The fixing of a system of vielbein $e^{\left( \alpha \right) }$ induces a
metric $g$ which, by definition, admits $e^{\left( \alpha \right) }$ as
orthonormal frame. The metric $g$ is global because the $e^{\left( \alpha
\right) }$ are the vielbein.

	The analogy between spin frames and vielbein is really very strict, and it is
one of the reasons which justify the choice of the name ''spin frames''. It should be stressed, however, that the two structures are not equivalent. The existence of the spin frames requires topological
conditions which are stronger than those ensuring for the existence of the vielbein.
These supplementary conditions define the coupling of spinors
with gravity in a canonical, geometric and global way.

	In other words, if we choose some vielbein over the manifold $M$,
we can not generally reconstruct the spin frame that induces the vielbein\footnote{This can be done by following
proposition (4.4.2.1).}. This is possible only if we lift the cocycle $%
a_{\left( \alpha \beta \right) }$ of the transition functions of the
vielbein to the group of $Spin\left( \eta \right) $ (see Ref.[1]). This
implies some topological constraints. Besides, even this is possible, in general there is not only one way to do that. In any
case, the vielbein do not induce a spin frame canonically.

	The supplementary information necessary to build the spin frame are actually
encoded in the choice of the spin bundle $\Sigma $. Once a trivialization of
this bundle is chosen, one obtains the $Spin\left( \eta \right) $%
-valued transition functions. By composing with the covering map $\ell
:Spin\left( \eta \right) \longrightarrow SO\left( \eta \right) $, a cocycle 
$\ell \left( G_{\left( \alpha \beta \right) }\right)
:U_{\alpha \beta }\longrightarrow SO\left( \eta \right) $ is obtained. This defines a way to glue together the local sections of $L\left( M\right) $ \textsl{
coherently}. The families of local sections of $L\left( M\right) $ which \textsl{%
glue together} \textsl{coherently} are the global objects which we have
called spin frames. Thanks to these information, which are codified by $\Sigma $, the
global spin frames do exist. This is in contrast with the global vielbein, even
on non parallelizable manifolds.

	Following from these considerations, it can be claimed that the framework of spin frames allows to
globalize the vielbein, which are necessarily local. Each time the vielbein
are used as dynamical fields (for example in the theory of Dirac spinors),
it is natural to substitute the vielbein with the spin frames, to
obtain a global theory geometrically well-formulated. In this sense, the
vielbein formalism is the local version of the formalism of spin frames.

\subsection{The Lie derivative of spin frames}

	Let $\Xi =\xi ^{\mu }\partial _{\mu }+\xi ^{ab}\sigma _{\alpha \beta }$ be
an infinitesimal generator of automorphisms over $\Sigma $ and indicate with 
$\Xi _{\left( v\right) }=\xi _{\left( v\right) }^{ab}\sigma _{\alpha \beta
}=\left( \xi ^{ab}+\Gamma _{\mu }^{ab}\xi ^{\mu }\right) \sigma _{\alpha
\beta }$ its vertical part with respect to the spin connection.

	Being $\Sigma _{\lambda }$ a Gauge-Natural bundle associated to $\Sigma $,
we can associate to $\Xi $ an infinitesimal generator $\Xi _{\lambda }$ of
transformations of $\Sigma _{\lambda }$ in the following way:
\begin{equation}
\left\{ 
\begin{tabular}{l}
\vspace{0.1cm}$\left( x,S\right) \longmapsto \left( f\left( x\right) ,\varphi \cdot
S\right) \hspace{1cm}\hspace{1.35cm}\Xi =\xi ^{\mu }\partial _{\mu }+\xi ^{ab}\sigma _{\alpha \beta }$ \\ 
$\left( x,e\right) \longmapsto \left( f\left( x\right) ,J\cdot e\cdot \ell
\left( \varphi ^{-1}\right) \right) \hspace{1cm}$ $\Xi _{\lambda }=\xi ^{\mu
}\partial _{\mu }+\Xi _{a}^{\mu }\partial _{\mu }^{a}$%
\end{tabular}
\right.
\end{equation}
where 
\begin{equation}
\Xi _{a}^{\mu }=\partial _{\rho }\xi ^{\mu }e_{a}^{\rho }-e_{b}^{\mu
}\partial _{\alpha }^{\beta }\ell _{a}^{b}\left( \I\right) \dot{\varphi}%
_{\beta }^{\alpha }=\partial _{\rho }\xi ^{\mu }e_{a}^{\rho }-e_{b}^{\mu
}\xi _{~a}^{b\cdot }
\end{equation}
The Lie derivative of a spin frame is thence given by: 
\begin{eqnarray}
\pounds _{\Xi }e_{a}^{\mu } &=&d_{\sigma }e_{a}^{\mu }\xi ^{\sigma }-\Xi
_{a}^{\mu }=  \nonumber \\
&=&\left( d_{\sigma }e_{a}^{\mu }+\Gamma _{\nu \sigma }^{\mu }e_{a}^{\nu
}-\Gamma _{a\sigma }^{b}e_{b}^{\mu }\right) \xi ^{\sigma }-\nabla _{\nu }\xi
^{\mu }e_{a}^{\nu }+e_{b}^{\mu }\xi _{\left(v\right) a}^{b~~\cdot }= 
\nonumber \\
&=&-\nabla _{\nu }\xi ^{\mu }e_{a}^{\nu }+e_{b}^{\mu }\xi _{\left(v\right)
a}^{b\ \ \ \ \cdot }
\end{eqnarray}

\subsection{The spinor fields bundle}

Consider now a vector space $V$ and a representation over it of the group $%
Spin\left( \eta \right) $, which we denote by $\sigma :\Sigma \times
V\longrightarrow V.$

	Consider also the vector bundle $E_{\sigma }=\Sigma \times _{\sigma }V$
associated to $\Sigma $ with respect to the chosen representation $\sigma $;
the sections of this bundle are regarded as \textsl{spinor fields}. Being
defined on $E_{\sigma }$ the spin connection (\ref{spin connection}), we can
define the \textsl{(formal) covariant derivative} of the sections of $%
E_{\sigma }$, obtaining 
\begin{equation}
\Omega _{\mu }^{i}=\upsilon _{\mu }^{i}+\frac{1}{8}\Gamma _{\mu }^{ab}\left[
\gamma _{a},\gamma _{b}\right] _{\beta }^{\alpha }\partial _{\alpha }^{\beta
}\sigma _{j}^{i}\left( \I\right) \upsilon ^{j}
\end{equation}
We also define the Lie derivative of the sections of $E_{\sigma }$ with
respect to the vector fields of the base $M$: 
\begin{equation}
\pounds _{X}\upsilon ^{i}:=\xi ^{\mu }\upsilon _{\mu }^{i}-\frac{1}{8}\xi
^{ab}\left[ \gamma _{a},\gamma _{b}\right] _{\beta }^{\alpha }\partial
_{\alpha }^{\beta }\sigma _{j}^{i}\left( \I\right) \upsilon ^{j}
\label{lie spinori}
\end{equation}
However, one can define a canonical lift over $\Sigma $ induced by a spin
frame $e_{\mu }^{a}$ given by 
\begin{equation}
\xi _{\left( \upsilon \right) }^{ab}=e_{\mu }^{a}\nabla _{\nu }\xi ^{\mu
}e^{b\nu }
\end{equation}
The Lie derivative (\ref{lie spinori}) with respect to such a field becomes 
\begin{equation}
\pounds _{\hat{K}\left( X\right) }\upsilon ^{i}=\xi ^{\mu }\nabla _{\mu
}\upsilon ^{i}+\frac{1}{8}\nabla _{\mu }\xi ^{\nu }e^{a\mu }e_{\nu
}^{b}\left[ \gamma _{a},\gamma _{b}\right] _{\beta }^{\alpha }\partial
_{\alpha }^{\beta }\sigma _{j}^{i}\left( \I\right) \upsilon ^{j}
\end{equation}
Notice however that this Lie derivative is not natural (i.e. it does not
preserve the commutators) unless $X$ is a Killing vector.

	We stress that in the context of Gauge-Natural theories, it is 
\textsl{not} necessary to define the Lie derivatives with respect to
the vector fields of the base. This is done in natural theories.
Gauge-Natural theories, indeed, have been formulated to study fields which
are not natural objects from a geometrical viewpoint (i.e., they are
not sections of natural bundles).

	We thus believe that the \textsl{true} analogy is with gauge theories, where it is necessary to renounce to
the naturality of the fields. Therefore one might say that Gauge-Natural theories rise from the need to provide a general
context to develop field theories with non natural objects.


\chapter{Supersymmetry}\label{chapt:susy}

\section{Introduction}

The formalism presented in the previous chapters can be now applied to supersymmetric theories. Supersymmetries consist of transformations exchanging bosons with fermions; when made local and embedded in a gravitational context, they originate the theory of Supergravity (SUGRA).

	SUGRA can be viewed as the theory of the gravitational field, associated to a
spin $2$ boson called the \textsl{graviton}, that interacts
with a $3/2$ spin fermion, the \textsl{gravitino}. In the simplest case, namely in d=4, N=1 Supergravity,
there are $4$ space-time dimensions (as in usual General Relativity), and
only one fermionic dimension (i.e., supersymmetries have only one generator).

	In this chapter we shall attempt to formulate Supergravity as a Gauge Natural field theory. This should be alternative to the framework based on
supermanifolds, and it is motivated by the following facts:

	- If space-time has to be modeled as a supermanifold (with fermionic
dimensions) at least one should clarify which extent is a notational trick
and which extent is fundamental.

	- Most of the motivations that bring to supermanifolds are based on quantum
considerations. At least at an early stage the quantum and the classical
formulation of the theory should be kept separated (if possible).

	The tools discussed in so far should allow to describe Supergravity (as a
first step) in a purely classic form. The procedure is the following: given
a manifold $M$, the supersymmetries are the automorphisms of a spin bundle $%
\Sigma $, whose structure group is a supergroup but with an ordinary base.
The configuration bundle is associated to the structure group, so that the
natural action of automorphisms of the structure bundle acting on
configurations reproduces supersymmetry transformations.

	The first step is achieved in the case of the Wess-Zumino model, where the SUSY transformations are independent of the space-time coordinates. In this case, just a subset of automorphisms is taken into consideration. On the other hand, the Rarita-Schwinger theory deals with \textit{local} supersymmetry. This introduces some nontrivial issues, as it is shown in the next chapter.

\section{The Wess-Zumino model}

In this section we discuss the simplest application of supersymmetries to a
physical system endowed with a gravitational background. This constitutes a
first step towards Supergravity. As it will be shown, here the
transformations are point-independent; the link between their algebra and
the Gauge-Natural theory that can be constructed is given by the Lie
derivatives of the fields. However, the following is not a complete discussion of the Wess-Zumino
model, but it is just an example of how the geometrical framework discussed so far can be applied to physical systems.

\subsection{Dirac matrices and Majorana spinors}

Let $\lambda :Spin\left( \eta \right) \times W\longrightarrow W$ be a
representation induced by a complex representation of the group $Spin\left(
\eta \right) $ over $\mathbf{C}^{k}$, given by the $k\times k$ Dirac matrices $%
\gamma _{a}$ such that $\left( a=0,1,2,3\right) $%
\begin{equation}
\left\{ \gamma _{a},\gamma _{b}\right\} =2\eta _{ab}\I
\end{equation}
as a consequence 
\begin{equation}
\left\{ \gamma _{a},\gamma _{5}\right\} =0
\end{equation}
where $\eta _{ab}$ is the canonical diagonal matrix of signature $\eta
=\left( r,s\right) $ and 
\begin{equation}
\gamma _{5}=-i\gamma _{0}\gamma _{1}\gamma _{2}\gamma _{3}
\end{equation}
Here and hereafter we will consider only the Dirac representation.

	In dimension $k=4$ and with Lorentz signature $\eta =\left( 1,3\right) $, we
choose the set of Dirac matrices 
\begin{equation}
\gamma _{0}=\left( 
\begin{array}{ll}
0 & \I \\ 
\I & 0
\end{array}
\right) ~\;\gamma _{i}=\left( 
\begin{array}{ll}
0 & \mathbf{-\sigma }^{i} \\ 
\mathbf{\sigma }^{i} & 0
\end{array}
\right)
\end{equation}
where the $2\times 2$ matrices $\mathbf{\sigma }^{i}$ denote the standard Pauli
matrices.\medskip\newline
\textbf{Definition (5.2.1):} a spinor field $\psi $ is called \textsl{Majorana spinor} if it satisfies the condition 
\begin{equation}
\psi =C\bar{\psi}^{\dagger }
\end{equation}
where $\ C$ is the \textsl{charge conjugation matrix}.
In Dirac representation, we have 
\begin{equation}
C=\left( 
\begin{array}{ll}
-i\mathbf{\sigma }^{2} & 0 \\ 
0 & i\mathbf{\sigma }^{2}
\end{array}
\right) \;~\psi _{Maj}=\left( 
\begin{array}{l}
\;\;\;\alpha \\ 
i\mathbf{\sigma }^{2~t}\alpha ^{\dagger }
\end{array}
\right)
\end{equation}
here $\alpha $ is any two-component anticommuting spinor.
In the following, the \textsl{Majorana flip identities }will be considerably
important: 
\begin{equation}
\left\{ 
\begin{tabular}{l}
\vspace{0.1cm}$\bar{\psi}\varphi =\bar{\varphi}\psi $ \\ 
\vspace{0.1cm}$\bar{\psi}\gamma ^{a}\varphi =-\bar{\varphi}\gamma ^{a}\psi $ \\ 
\vspace{0.1cm}$\bar{\psi}\gamma ^{5}\varphi =\bar{\varphi}\gamma ^{5}\psi $ \\ 
\vspace{0.1cm}$\bar{\psi}\gamma ^{a}\gamma ^{b}\varphi =-\bar{\varphi}\gamma ^{a}\gamma
^{b}\psi $ \\ 
$\bar{\psi}\gamma ^{5}\gamma ^{a}\varphi =\bar{\varphi}\gamma ^{5}\gamma
^{a}\psi $%
\end{tabular}
\right.
\end{equation}

\subsection{Covariance of the Lagrangian}

We consider as fields a Majorana anticommuting spinor $\psi $ and four
scalars $\left( A,B,C,D\right) $ on a space-time manifold $%
M$ with a spin frame (i.e. a vielbein) $e_{a}^{\mu }$.
After defining a structure bundle $\Sigma \longrightarrow M$, which is indeed a
spin bundle (see Chapter 4), the Lagrangian is 
\begin{eqnarray}
L_{WZ} &=&\frac{1}{2}\left( \nabla _{\mu }A\nabla ^{\mu }A+D^{2}+2mAD\right)
e\mathbf{ds}-\bar{\psi}\left( i\gamma ^{a}\nabla _{a}\psi +m\psi \right) e%
\mathbf{ds}+  \nonumber \\
&&+\frac{1}{2}\left( \nabla _{\mu }B\nabla ^{\mu }B+C^{2}-2mBC\right) e^{3}%
\mathbf{ds}
\end{eqnarray}
($e$ is the determinant of the vielbein $e_{a}^{\mu }$). The covariant
derivatives of the scalar fields are considered with respect to the
Levi-Civita connection induced by the metric, which is in turn induced by
the spin frame. Instead, the covariant derivatives of spinors are evaluated
with respect to the spin connection (\ref{spin connection}), namely: 
\begin{equation}
\nabla _{a}\psi =e_{a}^{\mu }\left( d_{\mu }\psi +\frac{1}{8}\Gamma _{\mu
}^{ab}\left[ \gamma _{a},\gamma _{b}\right] \psi \right) \;,\;\Gamma _{\mu
}^{ab}=e_{\lambda }^{a}\left( \Gamma _{\sigma \mu }^{\lambda }e^{\sigma
b}+d_{\mu }e^{\lambda b}\right)
\end{equation}
This Lagrangian is invariant (modulo divergence terms) under the
infinitesimal transformations 
\begin{equation}
\left\{ 
\begin{tabular}{l}
\vspace{0.1cm}$\delta A=\frac{1}{2}\left( \bar{\epsilon}\psi \right) $ \\ 
\vspace{0.1cm}$\delta B=-\frac{i}{2}\left( \bar{\epsilon}\gamma ^{5}\psi \right) e^{-1}$
\\ 
\vspace{0.1cm}$\delta C=-\frac{1}{2}\left( \bar{\epsilon}\gamma ^{5}\gamma ^{a}\nabla
_{a}\psi \right) e^{-1}$ \\ 
\vspace{0.1cm}$\delta D=\frac{i}{2}\left( \bar{\epsilon}\gamma ^{a}\nabla _{a}\psi
\right) $ \\ 
\vspace{0.1cm}$\delta \psi =\frac{1}{2}\left[ i\left( \gamma ^{a}\epsilon \right) \nabla
_{a}A+e\left( \gamma ^{5}\gamma ^{a}\epsilon \right) \nabla _{a}B+ie\left(
\gamma ^{5}\epsilon \right) C+\epsilon D\right] $ \\ 
$\delta \bar{\psi}=\frac{1}{2}\left[ -i\nabla _{a}A\left( \bar{\epsilon}%
\gamma ^{a}\right) -e\nabla _{a}B\left( \bar{\epsilon}\gamma ^{5}\gamma
^{a}\right) +ie\left( \bar{\epsilon}\gamma ^{5}\right) C+\bar{\epsilon}%
D\right] $%
\end{tabular}
\right.
\end{equation}
The transformation parameter $\epsilon $ is a Majorana anticommuting spin $1/2$ spinor which is assumed to be covariantly conserved: $\nabla
_{\mu }\epsilon =0.$ This condition is very strong, because it corresponds
to a point-independent transformation which defines gauge theories. If $%
\nabla _{\mu }\epsilon \neq 0,$ the transformations would be point dependent
and this happens in Supergravity, as we shall see in the next chapter.

	One can define the \textsl{infinitesimal generator of supersymmetries}
\begin{equation}
\Xi =\left( \delta A\right) \frac{\partial }{\partial A}+\left( \delta
B\right) \frac{\partial }{\partial B}+\left( \delta C\right) \frac{\partial 
}{\partial C}+\left( \delta D\right) \frac{\partial }{\partial D}+\left(
\delta \psi \right) \frac{\partial }{\partial \psi }+\left( \delta \bar{\psi}%
\right) \frac{\partial }{\partial \bar{\psi}}
\end{equation}
It can be shown that this leaves the Lagrangian invariant, modulo the following
divergence term,
\begin{equation}
\delta L_{WZ}=Div\left( \alpha \right)
\end{equation}
where 
\begin{eqnarray}
\alpha &=&\frac{1}{4}[\left( 2imA\left( \bar{\epsilon}\gamma ^{\mu }\psi
\right) +2\nabla ^{\mu }A\left( \bar{\epsilon}\psi \right) -\nabla _{\nu
}A\left( \bar{\epsilon}\gamma ^{\nu }\gamma ^{\mu }\psi \right) +iD\left( 
\bar{\epsilon}\gamma ^{\mu }\psi \right) \right) e+  \nonumber \\
&&+(2mB\left( \bar{\epsilon}\gamma ^{5}\gamma ^{\mu }\psi \right) -2i\nabla
^{\mu }B\left( \bar{\epsilon}\gamma ^{5}\psi \right) +i\nabla _{\nu }B\left( 
\bar{\epsilon}\gamma ^{5}\gamma ^{\nu }\gamma ^{\mu }\psi \right) + 
\nonumber \\
&&+C\left( \bar{\epsilon}\gamma ^{\mu }\gamma ^{5}\psi \right) )e^{2}]%
ds_{\mu }
\end{eqnarray}
Thus the infinitesimal generator $\Xi $ can be seen as a \textsl{generalized
symmetry} (see section 3.6).

\subsection{Closure of the algebra}

Here we calculate the commutator of two supersymmetries on the fields, in order to
check the closure of the algebra and to formulate the theory from a
Gauge-Natural point of view.

\subsubsection{Closure on the scalar fields}

We begin with the scalar fields: 
\begin{eqnarray}
\left[ \delta _{1},\delta _{2}\right] A&=& \frac{1}{4}[i\left( \bar{%
\epsilon}_{2}\gamma ^{a}\epsilon _{1}-\bar{\epsilon}_{1}\gamma ^{a}\epsilon
_{2}\right) \nabla _{a}A+ \nonumber \\ 
&&+e^{-1}\nabla _{a}B\left( \bar{\epsilon}_{2}\gamma ^{a}\gamma
^{5}\epsilon _{1}-\bar{\epsilon}_{1}\gamma ^{a}\gamma ^{5}\epsilon
_{2}\right) +D\left( \bar{\epsilon}_{2}\epsilon _{1}-\bar{\epsilon}%
_{1}\epsilon _{2}\right) + \nonumber\\ 
&&+iC\left( \bar{\epsilon}_{2}\epsilon _{1}-\bar{\epsilon}_{1}\epsilon
_{2}\right) =i\frac{1}{2}\left( \bar{\epsilon}_{2}\gamma ^{a}\epsilon
_{1}\right) \nabla _{a}A
\end{eqnarray}
this result holds by virtue of the Majorana flip identities 
\begin{equation}
\left\{ 
\begin{tabular}{l}
$\bar{\epsilon}_{2}\gamma ^{a}\gamma ^{5}\epsilon _{1}=\bar{\epsilon}%
_{1}\gamma ^{a}\gamma ^{5}\epsilon _{2}$ \\ 
$\bar{\epsilon}_{2}\epsilon _{1}=\bar{\epsilon}_{1}\epsilon _{2}$ \\ 
$\bar{\epsilon}_{1}\gamma ^{a}\epsilon _{2}=-\bar{\epsilon}_{2}\gamma
^{a}\epsilon _{1}$%
\end{tabular}
\right.
\end{equation}
Similarly, for the other scalar fields we obtain 
\begin{equation}
\left[ \delta _{1},\delta _{2}\right] B=\frac{i}{2}\left( \bar{\epsilon}%
_{2}\gamma ^{a}\epsilon _{1}\right) \nabla _{a}B
\end{equation}
\begin{equation}
\left[ \delta _{1},\delta _{2}\right] C=\frac{i}{2}\left( \bar{\epsilon}%
_{2}\gamma ^{a}\epsilon _{1}\right) \nabla _{a}C
\end{equation}
\begin{equation}
\left[ \delta _{1},\delta _{2}\right] D=\frac{i}{2}\left( \bar{\epsilon}%
_{2}\gamma ^{a}\epsilon _{1}\right) \nabla _{a}D
\end{equation}
These objects can be reinterpreted as the Lie derivatives with respect to an
appropriate vector field 
\begin{equation}
\xi =\frac{i}{2}\left( \bar{\epsilon}_{2}\gamma ^{\mu }\epsilon _{1}\right)
\partial _{\mu }
\end{equation}
of a scalar density of weight $k$: 
\begin{equation}
\pounds _{\xi }A=\xi ^{\mu }\nabla _{\mu }A+k\nabla _{\mu }\xi ^{\mu }A\;,\;%
\text{being} \:\nabla _{\nu }\xi ^{\mu }=0
\end{equation}

\subsubsection{Closure on the vielbein and on the spinor field}

A vector $\Xi =\xi ^{\mu }\partial _{\mu }+\xi ^{ab}\sigma _{ab}$ acts on $%
e_{a}^{\mu }$ as: 
\begin{equation}
\pounds _{\Xi }e_{a}^{\mu }=-\nabla _{\nu }\xi ^{\mu }e_{a}^{\nu
}+e_{b}^{\mu }\xi _{\left( \upsilon \right) a}^{b}
\end{equation}
Being 
\begin{equation}
\delta e_{a}^{\mu }=0
\end{equation}
the commutator on the vielbein is 
\begin{equation}
\left[ \delta _{1},\delta _{2}\right] e_{a}^{\mu }=0
\end{equation}
so if we want to obtain 
\begin{equation}
\left[ \delta _{1},\delta _{2}\right] e_{a}^{\mu }=\pounds _{\Xi }e_{a}^{\mu
}
\end{equation}
we have to choose a vertical field such that 
\begin{equation}
\pounds _{\hat{\xi}}e_{a}^{\mu }=0
\end{equation}
This happens if and only if 
\begin{equation}
\xi _{\left( \upsilon \right) }^{ab}=e_{\mu }^{a}\nabla _{\nu }\xi ^{\mu
}e^{b\nu }
\end{equation}
which is the \textsl{Kosmann lift}. This leads to 
\begin{equation}
\left[ \delta _{1},\delta _{2}\right] e_{a}^{\mu }=\pounds _{\hat{\xi}%
}e_{a}^{\mu }
\end{equation}
defined on the structure bundle $\Sigma :$%
\begin{equation}
\hat{\xi}=\xi ^{\mu }\left( \partial _{\mu }-\Gamma _{\mu }^{ab}\sigma
_{ab}\right) \oplus \left( e_{\mu }^{a}\nabla _{\nu }\xi ^{\mu }e^{b\nu
}\right) \sigma _{ab}\;,\;\xi ^{\mu }=\frac{i}{2}\left( \bar{\epsilon}%
_{2}\gamma ^{\mu }\epsilon _{1}\right)  \label{vettore}
\end{equation}
We have now to check if this vector is suited also for the spinor field $%
\psi $: the commutator is
\begin{eqnarray}
\left[ \delta _{1},\delta _{2}\right] \psi &=& \frac{i}{4}[\gamma
^{a}\left( \bar{\epsilon}_{1}\epsilon _{2}-\bar{\epsilon}_{2}\epsilon
_{1}\right) \nabla _{a}\psi -\gamma ^{a}\gamma ^{5}\left( \bar{\epsilon}
_{1}\epsilon _{2}-\bar{\epsilon}_{2}\epsilon _{1}\right) \gamma ^{5}\nabla
_{a}\psi + \nonumber\\ 
&&+\left( \bar{\epsilon}_{1}\epsilon _{2}-\bar{\epsilon}_{2}\epsilon
_{1}\right) \gamma ^{a}\nabla _{a}\psi -\gamma ^{5}\left( \bar{\epsilon}%
_{1}\epsilon _{2}-\bar{\epsilon}_{2}\epsilon _{1}\right) \gamma ^{5}\gamma
^{a}\nabla _{a}\psi ]= \nonumber \\ 
&&=\frac{i}{4}\left[ -\frac{1}{2}\left( \bar{\epsilon}_{1}\gamma
_{b}\epsilon _{2}\right) \right] \left( \gamma ^{a}\gamma ^{b}+\gamma
^{a}\gamma ^{5}\gamma ^{b}\gamma ^{5}+\gamma ^{b}\gamma ^{a}-\gamma
^{5}\gamma ^{b}\gamma ^{5}\gamma ^{a}\right) \nabla _{a}\psi + \nonumber \\ 
&&+\frac{i}{4}\left( \bar{\epsilon}_{1}\gamma _{b}\gamma _{c}\epsilon
_{2}\right) \left( \gamma ^{a}\gamma ^{b}\gamma ^{c}-\gamma ^{a}\gamma
^{5}\gamma ^{b}\gamma ^{c}\gamma ^{5}+\gamma ^{b}\gamma ^{c}\gamma
^{a}-\gamma ^{5}\gamma ^{b}\gamma ^{c}\gamma ^{5}\gamma ^{a}\right) \nabla
_{a}\psi = \nonumber \\ 
&&=\frac{i}{4}\left[ -\left( \bar{\epsilon}_{1}\gamma _{b}\epsilon
_{2}\right) \left\{ \gamma ^{a},\gamma ^{b}\right\} \right] \nabla _{a}\psi
= \nonumber \\ 
&&=\frac{i}{4}\left( \bar{\epsilon}_{2}\gamma ^{a}\epsilon _{1}\right)
\nabla _{a}\psi
\end{eqnarray}
by using the Majorana flip identities.
Now, the Lie derivative of the spinor $\psi $ with respect to a generic
infinitesimal generator defined on $\Sigma $ can be written as (\ref{lie spinori}): 
\begin{equation}
\pounds _{\Xi }\psi =\xi ^{\mu }\nabla _{\mu }\psi -\frac{1}{8}\left[ \gamma
_{a},\gamma _{b}\right] \psi \xi _{\left( \upsilon \right) }^{ab}
\end{equation}
Calculated with respect to (\ref{vettore}), this becomes 
\begin{equation}
\pounds _{\hat{\xi}}\psi =\xi ^{\mu }\nabla _{\mu }\psi -\frac{1}{8}\left[
\gamma _{a},\gamma _{b}\right] \psi \nabla ^{b}\xi ^{a}
\end{equation}
This is indeed the commutator of two supersymmetries on the spinor field 
$\psi $, because $\nabla ^{b}\xi ^{a}=0$ and $\left[ \delta _{1},\delta
_{2}\right] \psi =\xi ^{\mu }\nabla _{\mu }\psi $.

	In conclusion, once we take into account both the vector fields (\ref
{vettore}) and the supersymmetry generators, these form an algebra with the
following commutation rules: 
\begin{equation}
\left[ \delta _{1},\delta _{2}\right] =\pounds _{\hat{\xi}},\;\;\left[
\delta _{1},\pounds _{\hat{\xi}}\right] =0,\;\;\left[ \pounds _{\hat{%
\varsigma}},\pounds _{\hat{\xi}}\right] =0
\end{equation}
	These vector fields do not span an ordinary Lie algebra, since
some of the parameters are actually anticommuting. In contrast, the parameters of the ordinary Lie
algebras are scalars. They can be seen as generators of a
graded Lie algebra, a \textsl{superalgebra}, which will be analyzed in the Appendix.


\chapter{Supergravity}\label{chapt:rs}

\section{Introduction}

As we already remarked, Supergravity is the theory of the
gravitational field interacting with a spinor. Therefore we will consider as
fields the following:

	- a vielbein $e_{\mu }^{a}$ (or a spin frame in our formalism), with determinant $e$. It is defined on the structure bundle $\Sigma$.

	- the $\mu $-component of a 4-component spin $3/2$
Majorana (anticommuting) spinor $\psi _{\mu }$.

	- the principal spin connection on the bundle $\Sigma $, which we call $\omega $.

	The configuration bundle $B\longrightarrow M$ projects
into an ordinary 4-dimensional manifold $M$, with supermanifolds as fibers.
Regarding the spin connection $\omega $, we will use
the so-called standard approach. This consists of imposing the \textsl{null
(super)torsion constraint} 
\begin{equation}
T^{a}=\left( d_{\mu }e_{\nu }^{a}+\omega _{\;\; b\mu}^{a\,\cdot }e_{\nu }^{b}-%
\frac{i}{2}\bar{\psi}_{\mu }\gamma^{a}\psi _{\nu }\right) dx^{\mu }\wedge
dx^{\nu }=0  \label{torsione nulla}
\end{equation}
This condition is fixed \textsl{a priori} and it is kinematical, in the
sense that it does not affect the Lagrangian and accordingly, the dynamics
of fields. As it will be shown later, this simplifies the theory, because by
its means the connection is no longer an independent field. It then becomes expressed
as a function of the vielbein and of the gravitino. Thus there are only two independent fields: $e_{\mu }^{a}$ and $\psi _{\mu }$.

	The Rarita-Schwinger Lagrangian was postulated long ago \cite{RS}, 
\begin{eqnarray}
L &=&\mathcal{L}ds=\left( -4R_{\mu \nu }^{ab}e_{a}^{\mu }e_{b}^{\nu }e+8\bar{%
\psi}_{\mu }\gamma _{5}\gamma _{a}\nabla _{\nu }\psi _{\rho }e_{\sigma
}^{a}\epsilon ^{\mu \nu \rho \sigma }\right) ds:=  \nonumber \\
&=&\left( \mathcal{L}_{\mathcal{H}}+\mathcal{L}_{\mathcal{S}}\right) ds
\label{Lagrangiana}
\end{eqnarray}
where we set: 
\begin{equation}
\mathcal{L}_{\mathcal{H}}:=-4R_{\mu \nu }^{ab}e_{a}^{\mu }e_{b}^{\nu }e
\label{Lagrangiana di Hilbert}
\end{equation}
and 
\begin{equation}
\mathcal{L}_{\mathcal{S}}:=8\bar{\psi}_{\mu }\gamma _{5}\gamma _{a}\nabla
_{\nu }\psi _{\rho }e_{\sigma }^{a}\epsilon ^{\mu \nu \rho \sigma }
\label{Lagrangiana spine}
\end{equation}
These are, respectively, the Hilbert-Einstein and the spin lagrangian
densities. $ds$ is the standard volume element, and $R_{\mu \nu }^{ab}$ is the Riemann tensor of the connection $\omega $: 
\begin{equation}
R_{\mu \nu }^{ab}=d_{\mu }\omega _{\hspace{0.3cm}\nu }^{ab}-d_{\nu }\omega _{%
\hspace{0.3cm}\mu }^{ab}+\omega _{\ c\mu }^{a\cdot }\omega _{\hspace{0.3cm}%
\nu }^{cb}-\omega _{\ c\nu }^{a\cdot }\omega _{\hspace{0.3cm}\mu }^{cb}
\end{equation}
Exactly as in the Wess-Zumino model, $\gamma _{a}$ belongs to the set of $%
4\times 4$ matrices in the Dirac representation.
The covariant derivative of the gravitino is calculated with respect to the
spin connection: 
\begin{equation}
\nabla _{\mu }\psi _{\nu }=d_{\mu }\psi _{\nu }-\frac{1}{8}\left[ \gamma
_{a},\gamma _{b}\right] \omega _{\hspace{0.1cm}\hspace{0.1cm}\hspace{0.1cm}%
\mu }^{ab}\psi _{\nu }-\Gamma _{\mu \nu }^{\lambda }\psi _{\lambda }
\end{equation}
The supersymmetry parameter is a 4-component spin $1/2$ Majorana
spinor $\varepsilon $, which acts on the fields as follows: 
\begin{equation}
\left\{ 
\begin{tabular}{l}
$\delta e_{\mu }^{a}=\bar{\varepsilon}\gamma ^{a}\psi _{\mu }$ \\ 
$\delta \psi _{\mu }=\nabla _{\mu }\varepsilon $%
\end{tabular}
\right.  \label{supersimmetrie}
\end{equation}
We now have all the necessary elements to study the Rarita-Schwinger
model; we begin by working out the field
equations for the vielbein and for the gravitino. Those of the connection are also taken into account (though this field is not independent), to verify that they are identically satisfied by
virtue of the null torsion constraint (\ref{torsione nulla}).

\subsection{Expression of the principal connection $\omega $}

Let us use the constraint (\ref{torsione nulla}): 
\begin{equation}
d_{[\mu }e_{\nu ]}^{a}+\omega _{~~b[\mu \,}^{a\,\cdot }e_{\nu ]}^{b}=\frac{i%
}{2}\bar{\psi}_{\mu }\gamma _{a}\psi _{\nu }
\end{equation}
We denote 
\begin{equation}
\omega _{\lambda \mu \nu }=\frac{i}{2}\bar{\psi}_{\mu }\gamma _{\lambda
}\psi _{\nu }-e_{a\lambda }d_{\nu }e_{\mu }^{a}
\end{equation}
and permute the indices as follows:
\begin{equation}
\left\{ 
\begin{tabular}{l}
\vspace{0.1cm}$\frac{1}{2}\left( -\omega _{\lambda \nu \mu }+\omega _{\lambda \mu \nu
}\right) =-\frac{i}{2}\bar{\psi}_{\mu }\gamma _{\lambda }\psi _{\nu }-\frac{1%
}{2}\left( -e_{a\lambda }d_{\mu }e_{\nu }^{a}+e_{a\lambda }d_{\nu }e_{\mu
}^{a}\right) $ \\ 
\vspace{0.1cm}$\frac{1}{2}\left( \omega _{\nu \mu \lambda }-\omega _{\nu \lambda \mu
}\right) =\frac{i}{2}\bar{\psi}_{\lambda }\gamma _{\nu }\psi _{\mu }-\frac{1%
}{2}\left( e_{a\nu }d_{\lambda }e_{\mu }^{a}-e_{a\nu }d_{\mu }e_{\lambda
}^{a}\right) $ \\ 
$\frac{1}{2}\left( \omega _{\mu \lambda \nu }-\omega _{\mu \nu \lambda
}\right) =\frac{i}{2}\bar{\psi}_{\nu }\gamma _{\mu }\psi _{\lambda }-\frac{1%
}{2}\left( e_{a\mu }d_{\nu }e_{\lambda }^{a}-e_{a\mu }d_{\lambda }e_{\nu
}^{a}\right) $%
\end{tabular}
\right.
\end{equation}
By adding together the three equations, we get
\begin{eqnarray}
\omega _{\lambda }^{ab}&=& e^{a\nu }e^{b\mu }(-\frac{i}{2}\bar{\psi}_{\mu
}\gamma _{\lambda }\psi _{\nu }+\frac{i}{2}\bar{\psi}_{\lambda }\gamma _{\nu
}\psi _{\mu }+\frac{i}{2}\bar{\psi}_{\nu }\gamma _{\mu }\psi _{\lambda }+ \nonumber
\\ 
&&+\frac{1}{2}d_{\mu }g_{\lambda \nu }-\frac{1}{2}d_{\nu }g_{\mu \lambda }-
\frac{1}{2}d_{\lambda }g_{\mu \nu }+e_{c\mu }d_{\lambda }e_{\nu }^{c})= \nonumber \\ 
&&e^{a\mu }e^{b\nu }(-\frac{i}{2}\bar{\psi}_{\nu }\gamma _{\lambda }\psi
_{\mu }+\frac{i}{2}\bar{\psi}_{\lambda }\gamma _{\mu }\psi _{\nu }+\frac{i}{2
}\bar{\psi}_{\mu }\gamma _{\nu }\psi _{\lambda }+ \nonumber \\ 
&&-g_{\nu \rho }\Gamma _{\mu \lambda }^{\rho }+e_{c\nu }d_{\lambda }e_{\mu
}^{c})= \nonumber \\ 
&&=e^{a\mu }e^{b\nu }\left( \frac{i}{2}\bar{\psi}_{\mu }\gamma _{\lambda
}\psi _{\nu }+i\bar{\psi}_{\lambda }\gamma _{[\mu }\psi _{\nu ]}\right)
-e_{\rho }^{b}\Gamma _{\mu \lambda }^{\rho }e^{a\mu }-e_{\mu }^{b}d_{\lambda
}e^{a\mu }= \nonumber \\ 
&&=e^{\mu [a}e^{b]\nu }\left( \frac{i}{2}\bar{\psi}_{\mu }\gamma _{\lambda
}\psi _{\nu }+i\bar{\psi}_{\lambda }\gamma _{[\mu }\psi _{\nu ]}\right)
-\Gamma _{\lambda }^{ba}
\end{eqnarray}
where we have used (\ref{spin connection}). Therefore the null torsion connection which will be used in the following has the expression 
\begin{equation}
\omega _{\lambda }^{ab}=\Gamma _{\lambda }^{ab}+H_{\lambda }^{ab}
\label{connessione}
\end{equation}
where
\begin{eqnarray}
H_{\lambda }^{ab}&=& e^{\mu [a}e^{b]\nu }\left( \frac{i}{2}\bar{\psi}%
_{\mu }\gamma _{\lambda }\psi _{\nu }+i\bar{\psi}_{\lambda }\gamma _{[\mu
}\psi _{\nu ]}\right) = \nonumber \\ 
&&\frac{i}{2}\left( e^{\mu a}e^{b\nu }-e^{\mu b}e^{a\nu }\right) \left( 
\bar{\psi}_{\lambda }\gamma _{\mu }\psi _{\nu }-\bar{\psi}_{\lambda }\gamma
_{\nu }\psi _{\mu }+\bar{\psi}_{\mu }\gamma _{\lambda }\psi _{\nu }\right) =\nonumber
\\ 
&&i\left( \bar{\psi}_{\lambda }\gamma ^{a}\psi ^{b}-\bar{\psi}_{\lambda
}\gamma ^{b}\psi ^{a}+\bar{\psi}^{a}\gamma _{\lambda }\psi ^{b}\right)
\label{h}
\end{eqnarray}
or, equivalently, 
\begin{equation}
H^{abc}=H_{\lambda }^{ab}e^{\lambda c}=i\left( \bar{\psi}^{c}\gamma ^{a}\psi
^{b}+\bar{\psi}^{a}\gamma ^{b}\psi ^{c}+\bar{\psi}^{a}\gamma ^{c}\psi
^{b}\right)
\end{equation}

\section{Field equations for the Rarita-Schwinger Lagrangian}

We now want to evaluate the equations for the three fields involved, i.e.
the vielbein $e_{\mu }^{a}$, the Majorana spinor $\psi $ and $\omega _{\mu
}^{ab}$, the principal connection, independent of $e_{\mu }^{a}$, defined on
the bundle $\Sigma $.

	According to the general theory (see Chapter 2), the field equations
corresponding to the Lagrangian (\ref{Lagrangiana}) are: 
\begin{equation}
\delta \mathcal{L}=E_{i}\delta y^{i}\otimes ds+Div(F_{i}^{\mu }\delta
y^{i}\otimes ds_{\mu })  \label{moto}
\end{equation}
where 
\begin{equation}
E_{i}\delta y^{i}=E_{a}^{\mu }\delta e_{\mu }^{a}+E^{\mu }\delta \psi _{\mu
}+E_{ab}^{\mu }\delta \omega _{\mu }^{ab}
\end{equation}

\subsection{Equations for the vielbein field}

We apply Eq.(\ref{moto}): 
\begin{equation}
E_{a}^{\mu }\delta e_{\mu }^{a}=\left[ \left( R_{\mu }^{a}-\frac{1}{2}e_{\mu
}^{a}R\right) e-2e_{\nu }^{a}\bar{\psi}_{\lambda }\gamma _{5}\gamma _{\mu
}\nabla _{\rho }\psi _{\sigma }\epsilon ^{\lambda \nu \rho \sigma }\right]
\delta e_{a}^{\mu }=0
\end{equation}
Hence we conclude that the field equations for the vielbein are: 
\begin{equation}
R_{\mu }^{a}-\frac{1}{2}Re_{\mu }^{a}=\frac{2}{e}e_{\nu }^{a}\bar{\psi}%
_{\lambda }\gamma _{5}\gamma _{\mu }\nabla _{\rho }\psi _{\sigma }\epsilon
^{\lambda \nu \rho \sigma }  \label{campo vielbein}
\end{equation}

\subsection{Equations for the gravitino field}

In this case Eq.(\ref{moto}) reduces to: 
\begin{eqnarray}
E^{\mu }\delta \psi _{\mu }&=& 8\delta \bar{\psi}_{\mu }\gamma _{5}\gamma
_{a}\nabla _{\nu }\psi _{\rho }e_{\sigma }^{a}\epsilon ^{\mu \nu \rho \sigma
}+8\bar{\psi}_{\mu }\gamma _{5}\gamma _{a}\nabla _{\nu }\delta \psi _{\rho
}e_{\sigma }^{a}\epsilon ^{\mu \nu \rho \sigma }= \nonumber\\ 
&& =8\delta \bar{\psi}_{\mu }\gamma _{5}\gamma _{a}\nabla _{\nu }\psi _{\rho
}e_{\sigma }^{a}\epsilon ^{\mu \nu \rho \sigma }+8\nabla _{\nu }(\bar{\psi}
_{\mu }\gamma _{5}\gamma _{a}\delta \psi _{\rho }e_{\sigma }^{a}\epsilon
^{\mu \nu \rho \sigma })+ \nonumber \\ 
&& -8\nabla _{\nu }\bar{\psi}_{\mu }\gamma _{5}\gamma _{a}\delta \psi _{\rho
}e_{\sigma }^{a}\epsilon ^{\mu \nu \rho \sigma }-8\bar{\psi}_{\mu }\gamma
_{5}\gamma _{a}\delta \psi _{\rho }\nabla _{[\nu }e_{\sigma ]}^{a}\epsilon
^{\mu \nu \rho \sigma }+ \nonumber \\ 
&& -8i\bar{\psi}_{\mu }\gamma _{5}\gamma _{a}\delta \psi _{\rho }(\bar{\psi}%
_{\nu }\gamma _{a}\psi _{\sigma })\epsilon ^{\mu \nu \rho \sigma }=8\delta 
\bar{\psi}_{\mu }\gamma _{5}\gamma _{a}\nabla _{\nu }\psi _{\rho }e_{\sigma
}^{a}\epsilon ^{\mu \nu \rho \sigma }+ \nonumber \\ 
&& +8\nabla _{\nu }(\bar{\psi}_{\mu }\gamma _{5}\gamma _{a}\delta \psi
_{\rho }e_{\sigma }^{a}\epsilon ^{\mu \nu \rho \sigma })-8\delta \bar{\psi}%
_{\rho }\gamma _{5}\gamma _{a}\nabla _{\nu }\psi _{\mu }e_{\sigma
}^{a}\epsilon ^{\mu \nu \rho \sigma }= \nonumber \\ 
&&=8\delta \bar{\psi}_{\mu }\gamma _{5}\gamma _{a}\nabla _{\nu }\psi _{\rho
}e_{\sigma }^{a}\epsilon ^{\mu \nu \rho \sigma }+8\nabla _{\nu }(\bar{\psi}%
_{\mu }\gamma _{5}\gamma _{a}\delta \psi _{\rho }e_{\sigma }^{a}\epsilon
^{\mu \nu \rho \sigma })+ \nonumber \\ 
&&+8\delta \bar{\psi}_{\mu }\gamma _{5}\gamma _{a}\nabla _{\nu }\psi _{\rho
}e_{\sigma }^{a}\epsilon ^{\mu \nu \rho \sigma }= \nonumber \\ 
&&=16\delta \bar{\psi}_{\mu }\gamma _{5}\gamma _{a}\nabla _{\nu }\psi
_{\rho }e_{\sigma }^{a}\epsilon ^{\mu \nu \rho \sigma }+8\nabla _{\nu }(\bar{%
\psi}_{\mu }\gamma _{5}\gamma _{a}\delta \psi _{\rho }e_{\sigma
}^{a}\epsilon ^{\mu \nu \rho \sigma })
\end{eqnarray}
The bilinear $i\bar{\psi}_{\mu }\gamma _{5}\gamma _{a}\delta \psi _{\rho }(%
\bar{\psi}_{\nu }\gamma _{a}\psi _{\sigma })\epsilon ^{\mu \nu \rho \sigma }$
is equal to zero because among $\bar{\psi}_{\mu }$ and $\psi _{\rho }$,
only $C$-symmetric matrices give no vanishing contribution, and $\gamma
_{5}\gamma _{a}$ is $C$-skewsymmetric.

	Recalling now the first variation formula, i.e. Eq.(\ref{first variation}): 
\[
\left\langle \delta L\circ j^{k}\rho \mid j^{k}X\right\rangle =\left\langle 
\E\left( L\right) \circ j^{2k}\rho \mid X\right\rangle +d\left[
\left\langle \F\left( L,\gamma \right) \circ j^{2k-1}\rho \mid
j^{k-1}X\right\rangle \right] 
\]
we see that 
\begin{equation}
\E\left( L\right) \circ j^{2k}\rho =16\gamma _{5}\gamma _{a}\nabla
_{\nu }\psi _{\rho }e_{\sigma }^{a}\epsilon ^{\mu \nu \rho \sigma }
\end{equation}
and 
\begin{equation}
\F\left( L,\gamma \right) \circ j^{2k-1}\rho =8\bar{\psi}_{\mu }\gamma
_{5}\gamma _{a}\delta \psi _{\rho }e_{\sigma }^{a}\epsilon ^{\mu \nu \rho
\sigma }
\end{equation}
Thus the field equations for the $\frac{3}{2}$ spin field of component 
$\psi _{\mu }$ are given by Eq.(\ref{euler}): 
\begin{equation}
\E\left( L\right) \circ j^{2k}\rho =\gamma _{5}\gamma _{a}\nabla _{\nu
}\psi _{\rho }e_{\sigma }^{a}\epsilon ^{\mu \nu \rho \sigma }=0
\label{eq.campo gravitino}
\end{equation}

\subsection{Equations for the connection}

By recalling that 
\begin{equation}
\left\{ 
\begin{tabular}{l}
$\delta R_{\mu \nu }^{ab}=\nabla _{\mu }(\delta \omega _{\nu }^{ab})-\nabla
_{\nu }(\delta \omega _{\mu }^{ab})$ \\ 
$\delta \nabla _{\nu }(\psi _{\rho })=\nabla _{\nu }(\delta \psi _{\rho })-%
\frac{1}{4}\delta \omega _{\nu }^{ab}\gamma _{a}\gamma _{b}\psi _{\rho }$%
\end{tabular}
\right.
\end{equation}
the field equations for the connection are 
\begin{eqnarray*}
E_{ab}^{\mu }\delta \omega _{\mu }^{ab}& = & \frac{\partial \mathcal{L}}{%
\partial \omega _{\mu }^{ab}}\delta \omega _{\mu }^{ab}=-4\delta R_{\mu \nu
}^{ab}e_{a}^{\mu }e_{b}^{\nu }e+8\bar{\psi}_{\mu }\gamma _{5}\gamma
_{a}\delta (\nabla _{\nu }\psi _{\rho })e_{\sigma }^{a}\epsilon ^{\mu \nu
\rho \sigma }=  \\ 
&& =-4\nabla _{\mu }(\delta \omega _{\nu }^{ab})e_{a}^{\mu }e_{b}^{\nu
}e+4\nabla _{\nu }(\delta \omega _{\mu }^{ab})e_{a}^{\mu }e_{b}^{\nu }e+  \\ 
&&-2\bar{\psi}_{\mu }\gamma _{5}\gamma _{a}\gamma _{r}\gamma _{s}e_{\sigma
}^{a}\epsilon ^{\mu \nu \rho \sigma }\delta \omega _{\nu }^{rs}=  \\ 
&&=-4\nabla _{\mu }(\delta \omega _{\nu }^{ab}e_{a}^{\mu }e_{b}^{\nu
}e)+4\delta \omega _{\nu }^{ab}\nabla _{\mu }e_{a}^{\mu }e_{b}^{\nu }e+  \\ 
&&+4\delta \omega _{\nu }^{ab}e_{a}^{\mu }\nabla _{\mu }e_{b}^{\nu
}e+4\delta \omega _{\nu }^{ab}e_{a}^{\mu }e_{b}^{\nu }\nabla _{\mu
}e+4\nabla _{\mu }(\delta \omega _{\nu }^{ab}e_{b}^{\mu }e_{a}^{\nu }e)+  \\ 
&&-4\delta \omega _{\nu }^{ab}\nabla _{\mu }e_{a}^{\nu }e_{b}^{\mu
}e-4\delta \omega _{\nu }^{ab}\nabla _{\mu }e_{b}^{\mu }e_{a}^{\nu
}e-4\delta \omega _{\nu }^{ab}e_{b}^{\mu }e_{a}^{\nu }\nabla _{\mu }e+  \\ 
&&-2\bar{\psi}_{\mu }\gamma _{5}\gamma _{a}\gamma _{r}\gamma _{s}e_{\sigma
}^{a}\epsilon ^{\mu \nu \rho \sigma }\delta \omega _{\nu }^{rs}=-8(\nabla
_{\mu }\delta \omega _{\nu }^{ab}e_{a}^{\mu }e_{b}^{\nu }e)+  \\ 
&&+8\delta \omega _{\nu }^{ab}(\nabla _{\mu }e_{a}^{\mu }e_{b}^{\nu
}e+e_{a}^{\mu }\nabla _{\mu }e_{b}^{\nu }e+e_{a}^{\mu }e_{b}^{\nu }\nabla
_{\mu }e)+  \\ 
&&-2\bar{\psi}_{\mu }\gamma _{5}\gamma _{a}\gamma _{r}\gamma _{s}e_{\sigma
}^{a}\epsilon ^{\mu \nu \rho \sigma }\delta \omega _{\nu }^{rs}=-8(\nabla
_{\mu }\delta \omega _{\nu }^{ab}e_{a}^{\mu }e_{b}^{\nu }e)+A+B
\end{eqnarray*}
where we have set 
\begin{equation}
A=8\delta \omega _{\nu }^{ab}(\nabla _{\mu }e_{a}^{\mu }e_{b}^{\nu
}e+e_{a}^{\mu }\nabla _{\mu }e_{b}^{\nu }e+e_{a}^{\mu }e_{b}^{\nu }\nabla
_{\mu }e)
\end{equation}
and 
\begin{equation}
B=-2\bar{\psi}_{\mu }\gamma _{5}\gamma _{a}\gamma _{r}\gamma _{s}e_{\sigma
}^{a}\epsilon ^{\mu \nu \rho \sigma }\delta \omega _{\nu }^{rs}
\end{equation}
Let us now evaluate the covariant derivative of $e_{a}^{\mu }$: 
\[
\nabla _{\mu }(\delta _{\sigma }^{\nu })=\nabla _{\mu }(e_{c}^{\nu
}e_{\sigma }^{c})=\nabla _{\mu }e_{a}^{\nu }e_{\sigma }^{a}+e_{c}^{\nu
}\nabla _{\mu }e_{\sigma }^{c}=0 
\]
Therefore 
\begin{equation}
\nabla _{\mu }e_{a}^{\nu }=-e_{a}^{\sigma }e_{d}^{\nu }\nabla _{\mu
}e_{\sigma }^{d}  \label{derivata covariante vielbein}
\end{equation}
A similar formula holds for the covariant derivative of $e$: 
\begin{equation}
\nabla _{\mu }e=\nabla _{\mu }(dete_{\sigma }^{a})=ee_{a}^{\sigma }\nabla
_{\mu }e_{\sigma }^{a}  \label{derivata covariante determinante vielbein}
\end{equation}
By substituting (\ref{derivata covariante vielbein}) and (\ref{derivata
covariante determinante vielbein}) in $A$, we find: 
\begin{eqnarray}
A &=&8\delta \omega _{\nu }^{ab}(-e_{b}^{\sigma }e_{d}^{\nu }\nabla _{\mu
}e_{\sigma }^{d}e_{a}^{\mu }e-e_{a}^{\sigma }e_{d}^{\mu }\nabla _{\mu
}e_{\sigma }^{d}e_{b}^{\nu }e+e_{a}^{\mu }e_{b}^{\nu }\nabla _{\mu }e)= 
\nonumber \\
&=&8\delta \omega _{\nu }^{ab}[-e\nabla _{\mu }e_{\sigma }^{d}(e_{b}^{\sigma
}e_{d}^{\nu }e_{a}^{\mu }+e_{a}^{\sigma }e_{d}^{\mu }e_{b}^{\nu }-e_{a}^{\mu
}e_{b}^{\nu }e_{d}^{\sigma })]=  \nonumber \\
&=&8\delta \omega _{\nu }^{ab}[-e\nabla _{\mu }e_{\sigma }^{d}(e_{b}^{\sigma
}e_{d}^{\nu }e_{a}^{\mu }+2e_{a}^{\sigma }e_{d}^{\mu }e_{b}^{\nu })]
\end{eqnarray}
where we have used also the skewsymmetry of $\nabla _{\mu }e_{\sigma }^{d}$ in $%
\mu $ and $\sigma $, given by the skewsymmetry of $\omega _{\nu }^{ab}$ in $%
a $ and $b$.

	As far as $B$ is concerned, we use the formulas 
\begin{equation}
\gamma _{5}\gamma _{a}\gamma _{r}\gamma _{s}=i\gamma ^{d}\epsilon
_{rsad}+2\gamma _{5}\eta _{d[a}\gamma _{b]}  \label{anticommutatore gamma}
\end{equation}
and 
\begin{equation}
e_{\sigma }^{a}\epsilon ^{\mu \nu \rho \sigma }=e_{l}^{\mu }e_{m}^{\nu
}e_{n}^{\rho }\epsilon ^{lmna}
\end{equation}
which give 
\begin{eqnarray}
B &=&2\bar{\psi}_{\mu }\gamma _{5}\gamma _{a}\gamma _{r}\gamma _{s}e_{\sigma
}^{a}\epsilon ^{\mu \nu \rho \sigma }\delta \omega _{\nu }^{rs}=-2i\bar{\psi}%
_{\mu }\gamma ^{d}\psi _{\rho }\epsilon _{rsad}\epsilon ^{\mu \nu \rho
\sigma }e_{\sigma }^{a}\delta \omega _{\nu }^{rs}=  \nonumber \\
&=&-4ei(\bar{\psi}_{\mu }\gamma ^{\rho }\psi _{\rho }e_{a}^{\mu }e_{b}^{\nu
}-\bar{\psi}_{\mu }\gamma ^{\nu }\psi _{\rho }e_{a}^{\mu }e_{b}^{\rho }-\bar{%
\psi}_{\mu }\gamma ^{\mu }\psi _{\rho }e_{a}^{\nu }e_{b}^{\rho })\delta
\omega _{\nu }^{ab}=  \nonumber \\
&=&-4ei(2\bar{\psi}_{\mu }\gamma ^{\rho }\psi _{\rho }e_{a}^{\mu }e_{b}^{\nu
}-\bar{\psi}_{\mu }\gamma ^{\nu }\psi _{\rho }e_{a}^{\mu }e_{b}^{\rho
})\delta \omega _{\nu }^{ab}
\end{eqnarray}
So we conclude that
\begin{eqnarray}
E_{ab}^{\nu }\delta \omega _{\nu }^{ab}&=& -8(\nabla _{\mu }\delta \omega
_{\nu }^{ab})e_{a}^{\mu }e_{b}^{\nu }e+8\delta \omega _{\nu }^{ab}[-e\nabla
_{\mu }e_{\sigma }^{d}(e_{b}^{\sigma }e_{d}^{\nu }e_{a}^{\mu
}+2e_{a}^{\sigma }e_{d}^{\mu }e_{b}^{\nu })]+ \nonumber \\ 
&&-4ei\delta \omega _{\nu }^{ab}(2\bar{\psi}_{\mu }\gamma ^{\rho }\psi
_{\rho }e_{a}^{\mu }e_{b}^{\nu }-\bar{\psi}_{\mu }\gamma ^{\nu }\psi _{\rho
}e_{a}^{\mu }e_{b}^{\rho })=0
\end{eqnarray}
This formula leads to the required field equations for the connection 
\begin{equation}
E_{ab}^{\nu }=\nabla _{[\mu }e_{\sigma ]}^{d}(e_{b}^{\sigma }e_{d}^{\nu
}e_{a}^{\mu }+2e_{a}^{\sigma }e_{d}^{\mu }e_{b}^{\nu })+(2\bar{\psi}_{[\mu
}\gamma ^{\sigma }\psi _{\sigma ]}e_{a}^{\mu }e_{b}^{\nu }-\bar{\psi}_{[\mu
}\gamma ^{\nu }\psi _{\sigma ]}e_{a}^{\mu }e_{b}^{\sigma })=0
\label{equazioni connessione}
\end{equation}
which can be recast as 
\begin{equation}
(T_{[\mu \sigma ]}^{\,d}\delta _{\rho }^{\sigma }-2T_{[\mu \sigma
]}^{\,d}\delta _{\mu }^{\sigma })e_{a}^{\mu }e_{d}^{\nu }e_{b}^{\rho }=0
\end{equation}
where 
\begin{equation}
T_{[\mu \sigma ]}^{\,d}=\nabla _{[\mu }e_{\sigma ]}^{d}-\frac{i}{2}\bar{\psi}%
_{[\mu }\gamma ^{d}\psi _{\sigma ]}  \label{supertorsione antisimmetrizzata}
\end{equation}
is the \textsl{supertorsion} of the connection $\omega $. Now, if we require
the supertorsion to vanish, the field equations (\ref{equazioni connessione}%
) are identically satisfied, as expected. For this reason, at the beginning we have assumed the constraint of null torsion. This will simplify the treatment: in the condition of
covariance of the Lagrangian, the terms containing the field equations of
the connection will not contribute.

\section{Transformation of the connection under supersymmetries}

Since the connection is a function on $e_{\mu }^{a}$ and $\psi _{\mu }$,
the action of the supersymmetries on $e_{\mu }^{a}$ and $\psi _{\mu }$
induces the action on $\omega $. First of all, we recall the null torsion
constraint (\ref{torsione nulla}) 
\begin{equation}
T_{[\mu \nu ]}^{\,a}=\nabla _{[\mu }e_{\nu ]}^{a}-\frac{i}{2}\bar{\psi}%
_{[\mu }\gamma ^{a}\psi _{\nu ]}=0
\end{equation}
which can be recast as 
\begin{equation}
d_{[\mu }e_{\nu ]}^{a}+\omega _{~b[\mu }^{a\cdot }e_{\nu ]}^{b}-\Gamma
_{[\mu \nu ]}^{\lambda }e_{\lambda }^{a}-\frac{i}{2}\bar{\psi}_{[\mu }\gamma
^{a}\psi _{\nu ]}=0
\end{equation}
By varying this with respect to supersymmetries, one gets
\begin{eqnarray}
&&d_{[\mu }\delta e_{\nu ]}^{a}+\delta \omega _{~b[\mu }^{a\cdot }e_{\nu
]}^{b}+\omega _{~b[\mu }^{a\cdot }\delta e_{\nu ]}^{b}-\frac{i}{2}\delta 
\bar{\psi}_{[\mu }\gamma ^{a}\psi _{\nu ]}-\frac{i}{2}\bar{\psi}_{[\mu
}\gamma ^{a}\delta \psi _{\nu ]}=  \nonumber \\
&&=\nabla _{[\mu }\delta e_{\nu ]}^{a}+\delta \omega _{~b[\mu }^{a\cdot
}e_{\nu ]}^{b}-i\delta \bar{\psi}_{[\mu }\gamma ^{a}\psi _{\nu ]}=0
\end{eqnarray}
where we have used the symmetry of the Christoffel symbol in the lower
indices, the Majorana flip identity 
\begin{equation}
\bar{\psi}_{[\mu }\gamma ^{a}\delta \psi _{\nu ]}=-\delta \bar{\psi}_{[\nu
}\gamma ^{a}\psi _{\mu ]}
\end{equation}
and the fact that this bilinear is skewsymmetric in $\mu $ and $\nu $.
Therefore
\begin{eqnarray*}
\delta \omega _{~b[\mu }^{a\cdot }e_{\nu ]}^{b} &=& -\nabla _{[\mu }\delta
e_{\nu ]}^{a}+i\delta \bar{\psi}_{[\mu }\gamma ^{a}\psi _{\nu ]}=-i\nabla
_{[\mu }(\bar{\varepsilon}\gamma ^{a}\psi _{\nu ]})+i\nabla _{[\mu }\bar{%
\varepsilon}\gamma ^{a}\psi _{\nu ]}= \\ 
&&=-i\nabla _{[\mu }\bar{\varepsilon}\gamma ^{a}\psi _{\nu ]}-i\bar{%
\varepsilon}\gamma ^{a}\nabla _{[\mu }\psi _{\nu ]}+i\nabla _{[\mu }\bar{%
\varepsilon}\gamma ^{a}\psi _{\nu ]}=-i\bar{\varepsilon}\gamma ^{a}\nabla
_{[\mu }\psi _{\nu ]}
\end{eqnarray*}
Now, to isolate $\delta \omega _{~b\mu }^{a\cdot }$, we define 
\begin{equation}
\delta \omega _{ab[\mu }^{\,\cdot \,\cdot }e_{\nu ]}^{b}:=-(k_{a\mu \nu
}-k_{a\nu \mu })
\end{equation}
and 
\begin{equation}
-i\bar{\varepsilon}\gamma _{a}\nabla _{[\mu }\psi _{\nu ]}:=c_{a\mu \nu }
\end{equation}
An easy calculation shows that 
\begin{equation}
c_{a\mu \nu }+c_{\mu \nu a}-c_{\nu a\mu }=-k_{a\mu \nu }
\end{equation}
which is 
\begin{equation}
\delta \omega _{ab\nu }^{\,\cdot \,\cdot }e_{\mu }^{b}=i(\bar{\varepsilon}%
\gamma _{a}\nabla _{[\mu }\psi _{\nu ]}+\bar{\varepsilon}\gamma _{\mu
}\nabla _{[\nu }\psi _{a]}-\bar{\varepsilon}\gamma _{\nu }\nabla _{[a}\psi
_{\mu ]})
\end{equation}
or 
\[
\delta \omega _{ab\nu }^{\,\cdot \,\cdot }=i(\bar{\varepsilon}\gamma
_{a}\nabla _{[\mu }\psi _{\nu ]}+\bar{\varepsilon}\gamma _{\mu }\nabla
_{[\nu }\psi _{a]}-\bar{\varepsilon}\gamma _{\nu }\nabla _{[a}\psi _{\mu
]})e_{b}^{\mu } 
\]
Changing $\mu $ in $\rho $, $\nu $ in $\mu $ and raising the indices $a$ and 
$b$ gives: 
\begin{equation}
\delta \omega _{\mu }^{ab}=i(\bar{\varepsilon}\gamma _{\rho }\nabla _{[\nu
}\psi _{\mu ]}+\bar{\varepsilon}\gamma _{\nu }\nabla _{[\mu }\psi _{\rho ]}+%
\bar{\varepsilon}\gamma _{\mu }\nabla _{[\nu }\psi _{\rho ]})e^{\rho
a}e^{b\nu }
\end{equation}
which can be recast as 
\begin{equation}
\delta \omega _{\mu }^{ab}=-i(\bar{\varepsilon}\gamma _{\mu }\nabla _{\rho
}\psi _{\nu }+\bar{\varepsilon}\gamma _{\rho }\nabla _{\mu }\psi _{\nu }-%
\bar{\varepsilon}\gamma _{\rho }\nabla _{\nu }\psi _{\mu })e^{\,\rho
[a}e^{b]\nu }  \label{supersim conness}
\end{equation}

\section{Covariance of the Lagrangian}

Our next purpose is to check the covariance of the Lagrangian (\ref
{Lagrangiana}) with respect to automorphisms of $\Sigma $ and
supersymmetries.

\subsection{Covariance with respect to automorphisms}

The infinitesimal generator of automorphisms on $\Sigma $ is the right
invariant vector field: 
\[
\Xi =\xi ^{\mu }(x)\partial _{\mu }+\xi _{\left( \upsilon \right)
}^{ab}(x)\sigma _{ab} 
\]
The flow is defined by: 
\begin{equation}
\left\{ 
\begin{array}{ll}
x^{\prime }=\phi (x) &  \\ 
g^{\prime }=S(x)\cdot g & 
\end{array}
\right.
\end{equation}
therefore the components of $\Xi $ can be written as: 
\begin{equation}
\xi ^{\mu }=\dot{\phi}^{\mu }(x),\xi _{\left( \upsilon \right) }^{ab}=%
\dot{S}_{\beta }^{\alpha }\partial _{\alpha }^{\beta }\rho _{c}^{[a}(e)\eta
^{c]b}
\end{equation}
The infinitesimal generator $\Xi $ acts on the fields, by means of the
corresponding Lie derivatives, as follows (see Chapter 3): 
\begin{eqnarray}
&&\pounds _{\Xi }e_{\mu }^{a}=\nabla _{\mu }\xi ^{\nu }e_{\nu }^{a}+\nabla
_{\nu }e_{\mu }^{a}\xi ^{\nu }-e_{\mu }^{b}\xi _{(\upsilon )b}^{a}
\label{derivata di Lie vielbein}\\
&&\pounds _{\Xi }\psi _{\mu }=\xi ^{\rho }\nabla _{\rho }\psi _{\mu }+\nabla
_{\mu }\xi ^{\nu }\psi _{\nu }-\frac{1}{8}[\gamma _{a},\gamma _{b}]\psi
_{\mu }\xi _{(\upsilon )}^{\,{ab}}  \label{derivata di Lie gravitino}\\
&&\pounds _{\Xi }\omega _{\mu }^{ab}=R_{\nu \mu }^{ab}\xi ^{\nu }+\nabla _{\mu
}\xi _{(\upsilon )}^{\,ab}  \label{derivata di Lie connessione}
\end{eqnarray}
where 
\begin{equation}
\xi _{(\upsilon )}^{\,{ab}}=\xi ^{ab}+\omega _{\mu }^{ab}\xi ^{\mu }
\label{parte verticale}
\end{equation}
is the vertical part of $\Xi $. We remark that from (\ref{derivata di Lie
vielbein}) we obtain: 
\begin{equation}
\pounds _{\Xi }e_{a}^{\mu }=\nabla _{\nu }e_{a}^{\mu }\xi ^{\nu }-\nabla
_{\nu }\xi ^{\mu }e_{a}^{\nu }+e_{b}^{\mu }\xi _{(\upsilon )a}^{\,b}
\label{derivata di Lie vielbein inversa}
\end{equation}
Under (\ref{derivata di Lie vielbein}), (\ref{derivata di Lie gravitino}), (%
\ref{derivata di Lie connessione}) and (\ref{derivata di Lie vielbein inversa}%
), the two Lagrangian densities $\mathcal{L_{H}}$ and $\mathcal{L_{S}}$ are
separately covariant.

\subsubsection{Hilbert Lagrangian}

For the Hilbert Lagrangian, i.e.
\[
\mathcal{L_{H}}:=-4R_{\mu \nu }^{ab}e_{a}^{\mu }e_{b}^{\nu }e 
\]
the covariance condition is equivalent to the following identity: 
\begin{equation}
d_{\rho }(\mathcal{L_{H}}\xi ^{\rho })=p_{\mu }^{a}\pounds _{\Xi }e_{a}^{\mu
}+p_{ab}^{\mu \nu }\pounds _{\Xi }R_{\mu \nu }^{ab}
\label{covarianza automorfismi Hilbert}
\end{equation}
with the \textsl{naive momenta} of $\mathcal{L_{H}}$ 
\begin{eqnarray*}
&&p_{\mu }^{a}:=\frac{\partial \mathcal{L_{H}}}{\partial e_{a}^{\mu }}
=2e(R_{\mu }^{a}-\frac{1}{2}Re_{\mu }^{a}) \\
&&p_{ab}^{\mu \nu }:=\frac{\partial \mathcal{L_{H}}}{\partial R_{\mu \nu
}^{ab}}=ee_{[a}^{\mu }e_{b]}^{\nu }
\end{eqnarray*}
We remark that Eq.(\ref{covarianza automorfismi Hilbert}) can be written in
standard notation as 
\begin{equation}
\delta {\mathcal{L_{H}}}=p_{a}^{\mu }\delta e_{\mu }^{a}+p_{ab}^{\mu \nu
}\delta R_{\mu \nu }^{ab}=d_{\rho }(\mathcal{L_{H}}\xi ^{\rho })
\end{equation}
Now we want to evaluate the right hand side of (\ref{covarianza automorfismi
Hilbert}).
By applying the Lie derivative along $\Xi $ to the Riemann tensor 
\[
R_{\mu \nu }^{ab}=d_{\mu }\omega _{\nu }^{ab}-d_{\nu }\omega _{\mu
}^{ab}+\omega _{{\;\;c}\mu }^{a}\omega _{\nu }^{cb}-\omega _{{\;\;c}\nu
}^{a}\omega _{\mu }^{cb} 
\]
we obtain: 
\begin{equation}
\pounds _{\Xi }R_{\mu \nu }^{ab}=\nabla _{\mu }\pounds _{\Xi }\omega _{\nu
}^{ab}-\nabla _{\nu }\pounds _{\Xi }\omega _{\mu }^{ab}
\end{equation}
Now, using 
\begin{equation}
\lbrack \nabla _{\mu },\nabla _{\nu }]\xi _{(v)}^{\,ab}=R_{~{c\mu \nu }%
}^{a\cdot }\xi _{(v)}^{\,cb}+R_{c~{\mu \nu }}^{\cdot \,{b}}\xi _{(v)}^{\,ac}
\end{equation}
and the skewsymmetry of $R_{~{\mu \nu }}^{cb}$ in $b$ and $c$, we obtain the
following identity: 
\begin{equation}
2p_{ab}^{\mu \nu }\nabla _{\mu }\nabla _{\nu }\xi _{(v)}^{~ab}=p_{ab}^{\mu
\nu }[\nabla _{\mu },\nabla _{\nu }]\xi _{(v)}^{\,ab}=0
\end{equation}
For the skewsymmetry of $\xi _{(v)}^{\,{ab}}$ , and the definition of $%
p_{\mu }^{a}$,
\[
p_{\mu }^{a}e_{b}^{\mu }\xi _{(v)a}^{\,b}=0 
\]
So the r.h.s. of (\ref{covarianza automorfismi Hilbert}) becomes 
\begin{equation}
p_{\mu }^{a}\nabla _{\rho }e_{a}^{\mu }\xi ^{\rho }-p_{\mu }^{a}\nabla _{\nu
}\xi ^{\mu }e_{a}^{\nu }+2p_{ab}^{\mu \nu }\nabla _{\mu }R_{\rho \nu
}^{ab}\xi ^{\rho }+2p_{ab}^{\mu \nu }R_{\rho \nu }^{ab}\nabla _{\mu }\xi
^{\rho }
\end{equation}
Now we analyze the terms separately:
\begin{eqnarray}
p_{\mu }^{a}\nabla _{\rho }e_{a}^{\mu }\xi ^{\rho } &=& e(2R_{\mu
}^{a}-Re_{\mu }^{a})\xi ^{\rho }\nabla _{\rho }e_{a}^{\mu }= \nonumber \\ 
&=&e\xi ^{\rho }(-2R_{\mu }^{a}e_{a}^{\sigma }e_{d}^{\mu }\nabla _{\rho
}e_{\sigma }^{d}+Re_{\mu }^{a}e_{a}^{\sigma }e_{d}^{\mu }\nabla _{\rho
}e_{\sigma }^{d})= \nonumber \\ 
&=&e\xi ^{\rho }(-2R_{\mu }^{\sigma }e_{d}^{\mu }\nabla _{\rho }e_{\sigma
}^{d}+R\delta _{\mu }^{\sigma }e_{d}^{\mu }\nabla _{\rho }e_{\sigma }^{d})=\nonumber
\\ 
&=&e\xi ^{\rho }(-2R_{\mu }^{\sigma }e_{d}^{\mu }\nabla _{\rho }e_{\sigma
}^{d}+Re_{d}^{\sigma }\nabla _{\rho }e_{\sigma }^{d})
\label{covarianza 1}
\end{eqnarray}
where we used Eq.(\ref{derivata covariante vielbein}).
Moreover, 
\begin{equation}
-p_{\mu }^{a}\nabla _{\nu }\xi ^{\mu }e_{a}^{\nu }=-2e(R_{\mu }^{a}-Re_{\mu
}^{a})e_{a}^{\nu }\nabla _{\nu }\xi ^{\mu }=-2eR_{\mu }^{\nu }\nabla _{\nu
}\xi ^{\mu }+eR\delta _{\mu }^{\nu }\nabla _{\nu }\xi ^{\mu }
\label{covarianza 2}
\end{equation}
and
\begin{equation}
2p_{ab}^{\mu \nu }R_{\rho \nu }^{ab}\nabla _{\mu }\xi ^{\rho }=2ee_{a}^{\mu
}e_{b}^{\nu }R_{\rho \nu }^{ab}\nabla _{\mu }\xi ^{\rho }=2ee_{a}^{\mu
}R_{\rho }^{a}\nabla _{\mu }\xi ^{\rho }=2eR_{\mu }^{\nu }\nabla _{\nu }\xi
^{\mu }  \label{covarianza 3}
\end{equation}
Before studying the remaining term, we recall the Bianchi identities for the
curvature $R^{ab}$ of a generic principal connection $\Gamma $ (Eq.(\ref
{bianchi})): 
\begin{equation}
\nabla _{\mu }R_{\rho \nu }^{ab}+\nabla _{\rho }R_{\nu \mu }^{ab}+\nabla
_{\nu }R_{\mu \rho }^{ab}=0  \label{identità di Bianchi}
\end{equation}
which gives 
\begin{equation}
2e_{a}^{\mu }e_{b}^{\nu }\nabla _{\mu }R_{\rho \nu }^{ab}=e_{a}^{\mu
}e_{b}^{\nu }\nabla _{\rho }R_{\mu \nu }^{ab}
\end{equation}
Hence 
\begin{eqnarray}
2p_{ab}^{\mu \nu }\nabla _{\mu }R_{\rho \nu }^{ab}\xi ^{\rho }
&=&2ee_{a}^{\mu }e_{b}^{\nu }\nabla _{\mu }R_{\rho \nu }^{ab}\xi ^{\rho
}=ee_{a}^{\mu }e_{b}^{\nu }\nabla _{\rho }R_{\mu \nu }^{ab}\xi ^{\rho }= 
\nonumber \\
&=&\nabla _{\rho }(\mathcal{L_{H}}\xi ^{\rho })-e\nabla _{\rho }e_{\sigma
}^{a}e_{a}^{\sigma }R\xi ^{\rho }+2eR_{\mu }^{\sigma }e_{d}^{\mu }\nabla
_{\rho }e_{\sigma }^{d}\xi ^{\rho }-eR\nabla _{\rho }\xi ^{\rho }=  \nonumber
\\
&=&\nabla _{\rho }(\mathcal{L_{H}}\xi ^{\rho })-eR\nabla _{\rho }\xi ^{\rho
}+  \nonumber \\
&&+e\xi ^{\rho }(-\nabla _{\rho }e_{\sigma }^{a}e_{a}^{\sigma }R+eR_{\mu
}^{\sigma }e_{d}^{\mu }\nabla _{\rho }e_{\sigma }^{d})  \label{covarianza 4}
\end{eqnarray}
This equation has been evaluated integrating by parts $ee_{a}^{\mu
}e_{b}^{\nu }\nabla _{\rho }R_{\mu \nu }^{ab}\xi ^{\rho }$, and then using
the formulas (\ref{derivata covariante vielbein}) and (\ref{derivata
covariante determinante vielbein}).

	By adding now (\ref{covarianza 1}), (\ref{covarianza 2}), (\ref{covarianza 3}%
), (\ref{covarianza 4}), we verify the covariance of the Hilbert Lagrangian,
i.e. the condition (\ref{covarianza automorfismi Hilbert}) is satisfied.

\subsubsection{Spin Lagrangian}

As long as the spin Lagrangian density (\ref{Lagrangiana spine}), i.e. 
\[
\mathcal{L_{S}}:=8\bar{\psi}_{\mu }\gamma _{5}\gamma _{a}\nabla _{\nu }\psi
_{\rho }e_{\sigma }^{a}\epsilon ^{\mu \nu \rho \sigma } 
\]
is regarded, the covariance condition is 
\begin{equation}
d_{\rho }(\mathcal{L_{S}}\xi ^{\rho })=\pi _{a}^{\sigma }(\pounds _{\Xi
}e_{\sigma }^{a})+(\pounds _{\Xi }\bar{\psi}_{\mu })\pi ^{\mu }+\pi ^{\nu
\rho }\pounds _{\Xi }(\nabla _{\nu }\psi _{\rho })
\label{covarianza automorfismi spine}
\end{equation}
where the momenta are
\begin{equation}
\left\{ 
\begin{tabular}{l}
$\pi _{a}^{\sigma }:=\frac{\partial \mathcal{L_{S}}}{\partial e_{\sigma }^{a}%
}=\bar{\psi}_{\mu }\gamma _{5}\gamma _{a}\nabla _{\nu }\psi _{\rho }\epsilon
^{\mu \nu \rho \sigma }$ \\ 
$\pi ^{\mu }:=\frac{\partial \mathcal{L_{S}}}{\partial \bar{\psi}_{\mu }}%
=\gamma _{5}\gamma _{a}\nabla _{\nu }\psi _{\rho }e_{\sigma }^{a}\epsilon
^{\mu \nu \rho \sigma }$ \\ 
$\pi ^{\nu \rho }:=\frac{\partial \mathcal{L_{S}}}{\partial (\nabla _{\nu
}\psi _{\rho })}=\bar{\psi}_{\mu }\gamma _{5}\gamma _{a}e_{\sigma
}^{a}\epsilon ^{\mu \nu \rho \sigma }$%
\end{tabular}
\right.
\end{equation}
and the Lie derivatives hold as
\begin{equation}
\left\{ 
\begin{tabular}{l}
$\pounds _{\Xi }e_{\sigma }^{a}=\nabla _{\sigma }\xi ^{\nu }e_{\nu
}^{a}-e_{\sigma }^{b}\xi _{(v)b}^{a}$ \\ 
$\pounds _{\Xi }\bar{\psi}_{\mu }=\xi ^{\lambda }\nabla _{\lambda }\bar{\psi}%
_{\mu }+\nabla _{\mu }\xi ^{\lambda }\bar{\psi}_{\lambda }+\frac{1}{8}\bar{%
\psi}_{\mu }[\gamma _{l},\gamma _{m}]\xi _{(v)}^{lm}$ \\ 
$\pounds _{\Xi }(\nabla _{\nu }\psi _{\rho })=\xi ^{\lambda }\nabla
_{\lambda }\nabla _{\nu }\psi _{\rho }+\nabla _{\nu }\xi ^{\lambda }\nabla
_{\lambda }\psi _{\rho }+\nabla _{\rho }\xi ^{\lambda }\nabla _{\nu }\psi
_{\lambda }-\frac{1}{4}\gamma _{l}\gamma _{m}\xi _{(v)}^{lm}\nabla _{\nu
}\psi _{\rho }$%
\end{tabular}
\right.
\end{equation}
By substituting these expressions in (\ref{covarianza automorfismi spine}
), we obtain:
\begin{eqnarray}
d_{\rho }(\mathcal{L}_{\mathcal{S}}\xi ^{\rho })&=& \bar{\psi}_{\mu
}\gamma _{5}\gamma _{a}\nabla _{\nu }\psi _{\rho }\nabla _{\sigma }\xi
^{\lambda }e_{\lambda }^{a}\epsilon ^{\mu \nu \rho \sigma }-\xi ^{\lambda
}\nabla _{\lambda }\bar{\psi}_{\mu }\gamma _{5}\gamma _{a}\nabla _{\nu }\psi
_{\rho }e_{\sigma }^{a}\epsilon ^{\mu \nu \rho \sigma }+ \nonumber \\ 
&&+\nabla _{\mu }\xi ^{\lambda }\bar{\psi}_{\lambda }\gamma _{5}\gamma
_{a}\nabla _{\nu }\psi _{\rho }e_{\sigma }^{a}\epsilon ^{\mu \nu \rho \sigma
}+\bar{\psi}_{\mu }\gamma _{5}\gamma _{a}e_{\sigma }^{a}\nabla _{\nu }\xi
^{\lambda }\nabla _{\lambda }\psi _{\rho }\epsilon ^{\mu \nu \rho \sigma }+ \nonumber
\\ 
&&+\bar{\psi}_{\mu }\gamma _{5}\gamma _{a}e_{\sigma }^{a}\xi ^{\lambda
}\nabla _{\lambda }\nabla _{\nu }\psi _{\rho }\epsilon ^{\mu \nu \rho \sigma
}+\bar{\psi}_{\mu }\gamma _{5}\gamma _{a}e_{\sigma }^{a}\nabla _{\rho }\xi
^{\lambda }\nabla _{\nu }\psi _{\lambda }\epsilon ^{\mu \nu \rho \sigma }+ \nonumber
\\ 
&&-\bar{\psi}_{\mu }\gamma _{5}\gamma _{a}\nabla _{\nu }\psi _{\rho }\xi
_{(v)b}^{:a}e_{\sigma }^{b}\epsilon ^{\mu \nu \rho \sigma }+\bar{\psi}_{\mu
}\gamma _{l}\gamma _{m}\gamma _{5}\gamma _{a}\nabla _{\nu }\psi _{\rho }\xi
_{(v)}^{\,lm}e_{\sigma }^{a}\epsilon ^{\mu \nu \rho \sigma }+ \nonumber \\ 
&&-\bar{\psi}_{\mu }\gamma _{5}\gamma _{a}\gamma _{l}\gamma _{m}\nabla
_{\nu }\psi _{\rho }\xi _{(v)}^{\,lm}e_{\sigma }^{a}\epsilon ^{\mu \nu \rho
\sigma }
\label{covarianza spine estesa}
\end{eqnarray}
The addenda depending on the vertical field vanish altogether, provided the
following property of the gamma matrices: 
\begin{equation}
\lbrack \gamma _{l}\gamma _{m},\gamma _{5}\gamma _{a}]=-2\eta _{am}\gamma
_{l}\gamma _{5}+2\eta _{al}\gamma _{m}\gamma _{5}
\end{equation}
Eq.(\ref{covarianza spine estesa}) then deduces to: 
\begin{eqnarray*}
d_{\rho }(\mathcal{L_{S}}\xi ^{\rho }) &=&\bar{\psi}_{\mu }\gamma _{5}\gamma
_{a}\nabla _{\nu }\psi _{\rho }\nabla _{\sigma }\xi ^{\lambda }e_{\lambda
}^{a}\epsilon ^{\mu \nu \rho \sigma }+\xi ^{\lambda }\nabla _{\lambda }\bar{%
\psi}_{\mu }\gamma _{5}\gamma _{a}\nabla _{\nu }\psi _{\rho }e_{\sigma
}^{a}\epsilon ^{\mu \nu \rho \sigma } \\
&&+\nabla _{\mu }\xi ^{\lambda }\bar{\psi}_{\lambda }\gamma _{5}\gamma
_{a}\nabla _{\nu }\psi _{\rho }e_{\sigma }^{a}\epsilon ^{\mu \nu \rho \sigma
}+\bar{\psi}_{\mu }\gamma _{5}\gamma _{a}e_{\sigma }^{a}\nabla _{\nu }\xi
^{\lambda }\nabla _{\lambda }\psi _{\rho }\epsilon ^{\mu \nu \rho \sigma } \\
&&+\bar{\psi}_{\mu }\gamma _{5}\gamma _{a}e_{\sigma }^{a}\xi ^{\lambda
}\nabla _{\lambda }\nabla _{\nu }\psi _{\rho }\epsilon ^{\mu \nu \rho \sigma
}+\bar{\psi}_{\mu }\gamma _{5}\gamma _{a}e_{\sigma }^{a}\nabla _{\rho }\xi
^{\lambda }\nabla _{\nu }\psi _{\lambda }\epsilon ^{\mu \nu \rho \sigma }
\end{eqnarray*}
Now we want to calculate the right hand side of this equation. First of all
we integrate by parts the term $\bar{\psi}_{\mu }\gamma _{5}\gamma
_{a}e_{\sigma }^{a}\xi ^{\lambda }\nabla _{\lambda }\nabla _{\nu }\psi
_{\rho }\epsilon ^{\mu \nu \rho \sigma }$. Then we recall that
\begin{eqnarray}
\pounds _{\Xi }\epsilon ^{\mu \nu \rho \sigma } &=& \nabla _{\alpha }\xi
^{\mu }\epsilon ^{\alpha \nu \rho \sigma }+\nabla _{\alpha }\xi ^{\nu
}\epsilon ^{\mu \alpha \rho \sigma }+\nabla _{\alpha }\xi ^{\rho }\epsilon
^{\mu \nu \alpha \sigma }+ \nonumber \\ 
&&+\nabla _{\alpha }\xi ^{\sigma }\epsilon ^{\mu \nu \rho \alpha }-\nabla
_{\alpha }\xi ^{\alpha }\epsilon ^{\mu \nu \rho \sigma }=0
\label{derivata di lie epsilon}
\end{eqnarray}
and consequently the r.h.s. of (\ref{covarianza spine estesa}) becomes: 
\[
\nabla _{\mu }(\xi ^{\mu }\mathcal{L_{S}}) 
\]
The quantity $\xi ^{\mu }\mathcal{L_{S}}$ is a vector density, therefore 
\[
\nabla _{\mu }(\xi ^{\mu }\mathcal{L_{S}})=d_{\mu }(\xi ^{\mu }\mathcal{L_{S}%
}) 
\]
which proves the claim (\ref{covarianza automorfismi spine}).
This allows us to conclude that the matter Lagrangian (\ref{Lagrangiana
spine}) is covariant.
	Recalling that for the gravity Lagrangian (\ref{Lagrangiana di Hilbert}) we
obtain the same result, the covariance of the Rarita-Schwinger Lagrangian (%
\ref{Lagrangiana}) with respect to automorphisms on $\Sigma $ has been
recovered.

\subsection{Covariance under supersymmetries}

In this case the generator is no longer a vector, but an anticommuting
Majorana $1/2$ spinor $\varepsilon $. This generator acts
on the vielbein, on the gravitino and on the connection as follows (we recall Eq.(\ref{supersimmetrie}) and (\ref{supersim conness})): 
\begin{equation}
\left\{ 
\begin{tabular}{l}
\vspace{0.1cm}$\delta \psi _{\mu }=\nabla _{\mu }\varepsilon $ \\ 
\vspace{0.1cm}$\delta e_{\mu }^{a}=i\bar{\varepsilon}\gamma ^{a}\psi _{\mu }$ \\ 
$\delta \omega _{\mu }^{ab}=-i(\bar{\varepsilon}\gamma _{\rho }\nabla
_{[\lambda }\psi _{\mu ]}+\bar{\varepsilon}\gamma _{\lambda }\nabla _{[\mu
}\psi _{\rho ]}-\bar{\varepsilon}\gamma _{\mu }\nabla _{[\rho }\psi
_{\lambda ]})e^{\rho a}e^{b\lambda }$%
\end{tabular}
\right.
\end{equation}
The Rarita--Schwinger Lagrangian is: 
\[
\mathcal{L}=-4R_{\mu \nu }^{ab}e_{a}^{\mu }e_{b}^{\nu }e+8\bar{\psi}_{\mu
}\gamma _{5}\gamma _{a}\nabla _{\nu }\psi _{\rho }e_{\sigma }^{a}\epsilon
^{\mu \nu \rho \sigma }=(\mathcal{L_{H}})+(\mathcal{L_{S}}) 
\]
The condition of covariance, similar to (\ref{covarianza automorfismi
Hilbert}) and to (\ref{covarianza automorfismi spine}), is 
\begin{equation}
\left( \frac{\partial \mathcal{L}}{\partial e_{a}^{\mu }}-e_{\sigma }^{b}%
\frac{\partial \mathcal{L}}{\partial e_{\sigma }^{b}}e_{\mu }^{a}\right)
\delta e_{a}^{\mu }+\delta \bar{\psi}_{\mu }\frac{\partial \mathcal{L}}{%
\partial \bar{\psi}_{\mu }}+\frac{\partial \mathcal{L}}{\partial (\nabla
_{\nu }\psi _{\rho })}\delta (\nabla _{\nu }\psi _{\rho })=0
\label{covarianza per supersimmetrie}
\end{equation}
where 
\begin{eqnarray*}
&&\frac{\partial \mathcal{L}}{\partial e_{a}^{\mu }}=-8e(R_{\mu }^{a}-\frac{1%
}{2}Re_{\mu }^{a}) \\
&&\frac{\partial \mathcal{L}}{\partial e_{\sigma }^{a}}=8\bar{\psi}_{\mu
}\gamma _{5}\gamma _{a}\nabla _{\nu }\psi _{\rho }\epsilon ^{\mu \nu \rho
\sigma } \\
&&\frac{\partial \mathcal{L}}{\partial \bar{\psi}_{\mu }}=8\gamma _{5}\gamma
_{a}\nabla _{\nu }\psi _{\rho }e_{\sigma }^{a}\epsilon ^{\mu \nu \rho \sigma
} \\
&&\frac{\partial \mathcal{L}}{\partial (\nabla _{\nu }\psi _{\rho })}=8\bar{%
\psi}_{\mu }\gamma _{5}\gamma _{a}e_{\sigma }^{a}\epsilon ^{\mu \nu \rho
\sigma }
\end{eqnarray*}
and 
\begin{eqnarray}
&&\delta e_{a}^{\mu }=-e_{a}^{\sigma }e_{b}^{\mu }\delta e_{\sigma
}^{b}=-ie_{a}^{\sigma }\bar{\varepsilon}\gamma ^{\mu }\psi _{\sigma } \\ 
&&\delta \bar{\psi}_{\mu }=\nabla _{\mu }\bar{\varepsilon} \\ 
&&\delta \left(\nabla _{\nu }\psi _{\rho }\right)=\nabla _{\nu }(\delta \psi _{\rho })-
\frac{1}{4}\delta \omega _{\nu }^{ab}\gamma _{a}\gamma _{b}\psi _{\rho
}=\nabla _{\nu }\nabla _{\rho }\varepsilon -\frac{1}{4}\delta \omega _{\nu
}^{ab}\gamma _{a}\gamma _{b}\psi _{\rho}
\end{eqnarray}
We remark that, given the supersymmetry transformations, the two Lagrangians
(\ref{Lagrangiana di Hilbert}) and (\ref{Lagrangiana spine}) are not
separately covariant; for this reason, Eq.(\ref{covarianza per
supersimmetrie}) refers to the entire Lagrangian of the model $\mathcal{L_H}+\mathcal{L_S}$ given
above.
As we have already observed at the end of section 2 of this chapter, the
terms in this formula containing the field equations for the connection
identically vanish for the condition of null torsion (\ref{torsione nulla}).
These appear in (\ref{covarianza per supersimmetrie}) as the naive momenta of
the Lagrangian with respect to the connection $\omega $.

	We also remark that supersymmetries are vertical transformations, i.e. 
\[
\xi ^{\mu }=0 
\]
Therefore we have to prove the following identity: 
\begin{eqnarray}
&&8i(\overline{\varepsilon }\gamma ^{a}\psi _{\sigma })\bar{\psi}_{\mu
}\gamma _{5}\gamma _{a}\nabla _{\nu }\psi _{\rho }\epsilon ^{\mu \nu \rho
\sigma }+8\bar{\psi}_{\mu }\gamma _{5}\gamma _{a}\nabla _{\nu }\nabla _{\rho
}\varepsilon e_{\sigma }^{a}\epsilon ^{\mu \nu \rho \sigma }+ \nonumber\\ 
&&+8\nabla _{\mu }\bar{\varepsilon}\gamma _{5}\gamma _{a}\nabla _{\nu }\psi
_{\rho }e_{\sigma }^{a}\epsilon ^{\mu \nu \rho \sigma }+8ie\bar{\varepsilon}%
\gamma ^{\mu }\psi _{\sigma }R_{\mu }^{a}e_{a}^{\sigma }-4ie\bar{\varepsilon}%
\gamma ^{\mu }\psi _{\mu }R=0 
\label{covarianza per supersimmetrie estesa}
\end{eqnarray}
We begin with the integration by parts of $\nabla _{\mu }\bar{\varepsilon}%
\gamma _{5}\gamma _{a}\nabla _{\nu }\psi _{\rho }e_{\sigma }^{a}\epsilon
^{\mu \nu \rho \sigma }$: 
\begin{eqnarray}
&&\nabla _{\mu }\bar{\varepsilon}\gamma _{5}\gamma _{a}\nabla _{\nu }\psi
_{\rho }e_{\sigma }^{a}\epsilon ^{\mu \nu \rho \sigma }=\nabla _{\mu }(\bar{%
\varepsilon}\gamma _{5}\gamma _{a}\nabla _{\nu }\psi _{\rho }e_{\sigma
}^{a}\epsilon ^{\mu \nu \rho \sigma })+  \nonumber \\
&&-\bar{\varepsilon}\gamma _{5}\gamma _{a}\nabla _{\mu }\nabla _{\nu }\psi
_{\rho }e_{\sigma }^{a}\epsilon ^{\mu \nu \rho \sigma }-\bar{\varepsilon}%
\gamma _{5}\gamma _{a}\nabla _{\nu }\psi _{\rho }\nabla _{\mu }e_{\sigma
}^{a}\epsilon ^{\mu \nu \rho \sigma }
\end{eqnarray}
Now we use the constraint of null torsion (\ref{torsione nulla}): 
\[
T_{[\mu \sigma ]}^{\,d}=\nabla _{[\mu }e_{\sigma ]}^{d}-\frac{i}{2}\bar{\psi}%
_{[\mu }\gamma ^{d}\psi _{\sigma ]}=0 
\]
which implies 
\begin{equation}
\nabla _{[\mu }e_{\sigma ]}^{d}=\frac{i}{2}\bar{\psi}_{[\mu }\gamma ^{d}\psi
_{\sigma ]}  \label{torsione nulla antisimmetrizzata}
\end{equation}
Provided that, multiplying $\nabla _{\mu }e_{\sigma }^{d}$ by $\epsilon
^{\mu \nu \rho \sigma }$, only $\nabla _{[\mu }e_{\sigma ]}^{d}$ survives,
Eq.(\ref{covarianza per supersimmetrie estesa}) becomes:
\begin{eqnarray}
&&8[i(\overline{\varepsilon }\gamma ^{a}\psi _{\sigma })\bar{\psi}_{\mu
}\gamma _{5}\gamma _{a}\nabla _{\nu }\psi _{\rho }\epsilon ^{\mu \nu \rho
\sigma }+\bar{\psi}_{\mu }\gamma _{5}\gamma _{a}\nabla _{\nu }\nabla _{\rho
}\varepsilon e_{\sigma }^{a}\epsilon ^{\mu \nu \rho \sigma }+ \nonumber\\ 
&&-\bar{\varepsilon}\gamma _{5}\gamma _{a}\nabla _{\mu }\nabla _{\nu }\psi
_{\rho }e_{\sigma }^{a}\epsilon ^{\mu \nu \rho \sigma }-\frac{i}{2}\bar{%
\varepsilon}\gamma _{5}\gamma _{a}\nabla _{\nu }\psi _{\rho }(\bar{\psi}%
_{\mu }\gamma ^{a}\psi _{\sigma })\epsilon ^{\mu \nu \rho \sigma }]+ \nonumber \\ 
&&+8ie\bar{\varepsilon}\gamma ^{\mu }\psi _{\sigma }R_{\mu }^{a}e_{a}^{\sigma
}-4ie\bar{\varepsilon}\gamma ^{\mu }\psi _{\mu }R=0 
\end{eqnarray}
Now we consider the sum 
\begin{equation}
i(\bar{\varepsilon}\gamma ^{a}\psi _{\sigma })\bar{\psi}_{\mu }\gamma
_{5}\gamma _{a}\nabla _{\nu }\psi _{\rho }\epsilon ^{\mu \nu \rho \sigma }-%
\frac{i}{2}\bar{\varepsilon}\gamma _{5}\gamma _{a}\nabla _{\nu }\psi _{\rho
}(\bar{\psi}_{\mu }\gamma ^{a}\psi _{\sigma })\epsilon ^{\mu \nu \rho \sigma
}  \label{pippo}
\end{equation}
We start from the gravitino field equations (\ref{eq.campo gravitino}): 
\[
\gamma _{5}\gamma _{a}\nabla _{\nu }\psi _{\rho }e_{\sigma }^{a}\epsilon
^{\mu \nu \rho \sigma }=0 
\]
and take their covariant derivative: 
\begin{eqnarray}
&&\nabla _{\mu }\left( \gamma _{5}\gamma _{a}\nabla _{\nu }\psi _{\rho
}e_{\sigma }^{a}\right) \epsilon ^{\mu \nu \rho \sigma }=\gamma _{5}\gamma
_{a}\nabla _{\mu }\nabla _{\nu }\psi _{\rho }e_{\sigma }^{a}\epsilon ^{\mu
\nu \rho \sigma }+\gamma _{5}\gamma _{a}\nabla _{\nu }\psi _{\rho }\nabla
_{\mu }e_{\sigma }^{a}\epsilon ^{\mu \nu \rho \sigma }= \nonumber \\ 
&&=\gamma _{5}\gamma _{a}\nabla _{\mu }\nabla _{\nu }\psi _{\rho }e_{\sigma
}^{a}\epsilon ^{\mu \nu \rho \sigma }+\frac{i}{2}\gamma _{5}\gamma
_{a}\nabla _{\nu }\psi _{\rho }\left( \bar{\psi}_{\mu }\gamma ^{a}\psi
_{\sigma }\right) \epsilon ^{\mu \nu \rho \sigma }= \nonumber \\ 
&&=-\frac{1}{4}\gamma _{5}\gamma _{a}\gamma _{b}\gamma _{c}\psi _{\rho
}R_{\mu \nu }^{bc}e_{\sigma }^{a}\epsilon ^{\mu \nu \rho \sigma }+\frac{i}{2}%
\gamma _{5}\gamma _{a}\nabla _{\nu }\psi _{\rho }\left( \bar{\psi}_{\mu
}\gamma ^{a}\psi _{\sigma }\right) \epsilon ^{\mu \nu \rho \sigma }
\label{prop}
\end{eqnarray}
Now we recall the gamma matrices property 
\begin{equation}
\gamma _{5}\gamma _{a}\gamma _{b}\gamma _{c}=2\gamma _{5}\eta _{a[b}\gamma
_{c]}+i\epsilon _{abcd}\gamma ^{d}  \label{gamma}
\end{equation}
and the Fierz identity 
\begin{equation}
\gamma ^{a}\psi _{\mu }\bar{\psi}_{\nu }\gamma _{a}\psi _{\rho }\epsilon
^{\mu \nu \rho \sigma }ds=0
\end{equation}
which leads to 
\begin{equation}
\nabla _{\sigma }(\gamma ^{a}\psi _{\mu }\bar{\psi}_{\nu }\gamma _{a}\psi
_{\rho })\epsilon ^{\mu \nu \rho \sigma }ds=0  \label{deriv}
\end{equation}
By considering the total derivative and using the Majorana identity 
\begin{equation}
\nabla _{\sigma }\bar{\psi}_{\nu }\gamma _{a}\psi _{\rho }=-\bar{\psi}_{\rho
}\gamma _{a}\nabla _{\sigma }\bar{\psi}_{\nu }
\end{equation}
we get from Eq.(\ref{deriv}) 
\begin{equation}
\gamma ^{a}\nabla _{\nu }\psi _{\rho }\left( \bar{\psi}_{\mu }\gamma
_{a}\psi _{\sigma }\right) \epsilon ^{\mu \nu \rho \sigma }=-2\gamma
^{a}\psi _{\rho }\left( \bar{\psi}_{\mu }\gamma _{a}\nabla _{\nu }\psi
_{\sigma }\right) \epsilon ^{\mu \nu \rho \sigma }  \label{fierz}
\end{equation}
With the use of (\ref{gamma}) and (\ref{fierz}), Eq.(\ref{prop}) becomes: 
\begin{eqnarray}
&&\nabla _{\mu }\left( \gamma _{5}\gamma _{a}\nabla _{\nu }\psi _{\rho
}e_{\sigma }^{a}\right) \epsilon ^{\mu \nu \rho \sigma }=-\frac{1}{2}\gamma
_{5}\eta _{a[b}\gamma _{c]}\psi _{\rho }R_{\mu \nu }^{bc}e_{\sigma
}^{a}\epsilon ^{\mu \nu \rho \sigma }+ \nonumber \\ 
&&-\frac{i}{4}\gamma ^{d}\psi _{\rho }R_{\mu \nu }^{bc}e_{\sigma
}^{a}\epsilon _{abcd}\epsilon ^{\mu \nu \rho \sigma }-i\gamma _{5}\gamma
_{a}\psi _{\rho }\left( \bar{\psi}_{\mu }\gamma ^{a}\nabla _{\nu }\psi
_{\sigma }\right)
\end{eqnarray}
which, imposing the field equations of the vielbein (\ref{campo vielbein}) 
\begin{equation}
R_{\mu \nu }^{ab}e_{\sigma }^{c}\epsilon _{abcd}=-2\bar{\psi}_{\mu }\gamma
_{5}\gamma _{d}\nabla _{\nu }\psi _{\sigma }
\end{equation}
can be recast as
\begin{eqnarray}
&&\nabla _{\mu }\left( \gamma _{5}\gamma _{a}\nabla _{\nu }\psi _{\rho
}e_{\sigma }^{a}\right) \epsilon ^{\mu \nu \rho \sigma }=-\frac{1}{2}\gamma
_{5}\eta _{a[b}\gamma _{c]}\psi _{\rho }R_{\mu \nu }^{bc}e_{\sigma
}^{a}\epsilon ^{\mu \nu \rho \sigma }+ \nonumber \\ 
&&+\frac{i}{2}\gamma ^{d}\psi _{\rho }\left( \bar{\psi}_{\mu }\gamma
_{5}\gamma _{d}\nabla _{\nu }\psi _{\sigma }\right) \epsilon ^{\mu \nu \rho
\sigma }-i\gamma _{5}\gamma _{a}\psi _{\rho }\left( \bar{\psi}_{\mu }\gamma
^{a}\nabla _{\nu }\psi _{\sigma }\right) \epsilon ^{\mu \nu \rho \sigma }=\nonumber
\\ 
&&=\gamma _{5}\gamma _{a}\psi _{\rho }R_{\mu \nu }^{ab}e_{b\sigma }\epsilon
^{\mu \nu \rho \sigma }+\frac{i}{2}\gamma ^{a}\psi _{\rho }\left( \bar{\psi}
_{\mu }\gamma _{5}\gamma _{a}\nabla _{\nu }\psi _{\sigma }\right) \epsilon
^{\mu \nu \rho \sigma }+ \nonumber \\ 
&&-i\gamma _{5}\gamma _{a}\psi _{\rho }\left( \bar{\psi}_{\mu }\gamma
^{a}\nabla _{\nu }\psi _{\sigma }\right) \epsilon ^{\mu \nu \rho \sigma }
\end{eqnarray}
Finally, using the Bianchi identities 
\begin{equation}
R_{\mu \nu }^{ab}e_{b\sigma }=i\bar{\psi}_{\mu }\gamma ^{a}\nabla _{\nu
}\psi _{\sigma }
\end{equation}
(remember the constraint $T^{a}=0$), we get 
\begin{eqnarray}
&&\nabla _{\mu }\left( \gamma _{5}\gamma _{a}\nabla _{\nu }\psi _{\rho
}e_{\sigma }^{a}\right) \epsilon ^{\mu \nu \rho \sigma }=\frac{i}{2}\gamma
_{5}\gamma _{a}\psi _{\rho }\left( \bar{\psi}_{\mu }\gamma ^{a}\nabla _{\nu
}\psi _{\sigma }\right) \epsilon ^{\mu \nu \rho \sigma }+ \nonumber \\ 
&&+\frac{i}{2}\gamma ^{a}\psi _{\rho }\left( \bar{\psi}_{\mu }\gamma
_{5}\gamma _{a}\nabla _{\nu }\psi _{\sigma }\right) \epsilon ^{\mu \nu \rho
\sigma }-i\gamma _{5}\gamma _{a}\psi _{\rho }\left( \bar{\psi}_{\mu }\gamma
^{a}\nabla _{\nu }\psi _{\sigma }\right) \epsilon ^{\mu \nu \rho \sigma }\hspace{-0.1cm}=0
\end{eqnarray}
Which leads to 
\begin{equation}
-i\gamma _{5}\gamma _{a}\psi _{\rho }\left( \bar{\psi}_{\mu }\gamma
^{a}\nabla _{\nu }\psi _{\sigma }\right) \epsilon ^{\mu \nu \rho \sigma
}+i\gamma ^{a}\psi _{\rho }\left( \bar{\psi}_{\mu }\gamma _{5}\gamma
_{a}\nabla _{\nu }\psi _{\sigma }\right) \epsilon ^{\mu \nu \rho \sigma }=0
\label{proprietà}
\end{equation}
Now, using this property, Eq.(\ref{pippo}), i.e. 
\[
i(\bar{\varepsilon}\gamma ^{a}\psi _{\sigma })\left( \bar{\psi}_{\mu }\gamma
_{5}\gamma _{a}\nabla _{\nu }\psi _{\rho }\right) \epsilon ^{\mu \nu \rho
\sigma }-\frac{i}{2}\left( \bar{\varepsilon}\gamma _{5}\gamma _{a}\nabla
_{\nu }\psi _{\rho }\right) (\bar{\psi}_{\mu }\gamma ^{a}\psi _{\sigma
})\epsilon ^{\mu \nu \rho \sigma } 
\]
can be rewritten as 
\begin{equation}
-\left( \bar{\varepsilon}\gamma _{5}\gamma _{a}\psi _{\mu }\right) \left( 
\bar{\psi}_{\nu }\gamma ^{a}\nabla _{\rho }\psi _{\sigma }\right) \epsilon
^{\mu \nu \rho \sigma }-\frac{i}{2}\left( \bar{\varepsilon}\gamma _{5}\gamma
_{a}\nabla _{\nu }\psi _{\rho }\right) (\bar{\psi}_{\mu }\gamma ^{a}\psi
_{\sigma })\epsilon ^{\mu \nu \rho \sigma }
\end{equation}
We use again Eq.(\ref{fierz}), obtaining 
\begin{equation}
-\left( \bar{\varepsilon}\gamma _{5}\gamma _{a}\psi _{\mu }\right) \left( 
\bar{\psi}_{\nu }\gamma ^{a}\nabla _{\rho }\psi _{\sigma }\right) \epsilon
^{\mu \nu \rho \sigma }=+\frac{i}{2}\left( \bar{\varepsilon}\gamma
_{5}\gamma _{a}\nabla _{\nu }\psi _{\rho }\right) (\bar{\psi}_{\mu }\gamma
^{a}\psi _{\sigma })\epsilon ^{\mu \nu \rho \sigma }
\end{equation}
and the final result is 
\begin{equation}
i(\bar{\varepsilon}\gamma ^{a}\psi _{\sigma })\left( \bar{\psi}_{\mu }\gamma
_{5}\gamma _{a}\nabla _{\nu }\psi _{\rho }\right) \epsilon ^{\mu \nu \rho
\sigma }-\frac{i}{2}\left( \bar{\varepsilon}\gamma _{5}\gamma _{a}\nabla
_{\nu }\psi _{\rho }\right) (\bar{\psi}_{\mu }\gamma ^{a}\psi _{\sigma
})\epsilon ^{\mu \nu \rho \sigma }=0
\end{equation}
Therefore the covariance condition (\ref{covarianza per supersimmetrie})
reduces to
\begin{eqnarray*}
&&8\left( \bar{\psi}_{\mu }\gamma _{5}\gamma _{a}\nabla _{\nu }\nabla _{\rho
}\varepsilon \right) e_{\sigma }^{a}\epsilon ^{\mu \nu \rho \sigma }+8\left( 
\bar{\varepsilon}\gamma _{5}\gamma _{a}\nabla _{\mu }\nabla _{\nu }\psi
_{\rho }\right) e_{\sigma }^{a}\epsilon ^{\mu \nu \rho \sigma }+\\ 
&&8ie\left( \bar{\varepsilon}\gamma ^{\mu }\psi _{\sigma }\right) R_{\mu
}^{a}e_{a}^{\sigma }-4ie\left( \bar{\varepsilon}\gamma ^{\mu }\psi _{\mu
}\right) R+\nabla _{\mu }(\bar{\varepsilon}\gamma _{5}\gamma _{a}\nabla
_{\nu }\psi _{\rho }e_{\sigma }^{a}\epsilon ^{\mu \nu \rho \sigma })\hspace{-0.11cm}=\hspace{-0.07cm}0 \label{covarianza per supersimmetrie ridotta}
\end{eqnarray*}
We expand the double covariant derivatives of $\psi $ and $\varepsilon$: 
\begin{eqnarray*}
&&\nabla _{\nu }\nabla _{\mu }\psi _{\rho }=d_{\nu }\nabla _{\mu }\psi
_{\rho }-\frac{1}{8}[\gamma _{a},\gamma _{b}]\nabla _{\mu }\psi _{\rho
}\omega _{\nu }^{ab}-\Gamma _{\mu \nu }^{\lambda }\nabla _{\lambda }\psi
_{\rho }-\Gamma _{\rho \nu }^{\lambda }\nabla _{\mu }\psi _{\lambda }= \\
&&=d_{\mu \nu }\psi _{\rho }-\frac{1}{8}[\gamma _{a},\gamma _{b}]d_{\nu
}\psi _{\rho }\omega _{\mu }^{ab}-\frac{1}{8}[\gamma _{a},\gamma _{b}]\psi
_{\rho }d_{\nu }\omega _{\mu }^{ab}-\frac{1}{8}[\gamma _{a},\gamma _{b}%
]\nabla _{\mu }\psi _{\rho }\omega _{\nu }^{ab}= \\
&&=-\frac{1}{8}[\gamma _{a},\gamma _{b}](d_{\nu }\omega _{\mu }^{ab}+\frac{1%
}{8}[\gamma _{c},\gamma _{d}]\omega _{\nu }^{cd}\omega _{\mu }^{ab})\psi
_{\rho }= \\
&&-\frac{1}{8}[\gamma _{a},\gamma _{b}]d_{\nu }\omega _{\mu }^{ab}\psi
_{\rho }-\frac{1}{8}[\gamma _{b},\gamma _{c}]\omega _{\nu a}^{c\cdot }\omega
_{\mu }^{ab}\psi _{\rho }
\end{eqnarray*}
where we used the fact that $d_{\mu \nu }$ and $\Gamma _{\mu \nu }^{\lambda
} $, symmetric in $\mu $ and $\nu $, vanish if multiplied by $\epsilon ^{\mu
\nu \rho \sigma }$, and the property of the gamma matrices 
\begin{equation}
\lbrack \gamma _{a},\gamma _{b}][\gamma _{c},\gamma _{d}]=8\eta _{ad}[\gamma
_{b},\gamma _{c}]  \label{commutatori matrici gamma}
\end{equation}
Now 
\begin{eqnarray}
&&\nabla _{\nu }\nabla _{\mu }\psi _{\rho }\epsilon ^{\mu \nu \rho \sigma
}=(\nabla _{(\nu }\nabla _{\mu )}\psi _{\rho }+\nabla _{[\nu }\nabla _{\mu
]}\psi _{\rho })\epsilon ^{\mu \nu \rho \sigma }=  \nonumber \\
&&=\nabla _{[\nu }\nabla _{\mu ]}\psi _{\rho }\epsilon ^{\mu \nu \rho \sigma
}=-[\gamma _{a},\gamma _{b}]\psi _{\rho }R_{\nu \mu }^{ab}\epsilon ^{\mu \nu
\rho \sigma }
\end{eqnarray}
which implies 
\begin{equation}
\nabla _{\mu }\nabla _{\nu }\psi _{\rho }\epsilon ^{\mu \nu \rho \sigma
}=-[\gamma _{a},\gamma _{b}]\psi _{\rho }R_{\mu \nu }^{ab}\epsilon ^{\mu \nu
\rho \sigma }  \label{derivata doppia gravitino}
\end{equation}
Considering the spinor $\varepsilon $, the covariant derivative of which is
given by 
\[
\nabla _{\mu }\varepsilon =d_{\mu }\varepsilon -[\gamma _{a},\gamma
_{b}]\omega _{\mu }^{ab}\varepsilon 
\]
the calculation is the same as before, because the only difference between
the two covariant derivatives is given by the vanishing symmetric terms.

	Consequently, 
\begin{equation}
\nabla _{\nu }\nabla _{\rho }\varepsilon \epsilon ^{\mu \nu \rho \sigma
}=-[\gamma _{a},\gamma _{b}]R_{\nu \rho }^{ab}\varepsilon \epsilon ^{\mu \nu
\rho \sigma }  \label{derivata doppia epsilon}
\end{equation}
Hence, by using (\ref{derivata doppia gravitino}) and (\ref{derivata doppia
epsilon}), we obtain the following objects: 
\begin{eqnarray}
&&8\left( \bar{\psi}_{\mu }\gamma _{5}\gamma _{a}\nabla _{\nu }\nabla _{\rho
}\varepsilon \right) e_{\sigma }^{a}\epsilon ^{\mu \nu \rho \sigma }=-\frac{1%
}{2}\left( \bar{\psi}_{\mu }\gamma _{5}\gamma _{c}[\gamma _{a},\gamma
_{b}]\varepsilon \right) R_{\nu \rho }^{ab}e_{\sigma }^{a}\epsilon ^{\mu \nu
\rho \sigma }= \nonumber \\
&&=-\left( \bar{\psi}_{\mu }\gamma _{5}\gamma _{c}\gamma _{ab}\varepsilon
\right) R_{\nu \rho }^{ab}e_{\sigma }^{a}\epsilon ^{\mu \nu \rho \sigma
}=\left( \bar{\varepsilon}\gamma _{5}\gamma _{c}\gamma _{ab}\psi _{\rho
}\right) R_{\mu \nu }^{ab}e_{\sigma }^{c}\epsilon ^{\mu \nu \rho \sigma }
\end{eqnarray}
\begin{eqnarray}
&&-8\left( \bar{\varepsilon}\gamma _{5}\gamma _{a}\nabla _{\mu }\nabla _{\nu
}\psi _{\rho }\right) e_{\sigma }^{a}\epsilon ^{\mu \nu \rho \sigma }=\frac{1
}{2}\left( \bar{\varepsilon}\gamma _{5}\gamma _{c}[\gamma _{a},\gamma
_{b}]\psi _{\rho }\right) R_{\mu \nu }^{ab}e_{\sigma }^{c}\epsilon ^{\mu \nu
\rho \sigma }= \nonumber   \\ 
&&=\left( \bar{\varepsilon}\gamma _{5}\gamma _{c}\gamma _{ab}\psi _{\rho
}\right) R_{\mu \nu }^{ab}e_{\sigma }^{c}\epsilon ^{\mu \nu \rho \sigma
}=\left( \bar{\varepsilon}\gamma _{5}\gamma _{ab}\gamma _{c}\psi _{\rho
}\right) R_{\mu \nu }^{ab}e_{\sigma }^{c}\epsilon ^{\mu \nu \rho \sigma }  
\end{eqnarray}
which added together give: 
\begin{equation}
\left( \bar{\varepsilon}\gamma _{5}\{\gamma _{c},\gamma _{ab}\}\psi _{\rho
}\right) R_{\mu \nu }^{ab}e_{\sigma }^{c}\epsilon ^{\mu \nu \rho \sigma
}=2i\left( \bar{\varepsilon}\gamma ^{d}\psi _{\rho }\right) R_{\mu \nu
}^{ab}e_{\sigma }^{c}\epsilon _{abcd}\epsilon ^{\mu \nu \rho \sigma }
\label{pluto}
\end{equation}
Provided the property (\ref{anticommutatore gamma}), i.e. 
\begin{eqnarray*}
\gamma _{5}\{\gamma _{a},\gamma _{rs}\}=2i\gamma ^{d}\epsilon
_{rsad}+4\gamma _{5}\eta _{r[a}\gamma _{s]} 
\end{eqnarray*}
we can recast now (\ref{pluto}) recalling that 
\begin{eqnarray*}
\epsilon _{abcd}\epsilon ^{\mu \nu \rho \sigma }=-(4!)e_{[a}^{\mu
}e_{b}^{\nu }e_{c}^{\rho }e_{d]}^{\sigma }e 
\end{eqnarray*}
Therefore 
\begin{eqnarray*}
2i\left( \bar{\varepsilon}\gamma ^{d}\psi _{\rho }\right) R_{\mu \nu
}^{ab}e_{\sigma }^{c}\epsilon _{abcd}\epsilon ^{\mu \nu \rho \sigma
}=-8ie\left( \bar{\varepsilon}\gamma ^{\mu }\psi _{\sigma }\right) R_{\mu
}^{a}e_{a}^{\sigma }+4ie\left( \bar{\varepsilon}\gamma ^{\mu }\psi _{\mu
}\right) R 
\end{eqnarray*}
Finally, by substituting this equation in (\ref{covarianza per
supersimmetrie ridotta}), we obtain 
\begin{equation}
\delta \mathcal{L}=\nabla _{\mu }(\bar{\varepsilon}\gamma _{5}\gamma
_{a}\nabla _{\nu }\psi _{\rho }e_{\sigma }^{a}\epsilon ^{\mu \nu \rho \sigma
})
\end{equation}
which is the fundamental identity (\ref{identità fondamentale}). So we
conclude that the Rarita-Schwinger Lagrangian is covariant under the
supersymmetry transformations generated by the vector field $\Xi $.

	The fact that the Lagrangian is covariant only \textsl{on-shell} (i.e.
assuming the vielbein field equations) is very important. It gives
rise to serious problems in the construction of the theory, from our specific point of view. These
are related to the supersymmetry algebra, and are the same which arise in
General Relativity when dealing with the covariance of the Lagrangian under
the transformations of the Poincar\'{e} group [3]. In this case it can be shown that the
Hilbert-Einstein action is invariant under diffeomorphisms and Lorentz
rotations, but not under Poincar\'{e} translations [3]. To recover covariance
under the whole group, a torsion free condition similar to (\ref{torsione
nulla}) is introduced. The price to pay is to modify the Poincar\'{e}
algebra, which now closes only on-shell.

	In Supergravity, this happens as well. Without imposing the (super)torsion free condition (\ref{torsione nulla}), the action is not covariant under the translations of the Poincar\'{e}
supergroup; with the constraint, the theory is covariant, but the
Poincar\'{e} super algebra closes only on-shell [3]. The non closure of the
algebra is shown in section \ref{nonclosure}

	The on-shell covariance of the Rarita-Schwinger Lagrangian is a first warning of these problems. It shows indeed that the theory, if considered in the standard approach, is not consistent as a
Gauge-Natural theory.

	For these and other reasons, theoretical physicists have introduced the concept of \textsl{%
rehonomy}, and  Superstrings [2].
The formulation of Supergravity with a Gauge-Natural framework might
require another approach, which is introduced in the last section of this chapter.

\section{Closure of the supersymmetric algebra}\label{nonclosure}

After proving the covariance of the Lagrangian, we now obtain
the algebra of the supersymmetries. We shall check also if it is possible to
regard the action of the commutators on the fields as their Lie derivative
with respect to a suitable infinitesimal generator 
\begin{equation}
\Xi =\xi ^{\mu }(x)\partial _{\mu }+\xi ^{ab}(x)\sigma _{ab}
\end{equation}
We recall the supersymmetry transformations
\begin{equation}
\left\{
\begin{tabular}{l}
\vspace{0.1cm}$\delta \psi _{\mu }=\nabla _{\mu }\varepsilon $ \\ 
$\delta e_{\mu }^{a}=i\bar{\varepsilon}\gamma ^{a}\psi _{\mu }$%
\end{tabular}
\label{super}
\right.
\end{equation}
These can be viewed as vector fields along a configuration
(see Section 3.6 for the definition of generalized vector fields): 
\begin{eqnarray}
X &=&\left( \delta _{1}e_{\mu }^{a}\right) \frac{\partial }{\partial e_{\mu
}^{a}}+\left( \delta _{1}\psi _{\mu }\right) \frac{\partial }{\partial \psi
_{\mu }}=X^{i}\partial _{i} \\
Y &=&\left( \delta _{2}e_{\mu }^{a}\right) \frac{\partial }{\partial e_{\mu
}^{a}}+\left( \delta _{2}\psi _{\mu }\right) \frac{\partial }{\partial \psi
_{\mu }}=Y^{i}\partial _{i}
\end{eqnarray}
Their commutator must be evaluated on the infinite jet bundle, and then
projected down again on the configuration bundle. Since the components $
X^{i} $, $Y^{i}$ depend on the first derivative, we obtain: 
\begin{eqnarray}
\left[ X,Y\right] &=&\left( X^{k}\partial _{k}Y^{i}+X_{\sigma }^{k}\partial
_{k}^{\sigma }Y^{i}-Y^{k}\partial _{k}X^{i}-Y_{\sigma }^{k}\partial
_{k}^{\sigma }X^{i}\right) \partial _{i}=  \nonumber \\
&=&[\delta _{1}e_{\rho }^{c}\frac{\partial \left( \delta _{2}e_{\mu
}^{a}\right) }{\partial e_{\rho }^{c}}+\delta _{1}\psi _{\rho }\frac{%
\partial \left( \delta _{2}e_{\mu }^{a}\right) }{\partial \psi _{\rho }}+ 
\nonumber \\
&&+d_{\sigma }\left( \delta _{1}e_{\rho }^{c}\right) \frac{\partial \left(
\delta _{2}e_{\mu }^{a}\right) }{\partial \left( d_{\sigma }e_{\rho
}^{c}\right) }+d_{\sigma }\left( \delta _{1}\psi _{\rho }\right) \frac{%
\partial \left( \delta _{2}e_{\mu }^{a}\right) }{\partial \left( d_{\sigma
}\psi _{\rho }\right) }+  \nonumber \\
&&-\left( 1\longleftrightarrow 2\right) ]\frac{\partial }{\partial e_{\mu
}^{a}}+  \nonumber \\
&&+[\delta _{1}e_{\rho }^{c}\frac{\partial \left( \delta _{2}\psi _{\mu
}\right) }{\partial e_{\rho }^{c}}+\delta _{1}\psi _{\rho }\frac{\partial
\left( \delta _{2}\psi _{\mu }\right) }{\partial \psi _{\rho }}+  \nonumber
\\
&&+d_{\sigma }\left( \delta _{1}e_{\rho }^{c}\right) \frac{\partial \left(
\delta _{2}\psi _{\mu }\right) }{\partial \left( d_{\sigma }e_{\rho
}^{c}\right) }+d_{\sigma }\left( \delta _{1}\psi _{\rho }\right) \frac{%
\partial \left( \delta _{2}\psi _{\mu }\right) }{\partial \left( d_{\sigma
}\psi _{\rho }\right) }+  \nonumber \\
&&-\left( 1\longleftrightarrow 2\right) ]\frac{\partial }{\partial \psi
_{\mu }}  \label{commutatore}
\end{eqnarray}

\subsection{Commutator on the vielbein}

Taking Eq.(\ref{super}) into account, it is easy to verify that 
\begin{equation}
\frac{\partial \left( \delta _{2}e_{\mu }^{a}\right) }{\partial e_{\rho }^{c}%
}=\frac{\partial \left( \delta _{2}e_{\mu }^{a}\right) }{\partial \left(
d_{\sigma }e_{\rho }^{c}\right) }=\frac{\partial \left( \delta _{2}e_{\mu
}^{a}\right) }{\partial \left( d_{\sigma }\psi _{\rho }\right) }=0
\end{equation}
Therefore we are left with: 
\begin{equation}
\left( \delta _{1}\psi _{\rho }\right) \frac{\partial \left( \delta
_{2}e_{\mu }^{a}\right) }{\partial \psi _{\rho }}=i\bar{\varepsilon}%
_{2}\gamma ^{a}\delta _{\mu }^{\rho }\nabla _{\rho }\varepsilon _{1}=i\bar{%
\varepsilon}_{2}\gamma ^{a}\nabla _{\mu }\varepsilon _{1}
\end{equation}
\begin{equation}
\left( \delta _{2}\psi _{\rho }\right) \frac{\partial \left( \delta
_{1}e_{\mu }^{a}\right) }{\partial \psi _{\rho }}=i\bar{\varepsilon}%
_{1}\gamma ^{a}\nabla _{\mu }\varepsilon _{2}=-i\nabla _{\mu }\bar{%
\varepsilon}_{2}\gamma ^{a}\varepsilon _{1}
\end{equation}
By summing the above equations, one finds 
\begin{equation}
i\bar{\varepsilon}_{2}\gamma ^{a}\nabla _{\mu }\varepsilon _{1}+i\nabla
_{\mu }\bar{\varepsilon}_{2}\gamma ^{a}\varepsilon _{1}=i\nabla _{\mu
}\left( \bar{\varepsilon}_{2}\gamma ^{a}\varepsilon _{1}\right)
\end{equation}
If the commutator is applied to the vielbein, we then get 
\begin{equation}
\lbrack \delta _{1},\delta _{2}]e_{\mu }^{a}=i(\bar{\varepsilon}_{2}\gamma
^{a}\nabla _{\mu }\varepsilon _{1}-\bar{\varepsilon}_{1}\gamma ^{a}\nabla
_{\mu }\varepsilon _{2})  \label{commuta}
\end{equation}
because
\begin{equation}
\delta _{1}\delta _{2}e_{\mu }^{a}=\delta _{1}(i\bar{\varepsilon}_{2}\gamma
^{a}\psi _{\mu })=i\bar{\varepsilon}_{2}\gamma ^{a}\delta _{1}\psi _{\mu }=i%
\bar{\varepsilon}_{2}\gamma ^{a}\nabla _{\mu }\varepsilon _{1}
\end{equation}
and 
\begin{equation}
\delta _{2}\delta _{1}e_{\mu }^{a}=\delta _{2}(i\bar{\varepsilon}_{1}\gamma
^{a}\psi _{\mu })=i\bar{\varepsilon}_{1}\gamma ^{a}\delta _{2}\psi _{\mu }=i%
\bar{\varepsilon}_{1}\gamma ^{a}\nabla _{\mu }\varepsilon _{2}
\end{equation}
Using now the Majorana flip identities, we can rewrite Eq.(\ref{commuta}) as
follows: 
\begin{equation}
\lbrack \delta _{1},\delta _{2}]e_{\mu }^{a}=i\nabla _{\mu }(\bar{\varepsilon%
}_{2}\gamma ^{a}\varepsilon _{1})  \label{com}
\end{equation}
The Lie derivative of $e_{\mu }^{a}$ is given by (\ref{derivata di Lie
vielbein}): 
\begin{equation}
\pounds _{\Xi }e_{\mu }^{a}=\nabla _{\mu }\xi ^{\nu }e_{\nu }^{a}+\nabla
_{\nu }e_{\mu }^{a}\xi ^{\nu }-e_{\mu }^{b}\xi _{(v)b}^{\,a}  \label{lie}
\end{equation}
If we want to interpret the commutator (\ref{com}) as the Lie derivative (%
\ref{lie}) for some suitable vector field $\Xi $, we should require 
\begin{eqnarray}
\pounds _{\Xi }e_{\mu }^{a}&:=&[\delta _{1},\delta _{2}]e_{\mu
}^{a}=i\nabla _{\mu }(\bar{\epsilon _{2}}\gamma ^{a}\varepsilon
_{1})=i\nabla _{\mu }(\bar{\epsilon _{2}}\gamma ^{\rho }\varepsilon
_{1}e_{\rho }^{a})= \nonumber \\ 
&=&i\nabla _{\mu }(\bar{\epsilon _{2}}\gamma ^{\rho }\varepsilon
_{1})e_{\rho }^{a}+i(\bar{\epsilon _{2}}\gamma ^{\rho }\varepsilon
_{1})\nabla _{\mu }e_{\rho }^{a}=\nabla _{\mu }\xi ^{\rho }e_{\rho }^{a}+\xi
^{\rho }\nabla _{\mu }e_{\rho }^{a}= \nonumber \\ 
&=&\nabla _{\mu }\xi ^{\rho }e_{\rho }^{a}+\xi ^{\rho }\nabla _{\mu
}e_{\rho }^{a}+\xi ^{\rho }\nabla _{\rho }e_{\mu }^{a}-\xi ^{\rho }\nabla
_{\rho }e_{\mu }^{a}= \nonumber \\ 
&=&\nabla _{\mu }\xi ^{\rho }e_{\rho }^{a}+\xi ^{\rho }\nabla _{\rho
}e_{\mu }^{a}-e_{\mu }^{b}\xi _{(v)b}^{\,a}
\end{eqnarray}
Thus $\Xi $ can be defined as 
\[
\Xi =\xi ^{\rho }(x)\partial _{\rho }+\xi _{(v)}^{ab}(x)\sigma _{ab} 
\]
so that 
\begin{equation}
\xi ^{\rho }=i(\bar{\varepsilon}_{2}\gamma ^{\rho }\varepsilon _{1})
\label{campo orizz}
\end{equation}
and 
\begin{equation}
\xi _{(v)b}^{\,a}=(\xi ^{\rho }\nabla _{\rho }e_{\lambda }^{a}-\xi ^{\rho
}\nabla _{\lambda }e_{\rho }^{a})e_{b}^{\lambda }=2\xi ^{\rho
}e_{b}^{\lambda }\nabla _{[\rho }e_{\lambda ]}^{a}
\end{equation}

\subsection{Commutator on the gravitino}

We begin by expanding Eq.(\ref{connessione}: 
\begin{equation}
\delta \psi _{\mu }=\nabla _{\mu }\varepsilon =d_{\mu }\varepsilon -\frac{1}{%
4}\gamma _{a}\gamma _{b}\left( \Gamma _{\mu }^{ab}+H_{\mu }^{ab}\right)
\varepsilon
\end{equation}
Then we calculate each single term appearing in the r.h.s. of
Eq.(\ref{commutatore}):
\begin{equation}
\frac{\partial \left( \delta \psi _{\mu }\right) }{\partial e_{\rho }^{c}}=-%
\frac{1}{4}\gamma _{a}\gamma _{b}\left( \frac{\partial \Gamma _{\mu }^{ab}}{%
\partial e_{\rho }^{c}}+\frac{\partial H_{\mu }^{ab}}{\partial e_{\rho }^{c}}%
\right) \varepsilon
\end{equation}
Now, 
\begin{equation}
\frac{\partial \Gamma _{\mu }^{ab}}{\partial e_{\rho }^{c}}=\frac{\partial
\left( e_{\sigma }^{b}\Gamma _{\lambda \mu }^{\sigma }e^{a\lambda }\right) }{%
\partial e_{\rho }^{c}}+\frac{\partial \left( e_{\lambda }^{b}d_{\mu
}e^{a\lambda }\right) }{\partial e_{\rho }^{c}}
\end{equation}
The first term can be recast as follows:
\begin{eqnarray}
\frac{\partial \left( e_{\sigma }^{b}\Gamma _{\lambda \mu }^{\sigma
}e^{a\lambda }\right) }{\partial e_{\rho }^{c}} &=&\delta _{c}^{b}\delta
_{\sigma }^{\rho }\Gamma _{\lambda \mu }^{\sigma }e^{a\lambda }+g^{\rho
\lambda }\delta _{c}^{a}e_{\sigma }^{b}\Gamma _{\lambda \mu }^{\sigma
}+e_{\sigma }^{b}e^{a\lambda }\frac{\partial \Gamma _{\lambda \mu }^{\sigma }%
}{\partial e_{\rho }^{c}}=  \nonumber \\
&=&\delta _{c}^{b}\Gamma _{\lambda \mu }^{\rho }e^{a\lambda }+\delta
_{c}^{a}g^{\rho \lambda }e_{\sigma }^{b}\Gamma _{\lambda \mu }^{\sigma
}+e_{\sigma }^{b}e^{a\lambda }\frac{\partial \Gamma _{\lambda \mu }^{\sigma }%
}{\partial e_{\rho }^{c}}
\end{eqnarray}
$\Gamma _{\lambda \mu }^{\sigma }$ is a Christoffel symbol, so the Palatini
formula 
\begin{equation}
\Gamma _{\lambda \mu }^{\sigma }=\frac{1}{2}g^{\sigma \nu }\left( \partial
_{\lambda }g_{\mu \nu }+\partial _{\mu }g_{\lambda \nu }-\partial _{\nu
}g_{\lambda \mu }\right)
\end{equation}
must hold; this leads to 
\begin{eqnarray}
\frac{\partial \Gamma _{\lambda \mu }^{\sigma }}{\partial e_{\rho }^{c}} &=&%
\frac{\partial }{\partial e_{\rho }^{c}}\left[ \frac{1}{2}g^{\sigma \nu
}\left( \partial _{\lambda }g_{\mu \nu }+\partial _{\mu }g_{\lambda \nu
}-\partial _{\nu }g_{\lambda \mu }\right) \right] = \nonumber \\
&=&\frac{1}{2}\frac{\partial g^{\sigma \nu }}{\partial e_{\rho }^{c}}\left(
\partial _{\lambda }g_{\mu \nu }+\partial _{\mu }g_{\lambda \nu }-\partial
_{\nu }g_{\lambda \mu }\right) + \nonumber\\
&+&\frac{1}{2}g^{\sigma \nu }\frac{\partial }{
\partial e_{\rho }^{c}}\left( \partial _{\lambda }g_{\mu \nu }+\partial
_{\mu }g_{\lambda \nu }-\partial _{\nu }g_{\lambda \mu }\right) \label{Christoffel}
\end{eqnarray}
By recalling that 
\begin{equation}
g_{\mu \nu }=e_{\mu }^{m}e_{\nu }^{n}\eta _{mn}  \label{g}
\end{equation}
we get 
\begin{eqnarray}
&&\left( \partial _{\lambda }g_{\mu \nu }+\partial _{\mu }g_{\lambda \nu
}-\partial _{\nu }g_{\lambda \mu }\right) =\partial _{\lambda }\left( e_{\mu
}^{m}e_{\nu }^{n}\eta _{mn}\right) +\nonumber\\
&&+\partial _{\mu }\left( e_{\lambda
}^{m}e_{\nu }^{n}\eta _{mn}\right)
-\partial _{\nu }\left( e_{\lambda
}^{m}e_{\mu }^{n}\eta _{mn}\right) = e_{\mu }^{m}\left( \partial _{\lambda }e_{\nu m}
-\partial _{\nu }e_{\lambda
m}\right)+\nonumber\\
&&+e_{\nu }^{m}\left( \partial _{\mu }e_{\lambda m}+\partial
_{\lambda }e_{\mu m}\right) +e_{\lambda }^{m}\left( \partial _{\mu }e_{\nu
m}+\partial _{\nu }e_{\mu m}\right)
\end{eqnarray}
This gives 
\begin{eqnarray}
&&\frac{\partial }{\partial e_{\rho }^{c}}\left( \partial _{\lambda }g_{\mu
\nu }+\partial _{\mu }g_{\lambda \nu }-\partial _{\nu }g_{\lambda \mu
}\right) = \delta _{c}^{m}\delta _{\mu }^{\rho }\left( \partial _{\lambda
}e_{\nu m}-\partial _{\nu }e_{\lambda m}\right)+\nonumber\\
&&\delta _{c}^{m}\delta _{\nu }^{\rho }\left( \partial _{\mu }e_{\lambda
m}+\partial _{\lambda }e_{\mu m}\right) +\delta _{c}^{m}\delta _{\lambda
}^{\rho }\left( \partial _{\mu }e_{\nu m}+\partial _{\nu }e_{\mu m}\right)=\nonumber\\
&&=\delta _{\mu }^{\rho }\left( 
\partial _{\lambda }e_{\nu c}-\partial _{\nu
}e_{\lambda c}\right) +
\delta _{\nu }^{\rho }\left( \partial _{\mu
}e_{\lambda c}+\partial _{\lambda }e_{\mu c}\right)+\nonumber\\
&&+\delta _{\lambda
}^{\rho }\left( \partial _{\mu }e_{\nu c}+\partial _{\nu }e_{\mu c}\right)
\end{eqnarray}
By deriving Eq.( \ref{g}), one gets 
\begin{equation}
\frac{\partial g^{\sigma \nu }}{\partial e_{\rho }^{c}}=\frac{\partial }{%
\partial e_{\rho }^{c}}\left( e_{m}^{\sigma }e_{n}^{\nu }\right) \eta
^{mn}=e_{c}^{\nu }g^{\rho \sigma }+e_{c}^{\sigma }g^{\rho \nu }
\end{equation}
and by consequence 
\begin{eqnarray}
&&\frac{1}{2}e_{\sigma }^{b}e^{a\lambda }\frac{\partial g^{\sigma \nu }}{%
\partial e_{\rho }^{c}}\left( \partial _{\lambda }g_{\mu \nu }+\partial
_{\mu }g_{\lambda \nu }-\partial _{\nu }g_{\lambda \mu }\right) = \nonumber \\
&&=\frac{1}{2}e_{\sigma }^{b}e^{a\lambda }\left( e_{c}^{\nu }g^{\rho \sigma
}+e_{c}^{\sigma }g^{\rho \nu }\right) \left( \partial _{\lambda }g_{\mu \nu
}+\partial _{\mu }g_{\lambda \nu }-\partial _{\nu }g_{\lambda \mu }\right) 
\nonumber \\
&&=e_{\sigma }^{b}e^{a\lambda }\left( e_{c}^{\sigma }\Gamma _{\lambda \mu
}^{\rho }+e_{c}^{\nu }\Gamma _{\lambda \mu }^{\theta }g^{\rho \sigma }g_{\nu
\theta }\right) =\left( \delta _{c}^{b}\Gamma _{\lambda \mu }^{\rho
}+e^{b\rho }e_{c\sigma }\Gamma _{\lambda \mu }^{\sigma }\right) e^{a\lambda }
\label{primo}
\end{eqnarray}
This is the first term in the r.h.s of (\ref{Christoffel}). The second term
becomes 
\begin{eqnarray}
&&\frac{1}{2}e^{b\nu }e^{a\lambda }\frac{\partial }{\partial e_{\rho }^{c}}%
\left( \partial _{\lambda }g_{\mu \nu }+\partial _{\mu }g_{\lambda \nu
}-\partial _{\nu }g_{\lambda \mu }\right) =\nonumber\\
&&=\frac{1}{2}e^{b\nu }e^{a\lambda }[\delta _{\mu }^{\rho }\left( \partial
_{\lambda }e_{\nu c}-\partial _{\nu }e_{\lambda c}\right) +\delta _{\nu
}^{\rho }\left( \partial _{\mu }e_{\lambda c}+\partial _{\lambda }e_{\mu
c}\right) +\nonumber\\
&&+\delta _{\lambda }^{\rho }\left( \partial _{\mu }e_{\nu c}+\partial _{\nu
}e_{\mu c}\right) ]=e^{b\nu }e^{a\lambda }\left( \partial _{\lambda }e_{\nu
c}\delta _{\mu }^{\rho }-\partial _{\nu }e_{\mu c}\delta _{\lambda }^{\rho
}\right)  \label{secondo}
\end{eqnarray}
because, being multiplied by $\gamma _{a}\gamma _{b}$, it is antisymmetric
in $\nu $ and $\lambda$.

	Now, by adding (\ref{primo}) to (\ref{secondo}), Eq.(\ref{Christoffel}) can be
recast as 
\begin{eqnarray}
\frac{\partial \Gamma _{\mu }^{ab}}{\partial e_{\rho }^{c}} &=&\delta
_{c}^{b}d_{\mu }e^{a\rho }+\left( \delta _{c}^{b}\Gamma _{\lambda \mu
}^{\rho }+e^{b\rho }e_{c\sigma }\Gamma _{\lambda \mu }^{\sigma }\right)
e^{a\lambda }+\delta _{c}^{b}\Gamma _{\lambda \mu }^{\rho }e^{a\lambda }+ 
\nonumber \\
&&+\delta _{c}^{a}g^{\rho \lambda }e_{\sigma }^{b}\Gamma _{\lambda \mu
}^{\sigma }+e^{b\nu }e^{a\lambda }\left( \partial _{\lambda }e_{\nu c}\delta
_{\mu }^{\rho }-\partial _{\nu }e_{\mu c}\delta _{\lambda }^{\rho }\right)
\label{ojjo}
\end{eqnarray}
Let us now consider the elements of the second parenthesis of the commutator
(\ref{commutatore}) containing spinors. First of all, remember Eq.(\ref{h}): 
\begin{equation}
H_{\mu }^{ab}=i\left( \bar{\psi}_{\mu }\gamma ^{a}\psi ^{b}+\bar{\psi}%
^{a}\gamma ^{b}\psi _{\mu }+\bar{\psi}^{a}\gamma _{\mu }\psi ^{b}\right)
\end{equation}
This gives 
\begin{eqnarray}
\frac{\partial H_{\mu }^{ab}}{\partial e_{\rho }^{c}} &=&i\frac{\partial }{%
\partial e_{\rho }^{c}}\left( \bar{\psi}_{\mu }\gamma ^{a}\psi _{\nu
}e^{b\nu }+\bar{\psi}_{\nu }\gamma ^{b}\psi _{\mu }e^{a\nu }+\bar{\psi}_{\nu
}\gamma _{d}\psi _{\sigma }e^{a\nu }e^{b\sigma }e_{\mu }^{d}\right) = 
\nonumber \\
&=&i(\bar{\psi}_{\mu }\gamma ^{a}\psi _{\nu }g^{\rho \nu }\delta _{c}^{b}+%
\bar{\psi}_{\nu }\gamma ^{b}\psi _{\mu }g^{\rho \nu }\delta _{c}^{a}+\bar{%
\psi}_{\nu }\gamma _{d}\psi _{\sigma }g^{\rho \nu }\delta _{c}^{a}e^{b\sigma
}e_{\mu }^{d}+  \nonumber \\
&&+\bar{\psi}_{\nu }\gamma _{d}\psi _{\sigma }g^{\rho \sigma }\delta
_{c}^{b}e^{a\nu }e_{\mu }^{d}+\bar{\psi}_{\nu }\gamma _{d}\psi _{\sigma
}e^{a\nu }e^{b\sigma }\delta _{c}^{d}\delta _{\mu }^{\rho })=  \nonumber \\
&=&i(\bar{\psi}_{\mu }\gamma ^{a}\psi ^{\rho }\delta _{c}^{b}+\bar{\psi}%
^{\rho }\gamma ^{b}\psi _{\mu }\delta _{c}^{a}+\bar{\psi}^{\rho }\gamma
_{\mu }\psi ^{b}\delta _{c}^{a}+\bar{\psi}^{a}\gamma _{\mu }\psi ^{\rho
}\delta _{c}^{b}+  \nonumber \\
&&+\bar{\psi}^{a}\gamma _{c}\psi ^{b}\delta _{\mu }^{\rho })
\end{eqnarray}
which leads to 
\begin{eqnarray*}
-\frac{1}{4}\gamma _{a}\gamma _{b}\frac{\partial H_{\mu }^{ab}}{%
\partial e_{\rho }^{c}}\varepsilon _{1} &=&-\frac{i}{4}\gamma _{a}\gamma
_{b}\left( 2\bar{\psi}_{\mu }\gamma ^{a}\psi ^{\rho }\delta _{c}^{b}+2\bar{%
\psi}^{a}\gamma _{\mu }\psi ^{\rho }\delta _{c}^{b}+\bar{\psi}^{a}\gamma
_{c}\psi ^{b}\delta _{\mu }^{\rho }\right) \varepsilon _{1}= \\
&=&-\frac{i}{4}\left[ 2\gamma _{a}\gamma _{c}\varepsilon _{1}\left( \bar{\psi%
}_{\mu }\gamma ^{a}\psi ^{\rho }+\bar{\psi}^{a}\gamma _{\mu }\psi ^{\rho
}\right) +\gamma _{a}\gamma _{b}\varepsilon _{1}\left( \bar{\psi}^{a}\gamma
_{c}\psi ^{b}\right) \delta _{\mu }^{\rho }\right]
\end{eqnarray*}
where Majorana flip identities and the fact that, multiplied by $\gamma
_{a}\gamma _{b}$, $\frac{\partial H_{\mu }^{ab}}{\partial e_{\rho }^{c}}$ is
antisymmetric in $a$ and $b$, have been used.

	The term containing the principal connection $\Gamma _{\mu }^{ab}$ can be
recast with the use of (\ref{ojjo}): 
\begin{eqnarray}
-\frac{1}{4}\gamma _{a}\gamma _{b}\frac{\partial \Gamma _{\mu }^{ab}%
}{\partial e_{\rho }^{c}}\varepsilon _{1} &=&-\frac{1}{4}[\gamma _{a}\gamma
_{c}\varepsilon _{1}d_{\mu }e^{a\rho }+2\gamma ^{\lambda }\gamma
_{c}\varepsilon _{1}\Gamma _{\lambda \mu }^{\rho }+\gamma _{c}\gamma
_{\sigma }\varepsilon _{1}g^{\rho \lambda }\Gamma _{\lambda \mu }^{\sigma }+
\nonumber \\
&&+\gamma ^{\lambda }\gamma ^{\rho }\varepsilon _{1}e_{c\sigma }\Gamma
_{\lambda \mu }^{\sigma }+\gamma ^{\nu }\gamma ^{\lambda }\varepsilon
_{1}\left( \partial _{\lambda }e_{\nu c}\delta _{\mu }^{\rho }-\partial
_{\nu }e_{\mu c}\delta _{\lambda }^{\rho }\right) ]
\end{eqnarray}
Therefore the first element of the commutator acting on the gravitino can
be written as 
\begin{eqnarray*}
\delta _{1}e_{\rho }^{c}\frac{\partial \left( \delta _{2}\psi _{\mu }\right) 
}{\partial e_{\rho }^{c}} &=&-\frac{i}{4}\{\left( \bar{\varepsilon}%
_{1}\gamma ^{c}\psi _{\rho }\right) d_{\mu }e^{a\rho }\gamma _{a}\gamma
_{c}\varepsilon _{2}+2\left( \bar{\varepsilon}_{1}\gamma ^{c}\psi _{\rho
}\right) \gamma ^{\lambda }\gamma _{c}\varepsilon _{2}\Gamma _{\lambda \mu
}^{\rho }+ \\
&&+\left( \bar{\varepsilon}_{1}\gamma ^{c}\psi ^{\lambda }\right) \gamma
_{c}\gamma _{\sigma }\varepsilon _{2}\Gamma _{\lambda \mu }^{\sigma }+\left( 
\bar{\varepsilon}_{1}\gamma ^{c}\psi _{\rho }\right) \gamma ^{\lambda
}\gamma ^{\rho }\varepsilon _{2}e_{c\sigma }\Gamma _{\lambda \mu }^{\sigma }+
\\
&&+\left( \bar{\varepsilon}_{1}\gamma ^{c}\psi _{\rho }\right) \gamma ^{\nu
}\gamma ^{\lambda }\varepsilon _{2}\left( \partial _{\lambda }e_{\nu
c}\delta _{\mu }^{\rho }-\partial _{\nu }e_{\mu c}\delta _{\lambda }^{\rho
}\right) + \\
&&+2i\left( \bar{\varepsilon}_{1}\gamma ^{c}\psi _{\rho }\right) \left( \bar{%
\psi}_{\mu }\gamma ^{a}\psi ^{\rho }+\bar{\psi}^{a}\gamma _{\mu }\psi ^{\rho
}\right) \gamma _{a}\gamma _{c}\varepsilon _{2}+ \\
&&+i\left( \bar{\varepsilon}_{1}\gamma ^{c}\psi _{\rho }\right) \left( \bar{%
\psi}^{a}\gamma _{c}\psi ^{b}\right) \gamma _{a}\gamma _{b}\varepsilon
_{2}\delta _{\mu }^{\rho }\}
\end{eqnarray*}
We have now to evaluate the other two terms: 
\begin{eqnarray}
\frac{\partial \left( \delta \psi _{\mu }\right) }{\partial \left( d_{\sigma
}e_{\rho }^{c}\right) }&=&\frac{\partial }{\partial \left( d_{\sigma }e_{\rho
}^{c}\right) }\left[ -\frac{1}{4}\gamma _{a}\gamma _{b}\left(
\Gamma _{\mu }^{ab}+H_{\mu }^{ab}\right) \varepsilon \right] = \nonumber\\
&=&-\frac{1}{4}\gamma _{a}\gamma _{b}\left( \frac{\partial \Gamma
_{\mu }^{ab}}{\partial \left( d_{\sigma }e_{\rho }^{c}\right) }+\frac{
\partial H_{\mu }^{ab}}{\partial \left( d_{\sigma }e_{\rho }^{c}\right)}
\right) \varepsilon
\end{eqnarray}
Since $\frac{\partial H_{\mu }^{ab}}{\partial \left( d_{\sigma }e_{\rho
}^{c}\right) }=0$, we are left with 
\begin{eqnarray}
\frac{\partial \Gamma _{\mu }^{ab}}{\partial \left( d_{\sigma }e_{\rho
}^{c}\right) } &=&\frac{\partial \left( e_{\sigma }^{b}\Gamma _{\lambda \mu
}^{\sigma }e^{a\lambda }\right) }{\partial \left( d_{\sigma }e_{\rho
}^{c}\right) }+\frac{\partial \left( e_{\lambda }^{b}d_{\mu }e^{a\lambda
}\right) }{\partial \left( d_{\sigma }e_{\rho }^{c}\right) }=  \nonumber \\
&=&e_{\lambda }^{b}\delta _{\mu }^{\sigma }\delta _{c}^{a}g^{\rho \lambda
}=e^{b\rho }\delta _{\mu }^{\sigma }\delta _{c}^{a}
\end{eqnarray}
which leads to 
\begin{equation}
\frac{\partial \left( \delta \psi _{\mu }\right) }{\partial \left( d_{\sigma
}e_{\rho }^{c}\right) }=-\frac{1}{4}\gamma _{a}\gamma _{b}e^{b\rho }\delta
_{\mu }^{\sigma }\delta _{c}^{a}\varepsilon =-\frac{1}{4}\gamma _{c}\gamma
^{\rho }\delta _{\mu }^{\sigma }\varepsilon
\end{equation}
and finally to 
\begin{eqnarray}
d_{\sigma }\left( \delta _{1}e_{\rho }^{c}\right) \frac{\partial \left(
\delta _{2}\psi _{\mu }\right) }{\partial \left( d_{\sigma }e_{\rho
}^{c}\right) } &=&-\frac{i}{4}d_{\sigma }\left( \bar{\varepsilon}_{1}\gamma
^{c}\psi _{\rho }\right) \gamma _{c}\gamma ^{\rho }\delta _{\mu }^{\sigma
}\varepsilon _{2}=  \nonumber \\
&=&-\frac{i}{4}d_{\mu }\left( \bar{\varepsilon}_{1}\gamma ^{c}\psi _{\rho
}\right) \gamma _{c}\gamma ^{\rho }\varepsilon _{2}
\end{eqnarray}
Only the quantity
\begin{equation}
\delta _{1}\psi _{\rho }\frac{\partial \left( \delta _{2}\psi _{\mu }\right) 
}{\partial \psi _{\rho }}  \label{thy}
\end{equation}
has still to be calculated. Expanding the expression of the supersymmetries
on the gravitino field gives 
\begin{eqnarray}
\frac{\partial \left( \delta _{2}\psi _{\mu }\right) }{\partial \psi _{\rho }%
}&=&\frac{\partial \left( \nabla _{\mu }\varepsilon _{2}\right) }{\partial
\psi _{\rho }}=\frac{\partial }{\partial \psi _{\rho }}\left[ d_{\mu
}\varepsilon _{2}-\frac{1}{4}\gamma _{a}\gamma _{b}\left( \Gamma _{\mu
}^{ab}+H_{\mu }^{ab}\right) \varepsilon _{2}\right] = \nonumber\\
&=&-\frac{1}{4}\gamma _{a}\gamma _{b}\frac{\partial H_{\mu }^{ab}}{\partial
\psi _{\rho }}\varepsilon _{2}=-\frac{i}{4}\gamma _{a}\gamma _{b}\frac{%
\partial }{\partial \psi _{\rho }}\left( \bar{\psi}_{\mu }\gamma ^{a}\psi
^{b}+\bar{\psi}^{a}\gamma ^{b}\psi _{\mu }+\bar{\psi}^{a}\gamma _{\mu }\psi
^{b}\right) =\nonumber\\
&=&-\frac{i}{4}\gamma _{a}\gamma _{b}\frac{\partial }{\partial \psi _{\rho }}%
\left( \bar{\psi}_{\mu }\gamma ^{a}\psi _{\nu }e^{b\nu }+\bar{\psi}%
^{a}\gamma ^{b}\psi _{\mu }+\bar{\psi}^{a}\gamma _{\mu }\psi _{\nu }e^{b\nu
}\right) =  \nonumber \\
&=&-\frac{i}{4}\gamma _{a}\gamma _{b}\left( \bar{\psi}_{\mu }\gamma ^{a}%
\frac{\partial \psi _{\nu }}{\partial \psi _{\rho }}e^{b\nu }+\bar{\psi}%
^{a}\gamma ^{b}\frac{\partial \psi _{\mu }}{\partial \psi _{\rho }}+\bar{\psi%
}^{a}\gamma _{\mu }\frac{\partial \psi _{\nu }}{\partial \psi _{\rho }}%
e^{b\nu }\right) =  \nonumber \\
&=&-\frac{i}{4}\gamma _{a}\gamma _{b}\left[ \left( \bar{\psi}_{\mu }\gamma
^{a}+\bar{\psi}^{a}\gamma _{\mu }\right) e^{b\rho }+\bar{\psi}^{a}\gamma
^{b}\delta _{\mu }^{\rho }\right]
\end{eqnarray}
So (\ref{thy}) can be recast as 
\begin{eqnarray*}
\delta _{1}\psi _{\rho }\frac{\partial \left( \delta _{2}\psi _{\mu }\right) 
}{\partial \psi _{\rho }} &=&-\frac{i}{4}\gamma _{a}\gamma _{b}\left[ \left( 
\bar{\psi}_{\mu }\gamma ^{a}+\bar{\psi}^{a}\gamma _{\mu }\right) e^{b\rho }+%
\bar{\psi}^{a}\gamma ^{b}\delta _{\mu }^{\rho }\right] \varepsilon
_{2}\nabla _{\rho }\varepsilon _{1}= \\
&=&-\frac{i}{4}\left[ \gamma _{a}\gamma ^{\rho }\left( \bar{\psi}_{\mu
}\gamma ^{a}\varepsilon _{2}+\bar{\psi}^{a}\gamma _{\mu }\varepsilon
_{2}\right) \nabla _{\rho }\varepsilon _{1}+\gamma _{a}\gamma _{b}\bar{\psi}%
^{a}\gamma ^{b}\varepsilon _{2}\nabla _{\mu }\varepsilon _{1}\right]
\end{eqnarray*}
This allows to conclude that the commutator of supersymmetries
applied to the gravitino is 
\begin{eqnarray}
&&[\delta _{1}e_{\rho }^{c}\frac{\partial \left( \delta _{2}\psi _{\mu
}\right) }{\partial e_{\rho }^{c}}+\delta _{1}\psi _{\rho }\frac{\partial
\left( \delta _{2}\psi _{\mu }\right) }{\partial \psi _{\rho }}
+d_{\sigma }\left( \delta _{1}e_{\rho }^{c}\right) \frac{\partial \left(
\delta _{2}\psi _{\mu }\right) }{\partial \left( d_{\sigma }e_{\rho
}^{c}\right) }+\nonumber\\
&&+d_{\sigma }\left( \delta _{1}\psi _{\rho }\right) \frac{%
\partial \left( \delta _{2}\psi _{\mu }\right) }{\partial \left( d_{\sigma
}\psi _{\rho }\right) }
-\left( 1\leftrightarrow 2\right)]\frac{\partial }{\partial \psi
_{\mu }}=\nonumber\\
&&=-\frac{i}{4}\{\left( \bar{\varepsilon}_{1}\gamma ^{c}\psi _{\rho }\right)
d_{\mu }e^{a\rho }\gamma _{a}\gamma _{c}\varepsilon _{2}+2\left( \bar{%
\varepsilon}_{1}\gamma ^{c}\psi _{\rho }\right) \gamma ^{\lambda }\gamma
_{c}\varepsilon _{2}\Gamma _{\lambda \mu }^{\rho }+ \nonumber\\
&&+\left( \bar{\varepsilon}_{1}\gamma ^{c}\psi ^{\lambda }\right) \gamma
_{c}\gamma _{\sigma }\varepsilon _{2}\Gamma _{\lambda \mu }^{\sigma }+\left( 
\bar{\varepsilon}_{1}\gamma ^{c}\psi _{\rho }\right) \gamma ^{\lambda
}\gamma ^{\rho }\varepsilon _{2}e_{c\sigma }\Gamma _{\lambda \mu }^{\sigma }+
\nonumber\\
&&+\left( \bar{\varepsilon}_{1}\gamma ^{c}\psi _{\rho }\right) \gamma ^{\nu
}\gamma ^{\lambda }\varepsilon _{2}\left( \partial _{\lambda }e_{\nu
c}\delta _{\mu }^{\rho }-\partial _{\nu }e_{\mu c}\delta _{\lambda }^{\rho
}\right) + \nonumber\\
&&+2i\left( \bar{\varepsilon}_{1}\gamma ^{c}\psi _{\rho }\right) \left( \bar{%
\psi}_{\mu }\gamma ^{a}\psi ^{\rho }+\bar{\psi}^{a}\gamma _{\mu }\psi ^{\rho
}\right) \gamma _{a}\gamma _{c}\varepsilon _{2}+ \nonumber\\
&&+i\left( \bar{\varepsilon}_{1}\gamma ^{c}\psi _{\rho }\right) \left( \bar{%
\psi}^{a}\gamma _{c}\psi ^{b}\right) \gamma _{a}\gamma _{b}\varepsilon
_{2}\delta _{\mu }^{\rho }+\gamma _{a}\gamma ^{\rho }\left( \bar{\psi}_{\mu }\gamma ^{a}\varepsilon
_{2}+\bar{\psi}^{a}\gamma _{\mu }\varepsilon _{2}\right) \nabla _{\rho
}\varepsilon _{1}+\nonumber\\
&&+\gamma _{a}\gamma _{b}\left( \bar{\psi}^{a}\gamma
^{b}\varepsilon _{2}\right) \nabla _{\mu }\varepsilon _{1}
+d_{\mu }\left( \bar{\varepsilon}_{1}\gamma ^{c}\psi _{\rho }\right) \
\gamma_{c}\gamma^{\rho }\varepsilon _{2}-\left(
1\leftrightarrow 2\right) \}\frac{\partial }{\partial \psi _{\mu }}
\end{eqnarray}
The right hand side of this equation becomes, with some easy calculations, 
\begin{eqnarray}
&&-\frac{i}{2}\{[2\left( \bar{\varepsilon}_{[1}\gamma ^{b}\psi _{c}\right)
\Gamma _{\mu }^{ac}+\left( \bar{\varepsilon}_{[1}\gamma ^{a}\psi _{c}\right)
\left( e^{c\lambda }d_{\mu }e_{\lambda }^{b}+e^{c\lambda }e_{\sigma
}^{b}\Gamma _{\lambda \mu }^{\sigma }\right) +  \nonumber \\
&&+\left( \bar{\varepsilon}_{[1}\gamma ^{c}\psi _{\rho }\right) e^{a\lambda
}e^{\rho b}e_{c\sigma }\Gamma _{\lambda \mu }^{\sigma }+  \nonumber \\
&&+\left( \bar{\varepsilon}_{[1}\gamma ^{c}\psi _{\rho }\right) \left(
\partial _{\lambda }e_{\nu c}\delta _{\mu }^{\rho }-\partial _{\nu }e_{\mu
c}\delta _{\lambda }^{\rho }\right) e^{a\nu }e^{\lambda b}+  \nonumber \\
&&+2i\left( \bar{\varepsilon}_{[1}\gamma ^{b}\psi _{\rho }\right) \left( 
\bar{\psi}_{\mu }\gamma ^{a}\psi ^{\rho }+\bar{\psi}^{a}\gamma _{\mu }\psi
^{\rho }\right) +  \nonumber \\
&&+i\left( \bar{\varepsilon}_{[1}\gamma ^{c}\psi _{\rho }\right) \left( \bar{%
\psi}^{a}\gamma _{c}\psi ^{b}\right) \delta _{\mu }^{\rho }+d_{\mu }\left( 
\bar{\varepsilon}_{[1}\gamma ^{a}\psi _{\lambda }\right) e^{\lambda
b}]\gamma _{a}\gamma _{b}\varepsilon _{2]}+  \nonumber \\
&&-\left[ \left( \bar{\psi}_{\mu }\gamma ^{a}\varepsilon _{[1}+\bar{\psi}%
^{a}\gamma _{\mu }\varepsilon _{[1}\right) e^{\rho b}+\bar{\psi}^{a}\gamma
^{b}\varepsilon _{[1}\delta _{\mu }^{\rho }\right] \gamma _{a}\gamma
_{b}\nabla _{\rho }\varepsilon _{2]}\}\frac{\partial }{\partial \psi _{\mu }}
\end{eqnarray}
So we finally obtain 
\begin{eqnarray}
&&\left[ X,Y\right] \psi _{\mu }=-\frac{i}{2}\{[2\left( \bar{\varepsilon}%
_{[1}\gamma ^{b}\psi _{c}\right) \Gamma _{\mu }^{ac}+  \nonumber \\
&&\left( \bar{\varepsilon}_{[1}\gamma ^{a}\psi _{c}\right) \left(
e^{c\lambda }d_{\mu }e_{\lambda }^{b}+e^{c\lambda }e_{\sigma }^{b}\Gamma
_{\lambda \mu }^{\sigma }\right) +\left( \bar{\varepsilon}_{[1}\gamma
^{c}\psi _{\rho }\right) e^{a\lambda }e^{\rho b}e_{c\sigma }\Gamma _{\lambda
\mu }^{\sigma }+  \nonumber \\
&&\left( \bar{\varepsilon}_{[1}\gamma ^{c}\psi _{\rho }\right) \left(
\partial _{\lambda }e_{\nu c}\delta _{\mu }^{\rho }-\partial _{\nu }e_{\mu
c}\delta _{\lambda }^{\rho }\right) e^{a\nu }e^{\lambda b}+  \nonumber \\
&&2i\left( \bar{\varepsilon}_{[1}\gamma ^{b}\psi _{\rho }\right) \left( \bar{%
\psi}_{\mu }\gamma ^{a}\psi ^{\rho }+\bar{\psi}^{a}\gamma _{\mu }\psi ^{\rho
}\right) +  \nonumber \\
&&i\left( \bar{\varepsilon}_{[1}\gamma ^{c}\psi _{\rho }\right) \left( \bar{%
\psi}^{a}\gamma _{c}\psi ^{b}\right) \delta _{\mu }^{\rho }+d_{\mu }\left( 
\bar{\varepsilon}_{[1}\gamma ^{a}\psi _{\lambda }\right) e^{\lambda
b}]\gamma _{a}\gamma _{b}\varepsilon _{2]}+  \nonumber \\
&&-\left[ \left( \bar{\psi}_{\mu }\gamma ^{a}\varepsilon _{[1}+\bar{\psi}%
^{a}\gamma _{\mu }\varepsilon _{[1}\right) e^{\rho b}+\bar{\psi}^{a}\gamma
^{b}\varepsilon _{[1}\delta _{\mu }^{\rho }\right] \gamma _{a}\gamma
_{b}\nabla _{\rho }\varepsilon _{2]}\}
\end{eqnarray}
This form is very complicated. Now we want to rewrite this commutator as a
function of the field equations of the gravitino, in order to check that the
algebra closes only on-shell, as the on-shell covariance of the Lagrangian
seems to suggest.

	We start from (\ref{supersim conness})
\[
\delta \omega _{\mu }^{ab}=-i(\bar{\varepsilon}\gamma _{\mu }\nabla _{\rho
}\psi _{\nu }+\bar{\varepsilon}\gamma _{\rho }\nabla _{\mu }\psi _{\nu }-%
\bar{\varepsilon}\gamma _{\rho }\nabla _{\nu }\psi _{\mu })e^{\,\rho
[a}e^{b]\nu } 
\]
which can be recast as 
\begin{equation}
\delta \omega _{\mu }^{ab}=i(\bar{\varepsilon}\gamma _{\mu }\nabla _{[\nu
}\psi _{\rho ]}+2\bar{\varepsilon}\gamma _{[\rho }\psi _{\nu ]\mu
})e^{\,\rho a}e^{b\nu }
\end{equation}
where 
\begin{equation}
\psi _{\nu \mu }=\nabla _{[\nu }\psi _{\mu ]}
\end{equation}
We now use the property 
\begin{equation}
\gamma _{\mu }\nabla _{[\rho }\psi _{\nu ]}=-\frac{1}{2}\epsilon _{l\rho \nu
\mu }\gamma _{5}E^{l}-2\gamma _{[\rho }\psi _{\nu ]\mu }
\end{equation}
with 
\begin{equation}
E^{l}=\gamma _{5}\gamma _{a}\nabla _{\nu }\psi _{\rho }e_{\sigma
}^{a}\epsilon ^{l\nu \rho \sigma }
\end{equation}
which are the gravitino field equations. Therefore 
\begin{equation}
\delta \omega _{\mu }^{ab}=i(4\bar{\varepsilon}\gamma _{[\rho }\psi _{\nu
]\mu }+\frac{1}{2}\bar{\varepsilon}\gamma _{5}E^{l}\epsilon _{l\rho \nu \mu
})e^{\,\rho a}e^{b\nu }
\end{equation}
Now the commutator of two supersymmetries applied to the gravitino is 
\begin{eqnarray}
\left[ \delta _{1},\delta _{2}\right] \psi _{\mu }&=&\frac{1}{4}\delta
_{1}\omega _{\mu }^{ab}\gamma _{a}\gamma _{b}\varepsilon _{2}-\frac{1}{4}%
\delta _{2}\omega _{\mu }^{ab}\gamma _{a}\gamma _{b}\varepsilon _{1}= \nonumber \\ 
&=&\frac{i}{4}(4\bar{\varepsilon}_{1}\gamma _{[\rho }\psi _{\nu ]\mu }+
\frac{1}{2}\bar{\varepsilon}_{1}\gamma _{5}E^{l}\epsilon _{l\rho \nu \mu
})e^{\,\rho a}e^{b\nu }\gamma _{a}\gamma _{b}\varepsilon _{2}+ \nonumber \\ 
&-&\frac{i}{4}(4\bar{\varepsilon}_{2}\gamma _{[\rho }\psi _{\nu ]\mu }+
\frac{1}{2}\bar{\varepsilon}_{2}\gamma _{5}E^{l}\epsilon _{l\rho \nu \mu
})e^{\,\rho a}e^{b\nu }\gamma _{a}\gamma _{b}\varepsilon _{1}= \nonumber \\ 
&=&2i\gamma ^{\rho }\gamma ^{\nu }\varepsilon _{[2}\bar{\varepsilon}
_{1]}\gamma _{\rho }\psi _{\nu \mu }+\frac{i}{4}\gamma ^{\rho }\gamma ^{\nu
}\varepsilon _{[2}\bar{\varepsilon}_{1]}\gamma _{5}E^{l}\epsilon _{l\rho \nu
\mu }
\end{eqnarray}
With the Fierz formula 
\begin{equation}
\varepsilon _{[2}\bar{\varepsilon}_{1]}=-\frac{1}{4}\gamma ^{\lambda }\bar{%
\varepsilon}_{1}\gamma _{\lambda }\varepsilon _{2}+\frac{1}{8}\gamma
^{\lambda }\gamma ^{\theta }\bar{\varepsilon}_{1}\gamma _{\lambda }\gamma
_{\theta }\varepsilon _{2}
\end{equation}
we obtain 
\begin{eqnarray}
&&\left[ \delta _{1},\delta _{2}\right] \psi _{\mu }=2i\gamma ^{\rho }\gamma
^{\nu }\left( -\frac{1}{4}\gamma ^{\lambda }\bar{\varepsilon}_{1}\gamma
_{\lambda }\varepsilon _{2}+\frac{1}{8}\gamma ^{\lambda }\gamma ^{\theta }
\bar{\varepsilon}_{1}\gamma _{\lambda }\gamma _{\theta }\varepsilon
_{2}\right) \gamma _{\rho }\psi _{\nu \mu }+ \nonumber \\ 
&&+\frac{i}{4}\gamma ^{\rho }\gamma ^{\nu }\left( -\frac{1}{4}\gamma
^{\lambda }\bar{\varepsilon}_{1}\gamma _{\lambda }\varepsilon _{2}+\frac{1}{8
}\gamma ^{\lambda }\gamma ^{\theta }\bar{\varepsilon}_{1}\gamma _{\lambda
}\gamma _{\theta }\varepsilon _{2}\right) \gamma _{5}E^{l}\epsilon _{l\rho
\nu \mu }
\label{commutator}
\end{eqnarray}
Recalling Eq.(\ref{campo orizz}), i.e. 
\begin{equation}
\xi ^{\rho }=i(\bar{\varepsilon}_{2}\gamma ^{\rho }\varepsilon _{1})=-i(\bar{%
\varepsilon}_{1}\gamma ^{\rho }\varepsilon _{2})
\end{equation}
and the gamma matrices properties 
\begin{eqnarray}
&&\gamma ^{\rho }\gamma ^{\nu }\gamma ^{\lambda }\gamma _{\rho }=4\eta ^{\nu
\lambda }\\
&&\gamma ^{\rho }\gamma ^{\nu }\gamma ^{\lambda }\gamma ^{\theta }\gamma
_{\rho }=\gamma ^{\lambda }\gamma ^{\theta }\gamma ^{\nu }
\end{eqnarray}
the four terms of (\ref{commutator}) can be rewritten as: 
\begin{eqnarray}
1+3&=&\frac{1}{2}\xi _{\lambda }\gamma ^{\rho }\gamma ^{\nu }\gamma
^{\lambda }\gamma _{\rho }\psi _{\nu \mu }+\frac{1}{16}\xi _{\lambda }\gamma
^{\rho }\gamma ^{\nu }\gamma ^{\lambda }\gamma _{5}E^{l}\epsilon _{l\rho \nu
\mu }= \nonumber\\ 
&=&2\xi ^{\nu }\psi _{\nu \mu }+\frac{1}{16}\xi _{\lambda }\gamma ^{\rho
}\gamma ^{\nu }\gamma ^{\lambda }\gamma _{5}E^{l}\epsilon _{l\rho \nu \mu }
\end{eqnarray}
\begin{eqnarray}
2+4&=&\frac{i}{4}\left( \bar{\varepsilon}_{1}\gamma _{\lambda }\gamma
_{\theta }\varepsilon _{2}\right) \gamma ^{\rho }\gamma ^{\nu }\gamma
^{\lambda }\gamma ^{\theta }\gamma _{\rho }\psi _{\nu \mu }
+\frac{i}{32}\left( \bar{\varepsilon}_{1}\gamma _{\lambda }\gamma
_{\theta }\varepsilon _{2}\right) \gamma ^{\rho }\gamma ^{\nu }\gamma
^{\lambda }\gamma ^{\theta }\gamma _{5}E^{l}\epsilon _{l\rho \nu \mu }= \nonumber\\ 
&=&\left( \bar{\varepsilon}_{1}\gamma _{\lambda }\gamma _{\theta
}\varepsilon _{2}\right) (-\frac{i}{2}\gamma ^{\theta }\gamma ^{\lambda
}\gamma ^{\nu }\psi _{\nu \mu } 
+\frac{i}{32}\gamma ^{\rho }\gamma ^{\nu }\gamma ^{\lambda }\gamma
^{\theta }\gamma _{5}E^{l}\epsilon _{l\rho \nu \mu })
\end{eqnarray}
Using now the properties
\begin{eqnarray}
&&\gamma ^{\rho }\gamma ^{\nu }\gamma ^{\lambda }\gamma _{5}\epsilon _{l\rho
\nu \mu }=-2i\gamma _{l}\gamma _{\mu }\gamma ^{\lambda } \\ 
&&\gamma ^{\nu }\psi _{\nu \mu }=-\frac{i}{4}\gamma _{\mu }\gamma ^{l}E_{l}+
\frac{i}{2}E_{\mu }\\ 
&&\gamma ^{\rho }\gamma ^{\nu }\gamma ^{\lambda }\gamma ^{\theta }\gamma
_{5}\epsilon _{l\rho \nu \mu }=2i\gamma _{l}\gamma _{\mu }\gamma ^{\lambda
}\gamma ^{\theta }
\end{eqnarray}
We finally recover 
\begin{equation}
1+3=2\xi ^{\nu }\psi _{\nu \mu }-\frac{i}{8}\xi _{\lambda }\gamma _{l}\gamma
_{\mu }\gamma ^{\lambda }E^{l}
\end{equation}
and 
\begin{eqnarray}
2+4 &=&\left( \bar{\varepsilon}_{1}\gamma _{\lambda }\gamma _{\theta
}\varepsilon _{2}\right) \left( \frac{i}{4}\gamma ^{\lambda }\gamma ^{\theta
}\gamma ^{\nu }\psi _{\nu \mu }-\frac{1}{16}\gamma _{l}\gamma _{\mu }\gamma
^{\lambda }\gamma ^{\theta }E^{l}\right) =  \nonumber \\
&=&\left( \bar{\varepsilon}_{1}\gamma _{\lambda }\gamma _{\theta
}\varepsilon _{2}\right) \left( \frac{1}{16}\gamma ^{\lambda }\gamma
^{\theta }\gamma _{\mu }\gamma ^{l}E_{l}-\frac{1}{8}\gamma ^{\lambda }\gamma
^{\theta }E_{\mu }-\frac{1}{16}\gamma _{l}\gamma _{\mu }\gamma ^{\lambda
}\gamma ^{\theta }E^{l}\right) =  \nonumber \\
&=&\frac{1}{16}\left( \bar{\varepsilon}_{1}\gamma _{\lambda }\gamma _{\theta
}\varepsilon _{2}\right) \left( \gamma ^{\lambda }\gamma ^{\theta }\gamma
_{\mu }\gamma _{l}E^{l}-2\gamma ^{\lambda }\gamma ^{\theta }E_{\mu }-\gamma
_{l}\gamma _{\mu }\gamma ^{\lambda }\gamma ^{\theta }E^{l}\right)
\end{eqnarray}
Now, the property of the gamma matrices 
\begin{equation}
\left[ \gamma ^{\lambda }\gamma ^{\theta },\gamma _{\mu }\gamma _{l}\right]
=-2\delta _{l}^{\theta }\gamma ^{\lambda }\gamma _{\mu }+2\delta
_{l}^{\lambda }\gamma ^{\theta }\gamma _{\mu }
\end{equation}
gives 
\begin{eqnarray}
2+4 &=&\frac{1}{8}\left( \bar{\varepsilon}_{1}\gamma _{\lambda }\gamma
_{\theta }\varepsilon _{2}\right) \left( -\gamma ^{\lambda }\gamma _{\mu
}E^{\theta }+\gamma ^{\theta }\gamma _{\mu }E^{\lambda }-\gamma _{l}\gamma
_{\mu }\gamma ^{\lambda }\gamma ^{\theta }E^{l}\right) =  \nonumber \\
&=&\frac{1}{8}\left( \bar{\varepsilon}_{1}\gamma _{\lambda }\gamma _{\theta
}\varepsilon _{2}\right) \left( -\gamma ^{\lambda }\gamma _{\mu }\delta
_{l}^{\theta }+\gamma ^{\theta }\gamma _{\mu }\delta _{l}^{\lambda }-\gamma
_{l}\gamma _{\mu }\gamma ^{\lambda }\gamma ^{\theta }\right) E^{l}
\end{eqnarray}
Thus the commutator of two supersymmetries on the gravitino is expressed
in function of the gravitino field equations as follows: 
\begin{eqnarray}
\left[ \delta _{1},\delta _{2}\right] \psi _{\mu } &=&2\xi ^{\nu }\psi _{\nu
\mu }+\frac{1}{8}[\left( \bar{\varepsilon}_{1}\gamma _{\lambda }\gamma
_{\theta }\varepsilon _{2}\right) (-\gamma ^{\lambda }\gamma _{\mu }\delta
_{l}^{\theta }+  \nonumber \\
&&+\gamma ^{\theta }\gamma _{\mu }\delta _{l}^{\lambda }-\gamma _{l}\gamma
_{\mu }\gamma ^{\lambda }\gamma ^{\theta })+i\xi _{\lambda }\gamma
_{l}\gamma _{\mu }\gamma ^{\lambda }]E^{l}
\end{eqnarray}
It is now clear that the algebra closes on-shell. Furthermore we see that
the commutator is a Lie derivative of the field, but with respect to a
vector field over $\Sigma $ which depends on the field. In fact, 
\begin{equation}
\lbrack \delta _{1},\delta _{2}]e_{\mu }^{a}=\pounds _{\Xi }e_{\mu }^{a}
\end{equation}
gives 
\begin{equation}
\Xi =i(\bar{\varepsilon}_{2}\gamma ^{\rho }\varepsilon _{1})\partial _{\rho
}+2(\bar{\varepsilon}_{2}\gamma ^{\rho }\varepsilon _{1})e_{b}^{\lambda
}\nabla _{[\rho }e_{\lambda ]}^{a}\sigma _{a}^{b}=i\xi ^{\rho }\partial
_{\rho }+2\xi ^{\rho }e_{b}^{\lambda }\nabla _{[\rho }e_{\lambda
]}^{a}\sigma _{a}^{b}
\end{equation}
For the gravitino, the condition 
\begin{equation}
\lbrack \delta _{1},\delta _{2}]\psi _{\mu }=\pounds _{\Xi }\psi _{\mu }
\end{equation}
provided Eq.(\ref{derivata di Lie gravitino}), i.e. 
\begin{eqnarray*}
\pounds _{\Xi }\psi _{\mu }=\xi ^{\rho }\nabla _{\rho }\psi _{\mu }+\nabla
_{\mu }\xi ^{\nu }\psi _{\nu }-\frac{1}{4}\gamma _{a}\gamma _{b}\psi _{\mu
}\xi _{(\upsilon )}^{\,{ab}} 
\end{eqnarray*}
gives 
\begin{equation}
\xi ^{\rho }=i(\bar{\varepsilon}_{2}\gamma ^{\rho }\varepsilon _{1})
\end{equation}
and 
\begin{eqnarray}
\gamma _{a}\gamma _{b}\psi _{\mu }\xi _{(\upsilon )}^{\,{ab}} &=&4\nabla
_{\mu }\left( \xi ^{\nu }\psi _{\nu }\right) -\frac{1}{2}[\left( \bar{%
\varepsilon}_{1}\gamma _{\lambda }\gamma _{\theta }\varepsilon _{2}\right)
(-\gamma ^{\lambda }\gamma _{\mu }\delta _{l}^{\theta }+  \nonumber \\
&&+\gamma ^{\theta }\gamma _{\mu }\delta _{l}^{\lambda }-\gamma _{l}\gamma
_{\mu }\gamma ^{\lambda }\gamma ^{\theta })+i\xi _{\lambda }\gamma
_{l}\gamma _{\mu }\gamma ^{\lambda }]E^{l}
\end{eqnarray}
This is fairly different from what happens in the Wess-Zumino model. It is not unexpected, since we found previously that the Rarita-Schwinger Lagrangian is covariant on-shell. Both results mean that it is not
clear, at least within the standard approach, how to apply the Gauge-Natural
formalism to Supergravity. The correspondence between Lie derivatives and
commutators is indeed fundamental in this sense.

	All troubles seem to arise from the choice of the null torsion constraint. Therefore if the algebra would close off-shell, it might be possible to define a Gauge-Natural theory of Supergravity.
The Grignani-Nardelli approach could suit to our purposes; it is addressed in the next section.

\section{The Grignani-Nardelli approach}\label{GN}

This framework was first introduced in the early 80s by Stelle and West for
the SO(3,2) group spontaneously broken to the Lorentz group \cite{sugra3}. Later, Grignani and Nardelli adapted this formalism to a toy model \cite{sugra4} and to the Poincar\'{e} group \cite{sugra5}. We will show here only the basic features of this alternative; a complete analysis will be given in a future
work.

	The key ingredient is the introduction of an extra field in the theory: the
set of \textsl{Poincar\'{e} coordinates} $q^{a}\left( x\right) $. These are
regarded as Higgs fields that transform as vectors ''under Poincar\'{e}
transformations''. In our perspective they are simply the sections of a bundle $%
\Sigma _{\sigma }$ associated to the spin bundle by means of the
representation 
\begin{equation}
\sigma :Spin\left( \eta \right) \times \R^{m}\longrightarrow \R%
^{m}:\left( S,q^{a}\right) \longmapsto \ell _{b}^{a}\left( S\right) q^{b}
\end{equation}
These are used to redefine the vielbein as follows: 
\begin{equation}
e_{\mu }^{a}\longmapsto V_{\mu }^{a}=e_{\mu }^{a}+\nabla _{\mu }q^{a}
\label{nuova vielbein}
\end{equation}
This equation is justified by the following argument: for any fixed
space-time point $x^{\mu }$, we can choose a frame in which the space-time is
locally Minkowski. In $x^{\mu }$ one can choose $\omega _{\mu }^{ab}=e_{\mu
}^{a}=0$, so that the vielbein becomes $V_{\mu }^{a}=\partial _{\mu }q^{a}$.
This means that the $q^{a}$ can be regarded as the local orthonormal
coordinates at a fixed point. By general covariance, one can then evaluate
the form of the vielbein at any other point by imposing $\partial _{\mu
}q^{a}\longmapsto \nabla _{\mu }q^{a}$, which is Eq.(\ref{nuova vielbein}).
For this reason the $q^{a}$ are called ''Poincar\'{e} coordinates''.

	With the Poincar\'{e} coordinates, the fields involved are now four: the new
vielbein $V_{\mu }^{a}$, the gravitino $\psi _{\mu }$, the spin connection $%
\omega _{\mu }^{ab}$ (which is now independent) and the $q^{a}\left(
x\right) $. The supersymmetries are generated again by the 4-component, spin $1/2
$ Majorana spinor $\varepsilon $ as follows \footnote{The covariant derivative is
calculated with respect to the spin connection $\omega _{\mu }^{ab}$}:
\begin{equation}
\left\{ 
\begin{tabular}{l}
\vspace{0.1cm}$\delta V_{\mu }^{a}=i\bar{\varepsilon}\gamma ^{a}\psi _{\mu }$ \\ 
\vspace{0.1cm}$\delta \psi _{\mu }=\nabla _{\mu }\varepsilon $ \\ 
\vspace{0.1cm}$\delta \omega _{\mu }^{ab}=0$ \\ 
$\delta q^{a}=0$%
\end{tabular}
\right.  \label{G-N}
\end{equation}
By means of the substitution (\ref{nuova vielbein}), the Rarita-Schwinger
Lagrangian is rewritten as 
\begin{eqnarray}
L &=&\mathcal{L}ds=\left( -4R_{\mu \nu }^{ab}V_{a}^{\mu }V_{b}^{\nu }V+8\bar{%
\psi}_{\mu }\gamma _{5}\gamma _{a}\nabla _{\nu }\psi _{\rho }V_{\sigma
}^{a}\epsilon ^{\mu \nu \rho \sigma }\right) ds:=  \nonumber \\
&=&\left( \mathcal{L}_{\mathcal{H}}+\mathcal{L}_{\mathcal{S}}\right) ds
\end{eqnarray}
where $V$ is the determinant of the new vielbein $V_{\mu }^{a}$.
The supersymmetries (\ref{G-N}) close off-shell as follows:
\begin{equation}
\left\{ 
\begin{tabular}{l}
\vspace{0.1cm}$\left[ \delta _{1},\delta _{2}\right] V_{\mu }^{a}=\left[ \delta
_{1},\delta _{2}\right] e_{\mu }^{a}=\nabla _{\mu }q^{a}=i\nabla _{\mu
}\left( \bar{\varepsilon}_{2}\gamma ^{a}\varepsilon _{1}\right) $ \\ 
\vspace{0.1cm}$\left[ \delta _{1},\delta _{2}\right] \psi _{\mu }=\frac{1}{4}\gamma
_{a}\gamma _{b}\left( \delta _{2}\omega _{\mu }^{ab}\varepsilon _{1}-\delta
_{1}\omega _{\mu }^{ab}\varepsilon _{2}\right) =0$ \\ 
\vspace{0.1cm}$\left[ \delta _{1},\delta _{2}\right] \omega _{\mu }^{ab}=0$ \\ 
$\left[ \delta _{1},\delta _{2}\right] q^{a}=0$%
\end{tabular}
\right.
\end{equation}
Therefore their algebra is well-defined. This is the possible starting
point of a Gauge-Natural theory. There is indeed a theorem claiming that the
commutator of two generalized symmetries is a generalized symmetry.
This leads to the closure of their algebra.
In conclusion, a formulation of Supergravity that is analogous to what has been done for the
Wess-Zumino model seems to be possible. It can be developed in the same way
we tried to do within the standard approach. The form of the generator of automorphisms
on $\Sigma $ can be obtained from the identification between
commutators and Lie derivatives, possibly leading to a reformulation of
Supergravity as a Gauge-Natural field theory.

\chapter*{Conclusions and perspectives}

In this thesis the Gauge-Natural framework has been discussed. First, we have gone through the mathematical basics of this formalism, namely we have defined i) principal, associated and jet bundles, ii) a geometrical setup for the Lagrangian formulation of field theories and iii) spin structures on the Gauge-Natural bundles.

	Chapters \ref{chapt:susy} and \ref{chapt:rs} focus instead on the applications of this model to some basic supersymmetric theories. Chapter \ref{chapt:susy} deals with the Wess-Zumino model, namely with global supersymmetries. It is shown that this theory can be consistently embedded into a Gauge-Natural formulation. In Chapter \ref{chapt:rs} we have extended the discussion to local SUSY, and analyzed the Rarita-Schwinger (R-S) model. We have calculated the on-shell covariance of the Lagrangian and the closure of the SUSY algebra. Since the SUGRA algebra closes on-shell, it turns out that the description of the R-S model is problematic for the Gauge-Natural formalism.

	Despite these difficulties, we have pointed out possible solutions. The choice of the Grignani-Nardelli model, that is sketched in Section \ref{GN}, may solve the problems encountered with the null torsion constraint. There is also another possibility: a definition of on-shell symmetries within the Gauge-Natural framework. These are topics of interest for future investigations.

	Let us finally comment on the differences between the Gauge-Natural model and the standard approach to gauge field theories. It seems that the former
allows a better control on \textsl{global }properties, by the global nature of the structure bundle. The concept of structure bundle provides the symmetries and the conserved quantities with additional information, which one could not obtain in the approach based on manifolds. This
has been shown for General Relativity, gauge and spinor theories by the Mathematical Physics group in Turin.

	It follows that the Gauge-Natural formalism can be, in some sense, innovative. Other global frameworks which describe the interactions between spinor fields and the gravitational field do not actually exist. Certainly, local formulations give the correct field equations. However, their locality makes it impossible to study the conserved quantities, which are non-local objects, in a fully consistent manner.

\subsubsection*{Acknowledgments}

	I would like to thank my family for support, encouragement and understanding, my advisor Prof. Mauro Francaviglia for his useful suggestions and Dr. Lorenzo Fatibene for his constant and irreplaceable help throughout this work.

\appendix 

\chapter{Group theory for supersymmetries}

\section{Superalgebras}

A super Lie algebra is a vector space $\A$ over the real or complex
field which is the direct product of the two subspaces $\Pbb$ and $\D
$: 
\begin{equation}
\A=\Pbb\oplus \D  \label{splitting}
\end{equation}
$\Pbb$ and $\D$ are called, respectively, the even and odd subspace.

	To provide $\A$ with the structure of algebra, one must define in addiction
to the sum and the product by a scalar a suitable Lie bracket, which we
denote by [ , $]_{\pm },$satisfying some additional properties.
The first property of the Lie bracket is: 
\begin{equation}
i)\;\forall X,Y\in \Pbb,\;[X,Y]_{\pm }\in \Pbb\;\wedge \;[X,Y]_{\pm
}=-[Y,X]_{\pm }
\end{equation}
i.e. $\Pbb$ is a subalgebra. Furthermore, on this subspace the properties
of the Lie bracket are the same of an ordinary Lie algebra. Consequently, $%
\Pbb$ is an ordinary Lie algebra. 
\begin{equation}
ii)\;\forall X\in \Pbb,\;\forall \Psi \in \D,\;[X,\Psi ]_{\pm }\in 
\D,\;[X,\Psi ]_{\pm }=-[\Psi ,X]_{\pm }
\end{equation}
\begin{equation}
\forall Y\in \Pbb,\hspace{0.1cm}[X,[Y,\Psi ]_{\pm }]_{\pm }+[Y,[\Psi
,X]_{\pm }]_{\pm }+[\Psi ,[X,Y]_{\pm }]_{\pm }=0  \label{tommaso}
\end{equation}
These equations state that the odd subspace $\D$ is a carrier space for
a representation of the Lie algebra $\Pbb$, the Lie bracket [ , $]_{\pm }$
defining the action of $\Pbb$ on $\D$.
Indeed, Eq.(\ref{tommaso}) can be rewritten as follows: 
\begin{equation}
\lbrack X,[Y,\Psi ]_{\pm }]_{\pm }-[Y,[X,\Psi ]_{\pm }]_{\pm }=[[X,Y]_{\pm
},\Psi ]_{\pm }
\end{equation}
which implies that the action of elements of $\Pbb$ is consistent with
the Lie bracket defined over $\Pbb$. 
\begin{equation}
iii)\;\forall \Xi ,\Psi ,\Lambda \in \D,\;[\Psi ,\Xi ]_{\pm }\in \Pbb%
\;\wedge \;[\Psi ,\Xi ]_{\pm }=[\Xi ,\Psi ]_{\pm }  \label{lo}
\end{equation}
\begin{equation}
\lbrack \Psi ,[\Xi ,\Lambda ]_{\pm }]_{\pm }+[\Xi ,[\Lambda ,\Psi ]_{\pm
}]_{\pm }+[\Lambda ,[\Psi ,\Xi ]_{\pm }]_{\pm }=0  \label{la}
\end{equation}
\begin{equation}
\forall X\in \Pbb,\;\forall \Xi ,\Psi ,\Lambda \in \D,\;[X,[\Psi
,\Xi ]_{\pm }]_{\pm }-[\Xi ,[X,\Psi ]_{\pm }]_{\pm }+[\Psi ,[\Xi ,X]_{\pm
}]_{\pm }=0  \label{li}
\end{equation}
Eqs.(\ref{lo}), (\ref{la}) and (\ref{li}) introduce a symmetric Lie bracket, that
is an anticommutator, over the odd space $\D$, and they state that the
anticommutator of two odd elements is an even one. The following property
\begin{equation}
iv)\;\forall \alpha ,\beta \in \C,\;\forall F,G,H\in \A,\;[\alpha
F+\beta G,H]_{\pm }=\alpha [F,H]_{\pm }+\beta [G,H]_{\pm }
\label{distributiva}
\end{equation}
states that the Lie bracket is distributive with respect to the vector space
operations.
By using this last property, we now introduce a more compact notation, and
define the concept of \textsl{grading}.

	Let $\Z_{2}$ be the set of integer numbers modulo 2; the two
equivalence classes are represented by 0 and 1. To each element $A \in \A$ 
we associate a degree
$a$ which is an element of $\Z_{2}$; $a$ is 1 if $A$ lies in the odd space, zero if $A$
lies in the even one: 
\begin{eqnarray}
A &\in &\Pbb\Longrightarrow a=0\;(mod\;2) \\
A &\in &\D\Longrightarrow a=1\;(mod\;2)
\end{eqnarray}
We can now rewrite the defining properties of the Lie bracket of two
arbitrary elements of the superalgebra. These elements, which in general do
not have a definite degree, because they are the sum of an even and an odd
part, for the property (\ref{distributiva}) can be decomposed into a sum of
terms which are Lie brackets of elements possessing a definite grading. Then
if A, B , C are elements of $\A$ endowed with this property, we can
write: 
\begin{equation}
\lbrack A,B]_{\pm }=(-1)^{(1+ab)}[B,A]_{\pm }
\end{equation}
\begin{equation}
\lbrack A,[B,C]_{\pm }]_{\pm }+(-1)^{a(b+c)}[B,[C,A]_{\pm }]_{\pm
}+(-1)^{b(a+c)}[C,[A,B]_{\pm }]_{\pm }=0
\end{equation}
If we define $T_{A}$ ($A = 1,\ldots,d = dim \A$) as the generators of $\A
$, they have a definite degree, since $\A$ is the direct sum of $\Pbb$
and $\D$. This means that the basis $\{T_{A}\}$ can be chosen
so that it is the union of a basis for $\Pbb$ and a basis for $\D$.%

	To completely specify the superalgebra, we have to give the Lie bracket of
two generators: 
\begin{equation}
\lbrack T_{A},T_{B}]_{\pm }=C_{AB}^{\;\cdot \;\cdot
\;F}T_{F},\;\;C_{AB}^{\;\cdot \;\cdot \;F}\in \R
\end{equation}
$C_{AB}^{\;\cdot \;\cdot \;F}$ are graded structure constants of the
superalgebra, and from (\ref{a}) and (\ref{b}) it follows that they have to
satisfy the two properties 
\begin{equation}
C_{AB}^{\;\cdot \;\cdot \;F}=(-1)^{(1+ab)}C_{BA}^{\;\cdot \;\cdot \;F}
\end{equation}
\begin{equation}
C_{AL}^{\;\cdot \;\cdot \;M}C_{BC}^{\;\cdot \;\cdot
\;L}+(-1)^{a(b+c)}C_{BL}^{\;\cdot \;\cdot \;M}C_{CA}^{\;\cdot \;\cdot
\;L}+(-1)^{b(a+c)}C_{CL}^{\;\cdot \;\cdot \;M}C_{AB}^{\;\cdot \;\cdot \;L}=0
\end{equation}
We now want to introduce a matrix representation of the Lie superalgebras
discussed up to now. We consider complex matrices in dimension 
\[
d=m+N 
\]
where $m$ and $N$ are two integer numbers. Any $d\times d$ matrix can be
written in block form, with complex entries, as follows: 
\begin{equation}
Q=\left( 
\begin{array}{ll}
A & B \\ 
C & D
\end{array}
\right)  \label{supermatrice}
\end{equation}
$A$ is $m\times m$, $D$ is $N\times N$ and $B$ and $C$ are $m\times N$ and $%
N\times m$ respectively. The space of $d\times d$ matrices is a $d^{2}$%
-dimensional vector space which can be split, according to (\ref{splitting}%
), into an even and odd subspace by defining: 
\begin{equation}
Q\in \Pbb\Longleftrightarrow B=C=0\Longrightarrow Q=\left( 
\begin{array}{ll}
A & 0 \\ 
0 & D
\end{array}
\right)
\end{equation}

\begin{equation}
Q\in \D\Longleftrightarrow A=D=0\Longrightarrow Q=\left( 
\begin{array}{ll}
0 & B \\ 
C & 0
\end{array}
\right)
\end{equation}
The Lie bracket of two ''supermatrices'' can now be derived from the grading
just introduced and the axioms of a superalgebra: 
\begin{equation}
\left\{ 
\begin{tabular}{l}
\vspace{0.1cm}$\forall Q_{1},Q_{2}\in \Pbb,\;\;\;\;[Q_{1},Q_{2}]_{\pm }=[Q_{1},Q_{2}]$
\\ 
\vspace{0.1cm}$\forall Q_{1}\in \Pbb,\forall Q_{2}\in \D\;\;\;\;[Q_{1},Q_{2}]_{\pm
}=[Q_{1},Q_{2}]$ \\ 
$\forall Q_{1},Q_{2}\in \D,\;\;\;\;[Q_{1},Q_{2}]_{\pm }=\{Q_{1},Q_{2}\}$%
\end{tabular}
\right.  \label{i}
\end{equation}
where [ , ] and $\{\;,\;\}$ denote the usual commutator and anticommutator
of two matrices. Finally, we can express Eq.(\ref{i}) by stating that the Lie bracket of two
''supermatrices'' is a new matrix of the same type:

\begin{equation}
\lbrack Q_{1},Q_{2}]_{\pm }=Q_{3}=\left( 
\begin{array}{ll}
A_{3} & B_{3} \\ 
C_{3} & D_{3}
\end{array}
\right)  \label{sonata}
\end{equation}
where: 
\begin{eqnarray*}
A_{3} &=&[A_{1},A_{2}]+B_{1}C_{2}+B_{2}C_{1} \\
D_{3} &=&[D_{1},D_{2}]+C_{1}B_{2}+C_{2}B_{1} \\
B_{3} &=&A_{1}B_{2}-B_{2}D_{1}-A_{2}B_{1}+B_{1}D_{2} \\
C_{3} &=&D_{1}C_{2}-C_{2}A_{1}-D_{2}C_{1}+C_{1}A_{2}
\end{eqnarray*}
This superalgebra is called the general graded Lie algebra $GL\left(
m/N\right) $; it is not simple. The simple algebras $Osp\left( 2p/N\right) $
and $SU\left( m/N\right) $ are obtained as superalgebras of $GL\left(
m/N\right) $ by imposing further conditions on the graded matrices (i.e.,
the supermatrices).

\newpage
\section{Grassmann algebras}

In order to exponentiate the superalgebras and obtain the corresponding
supergroups, we have to define these particular algebras. Their elements
will be the parameters of the supergroups.
A Grassmann algebra $GA_{n}$ is an extension of the complex field; its
generators are $n$ objects 
\begin{equation}
\pi _{i}\;\;\;\;i=1,2,\ldots,n
\end{equation}
which satisfy the following anticommutation relations: 
\begin{equation}
\left\{ \pi _{i},\pi _{j}\right\} =0\Longrightarrow \pi _{i}^{2}=0
\end{equation}
both if $i=j$ and $i\neq j$. Consider now all the possible monomials $\pi
_{i_{1}}\ldots\pi _{i_{k}}$: the number $N_{k}$ of different k-monomials is $%
N_{k}={n\choose k}$, and the total number of monomials is 
\begin{equation}
N=\sum_{s=0}^{n}{n\choose k}=2^{n}
\end{equation}
The Grassmann algebra $GA_{n}$ generated by $\{\pi _{i}\}$ is the $2^{n}$%
-dimensional complex vector space spanned by all the linear combinations of
the $2^{n}$ monomials $\pi _{i_{1}}\ldots\pi _{i_{k}}$.
We can therefore write an element $\alpha \in GA_{n}$ as 
\begin{equation}
\alpha =z+\alpha _{i}\pi ^{i}+\alpha _{ij}\pi ^{i}\pi ^{j}+\alpha _{ijk}\pi
^{i}\pi ^{j}\pi ^{k}+\ldots+\alpha _{12\ldots n}\pi ^{1}\pi ^{2}\ldots\pi ^{n}
\label{grass}
\end{equation}
where $z$, $\alpha _{i}$, $\alpha _{ij}$, $\alpha _{ijk},\ldots$ are complex
numbers.
In particular, if $\alpha _{i}=\alpha _{ij}=\alpha _{ijk}=\ldots=0$, $\alpha $
is an ordinary complex number. Moreover, $\alpha _{i_{1}\ldots i_{k}}$ is a
skewsymmetric tensor w. r. to linear changes of generators.

		The whole set $GA_{n}$ is an algebra because the product of the generators
induces, canonically, a product operation of the elements of $GA_{n}$. The
product operation in this algebra is associative and distributive, but it is
not commutative. Every even monomial commutes with any other monomial, odd
or even. Consequently, every element of the algebra should split into an
even and an odd part: 
\begin{equation}
\alpha =\alpha ^{(+)}+\alpha ^{{(-)}}\Longrightarrow
GA_{n}=GA_{n}^{(+)}\oplus GA_{n}^{(-)}
\end{equation}
where the even and the odd part are a linear combination, respectively, of
all the even monomials and of all the odd ones. So there is a $\Z_{2}$
grading of the Grassmann algebra, with the following properties: 
\begin{equation}
GA_{n}^{(+)}\cdot GA_{n}^{(+)}\subset GA_{n}^{(+)}
\end{equation}
\begin{equation}
GA_{n}^{(+)}\cdot GA_{n}^{(-)}\subset GA_{n}^{(-)}
\end{equation}
\begin{equation}
GA_{n}^{(-)}\cdot GA_{n}^{(-)}\subset GA_{n}^{(+)}
\end{equation}
While an even element commutes with every other element of $GA_{n}$, two odd
elements anticommute. Therefore, as we did for superalgebras, we define the
grading $a$ of an element $\alpha \in GA_{n}$ to be zero if it is even and
to be $1$ if it is odd: 
\begin{equation}
\alpha \beta =(-1)^{ab}\beta \alpha
\end{equation}
We can now use the formalism developed above to introduce the complex
conjugation on Grassmann algebras.
If $n=2p$, we assume as generators 
\begin{eqnarray*}
\pi _{\alpha }\hspace{1cm}(\alpha &=&1,2,\ldots,p) \\
\pi _{\beta }\hspace{1cm}(\beta &=&p+1,\ldots,2p)
\end{eqnarray*}
The action of the complex conjugation $*$ on the generators defined above is:
\begin{equation}
\left\{ 
\begin{tabular}{l}
$(\pi _{\alpha })^{*}=\pi _{\beta }$ \\ 
$(\pi _{\beta })^{*}=\pi _{\alpha }$ \\ 
$(\pi _{i}\pi _{j})^{*}=(\pi _{j})^{*}(\pi _{i})^{*}$%
\end{tabular}
\right.
\end{equation}
The mapping $*$ extends canonically to all the element of the algebra. If $%
\alpha \in GA_{(2p)}$ is given by (\ref{grass}), then 
\begin{equation}
\alpha ^{*}=z^{*}+\alpha _{i}^{*}(\pi ^{i})^{*}+\alpha _{ij}^{*}(\pi
^{j})^{*}(\pi ^{i})^{*}+\ldots
\end{equation}
Its formal properties are:
\begin{equation}
\left\{ 
\begin{tabular}{l}
\vspace{0.1cm}$\forall \alpha \in GA_{(2p)},\;\;\;\;\hspace{1cm}\hspace{0.5cm}\hspace{%
0.25cm}\!\!(\alpha ^{*})^{*}=\alpha $ \\ 
\vspace{0.1cm}$\forall \alpha _{1}\alpha _{2}\in GA_{(2p)},\;\;\;\;\hspace{0.5cm}\hspace{%
0.5cm}\,(\alpha _{1}\alpha _{2})^{*}=\alpha _{2}^{*}\alpha _{1}^{*}$ \\ 
$\forall a\in \C,\;\forall \alpha \in GA_{(2p)},\;\;\;\;(a\alpha
)^{*}=a^{*}\alpha ^{*}$%
\end{tabular}
\right.
\end{equation}
Given the complex conjugation, the notions of reality and of norm are
defined in the same way as for complex numbers: 
\[
\forall \alpha \in \mathbf{\R}\;\Longrightarrow \;\alpha ^{*}=\alpha 
\]
\[
\Vert \alpha \Vert ^{2}=\alpha ^{*}\alpha 
\]
We remark, however, that $\Vert \alpha \Vert ^{2}$ is NOT positive definite:
in fact, the norm of an imaginary odd element is always zero.
We are now ready to introduce the concept of an analytic function on a
Grassmann algebra $GA_{n}$ into itself: if 
\[
f\;:\;GA_{n}\longrightarrow GA_{n} 
\]
it can be defined via a power series expansion: 
\begin{equation}
\forall \alpha \in GA_{n},\;\;\;\;f(\alpha )=\sum_{m=0}^{\infty }f_{m}\alpha
^{m}\in GA_{n}  \label{serie}
\end{equation}
where $f_{m}$ are coefficients of a series with finite convergence radius.
If $\alpha $ is an even element, the series may extend to infinity; anyway,
if $\alpha $ is odd, the series stops after the first element since $\alpha
^{2}=0.$

\section{Supermanifolds}

Provided with Grassmann algebras, we can introduce the concept of
supermanifold. In order to give only a basic notion of this object, we
defined it as a smooth space whose point are labeled by two sets of
coordinates: bosonic and fermionic.

	The bosonic coordinates are chosen to be even elements of a Grassmann
algebra $GA_{\infty }$, while the odd ones are odd elements of the same
algebra. Being the concept of function well defined in $GA_{\infty }$, the
whole formalism of differential geometry can be used. Accordingly, we shall denote by $\mathcal{M}^{p/q}$ a supermanifold with $p$
bosonic dimensions and $q$ fermionic ones. The coordinates of a point $p\in 
\mathcal{M}^{p/q}$ will be denoted by $\{x^{a},\theta ^{\alpha }\}$, where $%
x^{a}$ $(a=1,2,\ldots,p)$ are bosonic and $\theta ^{\alpha }$ $(\alpha
=1,2,\ldots,q)$ are fermionic.\newline
A \textsl{superfield} is a function of several variables mapping $\mathcal{M}%
^{p/q}$ into $GA_{\infty }$: 
\begin{equation}
\phi \;:\mathcal{M}^{p/q}\longrightarrow GA_{\infty }
\end{equation}
Using the nilpotency of $\theta ^{\alpha }$, $\phi (x,\theta )$ can be
written as a polynomial in $\theta ^{\alpha }$, whose coefficients are
functions of the bosonic coordinates only: 
\begin{equation}
\phi (x,\theta )=\varphi (x)+\varphi _{\alpha }(x)\theta ^{\alpha }+\varphi
_{\alpha _{1}\alpha _{2}}(x)\theta ^{\alpha _{1}\alpha _{2}}+\ldots+\varphi
_{\alpha _{1}\ldots\alpha _{q}}(x)\theta ^{\alpha _{1}\ldots\alpha _{q}}
\end{equation}
We emphasize that all the $\varphi _{\alpha _{1}\ldots\alpha _{q}}(x)$ are
completely antisymmetric in their indices because of the anticommutativity
of the $\theta _{.s}$. In supersymmetric theories, where the fermionic
coordinates $\theta _{\alpha }$ are spinors, the fields in the collection
have different spins, bosons and fermions necessarily coexisting in the same
superfield.

	The space of superfields is called $\mathcal{C}(\mathcal{M}^{p/q})$. The
differential operators acting on it are linear combinations of the
fundamental derivatives 
\begin{equation}
\partial _{a}=\frac{\partial }{\partial x^{a}}
\end{equation}
and 
\begin{equation}
\partial _{\alpha }=\frac{\partial }{\partial \theta ^{\alpha }}
\end{equation}
which act on the superfield in the following way:
\begin{equation}
\left\{ 
\begin{tabular}{l}
\vspace{0.1cm}$\partial _{a}\phi (x,\theta )=\partial _{a}\varphi (x)+\partial _{a}\varphi
_{\alpha }(x)\theta ^{\alpha }+\ldots$ \\ 
$\partial _{\alpha }\phi (x,\theta )=\varphi _{\alpha }(x)+2\varphi _{\alpha
\beta }(x)\theta ^{\beta }+\varphi _{\alpha \beta \gamma }(x)\theta ^{\beta
}\theta ^{\gamma }+\ldots$%
\end{tabular}
\right.
\end{equation}
It is easy to verify the formal properties
\begin{equation}
\left\{ 
\begin{tabular}{l}
\vspace{0.1cm}$\lbrack \partial _{\alpha },\partial _{\beta }]=0$ \\ 
\vspace{0.1cm}$\lbrack \partial _{a},\partial _{\beta }]=0$ \\ 
\vspace{0.1cm}$\{\partial _{\alpha },\partial _{\beta }\}=0$ \\ 
$\partial _{\alpha }[\theta ^{\beta }\phi (x,\theta )]=\delta _{\alpha
}^{\beta }\phi -\theta ^{\beta }\partial _{\alpha }\phi $%
\end{tabular}
\right.
\end{equation}
The tangent space to $\mathcal{M}^{p/q}$, $T(\mathcal{M}^{p/q})$, can be
therefore defined; it is spanned by the differential operators 
\begin{equation}
t=t^{a}(x,\theta )\partial _{a}+t^{\alpha }(x,\theta )\partial _{\alpha }
\end{equation}
where $t^{a}$ and $t^{\alpha }$ are respectively bosonic and fermionic
superfields.

	At each point $p=$ $\left( x,\theta \right) $, $T(\mathcal{M}^{p/q})$ is a
graded vector space with $p$ bosonic and $q$ fermionic dimensions.
We can also define a graded vector space $V\left( n/m\right) $: let $%
\{e_{a},e_{\alpha }\}$ be a collection of $n$ elements $e_{a}$ $%
(a=1,2,\ldots,n) $ and $m$ elements $e_{\alpha }$ $(\alpha =1,2,\ldots,m)$
respectively called the \textsl{bosonic} and \textsl{fermionic fundamental
vectors}. An element $\nu \in V(n/m)$ is a linear combination 
\begin{equation}
\nu =\nu ^{a}e_{a}+\nu ^{\alpha }e_{\alpha }
\end{equation}
with $\nu ^{a}\in GA_{\infty }^{(+)}$ and $\nu ^{\alpha }\in GA_{\infty
}^{(-)}$.

	In complete analogy to ordinary vector space theory, one can introduce the
dual space $V^{*}(n/m)$ defining a basis of linear functionals $%
\{e^{a},e^{\alpha }\}$ and the following rules:
\begin{equation}
\left\{ 
\begin{tabular}{l}
\vspace{0.1cm}$\forall w\in V(n/m),\;\;(w,e^{*})\in GA_{\infty }$ \\ 
\vspace{0.1cm}$(e_{a},e^{b})=\delta _{a}^{b}\;\;,\;\;(e_{\alpha },e^{b})=0$ \\ 
\vspace{0.1cm}$(e_{\alpha },e^{\beta })=\delta _{\alpha }^{\beta }\;\;,\;\;(e_{a},e^{\beta
})=0$ \\ 
$\forall \nu ^{*}\in V^{*}(n/m),\;\;\nu ^{*}=\nu _{a}e^{a}+\nu _{\alpha
}e^{\alpha }$%
\end{tabular}
\right.
\end{equation}
Hence 
\begin{equation}
\forall \nu \in V(n/m),\;\;\forall w^{*}\in V^{*}(n/m),\;\;(\nu ,w^{*})=(\nu
^{a}w_{a}+\nu ^{\alpha }w_{\alpha })\in GA_{\infty }
\end{equation}
The differential 1-forms on ${\mathcal{M}}^{p/q}$ are elements of the dual
vector space $T^{*}({\mathcal{M}}^{p/q})$, where a basis is provided by the
differentials $dx^{a}$ and $d\theta ^{\alpha }$:
\begin{equation}
\left\{ 
\begin{tabular}{l}
\vspace{0.1cm}$(\partial _{a},dx^{b})=\delta _{a}^{b}\;\;,\;\;(\partial _{\alpha
},dx^{b})=0$ \\ 
$(\partial _{a},d\theta ^{\beta })=0\;\;,\;\;(\partial _{\alpha },d\theta
^{\beta })=\delta _{\alpha }^{\beta }$%
\end{tabular}
\right.
\end{equation}
so we are lead to define a 1-form $\omega \in T^{*}({\mathcal{M}}^{p/q})$
as: 
\begin{equation}
\omega =dx^{a}\omega _{a}(w,\theta )+d\theta ^{\alpha }\omega _{\alpha
}(x,\theta )
\end{equation}
where $\omega _{a}(w,\theta )$ and $\omega _{\alpha }(x,\theta )$ are
respectively bosonic and fermionic superfields. To extend this concept to $p$%
-forms, we must define some rules for the exterior product:
\begin{equation}
\left\{ 
\begin{tabular}{l}
\vspace{0.1cm}$dx^{a}\wedge dx^{b}=-dx^{b}\wedge dx^{a}$ \\ 
\vspace{0.1cm}$dx^{a}\wedge d\theta ^{\beta }=-d\theta ^{\beta }\wedge dx^{a}$ \\ 
$d\theta ^{\alpha }\wedge d\theta ^{\beta }=d\theta ^{\beta }\wedge d\theta
^{\alpha }$%
\end{tabular}
\right.
\end{equation}
and if we define
\begin{eqnarray}
\omega ^{(p)}&=& \omega _{a_{1}\ldots a_{p}}(x,\theta )dx^{a_{1}}\wedge
dx^{a_{2}}\wedge \ldots\wedge dx^{a_{p}}+ \nonumber \\ 
&&+\omega _{\alpha _{1}a_{2}\ldots a_{p}}(x,\theta )d\theta ^{\alpha
_{1}}\wedge dx^{a_{2}}\wedge \ldots\wedge dx^{a_{p}}+\ldots+ \nonumber \\ 
&&+\omega _{\alpha _{1}\ldots\alpha _{p}}d\theta ^{\alpha _{1}}\wedge
\ldots\wedge d\theta ^{\alpha _{p}}
\end{eqnarray}
where again $\omega _{\alpha _{1}\ldots\alpha _{p}a_{m+1}\ldots a_{p}}(x,\theta )$
are fermionic or bosonic superfields, depending on whether the number of
Greek indices is odd or even. In this way, the usual grading of the exterior
product of forms is respected: 
\begin{equation}
\omega ^{(p)}\wedge \omega ^{(q)}=(-1)^{pq}\omega ^{(q)}\wedge \omega ^{(p)}
\label{zappa}
\end{equation}
This equation can be generalized: indeed, the above choice of the bosonic or
fermionic character of $\omega _{\alpha _{1}\ldots\alpha
_{p}a_{m+1}\ldots a_{p}}(x,\theta )$ is the right one for a bosonic $p$-form $%
\omega ^{(p)}$. However, we can consider also fermionic $p$-forms, like the
coordinate differentials and, in general, all the $p$-forms carrying free
fermionic indices in an odd number.
Consequently, Eq.(\ref{zappa}) is replaced by 
\begin{equation}
\omega _{(a)}^{(p)}\wedge \omega _{(b)}^{(q)}=(-1)^{ab+pq}\omega
_{(b)}^{(q)}\wedge \omega _{(a)}^{(p)}
\end{equation}
With the notions of supermanifold, super Lie algebra and Grassmann algebra,
we are now ready to introduce the concept of supergroup.

\section{Supergroups}

Consider $GL\left( m/N\right) ,$ namely the algebra of $\left( m+N\right)
\times \left( m+N\right) $ complex matrices, closed under (\ref{sonata}),
and let $\{t_{a},t_{\alpha }\}$ be a basis of $GL\left( m/N\right) $.
$\{t_{a}\}$ $\left( a=1,2,\ldots,m^{2}+N^{2}\right) $ is a basis of the even
subspace: 
\begin{equation}
t_{a}=\left( 
\begin{array}{ll}
A_{a} & 0 \\ 
0 & D_{a}
\end{array}
\right)
\end{equation}
while $\{t_{\alpha }\}$ is a basis of the odd subspace: 
\begin{equation}
t_{\alpha }=\left( 
\begin{array}{ll}
0 & B_{\alpha } \\ 
C_{\alpha } & 0
\end{array}
\right)
\end{equation}
Any matrix $Q\in GL\left( m/n\right) $ can be written as 
\begin{equation}
Q=Q^{a}t_{a}+Q^{\alpha }t_{\alpha }
\end{equation}
where $Q^{a},Q^{\alpha }\in $ $\C$ and the Lie bracket, according to (%
\ref{sonata}), is 
\begin{equation}
\lbrack Q_{1},Q_{2}]_{\pm
}=Q_{1}^{a}Q_{2}^{b}[t_{a},t_{b}]+(Q_{1}^{a}Q_{2}^{\beta }-Q_{1}^{\beta
}Q_{2}^{a})[t_{a},t_{\beta }]+Q_{1}^{\alpha }Q_{2}^{\beta }\{t_{\alpha
},t_{\beta }\}
\end{equation}
Notice that the right hand side of this equation would be the ordinary
commutator of $Q_{1}$ and $Q_{2}$ 
\begin{equation}
\lbrack Q_{1},Q_{2}]=(Q_{1}Q_{2}-Q_{2}Q_{1})
\end{equation}
if $Q_{a}^{1}$, $Q_{a}^{2}$, $Q_{\alpha }^{1}$, $Q_{\alpha }^{2}$, instead
of being complex numbers, were, respectively, even and odd elements of a
Grassmann algebra $GA_{\infty }$. Therefore, to every superalgebra we
associate a graded vector space spanned by the linear combinations of the
even generators of $GA_{\infty }$ and of the odd generators with odd
elements of the same $GA_{\infty }$.

	The ordinary commutator of elements of the associated vector space provides
an isomorphic realization of the superalgebra.
This point of view is useful, because we can now define the supergroup
corresponding to a given superalgebra as the exponentiation of the graded
vector space $\A^{\prime }$ associated to $\A$. The difference
between the two spaces is that in $\A^{\prime }$ complex numbers are
replaced by elements of the Grassmann algebra.

	It follows that we can define the \textsl{supergroup associated to }$\A$
as 
\begin{equation}
\mathcal{G}=exp(\A^{\prime })
\end{equation}
In the case of $GL\left( m/N\right) $, an element of the associated vector
space is a graded matrix, whose entries are elements of the Grassmann
algebra: even in the diagonal blocks $\left( A,D\right) $, odd in the off
diagonal ones $\left( B,C\right) $. Such objects can be viewed as $%
GA_{\infty }$ - linear operators on graded vector spaces. The product
operation is the ordinary product of graded matrices, whose construction is
exactly the same as in the first section of this chapter: if a graded matrix
is 
\begin{equation}
Q=\left( 
\begin{array}{ll}
A & \Sigma \\ 
\Upsilon & D
\end{array}
\right)
\end{equation}
$A,D$ are $m\times m$ and $N\times N$ matrices, with commuting entries,
while $\Sigma $ and $\Upsilon $ are $m\times N$ and $N\times m$ matrices,
respectively, with anticommuting entries.

	The product $Q_{1}Q_{2}$ is defined as for ordinary matrices: 
\begin{equation}
Q_{1}Q_{2}=Q_{3}=\left( 
\begin{array}{ll}
A_{3} & \Sigma _{3} \\ 
\Upsilon _{3} & D_{3}
\end{array}
\right)
\end{equation}
where 
\begin{eqnarray*}
A_{3} &=&A_{1}A_{2}+\Sigma _{1}\Upsilon _{2} \\
\Sigma _{3} &=&A_{1}\Sigma _{2}+\Sigma _{1}D_{2} \\
\Upsilon _{3} &=&\Upsilon _{1}A_{2}+D_{1}\Upsilon _{2} \\
D_{3} &=&\Upsilon _{1}\Sigma _{2}+D_{1}D_{2}
\end{eqnarray*}
There follow the definitions of transposition, Hermitian conjugation,
supertrace and superdeterminant: 
\begin{equation}
Q^{T}=\left( 
\begin{array}{ll}
A^{T} & \Upsilon ^{T} \\ 
-\Sigma ^{T} & D^{T}
\end{array}
\right) \;\;,\;\;Q^{\dagger }=\left( 
\begin{array}{ll}
A^{\dagger } & \Upsilon ^{\dagger } \\ 
\Sigma ^{\dagger } & D^{\dagger }
\end{array}
\right)
\end{equation}
\begin{equation}
\left\{ 
\begin{tabular}{l}
\vspace{0.1cm}$StrQ=TrA-TrD$ \\ 
$SdetQ=(detA)(detD^{\prime })$%
\end{tabular}
\right.
\end{equation}
$D^{\prime }$ is defined by the inverse of $Q$: 
\begin{equation}
Q^{-1}=\left( 
\begin{array}{ll}
A^{\prime } & \Sigma ^{\prime } \\ 
-\Upsilon ^{\prime } & D^{\prime }
\end{array}
\right)
\end{equation}
With these definitions, the following properties valid for ordinary matrices
continue to hold true:
\begin{equation}
\left\{ 
\begin{tabular}{l}
\vspace{0.1cm}$(Q_{1}Q_{2})^{T}=Q_{2}^{T}Q_{1}^{T}$ \\ 
\vspace{0.1cm}$(Q_{1}Q_{2})^{\dagger }=Q_{2}^{\dagger }Q_{1}^{\dagger }$ \\ 
\vspace{0.1cm}$Str(Q_{1}Q_{2})=Str(Q_{2}Q_{1})$ \\ 
\vspace{0.1cm}$Sdet(Q_{1}Q_{2})=(SdetQ_{1})(SdetQ_{2})$ \\ 
$Sdet(expQ)=exp(StrQ)$%
\end{tabular}
\right.
\end{equation}

\section{$Osp\left( m/N\right) $ and the Poincar\'{e} supergroup}

First of all we introduce the orthosymplectic algebra $Osp\left( m/N\right) $%
, which exists only when $m=2p$ is even; it is a (super)subalgebra of $%
GL\left( 2p/N\right) $ characterized by the following conditions (see Eq.(%
\ref{supermatrice})): $\forall Q\in Osp\left( 2p/N\right) $, 
\begin{equation}
\left\{ 
\begin{tabular}{l}
\vspace{0.1cm}$A^{T}\Omega _{(2p)}+\Omega _{(2p)}A=0$ \\ 
\vspace{0.1cm}$D^{T}\Omega _{(N)}+\Omega _{(N)}D=0$ \\ 
$C=\Omega _{(N)}B^{T}\Omega _{(2p)}$%
\end{tabular}
\right.
\end{equation}
where the two matrices $\Omega _{(2p)}$ and $\Omega _{(N)}$ have been chosen
so that 
\begin{equation}
\begin{tabular}{l}
$\Omega _{(2p)}^{2}=\I\;\;,\;\;\Omega _{(2p)}^{T}=-\Omega _{(2p)}$ \\ 
$\Omega _{(N)}^{T}=\Omega _{(N)}$%
\end{tabular}
\label{cl}
\end{equation}
From these equations, one can see that since $\Omega _{(2p)}$ is
skewsymmetric, the matrices $A$ span a symplectic subalgebra $Sp\left( 2p,%
\C\right) $ of $Osp\left( 2p/N\right) $.
On the other hand, $\Omega _{(N)}$ is symmetric, thence it is an orthogonal
metric, and the submatrices $D$ span an orthogonal subalgebra $O\left( N,%
\C\right) $ of $Osp\left( 2p/N\right) $.
The ordinary Lie subalgebra of $Osp\left( 2p/N\right) $ is therefore 
\begin{equation}
\G=Sp(2p)\otimes O(N)
\end{equation}
The symplectic and orthogonal algebra acts on the off-diagonal matrices $B$
and $C$ transforming respectively in the defining representations of $%
Sp\left( 2p\right) $ and $O\left( N\right) $. For our purposes are important
only the algebras $Osp\left( 4/N\right) $, with $1\le N\le 8$.
In this case, the Lie algebra isomorphism 
\begin{equation}
Sp(4,C)\sim O(5,C)
\end{equation}
can be used, and imposing suitable reality conditions, one obtains a real
superalgebra $Osp\left( 4/N\right) $ whose Lie algebra is $SO\left(
2,3\right) $ $\otimes $ $SO\left( N\right) $.
$SO\left( 2,3\right) $ is the anti-de Sitter group, containing the Lorentz
generators $M_{ab}$ and the non commuting anti-de Sitter translations $P_{a}$. The off-diagonal generators transform as vectors under $SO\left( N\right) $
and as spinors under $SO\left( 2,3\right) $, playing a role analogue to the
role of the supersymmetry generators in the Poincar\'{e} superalgebra.

	Exploiting a method called Inon\"{u}--Wigner contraction [1], it is
possible to get the $N$-extended Poincar\'{e} algebra from $Osp\left(
4/N\right) $. But here we follow another procedure: suppose that
\begin{equation}
\Omega ^{\prime }=\left( 
\begin{array}{ll}
\Omega _{\left( m\right) } & 0 \\ 
0 & \Omega _{\left( N\right) }
\end{array}
\right)
\end{equation}
\bigskip \newline
This is a graded matrix of even type, whose diagonal blocks are defined in (%
\ref{cl}).
$\Omega ^{\prime }$ is called orthosymplectic metric: it can be utilized
to define a quadratic form on a graded vector space $V\left( m/N\right) $ of
which it is the generalization of a symplectic plus an orthogonal quadratic
form.

	Given two elements $\nu ,w\in V(m/N)$, we set 
\begin{equation}
\Omega ^{\prime }(\nu ,w)=\nu ^{T}\Omega ^{\prime }w=\nu ^{\alpha }w^{\beta
}\Omega _{(m)\alpha \beta }+\nu ^{a}w^{b}\Omega _{(N)ab}
\end{equation}
and we can define the complex orthosymplectic group $Osp\left( m/N;\C%
\right) $ as the group of graded matrices $O$ which preserve $\Omega
^{\prime }$: 
\begin{equation}
\Omega ^{\prime }(O\nu ,Ow)=\Omega ^{\prime }(\nu ,w)
\end{equation}
that is 
\begin{equation}
O^{T}\Omega ^{\prime }O=\Omega ^{\prime }  \label{as}
\end{equation}
Setting 
\begin{equation}
O=exp(\Lambda )
\end{equation}
and considering $\Lambda $ infinitesimal, we see that (\ref{as}) is
equivalent to 
\begin{equation}
\Omega ^{\prime }\Lambda \Omega ^{\prime -1}=-\Lambda ^{T}
\label{condizione}
\end{equation}
Now we can finally obtain the explicit form of the superalgebra $Osp\left(
4/N\right) $, making the following choice for the matrix $\Omega ^{\prime }$%
: 
\begin{equation}
\Omega ^{\prime }=\left( 
\begin{array}{ll}
C & 0 \\ 
0 & \I_{\left( N\right) }
\end{array}
\right)
\end{equation}
where $C$ is the charge conjugation matrix and $\I_{(N)}$ is the
identity matrix in $N$-dimensions. The most general graded matrix $\Lambda $ which satisfies (\ref{condizione})
is 
\begin{equation}
\Lambda =\left( 
\begin{array}{ll}
\vspace{0.1cm}-\frac{1}{4}\epsilon ^{ab}\gamma _{ab}+\frac{i}{2}\epsilon ^{a}\gamma _{a} & 
\xi ^{B} \\ 
\hspace{1.2cm}\bar{\xi}^{A} & \frac{1}{2}\epsilon ^{AB}
\end{array}
\right)  \label{de Sitter}
\end{equation}
where $\epsilon ^{ab}=-\epsilon ^{ba}$ are the parameters of the Lorentz
subalgebra and $\epsilon ^{a}$ may be interpreted as the parameters of the
anti-de Sitter boosts. Indeed the $4\times 4$ matrices 
\begin{equation}
L=\frac{1}{4}\epsilon ^{ab}\gamma _{ab}-\frac{i}{2}\epsilon ^{a}\gamma _{a}
\end{equation}
generate the anti-de Sitter group $SO\left( 2,3\right) $. Furthermore the
skewsymmetric parameters $\epsilon _{AB}$ correspond to the generators of $%
SO\left( N\right) $, while the $\xi _{A}$ are the supersymmetry parameters
and Majorana spinors.

	Given these interpretations to the objects in Eq.(\ref{de Sitter}), it is
clear that a matrix representation for the Poincar\'{e} superalgebra is
given by 
\begin{equation}
\Theta =\left( 
\begin{array}{ll}
\vspace{0.1cm}-\frac{1}{4}\epsilon ^{ab}\gamma _{ab} & \xi ^{B} \\ 
\hspace{0.8cm}\bar{\xi}^{A} & \frac{1}{2}\epsilon ^{AB}
\end{array}
\right)
\end{equation}
where we have imposed the $\epsilon _{a.s}$ to vanish.

	Finally, one can evaluate the algebra of the Poincar\'{e} supergroup:
writing $\Theta $ as a $GA_{\infty }$ linear combination of matrices: 
\begin{equation}
\Theta =-(\epsilon ^{a}P_{a}+\epsilon ^{ab}M_{ab}+\epsilon ^{AB}T_{AB}+\bar{Q%
}_{A}\xi ^{A})
\end{equation}
calculating the commutator 
\begin{equation}
\lbrack \Theta _{1},\Theta _{2}]=\Theta _{3}
\end{equation}
and expanding the result along the generators $M_{ab},P_{a},T_{AB},\bar{Q}%
_{A}$ (the $P_{a}$ are the generators of translations in Minkowski
space-time), the following relations hold: 
\begin{equation}
\left\{ 
\begin{tabular}{l}
\vspace{0.1cm}$\lbrack M_{ab},M_{cd}]=\frac{1}{2}(\eta _{bc}M_{ad}+\eta _{ad}M_{bc}-\eta
_{bd}M_{ac}-\eta _{ac}M_{bd})$ \\ 
\vspace{0.1cm}$\lbrack P_{a},P_{b}]=0$ \\ 
\vspace{0.1cm}$\lbrack M_{ab},P_{c}]=-\frac{1}{2}(\eta _{ac}P_{b}-\eta _{bc}P_{a})$ \\ 
\vspace{0.1cm}$\lbrack M_{ab},\bar{Q}_{A\beta }]=\frac{1}{4}\bar{Q}_{A\beta }(\gamma
_{ab})_{\alpha \beta }$ \\ 
\vspace{0.1cm}$\lbrack P_{a},\bar{Q}_{A\beta }]=0$ \\ 
\vspace{0.1cm}$\{\bar{Q}_{A\alpha },\bar{Q}_{B\beta }\}=i(C\gamma ^{a})_{\alpha \beta
}\delta _{AB}P_{a}-4C_{\alpha \beta }T_{AB}$ \\ 
\vspace{0.1cm}$\lbrack T_{AB},\bar{Q}_{C\alpha }]=0$ \\ 
$\lbrack T_{AB},T_{CD}]=0$%
\end{tabular}
\right.
\end{equation}

\samepage

\end{document}